\documentclass[12pt]{iopart}

\newcommand{\bl}[1]{{\color{blue} #1}}
\newcommand{\re}[1]{{\color{red} #1}}
\newcommand{\rb}[1]{\raisebox{1.5ex}[-1.5ex]{#1}} 
\usepackage{graphicx,dsfont}
\usepackage{color,url,amssymb,multirow,longtable,dsfont,longtable}

\def\rmb#1{{\bf #1}}

\def\beq{\begin{equation}}
\def\eeq{\end{equation}}

\def\slash#1{#1 \hskip -0.5em / } 
\def\beqy{\begin{eqnarray}}
\def\eeqy{\end{eqnarray}}

\def\hlf{{{1\over2}}}
\def\thlf{{{3\over2}}}
\def\fhlf{{{5\over2}}}
\def\shlf{{{7\over2}}}
\def\nhlf{{{9\over2}}}
\def\elhlf{{{11\over2}}}
\def\thhlf{{{13\over2}}}
\def\fihlf{{{15\over2}}}

\def\starfour{\tiny{\left(\hspace*{-7pt}\begin{array}{c}\ast \\[-5pt]\ast \\[-5pt] \ast \\[-5pt] \ast \end{array}\hspace*{-7pt}\right)}}
\def\starthree{\tiny{\left(\hspace*{-7pt}\begin{array}{c}\ast \\[-5pt]\ast \\[-5pt] \ast \end{array}\hspace*{-7pt}\right)}}
\def\startwo{\tiny{\left(\hspace*{-7pt}\begin{array}{c}\ast \\[-5pt]\ast \end{array}\hspace*{-7pt}\right)}}
\def\starone{\tiny{\left(\hspace*{-7pt}\begin{array}{c}\ast \end{array}\hspace*{-7pt}\right)}}
\def\starfour{\tiny{\left(****\right)}}
\def\starthree{\tiny{\left(***\right)}}
\def\startwo{\tiny{\left(**\right)}}
\def\starone{\tiny{\left(*\right)}}

\eqnobysec
\begin{document}



\review[Progress Toward Understanding Baryon Resonances]{Progress
  Toward Understanding Baryon Resonances}

\author{V. Crede and W. Roberts}

\address{Florida State University, Department of Physics, Tallahassee,
  FL 32306, USA}
\ead{\mailto{crede@fsu.edu}, \mailto{wroberts@fsu.edu}}
\begin{abstract}
The composite nature of baryons manifests itself in the
existence of a rich spectrum of excited states, in particular in the important 
mass region 1-2~GeV for the light-flavoured baryons. The properties of these resonances can be 
identified by systematic investigations using electromagnetic and strong 
probes, primarily with beams of electrons, photons, and pions. After decades
of research, the fundamental degrees of freedom underlying the baryon
excitation spectrum are still poorly understood. The search for hitherto
undiscovered but predicted resonances continues at many laboratories
around the world. Recent results from photo- and electroproduction 
experiments provide intriguing indications for new states and shed
light on the structure of some of the known nucleon excitations. The
continuing study of available data sets with consideration of new observables 
and improved analysis tools have also called into question some of the 
earlier findings in baryon spectroscopy. Other breakthrough measurements 
have been performed in the heavy-baryon sector, which has seen a fruitful 
period in recent years, in particular at the $B$~factories and the Tevatron. 
First results from the LHC indicate rapid progress in the field of bottom 
baryons. In this review, we discuss the recent experimental progress and 
give an overview of theoretical approaches.
\end{abstract}

\maketitle

\section{Introduction}
A better understanding of the nucleon as a bound state of quarks and
gluons as well as the spectrum and internal structure of excited baryons 
remains a fundamental challenge and goal in hadronic physics. In particular, 
the mapping of the nucleon excitations provides access to strong interactions 
in the domain of quark confinement. While the peculiar phenomenon of confinement 
is experimentally well established and believed to be true, it remains analytically 
unproven and the connection to quantum chromodynamics (QCD) -- the fundamental 
theory of the strong interactions -- is only poorly understood. In the early years 
of the 20th century, the study of the hydrogen spectrum has established without question
that the understanding of the structure of a bound state and of its excitation
spectrum need to be addressed simultaneously. The spectroscopy of excited baryon resonances 
and the study of their properties is thus complementary to understanding the 
structure of the nucleon in deep inelastic scattering experiments that provide 
access to the properties of its constituents in the ground state. However, the 
collective degrees of freedom in such experiments are lost. 

An extensive data set of observables in light-meson photo- and electro-production 
reactions has been accumulated over recent years at facilities worldwide such 
as Jefferson Laboratory in the United States, the ELectron Stretcher Accelerator 
(ELSA), the MAinz MIcrotron (MAMI), and the GRenoble Anneau Accelerateur Laser 
(GRAAL) facility in Europe as well as the 8 GeV Super Photon Ring (SPring-8) in 
Japan hosting the Laser Electron Photon Experiment (LEPS). The data set includes cross section data and polarisation 
observables for a large variety of final states, such as $\pi N$, $\eta N$, $\omega N$, 
$\pi\pi N$, $K\Lambda$, $K\Sigma$, etc. These data complement the earlier 
spectroscopy results from $\pi$- and $K$-induced reactions. For a long time, 
the lack of experimental data and the broad and overlapping nature of light-flavour 
baryon resonances has obscured our view of the nucleon excitation spectrum. 
These experiments therefore represent an important step toward the unambiguous 
extraction of the scattering amplitude in these reactions, which will allow the
identification of individual resonance contributions.

Hadronic spectroscopy cannot be interpreted by applying standard perturbation
theory:  phenomenology as well as QCD-based models have provided much
of the insight and theoretical guidance. The recent advances in lattice gauge theory and 
the availability of large-scale computing technology make it possible for the
first time to complement these approaches with numerical solutions of QCD. 
Spin-parity assignments for excited states have even been successfully worked out by some groups. Some collaborations have carried out simulations with pion masses as light as 156 MeV, but the resonance nature of the states, as well as the presence of thresholds complicate the extraction of information from such calculations. In addition, to the best of our knowledge, such simulations have focused on states with lower values of angular momentum. In other simulations with larger pion masses ($m_\pi\gtrsim 400$~MeV), a rich spectrum 
of excited states is obtained, and the low-lying states of some lattice-QCD
calculations 
have the
same quantum numbers as the states in models based on three quarks with wave functions based on the irreducible representations of SU(6)\,$\otimes$\,O(3).  The 
good qualitative agreement may be surprising since the connection between the 
relevant quark degrees of freedom, the constituent or dressed quarks, and those 
of the QCD Lagrangian is not well understood. The lattice results appear to answer
the long-standing question in hadron spectroscopy of whether the large number of 
excited baryons predicted by quark models, but experimentally not 
observed, is realised in nature.

The main goals of recent experiments are the determination of the excited 
baryon spectrum, the identification of possible new symmetries in the spectrum, 
and understanding the structure of states that appear to be built of three valence 
quarks at a microscopic level. The fundamental questions comprise the 
quest for the number of relevant degrees of freedom and a better understanding
of the mechanisms responsible for confinement and chiral symmetry breaking: 
How are the valence or dressed quarks with their clouds of gluons and 
quark-antiquark pairs related to the quark and gluon fields of the underlying 
Lagrangian of QCD? How does the chiral symmetry structure of QCD lead to
dressed quarks and produce the well-known long distance behaviour?
The new experimental and theoretical information bears directly on both the
search for, and our current understanding of baryon resonances. This paper 
will review the new data, present an overview of theoretical approaches and
a sampling of the phenomenological work that has been developed based 
on the new experimental results.

\subsection{Guide to the Literature}
The Particle Data Group~(PDG) regularly includes minireviews on a
large variety of topics in their Reviews of Particle Properties~(RPP),
which have been published biennially for many decades. Useful
minireviews on $N^\ast$ and $\Delta$~resonances as well as charmed
baryons can be found in the latest edition of the RPP~\cite{Beringer:1900zz}. 
Owing to the lack of  suitable $K$~beams, little progress has been reported 
in reviews on $\Lambda$ and $\Sigma$~resonances in the 2010 edition 
of the RPP~\cite{FERMILAB-PUB-10-665-PPD}. The latter also includes a
very brief note on $\Xi$~baryons based on a previous review by Meadows,
published in the proceedings of the {\it Baryon 1980} 
conference~\cite{Meadows:1980vr}.

A comprehensive review of baryon spectroscopy is contained in the 
2009 article by Klempt and Richard~\cite{Klempt:2009pi}, who discuss
prospects for photo- and electroproduction experiments. Some of the
open questions discussed in that review have already found answers 
from recent measurements. An older general article 
on baryon spectroscopy by Hey and Kelly~\cite{Hey:1982aj} still
provides useful information, particularly on some aspects of the
theoretical data analysis. Quark model developments have been discussed 
by Capstick and Roberts in 2000~\cite{Capstick:2000qj}.

Further reviews were dedicated to particular aspects of experiments
using electromagnetic probes. About a decade ago, Krusche and
Schadmand gave a nice summary on low-energy photoproduction~\cite{Krusche:2003ik}
and in 2007, Drechsel and Walcher looked at hadron structure at low
$Q^2$~\cite{Drechsel:2007sq}. The work of  Tiator {\it et 
al.}~\cite{Tiator:2011pw} and Aznauryan and Burkert~\cite{Aznauryan:2011qj} provide  more recent reviews on the 
electroexcitation of nucleon resonances.
\section{Baryon Spectroscopy}
Baryons are strongly interacting fermions and all established baryons
are consistent with a $qqq$~configuration, so that the baryon number
is $B = \frac{1}{3} + \frac{1}{3} + \frac{1}{3} = 1$. Other exotic baryons 
have been proposed, such as pentaquarks -- baryons made of four quarks 
and one antiquark $(B = \frac{1}{3} + \frac{1}{3} + \frac{1}{3} + \frac{1}{3} - \frac{1}{3} = 1)$, 
but their existence is not generally accepted. In quantum chromodynamics, 
heptaquarks (5 quarks, 2 antiquarks), nonaquarks (6 quarks, 3 antiquarks), etc. 
could also exist. Baryons which consist only of $u$~and $d$~quarks are called 
nucleons if they have isospin $I=\frac{1}{2}$ or $\Delta$~resonances if they have 
isospin $I=\frac{3}{2}$. Baryons containing $s$~quarks are called hyperons and 
are labelled as $\Lambda,~\Sigma,~\Xi$, and $\Omega$ depending on
the number of $s$~quarks and isospin (Table~\ref{Table:Baryons}). Baryons 
containing $c$ or $b$~quarks are labelled with an index. For example, the
$\Lambda^+_c$ has isospin zero and quark content $udc$, the $\Xi_c^+$ has 
quark content $usc$ and the $\Xi^{++}_{cc}$ has quark content $ucc$.

\begin{table}[b]
\begin{center}
\caption{\label{Table:Baryons} Summary of baryon quantum numbers.
  Antihyperons have positive Strangeness +1, +2, and +3.}
\begin{tabular}{@{}l|cc|cc|c|c}
\br
 & $N$ & $\Delta$ & $\Lambda$ & $\Sigma$ & $\Xi$ & $\Omega$\\
\mr
Isospin, $I$ & $\frac{1}{2}$ & $\frac{3}{2}$ & 0 & 1 & $\frac{1}{2}$ & 0\\
Strangeness, $S$ & \multicolumn{2}{c|}{0} & \multicolumn{2}{c|}{$-1$} & $-2$ & $-3$\\
Number of strange quarks & \multicolumn{2}{c|}{0} & \multicolumn{2}{c|}{1} & 2 & 3\\
\br
\end{tabular}
\end{center}
\end{table}
\subsection{Baryons Composed of $u$, $d$ and $s$ Quarks} \label{sec:flavours}
\begin{figure}[h]
\centerline{
\includegraphics[width=3.5in]{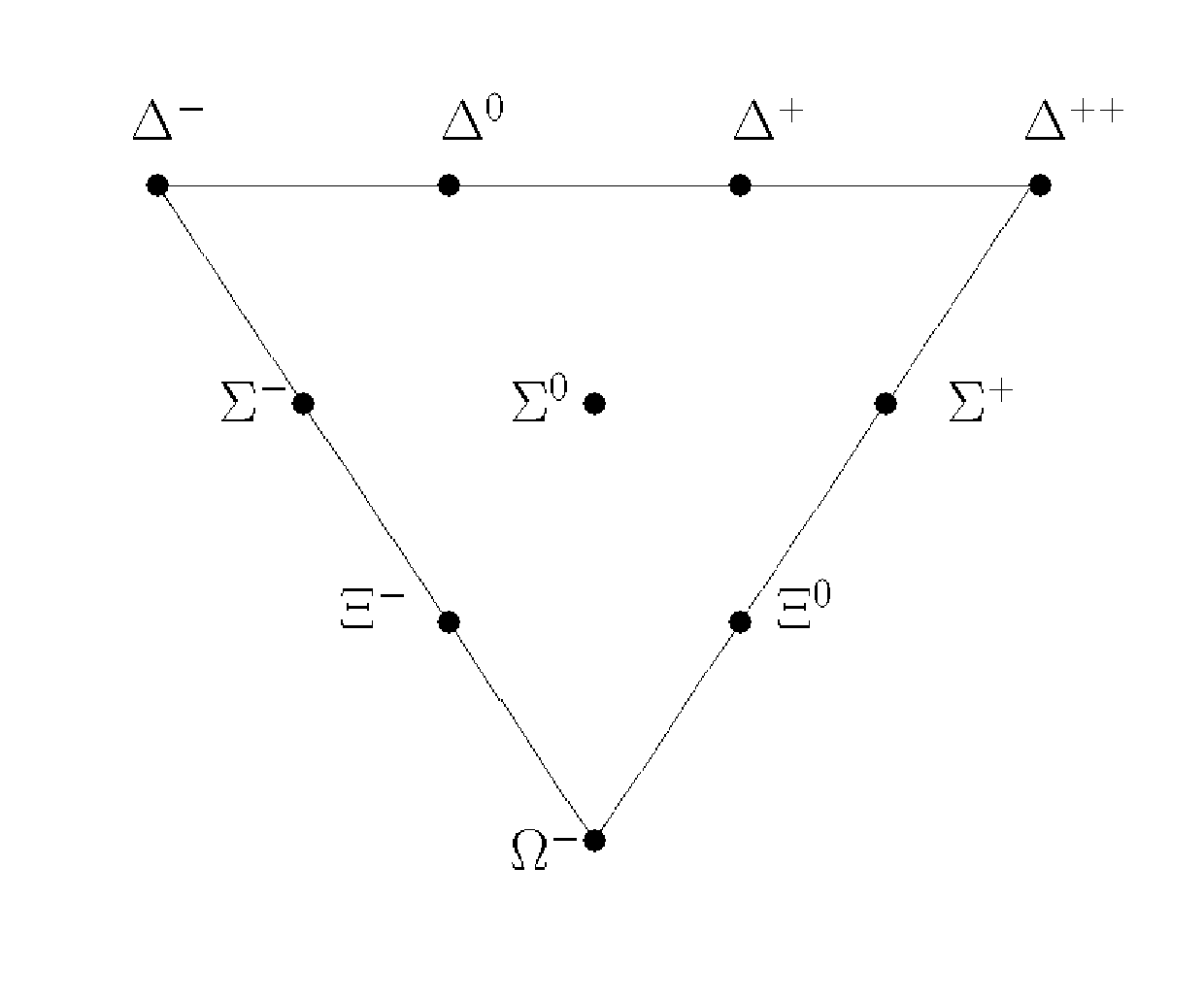}\hspace{-1cm}
\includegraphics[width=3.5in]{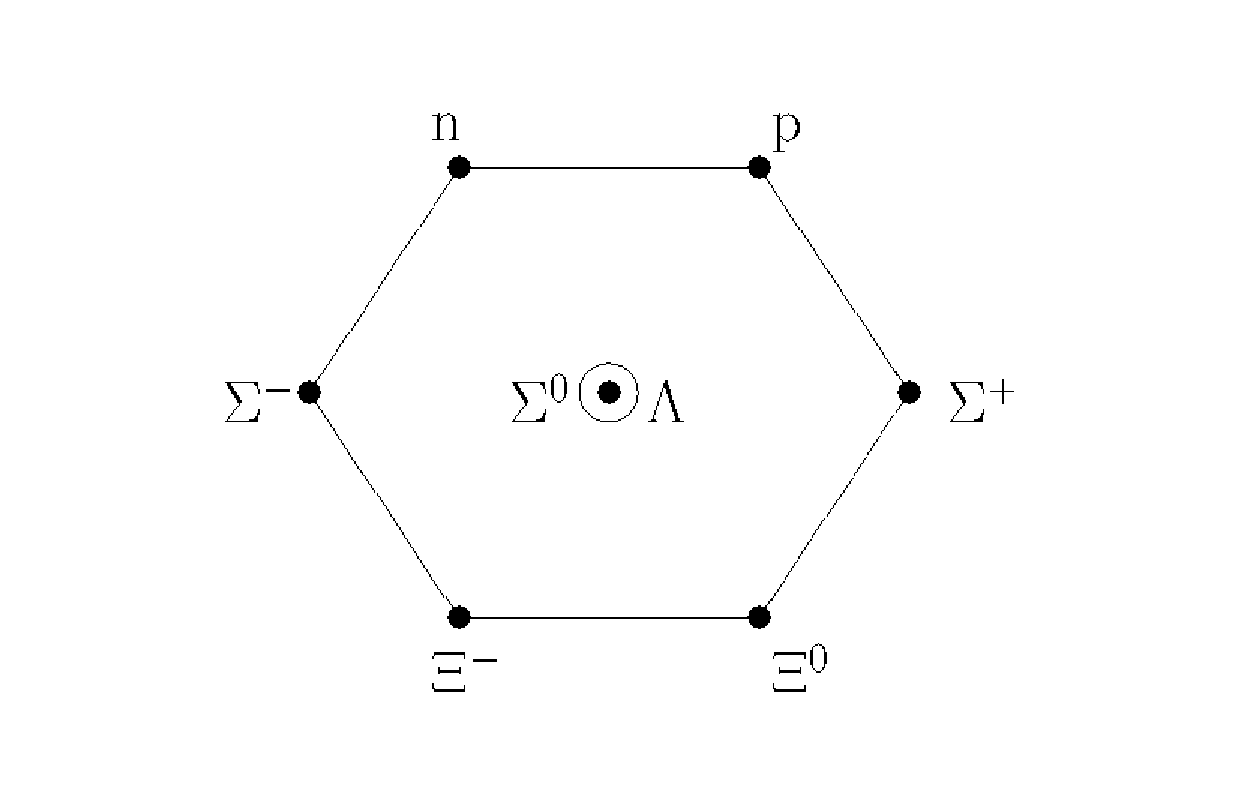}}
\caption{The symmetric {\bf 10} (left) and mixed symmetric {\bf 8} (right) of SU(3).}
\label{fig:multiplet10+8}
\end{figure}

As fermions baryons obey the Pauli principle, so the total wave function
\begin{eqnarray}
|qqq\rangle_A = |{\rm colour}\rangle_A\,\times\, |{\rm space,~spin,~flavour}\rangle_S
\end{eqnarray}
must be antisymmetric (denoted by the index $A$) under the interchange of any two equal-mass
quarks. Since all observed hadrons are colour
singlets, the colour component of the wave function must be completely
antisymmetric. For the 
light-flavour baryons, the three flavors $u$, $d$ and $s$ can be
treated in an approximate SU(3) framework, in which each quark is a
member of an SU(3) triplet. The flavour wave functions of  baryon states can then be constructed to be members of SU(3) multiplets as
\begin{equation}
{\bf 3}\,\otimes\, {\bf 3}\,\otimes\, {\bf 3}\,=\,{\bf 10}_S\,\oplus\,
{\bf 8}_M\,\oplus\, {\bf 8}_M\,\oplus\, {\bf 1}_A\,.
\end{equation}
The proton and neutron, of which all the matter that we see is composed, are members of both octets. The weight diagrams for the decuplet and octet representations of SU(3) are shown in Figure~\ref{fig:multiplet10+8}.

The total baryon spin has two possible values. The three 
quark spins ($s=\frac{1}{2}$) can yield a total baryon spin of either $S=\frac{1}{2}$ or $S=\frac{3}{2}$. The latter
configuration yields a completely symmetric spin wave function,
whereas the $S=\frac{1}{2}$~state  exhibits a mixed symmetry. The flavour and spin can be combined in an approximate 
spin-flavour SU(6), where the multiplets are
\begin{eqnarray}
{\bf 6}\,\otimes\, {\bf 6}\,\otimes\, {\bf 6}\,=\,{\bf 56}_S\,\oplus\,
{\bf 70}_M\,\oplus\, {\bf 70}_M\,\oplus\, {\bf 20}_A\,.
\end{eqnarray}
These can be decomposed into flavour SU(3) multiplets
\begin{eqnarray}
{\bf 56}\,=\, ^4{\bf 10}\,\oplus\, ^2{\bf 8}\\\label{Equation:70plet} {\bf 70}\,=\,
^2{\bf 10}\,\oplus\, ^4{\bf 8}\,\oplus\, ^2{\bf 8}\,\oplus\, ^2{\bf 1}\\\label{Equation:20plet}
{\bf 20}\,=\, ^2{\bf 8}\,\oplus\, ^4{\bf 1}\,,
\end{eqnarray}
where the superscript $(2S+1)$ gives the spin for each particle in the
SU(3)~multiplet. The proton and neutron belong to the ground-state
{\bf 56}, in which the orbital angular momentum between any pair of
quarks is zero, and are members of the octet with spin and parity $J^P
= \frac{1}{2}^+$. The $\Delta$~resonance is a member of the decuplet with spin
and parity $J^P = \frac{3}{2}^+$.
We refer the interested reader to the literature for the correctly
symmetrized wave functions for the SU(3)  multiplets. The wave 
functions of the {\bf 70} and {\bf 20} require some excitation of the
spatial part to make the overall non-colour
(spin\,$\times$\,space\,$\times$\,flavour) component of the wave
function symmetric. Orbital motion is accounted for by classifying
states in SU$(6)\,\otimes\,$O(3)
supermultiplets, with the O(3)~group describing the orbital motion. 

In addition to spin-flavour multiplets, it is convenient to classify
baryons into bands according to the harmonic oscillator model
with equal quanta of excitation, $N = 0,1,2, ...$. Each band consists 
of a number of supermultiplets, specified by $({\bf D},\,L^P_N)$, where 
${\bf D}$ is the dimensionality of the SU(6) representation, $L$ is the total 
quark orbital angular momentum, and $P$ is the parity. 
The first-excitation band contains only one supermultiplet, $({\bf 70},\,1_1^-)$, 
corresponding to states with one unit of orbital angular momentum and
negative parity, whereas the second-excitation band contains
five supermultiplets corresponding to states with positive parity and either two units of 
angular momentum or one unit of radial excitation. 
Table~\ref{Table:Classification} shows the supermultiplets contained 
in the first three bands and the known, well-established baryons that
are members of the ground-state {\bf 56}. Further quark-model
assignments for some of the known baryons will be discussed in 
Section~\ref{Section:Discussion}.

\begin{table}
\begin{center}
\caption{\label{Table:Classification}Supermultiplets, $({\bf D},\,L_N^P)$,  
  contained in the first three bands~\cite{Beringer:1900zz} and assignments
  for the ground-state octet and decuplet to known baryons.}
\begin{tabular}{@{}l|ccccc}
\br
 N & \multicolumn{5}{c}{Supermultiplets}\\
\mr
0 & \multicolumn{5}{c}{$({\bf 56},\,0_0^+)$}\\
   & $S = \frac{1}{2}^+$ & $N(939)$ & $\Lambda(1116)$ & $\Sigma(1193)$ & $\Xi(1318)$\\[0.5ex]
   & $S = \frac{3}{2}^+$ & $\Delta(1232)$ & $\Sigma(1385)$ & $\Xi(1530)$ & $\Omega(1672)$\\
\mr
1 & \multicolumn{5}{c}{$({\bf 70},\,1_1^-)$}\\
\mr
2 & $({\bf 56},\,0_2^+)$ & $({\bf 70},\,0_2^+)$ & $({\bf 56},\,2_2^+)$ & $({\bf 70},\,2_2^+)$ & $({\bf 20},\,1_2^+)$\\
\br
\end{tabular}
\end{center}
\end{table}

\subsection{Baryons Containing Heavy Quarks}
The baryons containing a single charm quark can be described in terms of SU(3) flavour multiplets, but these represent but a subgroup of the larger SU(4) group that includes all of the baryons containing zero, one, two or three charmed quarks. Furthermore, this multiplet structure is expected to be repeated for every combination of spin and parity, leading to a very rich spectrum of states. One can also construct SU(4) multiplets in which charm is replaced by beauty, as well as place the two sets of SU(4) structures within a larger SU(5) group to account for all the baryons that can be constructed from the five flavours of quark accessible at low to medium energies. It must be understood that the classification of states in SU(4) and SU(5) multiplets serves primarily for enumerating the possible states, as these symmetries are badly broken. Only at the level of the SU(3) ($u,\,\,d,\,\,s)$ and SU(2) $(u,\,\,d)$ subgroups can these symmetries be used in any quantitative way to understand the structure and properties of these states.

\begin{figure}[h]
\caption{(a) The symmetric {\bf 20} of SU(4), showing the SU(3)
  decuplet on the lowest layer. (b) The mixed-symmetric {\bf 20}s
  and (c) the antisymmetric {\bf 4} of SU(4). The mixed-symmetric {\bf
    20}s have the SU(3) octet on the lowest layer, while the {\bf 4} has
  the SU(3) singlet at the bottom. Note that there are two $\Xi_c^+$ and
  two $\Xi_c^0$~resonances on the middle layer of the mixed-symmetric {\bf 20}.}
\label{Figure:Multiplets}
\begin{center}
\includegraphics[scale=1.2]{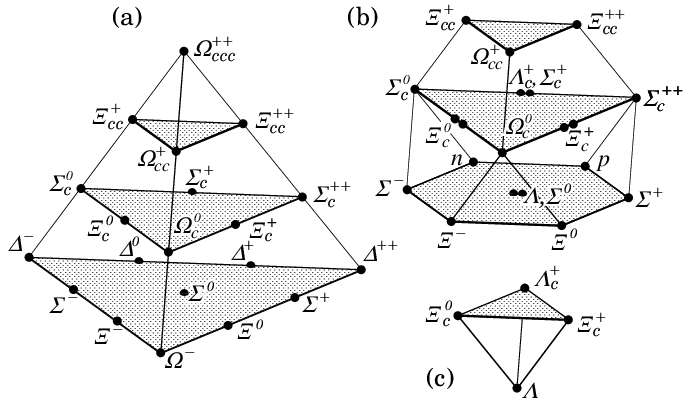}
\end{center}
\end{figure}

The multiplet structure for flavour SU(4) is ${\bf 4}\,\otimes\,{\bf 4}\,\otimes {\bf
  4} = {\bf 20}_S\,\oplus\,{\bf 20}_M\,\oplus\,{\bf 20}_M\,\oplus\,{\bf 4}_A$. The symmetric {\bf 20} contains the SU(3) decuplet as a subset, forming the `ground floor' of the weight diagram shown in Figure~\ref{Figure:Multiplets}~(a), and all the ground-state baryons in this multiplet have $J^P=\frac{3}{2}^+$. The mixed-symmetric {\bf 20}s (Figure~\ref{Figure:Multiplets}~(b)) contain the SU(3) octets on the lowest level, and all the ground-state baryons in this multiplet have $J^P=\frac{1}{2}^+$. The ground-floor state of the {\bf 4} (Figure~\ref{Figure:Multiplets}~(c)) is the SU(3) singlet $\Lambda$ with $J^P=\frac{1}{2}^-$.

Within the flavour SU(3) subgroups, the ground-state heavy baryons containing a single heavy quark belong either to a sextet of flavour symmetric states, or an antitriplet of flavour antisymmetric states, both of which sit on the second layer of the mixed-symmetric {\bf 20} of SU(4) of Figure~\ref{Figure:Multiplets}~(b). There is also expected to be a sextet of states with $J^P=\thlf^+$ sitting on the second floor of the symmetric {\bf 20}. The members of the two multiplets of singly-charmed baryons have flavour wave functions
\beqy
\Sigma_c^{++}&=&uuc,\,\,\,\,\Sigma_c^{+}=\frac{1}{\sqrt2}\left(ud+du\right)c,\,\,\,\,\Sigma_c^{0}=ddc\nonumber\\
\Xi_c^{'+}&=&\frac{1}{\sqrt2}\left(us+su\right)c,\,\,\,\,\Xi_c^{'0}=\frac{1}{\sqrt2}\left(ds+sd\right)c,
\nonumber\\
\Omega_c^{0}&=&ssc,
\eeqy
for the sextet and
\beqy
\Lambda_c^{+}=\frac{1}{\sqrt2}\left(ud-du\right)c,\,\,\,\,
\Xi_c^{+}=\frac{1}{\sqrt2}\left(us-su\right)c,\,\,\,\,
\Xi_c^{0}=\frac{1}{\sqrt2}\left(ds-sd\right)c,
\eeqy
for the antitriplet. There is a similar set of flavour wave functions for baryons containing a single $b$ quark. 

For the baryons containing two charm or two beauty quarks, the flavour wave functions are 
\beq
\Xi_{cc}^{++}=ccu,\,\,\,\,\Xi_{bb}^{0}=bbu,\,\,\,\,\Omega_{cc}^{+}=ccs,\,\,\,\,\Omega_{bb}^{-}=bbs.
\eeq
When the two heavy quarks are different, there are two ways of
constructing their flavour wave functions. One can imagine that the
two heavy quarks are members of a (pseudo-)symmetry group,
SU(2)$_{bc}$, and that the pair of heavy quarks form either triplet or
singlet representations of this group. Two members of this triplet
would then be the $\Xi_{cc}$ and $\Xi_{bb}$, with the third member
being the state 
\beq
\Xi_{bc}^+=\frac{1}{\sqrt2}\left(cb+bc\right)u.
\eeq
The singlet state would then be 
\beq
\Xi_{bc}^{'+}=\frac{1}{\sqrt2}\left(cb-bc\right)u.
\eeq 
Alternatively, the flavour wave functions of these two states may be
written as 
\beq
\Xi_{bc}^{'+}=\frac{1}{\sqrt2}\left(uc+cu\right)b
\eeq 
and 
\beq
\Xi_{bc}^{+}=\frac{1}{\sqrt2}\left(uc-cu\right)b.
\eeq
The same choices of wave function need to be made when the light quark in the baryon is a down quark or a strange quark. 

\subsection{Naming Scheme for Light Baryons}
An indispensable tool in hadron spectroscopy are of course the
listings by the PDG. The group lists established baryon resonances in
their Baryon Summary Table. Resonances are given a star rating based
on their overall status. The ratings range from $\ast\,\ast\,\ast\,\ \ast$ 
(existence is certain) and $\ast\,\ast\,\ast$ (existence is likely to
certain, but further confirmation is desirable) to $\ast\,\ast$
(evidence of existence is only fair) and $\ast$ (evidence of 
existence is poor)~\cite{Beringer:1900zz}.

In its latest 2012 edition~\cite{Beringer:1900zz}, the PDG has replaced the
historical naming scheme for observed light nonstrange baryons -- based on elastic 
$\pi N$~scattering -- with a more general scheme which applies to 
all baryons independent of the production mechanism, including
those resonances that are not produced in formation experiments, 
e.g. $\Xi$~or $\Omega$~resonances with $|S|>1$ and heavy baryons.
For example, the $\Delta$~resonance was observed originally in the 
$P_{33}$ partial wave with $L = 1$ referring to the angular momentum 
between the $\pi$ and the nucleon in $\pi N$~scattering experiments. 
With the plethora of new results coming from photoproduction experiments, 
the resonance label has switched from the incoming $\pi N$~partial wave, 
$L_{2I\,2J}$, to the spin-parity of the state. Consequently, the name of 
the $\Delta$~resonance has changed from $\Delta(1232)P_{33}$ to
$\Delta(1232)\,\frac{3}{2}^+$.
\section{Heavy-Quark Baryons}
Baryons containing heavy quarks have been the focus of much attention
recently, in particular since the experimental discovery of new resonances at 
Fermilab and the Large Hadron Collider~(LHC). The heavy quark provides a 
`flavour tag' that may give insights into the mechanism of confinement 
and the systematics of hadron resonances that cannot be obtained with light 
quarks. All of the states containing heavy quarks are expected to be somewhat 
narrow, for the most part, so that their detection and isolation is relatively easy,
and in general does not rely on the extensive partial-wave-analysis
machinery usually necessary for identifying light baryons (most states
found to date have widths of a few MeV, with the largest reported
width being a few tens of MeV, but with large uncertainties). Such
analyses may be required for determining the quantum numbers of the
states, but even then, the procedure may still be simpler than in the
case of light baryons, as it is expected to be largely free of the
various interferences that arise with nearby, broad and overlapping
resonances.

In addition, the heavy-quark symmetries provide a framework for
understanding and predicting the spectrum of one flavour of heavy
baryons, say those containing a $b$ quark, if the spectrum of baryons
containing a $c$ quark has been obtained \cite{Isgur:1991wq}. Used
judiciously, this heavy-quark symmetry can provide some qualitative
insight, and perhaps even quantitative, into the spectrum of light
baryons, particularly the hyperons.

Despite the wealth of information that they can provide, and many
theoretical treatises, surprisingly little is known experimentally
about the heavy baryons \cite{Beringer:1900zz}. This is largely because despite
the comments above, they are difficult to produce. Unlike the heavy
mesons, there are no known resonant production mechanisms, so these baryons
can only be obtained by continuum production, where cross sections are
small, as products in the decays of heavy mesons, or at hadron
colliders. Not surprisingly, the $B$~factories, and CLEO before that,
have been the main source of these baryons.

The first observations of beauty baryons were made at the Organisation 
Europ\'eenne pour la Recherche Nucl\'eaire (CERN) by the Large Electron 
Positron (LEP) experiments. These observations were mostly limited to a few 
decay modes of the $\Lambda_b^0$~baryon and some indications for other
ground-state baryons based on their semileptonic decays. In recent years, 
the Tevatron experiments at Fermilab have observed a large number of new
heavy-baryon states. In addition to direct observations of most of the
ground states, the published results even include precision lifetime
measurements, and several new decay modes. It is clear that future
results will again come from CERN. The LHC experiments have already
demonstrated their discovery potential. In 2012 alone, LHCb has announced 
two new $\Lambda_b^{\ast0}$~states and the Compact Muon Solenoid (CMS) 
Collaboration has reported a new $\Xi_b^{\ast0}$~resonance (Table~\ref{Table:HeavyBaryons}).

\subsection{The Known Baryons}

\begin{table}\small
\begin{center}
\caption{\label{Table:HeavyBaryons} (Colour online) The heavy baryons 
  with at least a  $\ast\,\ast\,\ast$~assignment as listed by the PDG indicating that 
  existence ranges from very likely to certain~\cite{Beringer:1900zz}. Only the 
  $\Lambda_c^+$ and the $\Sigma_c(2455)$ are considered certain with 
  $\ast\,\ast\,\ast\,\ \ast$~assignments. Two further states not listed in the 
  table are well established: $\Lambda_c(2880)^+$ with a possible $J^P=\frac{5}{2}^+$ 
  assignment and a mass of $2881.50\pm 0.35$~MeV and $\Xi_c^{+(0)}(3080)$ with 
  unknown quantum numbers and a mass of $3077.0\pm 0.4~(3079.9\pm 
  1.4)$~MeV. Masses which have changed slightly from the 2010 edition are marked
  with $(^\ast)$. Recent findings (not included in the 2012 edition) are highlighted 
  in red with statistical errors only.}  
\begin{tabular}{@{}l|c|c|c|c|c}
\br
 & Mass [\,MeV\,] & Mass [\,MeV\,] & Mass [\,MeV\,] & Mass [\,MeV\,] & Mass [\,MeV\,]\\
 & $J^P = \frac{1}{2}^+$ & $J^P = \frac{3}{2}^+$ & $J^P = \frac{1}{2}^-$ & $J^P = \frac{3}{2}^-$ & $J^P = ?^?$\\
\mr
$\Lambda_c^{+}$ & $2286.46\pm 0.14$ & & $2592.25\pm 0.28\,^\ast$ & $2628.11\pm 0.19\,^\ast$ & $2939.3^{+1.4}_{-1.5}$\\
$\Sigma_c^{++}$ & $2453.98\pm 0.16\,^\ast$ & $2517.9\pm 0.6\,^\ast$ & & & $2801^{+4}_{-6}$\\
$\Sigma_c^{+}$ & $2452.9\pm 0.4$ & $2517.5\pm 2.3$ & & & $2792^{+14}_{-5}$\\
$\Sigma_c^{0}$ & $2453.74\pm 0.16\,^\ast$ & $2518.8\pm 0.6\,^\ast$ & & & $2806^{+5}_{-7}\,^\ast$\\
$\Xi_c^+$ & $2467.6^{+0.4}_{-1.0}$ & $2645.9^{+0.5}_{-0.6}$ & $2789.1\pm 3.2$ & $2816.6\pm 0.9$ & $2971.4\pm 3.3$\\
$\Xi_c^0$ & $2471.09^{+ 0.35}_{-1.00}$ & $2645.9\pm 0.5$ & $2791.8\pm 3.3$ & $2819.6\pm 1.2$ & $2968.0\pm 2.6$\\
$\Xi_c^{\prime +}$ & $2575.6\pm 3.1$ & & & &\\
$\Xi_c^{\prime 0}$ & $2577.9\pm 2.9$ & & & &\\
$\Omega_c^0$ & $2695.2^{+1.8}_{-1.6}$ & $2765.9\pm 2.0$ & & &\\
\mr
$\Lambda_b^0$ & \multicolumn{2}{l}{{\color{red} $5619.53\pm 0.13\pm0.45$}~\cite{Aaij:2013ky}} &
\multicolumn{1}{l}{{\color{red} $5911.95\pm 0.12$}} & \multicolumn{2}{l}{{\color{red} $5919.76\pm 0.12$}~\cite{Aaij:2012da}}\\
$\Sigma_b^+$ & $5811.3^{+0.9}_{-0.8}\pm 1.7\,^\ast$ &
\multicolumn{2}{l}{$5832.1\pm 0.7^{+1.7}_{-1.8}\,^\ast$} & \multicolumn{1}{l}{} &\\
$\Sigma_b^-$ & $5815.5^{+0.6}_{-0.5}\pm 1.7\,^\ast$ & \multicolumn{2}{l|}{$5835.1\pm 0.6^{+1.7}_{-1.8}\,^\ast$} & &\\
$\Xi_b^-$ & \multicolumn{3}{l|}{$5791.1\pm 2.2\,^\ast$~({\color{red} $5795.8\pm 0.9\pm 0.4$}~\cite{Aaij:2013ky})} & &\\
$\Xi_b^0$ & \multicolumn{3}{l|}{$5787.8\pm 5.0\pm 1.3\,^\ast$\hspace{1mm}{\color{red} $5945.0\pm 0.7\pm 0.3$}~\cite{Chatrchyan:2012ni}} & &\\
$\Omega_b^-$ & \multicolumn{3}{l|}{$6071\pm 40$~({\color{red} $6046.0\pm 2.2\pm 0.5$}~\cite{Aaij:2013ky})} & &\\
\br
\end{tabular}
\end{center}
\end{table}

In the 2012 edition of the RPP, the PDG lists 23~singly-charmed, one doubly-charmed, 
and six~beauty baryons in its Baryon Summary Table~\cite{Beringer:1900zz}. The 
existence of only 23 of these ranges from likely to certain with a 3-star 
assignment by the PDG and only two baryons, the $\Lambda_c^+$ and 
$\Sigma_c(2455)$, have a 4-star assignment. The known $C = +1$
and $B = +1$ baryons with at least three stars are listed in
Table~\ref{Table:HeavyBaryons}. For most of the beauty baryons, with
the exception of the $\Lambda_b^0$, only one or two decay modes have
been observed. None of the $I$,~$J$, or $P$ quantum numbers have
actually been measured, but are based on quark model expectations.

\subsection{\label{Section:HeavyBaryonExperiments}Experimental Methods
  and Major Experiments}
Recent experimental results on heavy-flavour baryon spectroscopy have
been reported by several experiments, mostly by the so-called $B$~factories 
at $e^+e^-$ collider facilities and the experiments at the Tevatron
$p\bar{p}$ collider facility. Some earlier photoproduction experiments at 
Fermilab, E687, E691, E791, and Focus made important contributions;
technical details on these experiments can be found elsewhere in the literature. 
The SELEX experiment used a hadron beam and produced interesting
results on charmed baryons including the possible first observation of
a doubly-charmed baryon~\cite{Mattson:2002vu,Ocherashvili:2004hi}.

Charmed baryons have been studied intensively at the Cornell Electron-positron Storage Ring
(CESR) at Cornell University using the CLEO detector. Based on the CLEO-II
detector, the CLEO-III detector started operation in 1999~\cite{Kopp:1996kg}. 
CLEO consisted of drift chambers for tracking and $dE$/$dx$ measurements 
and a CsI~electromagnetic calorimeter based on 7800~modules inside a 1.5~T 
magnetic field. For CLEO-III, a silicon-strip vertex detector and a ring-imaging 
$\rm\check{C}$erenkov detector for particle identification were added. The 
integrated luminosity accumulated by the CLEO-III detector in 1999-2003 
was 16~fb$^{-1}$. In 2003, CLEO was upgraded to CLEO-c in order to study 
charmed mesons at high luminosities. The accelerator ran at significantly 
lower energies, mostly at the $\psi\,^{\prime}(3770)$ charmonium state, so that charmed baryons could 
no longer be produced. The study of $N^\ast$~states in $\psi\,^{\prime}(3770)\to p\bar{p}$
+ mesons remained a possibility, though. CLEO-c operations ended in Spring 2008.
Selected results of the CLEO collaboration can be found in~\cite{Avery:1988uh,Edwards:1994ar,
Avery:1995ps,Gibbons:1996yv,Jessop:1998wt,Alexander:1999ud,CroninHennessy:2000bz,
Csorna:2000hw,Artuso:2000xy,Ammar:2000uh,Athar:2004ni,Alexander:2010vd}.

In 1999, two $B$-factories started data-taking with the main
goal of studying time-dependent $CP$ asymmetries in the decay of 
$B$~mesons: BABAR at the Stanford Linear Accelerator Center (SLAC) and
Belle at the KEKB $e^+e^-$ collider in Tsukuba, Japan. The very high 
luminosities of the electron-positron colliders and the general-purpose 
character of the detectors also made them suitable places for the study 
of heavy baryons. The peak cross section at the $\Upsilon(4S)$ is about 
1.2~nb, which sits on approximately 3.5~nb of what is called the continuum
background. Most of the information on heavy baryons comes from the reaction 
$e^+e^-$~$\to$ continuum and, to a lesser extent, from $B$~decays.

BABAR was a cylindrically-shaped detector~\cite{Aubert:2001tu} with
the interaction region at its center. The electron beam of the
PEP-II facility collided with a positron beam of lower energy to 
produce a center-of-mass energy near or at 10.58~GeV, which corresponds
to the $\Upsilon(4S)$~resonance. This state decays almost
instantly into two $B$~mesons. About $10^9$~$B$~mesons were
recorded. The momenta of charged particles were measured with a
combination of a five-layer silicon vertex tracker and a 40-layer 
drift chamber in a solenoidal magnetic field of 1.5~T. The momentum
resolution was about $\sigma_{p_t}\approx 0.5\,\%$ at $p_t =
1.0$~GeV. A detector of internally reflected \v{C}erenkov
radiation was used for charged particle identification. The
electromagnetic calorimeter was a finely segmented array of CsI(Tl)
crystals with an energy resolution of $\sigma_E/E\approx 2.3\,\%\times
E^{-1/4} + 1.9\,\%$, where the energy is in GeV. The instrumented flux
return contained resistive plate chambers for muon and long-lived
neutral hadron identification. Some selected results are given 
in~\cite{Aubert:2005gt,Aubert:2006je,Aubert:2006qw,Aubert:2006sp,
Aubert:2007dt,:2008if,Aubert:2008ty}.

The Belle experiment operates at the KEKB $e^+e^-$ accelerator, the world's highest 
luminosity machine with a world record in luminosity of $1.7\times 10^{34}~{\rm cm}
^{-2}$s$^{-1}$ and with an integrated luminosity exceeding $700~{\rm fb}^{-1}$. The Belle
detector~\cite{Iijima:2000cq} is a large-solid-angle magnetic spectrometer based on 
a 1.5~T superconducting solenoid. Charged particle tracking is provided by a three-layer 
silicon vertex detector and a 50-layer central drift chamber (CDC) surrounding the 
interaction point. The charged particle acceptance covers laboratory polar angles between 
$\theta = 17^\circ$ and $150^\circ$, corresponding to about 92\,\% of the total solid 
angle in the center-of-mass system. 
Charged hadron identification and pion/kaon separation is provided by $dE/dx$~measurements 
in the CDC, an array of 1188 aerogel \v{C}erenkov counters (ACC), and a barrel-like array 
of 128~time-of-flight scintillation counters with rms resolution of 0.95 ps. Particles that produce electromagnetic 
showers are detected in an array of 8736 CsI(Tl) crystals of projective 
geometry that covers the same solid angle as the charged particle system. 
Results on charmed-baryon spectroscopy relevant to this review can be found 
in~\cite{Iijima:2000cq,Mizuk:2004yu,Abe:2004sr,Chistov:2006zj,Abe:2006rz,Solovieva:2008fw,:2008wz}.

The large data sets available from the Tevatron $p\bar{p}$ collider at
Fermilab at a center-of-mass energy of $\sqrt{s} = 1.96$~TeV have
rendered possible significant progress in the field of heavy baryon
spectroscopy, in particular for beautiful baryons. Both experiments,
the Collider Detector at Fermilab (CDF) and D\O, collected about 10~fb$^{-1}$ of data during the Run II
period from 2001 to 2011 and could trigger on dimuon pairs so that heavy
baryons decaying to $J/\psi$ can be studied. The most relevant components of the D\O~detector for the results
discussed in this review are the central tracking system and the muon
spectrometer. The central tracking system consists of a silicon
microstrip tracker (SMT) and a central fiber tracker (CFT) that are
surrounded by a 2~T superconducting solenoid. The muon spectrometer is
located outside the calorimeter system and covers the pseudorapidity
region $|\eta| < 2$ ($\eta = -{\rm ln} [{\rm tan} (\theta/2)]$ and
$\theta$ is the polar angle). It comprises a layer of drift tubes and
scintillator trigger counters in front of 1.8~T iron toroids followed
by two similar layers behind the toroids. Results on heavy-baryon
spectroscopy from the D\O~collaboration are summarized in
\cite{Abazov:2004bn,arXiv:0706.1690,arXiv:0704.3909,arXiv:0706.2358,arXiv:0808.4142,arXiv:1105.0690}. 

The tracking system of the CDF~II detector~\cite{Acosta:2004yw} lies
within a uniform, axial magnetic field of 1.4~T surrounded by calorimeters 
using lead-scintillator sampling (electromagnetic calorimeters) and iron-scintillator
sampling (hadron calorimeters). The inner tracking volume is comprised
of 6-7 layers of double-sided silicon microstrip detectors and the
remainder is occupied by an open-cell drift chamber (COT). The silicon 
detector provides a vertex resolution of approximately $15~\mu$m in the
transverse and $70~\mu$μm in the longitudinal direction and the COT a
transverse momentum resolution of $\sigma(p_T)/p^2_T \approx 0.1\,\% /
({\rm GeV} / c)$. CDF was able to trigger on hadronic decays of heavy
baryons which are identified by tracks displaced from the primary
vertex. Recent results relevant to this review can be found
in~\cite{hep-ex/0208035,hep-ex/0507067,hep-ex/0508022,
hep-ex/0609021,hep-ex/0601003,arXiv:0707.0589,arXiv:0706.3868,
arXiv:0810.3213,arXiv:0905.3123,arXiv:0812.4271,arXiv:0912.3566, 
Aaltonen:2011wd,Aaltonen:2011sf,CDF:2011ac}.

The LHC experiments, in particular LHCb and CMS, have just started to contribute
significantly to hadron spectroscopy. Their most recent results on 
$b$~baryons are based on the 2011 data-taking at a center-of-mass
energy of $\sqrt{s} = 7$~TeV and are of unprecedented statistical quality. 
We refer to~\cite{Alves:2008zz} for LHCb and~\cite{Chatrchyan:2008aa} for 
CMS for more details on the experimental setups. Results relevant for
this review are summarized in~\cite{Aaij:2013ky,Aaij:2012da,Chatrchyan:2012ni,
Aaij:2011ep}.

\subsection{Recent Results in the Spectroscopy of Charmed Baryons}
\subsubsection{The $\Lambda_c$ States}
\begin{figure}[t]
\caption{\label{Figure:Lambda_c}(Colour online) Left: The $m(pK^-\pi^+\pi^+\pi^-)
  - m(pK^-\pi^+)$~difference spectrum together with the fit (black solid line)
  from CDF~\cite{Aaltonen:2011sf}. The red dash-dotted line corresponds to 
  the sum of three background contributions. The two peaks corresponding to  
  $\Lambda_c(2595)^+ - \Lambda_c^+$ and $\Lambda_c(2625)^+ -\Lambda_c^+$
  candidates are visible. Right: Comparison of $\Lambda_c^{\ast +}$ mass difference 
  measurements by CLEO~\cite{Edwards:1994ar}, Fermilab E687~\cite{Frabetti:1993hg,
  Frabetti:1995sb}, ARGUS~\cite{Albrecht:1997qa}, and CDF~\cite{Aaltonen:2011sf}. 
  Pictures from~\cite{Aaltonen:2011sf}.}
\includegraphics[scale=0.38]{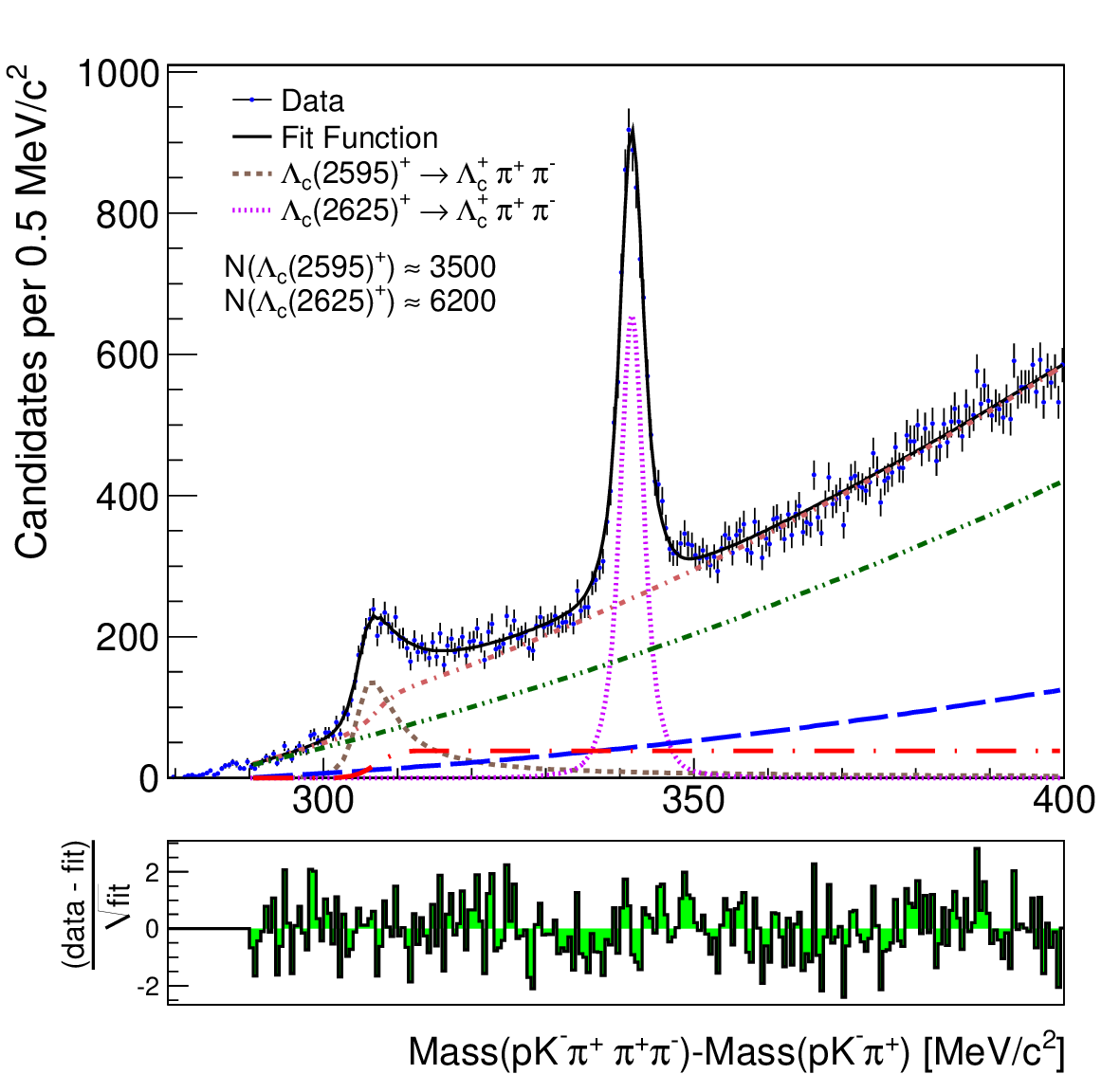}
\begin{minipage}{4cm}\vspace{-7cm}
\begin{tabular}{c}
\includegraphics[scale=0.38]{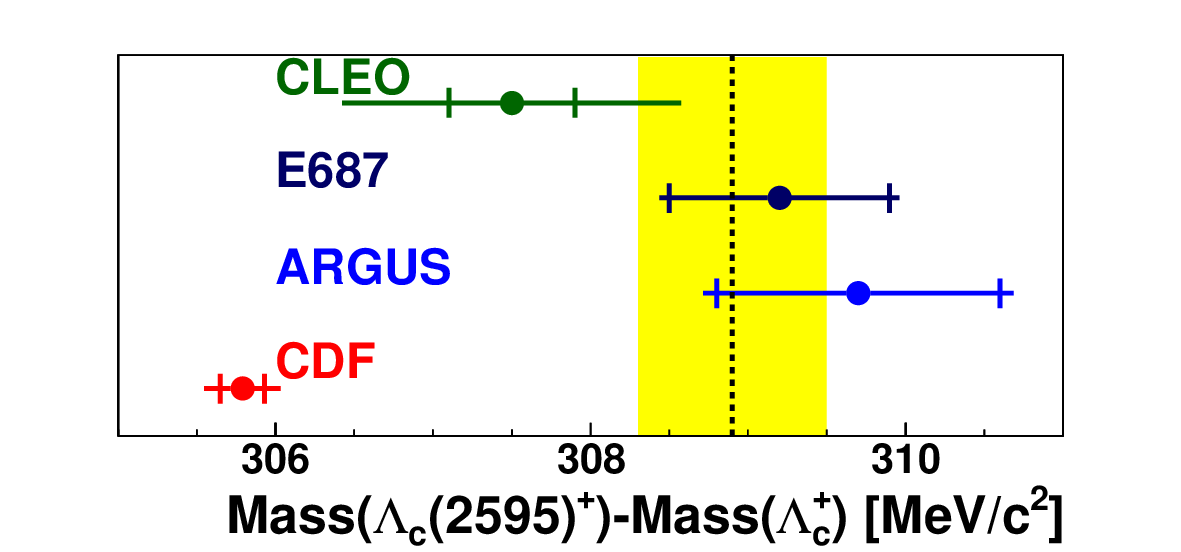}\\
\includegraphics[scale=0.38]{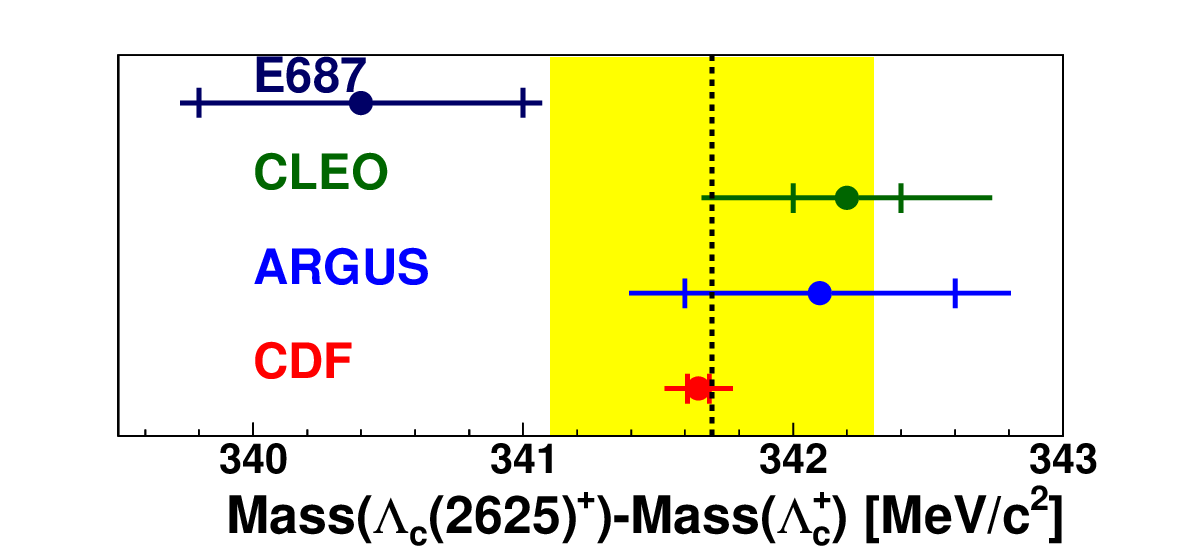}
\end{tabular}
\end{minipage}
\end{figure}

The $\Lambda_c^+$~resonance (quark content $udc$) was the first 
and is now the best known charmed baryon. It was reported at Fermilab in 
1976~\cite{Knapp:1976qw} shortly after the discovery of the $J/\psi$~meson. 
The PDG lists 85~different values for branching fractions for the $\Lambda_c^+$ 
including hadronic, semileptonic, and inclusive modes~\cite{Beringer:1900zz}. 
Most of the branching fractions are measured relative to the decay mode $\Lambda_c^+\to pK^-\pi^+$ 
and there are no completely model-independent measurements of the absolute branching 
fraction for $\Lambda_c^+\to pK^-\pi^+$. A minireview on the measurements 
that have been used to extract $\cal{B}$$(\Lambda_c^+\to pK^-\pi^+)$ can be 
found in~\cite{FERMILAB-PUB-10-665-PPD}. The most precise mass determination 
was reported by BABAR in 2005 based on its decay into $\Lambda K_S^0 K^+$ and 
$\Sigma^0 K_S^0 K^+$~\cite{Aubert:2005gt}
\begin{eqnarray}
    M_{\Lambda_c^+}\,=\,2286.4\pm 0.14~{\rm MeV}
\end{eqnarray}
and is the sole result used by the PDG for their average of the measurements. 
Although the spin $J$ has not been measured directly, a $J=\frac{1}{2}$~assignment 
for the lowest-mass $\Lambda_c^+$ appears likely and is consistent with an 
analysis of $p K^-\pi^+$~decays~\cite{Jezabek:1992vi}.

Additional charmed baryons were discovered long after the initial observation 
of the $\Lambda_c^+$~ground state. In 1993, the ARGUS collaboration at the 
$e^+e^-$~storage ring DORIS~II at the Deutsches Elektronen-Synchrotron (DESY) 
discovered a first $\Lambda_c^+$~excitation, the $\Lambda_c(2625)^+$~\cite{Albrecht:1993pt}. 
In 1995, this sighting was confirmed by the CLEO collaboration, who also reported on the 
first observation of the lower-mass $\Lambda_c(2595)^+$~\cite{Edwards:1994ar}. This 
second peak was later confirmed by the E687 collaboration~\cite{Frabetti:1993hg,
Frabetti:1995sb} at Fermilab and the ARGUS collaboration~\cite{Albrecht:1997qa}.
The most precise study of the $\Lambda_c(2595)/\Lambda_c(2625)$~system has been 
performed recently by the CDF collaboration~\cite{Aaltonen:2011sf}, who observed 
a number of excited charmed baryon~states. Figure~\ref{Figure:Lambda_c} (left) 
shows the $m(pK^-\pi^+\pi^+\pi^-)-m(pK^-\pi^+)$ mass difference distribution 
of $\Lambda_c^{\ast +}$~candidates decaying to $\Lambda_c^+\pi^+\pi^-$. The 
signals are described by a convolution of non-relativistic Breit-Wigner and Gaussian 
resolution functions, where the signal for the $\Lambda_c(2595)^+$ also takes into
account the mass dependence of the natural width owing to the proximity of the 
$\Sigma_c\pi$~threshold to the $\Lambda_c(2595)^+$~mass. These finite width 
effects were discussed in~\cite{Blechman:2003mq} and yield a lower $\Lambda_c(2595)^+$~mass 
measurement than observed by previous experiments. In Figure~\ref{Figure:Lambda_c} 
(left), a clear peak for the $\Lambda_c(2595)^+$ is visible at $\Delta m = 305.79\pm
0.14\pm 0.20$~MeV~\cite{Aaltonen:2011sf}. An additional,  even stronger peak 
is observed at $\Delta m = 341.65\pm 0.04\pm 0.12$~MeV which corresponds 
to the $\Lambda_c(2625)^+$. Neither state has been seen in $\Lambda_c^+\pi^0$, 
thus ruling out an interpretation as excited $\Sigma_c^+$~states. Figure~\ref{Figure:Lambda_c} 
(right) shows a comparison of $\Lambda_c^{\ast +}$ mass difference measurements 
by various experiments. The significant shift of the $\Lambda_c(2595)^+$ mass 
towards lower values is attributed to the fact that the improved statistics reported 
by CDF~\cite{Aaltonen:2011sf} allows for a more detailed study of the state's line 
shape including the effects of the $\Lambda_c(2595)^{\ast +}\to\Sigma_c(2455)\pi$~threshold, 
which was taken into account in the fit. The proximity to the $\Sigma_c\pi$~threshold 
and assuming $J^P = \frac{1}{2}^+$ for the $\Sigma_c(2455)$ (lowest-mass
$\Sigma_c$~state) makes a $J^P = \frac{1}{2}^-$~assignment for the 
$\Lambda_c(2595)^+$ likely.

\begin{figure}[t]
\caption{\label{Figure:Spin} (Colour online). Left: The yield of
  $\Lambda_c(2880)^+\to\Sigma_c^{0,++} \pi^\pm$~decays from Belle
  as a function of the helicity angle~\cite{Abe:2006rz}. The dotted,
  dashed, and solid curves correspond to a spin-$\frac{1}{2}$, spin-$\frac{3}{2}$, and
  spin-$\frac{5}{2}$ hypothesis, respectively. Right: The efficiency-corrected
  helicity-angle distribution for $\Sigma_c(2455)^0$~candidates from
  BABAR~\cite{:2008if}. The helicity angle cos\,$\theta_h$ is defined as
  the angle between the momentum vector of the $\Lambda_c^+$ and
  the momentum vector of the recoiling $B$-daughter $\bar{p}$ in the
  rest frame of the $\Sigma_c(2455)^0$. The curves denote the spin-$\frac{1}{2}$ 
  (solid line) and spin-$\frac{3}{2}$ (dashed line) hypothesis. Pictures
  from~\cite{Abe:2006rz} and~\cite{:2008if}, respectively.}
\includegraphics[scale=0.59]{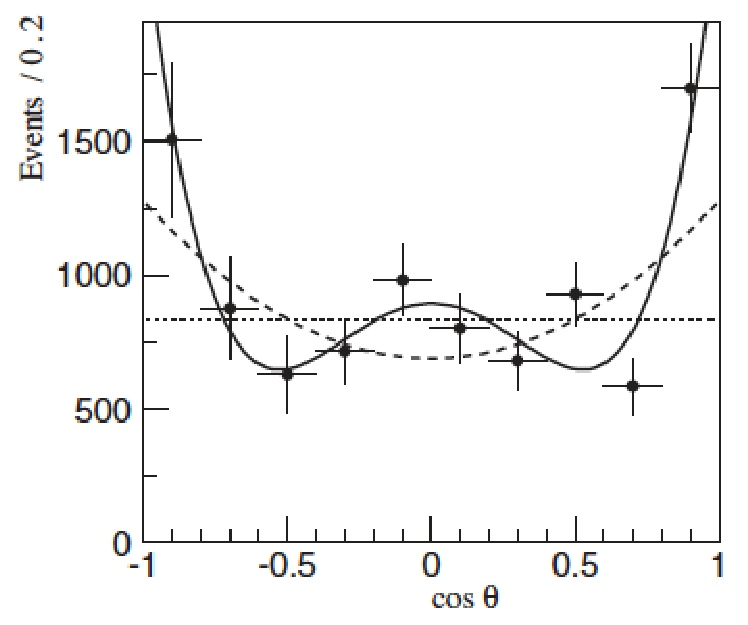}
\includegraphics[scale=0.4]{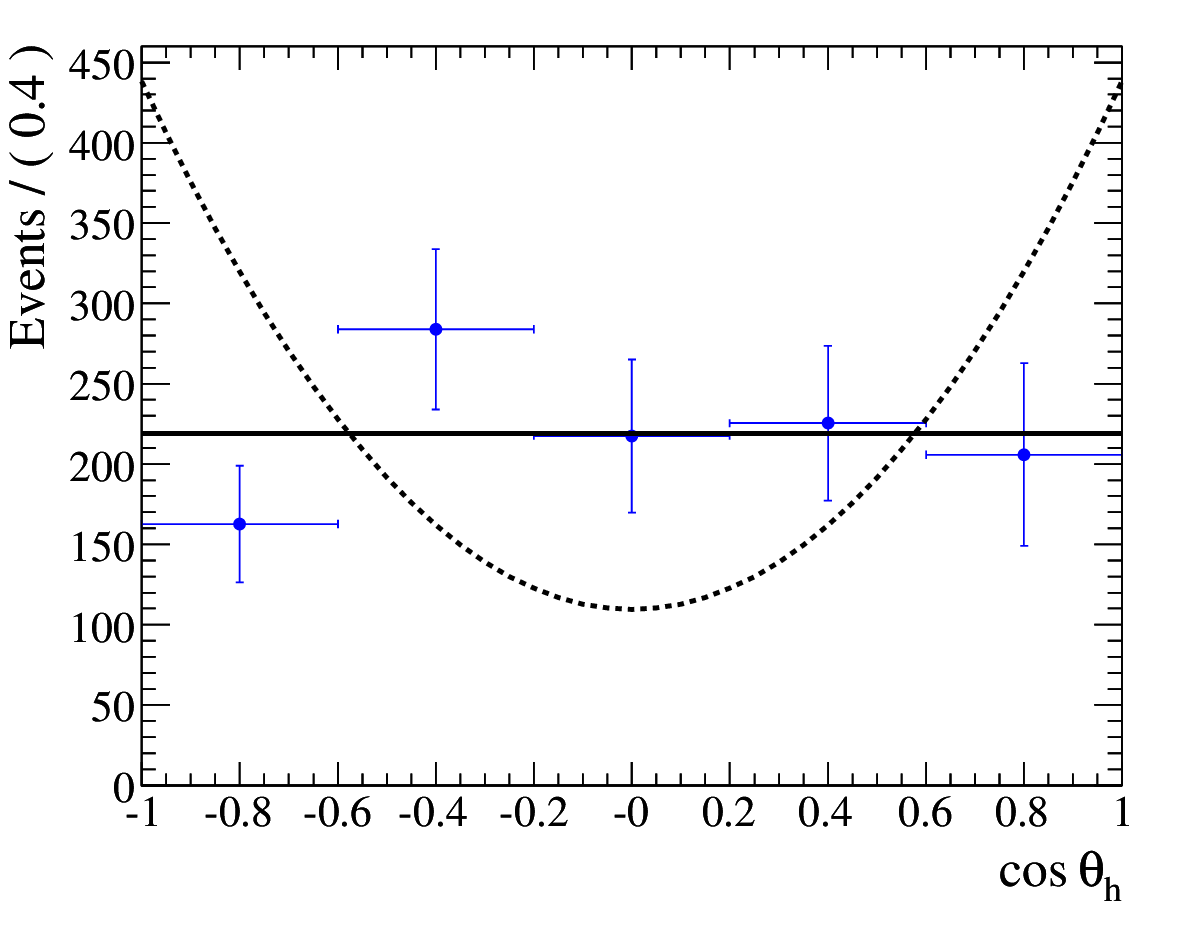}
\end{figure}

In 2007, BABAR reported two narrow signals observed in the invariant 
$D^0 p$~mass distribution which were absent in the corresponding $D^+ p$ 
mass spectra~\cite{Aubert:2006sp}. Consequently, the peaks were identified 
as $\Lambda_c^{\ast +}$ candidates. The lower-mass peak was initially 
discovered in 2001 by CLEO in the $\Lambda_c\pi^+\pi^-$ final state and 
called $\Lambda_c(2880)^+$~\cite{Artuso:2000xy}. Belle 
also studied the $\Lambda_c\pi^+\pi^-$~system and confirmed the 
$\Lambda_c(2880)^+$ as well as the higher-mass BABAR peak, now called
$\Lambda_c(2940)^+$~\cite{Abe:2006rz}. Table~\ref{Table:Lambda_c}
summarizes the mass and width results for both states. The decay angular 
distribution of $\Lambda_c(2880)^+\to\Sigma_c(2455)^{0,++}\pi^\pm$
depends on the spin of the $\Lambda_c(2880)^+$ and is shown in 
Figure~\ref{Figure:Spin}~(left) ~\cite{Abe:2006rz}. Assuming spin~$\frac{1}{2}^+$ 
for the lowest-mass $\Sigma_c$~state, a high-spin assignment for the
$\Lambda_c(2880)^+$ of $J=\frac{5}{2}$ is favoured over the spin-$\frac{1}{2}$ 
or $\frac{3}{2}$~hypothesis. The $\Lambda_c(2880)^+$~yields were 
determined from fits of the $\Lambda_c^+\pi^+\pi^-$ spectrum in
cos\,$\theta$~bins for the $\Sigma_c(2455)$~signal region and sidebands.
We note that the fit shown in~\cite{Abe:2006rz} systematically underestimates 
the data points between the $\Lambda_c(2880)^+$ and $\Lambda_c(2940)^+$
signals. The paper states that this might be due to the presence of an additional 
resonance or due to interference and takes these possibilities into account
as a systematic uncertainty. This issue needs to be further addressed in the 
future in order to reduce the uncertainty in this rare attempt of measuring the $J$~quantum 
number of a heavy baryon. The authors also determined the ratio of 
$\Lambda_c(2880)^+$ partial widths~\cite{Abe:2006rz}
\begin{eqnarray}
    \Gamma(\Sigma_c(2520)\pi)\,/\,\Gamma(\Sigma_c(2455)\pi) = 0.225\pm
    0.062\pm 0.025\,,
\end{eqnarray}
which favours a spin-parity assignment of $\frac{5}{2}^+$ over $\frac{5}{2}^-$
in various models~\cite{Isgur:1991wq,Capstick:1986bm,Cheng:2006dk}.
It is interesting to note that the favoured $J^P=\frac{5}{2}^+$ assignment 
for the $\Lambda_c(2880)^+$ requires a large decay angular momentum of $L=3$ for
all observed decays, into $\Sigma_c(2455)\pi$, $\Sigma_c(2520)\pi$ and $p D^0$. 
However, a lower angular momentum may be sufficient for the $\Lambda_c(2880)^+$ 
decay into $\Sigma_c(2520)\pi$ and it remains a mystery why the $L=1$~decay into 
$\Sigma_c(2520)$ should be suppressed. It is also worth pointing out that the original
CLEO analysis~\cite{Artuso:2000xy} observes $\Lambda_c(2880)^+$ decays via
$\Sigma_c\pi$ and nonresonantly to $\Lambda_c^+\pi^+\pi^-$, but not via 
$\Sigma_c^\ast$. This analysis suffers from significantly lower statistics and thus, 
does not contradict the Belle results.

We finally mention the $\Lambda_c(2765)^+$, which is listed as a ($\ast$)~state 
in the RPP~\cite{Beringer:1900zz}, but its quantum numbers are unknown. The 
state was observed as a broad, statistically significant peak by the CLEO 
collaboration in $\Lambda_c^+\pi^+\pi^-$~\cite{Artuso:2000xy}, but is not even 
fully identified as a $\Lambda_c$~resonance. It should be studied further in
$\Lambda_c^+\pi^+\pi^0$~events, for example by the BABAR or Belle collaboration.

\begin{table}
\begin{center}
\caption{\label{Table:Lambda_c}Measurements of mass and width of the
  $\Lambda_c(2800)^+$ and $\Lambda_c(2800)^+$.}
\begin{tabular}{l|c|c|c}
\br
 & Mass [\,MeV\,] & Width [\,MeV\,] & Decay Modes\\
\mr
$\Lambda_c(2880)^+$~\cite{Artuso:2000xy} & $2882.5\pm 1\pm 2$ & $< 8$ & $\Lambda_c^+\pi^+\pi^-$\\
$\Lambda_c(2880)^+$~\cite{Aubert:2006sp} & $2881.9\pm 0.1\pm 0.5$ & $5.8\pm 1.5\pm 1.1$ & $p D^0$\\
$\Lambda_c(2880)^+$~\cite{Abe:2006rz} & $2881.2\pm 0.2\pm 0.4$ & $5.8\pm 0.7\pm 1.1$ & $\Sigma_c(2455)^{0,++}\pi^\pm$\\
\mr
$\Lambda_c(2940)^+$~\cite{Aubert:2006sp} & $2939.8\pm 1.3\pm 1.0$ & $17.5\pm 5.2\pm 5.9$ & $p D^0$\\
$\Lambda_c(2940)^+$~\cite{Abe:2006rz} & $2938.0\pm 1.3^{+2.0}_{-4.0}$ & $13^{+8\,+27}_{-5\,-7}$ & $\Sigma_c(2455)^{0,++}\pi^\pm$\\
\br
\end{tabular}
\end{center}
\end{table}

\subsubsection{The $\Sigma_c$ States}
\begin{figure}[t]
\caption{\label{Figure:Sigma_c} (Colour online). The difference
  spectra $m(pK^-\pi^+\pi^-)
  - m(pK^-\pi^+)$~(right) and $m(pK^- \pi^+\pi^+) - m(pK^-\pi^+)$
  (left) together with the fits (black solid line)
  from CDF~\cite{Aaltonen:2011sf}. The red dash-dotted line
  corresponds to the sum of all three background contributions
  including reflections from $\Lambda_c^{\ast +}$~decays. The two
  peaks corresponding to $\Sigma_c^{0\,(++)} - \Lambda_c^+$ and
  $\Sigma_c^{\ast\,0\,(++)} -\Lambda_c^+$~candidates are visible. 
  Pictures from~\cite{Aaltonen:2011sf}.}
\includegraphics[scale=0.375]{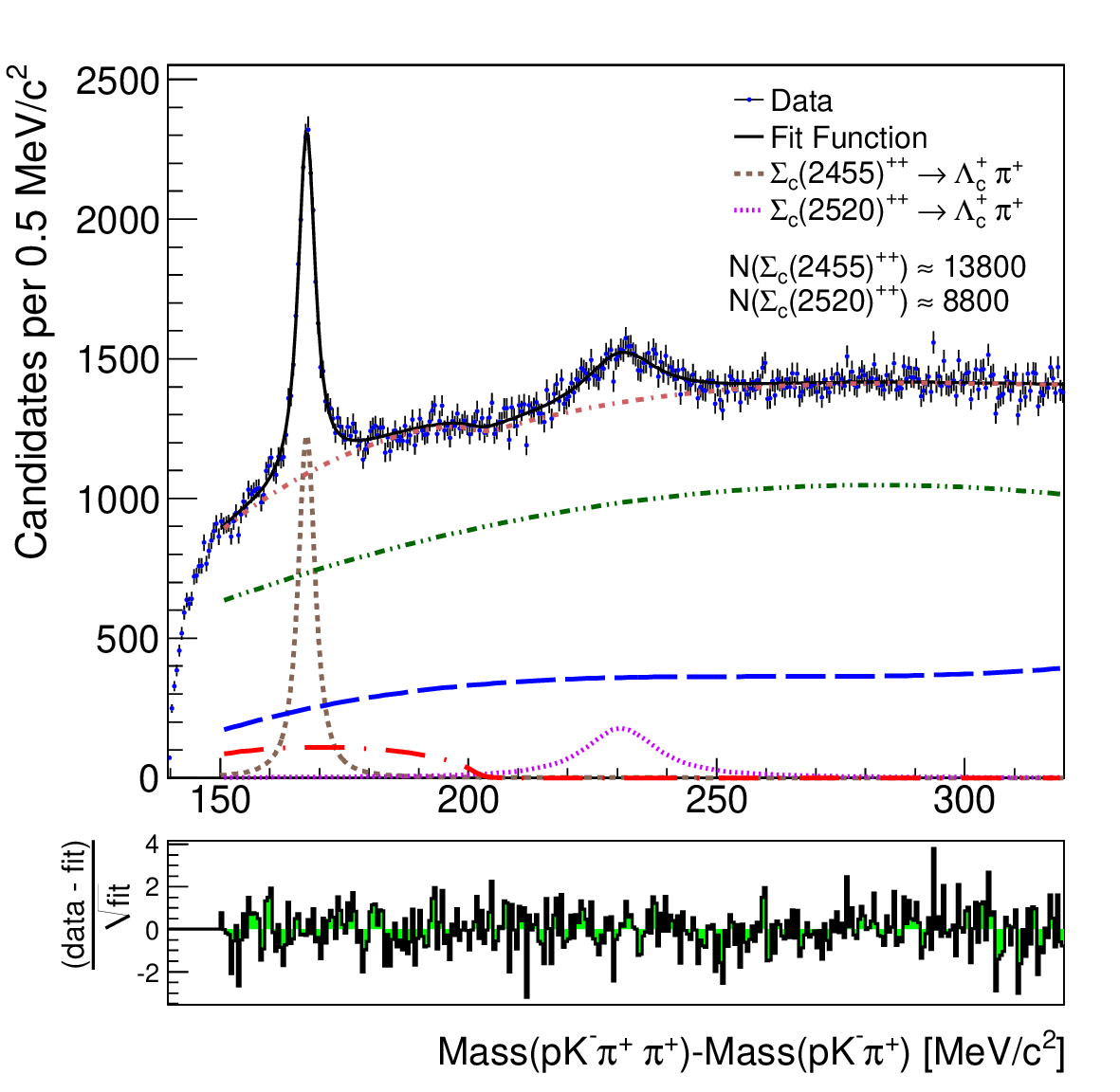}
\includegraphics[scale=0.375]{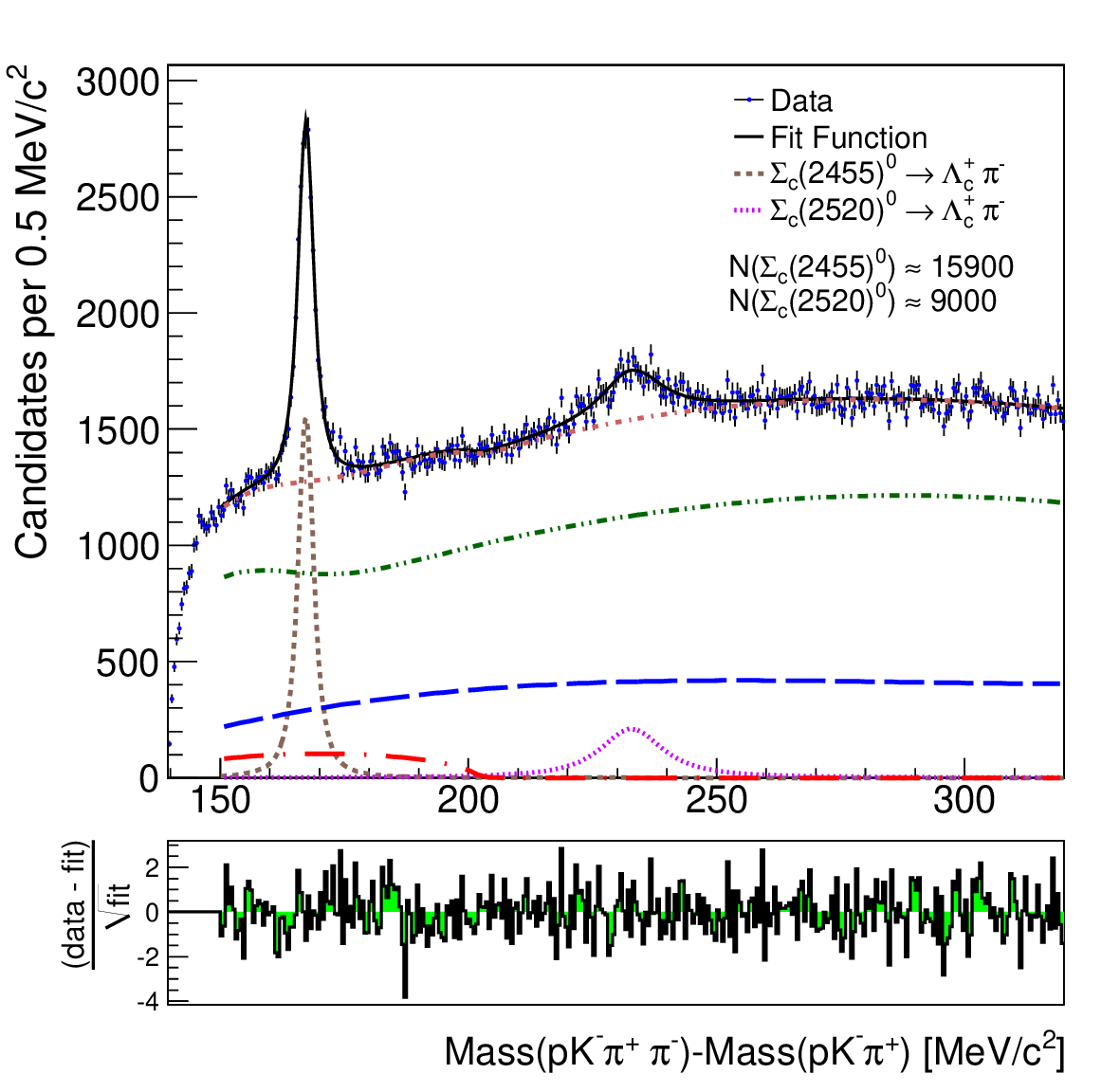}
\end{figure}

$\Sigma_c$ states are expected to occur in triplets: $\Sigma_c^{++}$, $\Sigma_c^+$ and $\Sigma_c^0$, with quark content $uuc$, $udc$ and $ddc$, respectively. The two lowest-lying $\Sigma_c$~resonances, $\Sigma_c(2455)$ and
$\Sigma_c(2520)$, are well established and both have been confirmed in
several experiments~\cite{Beringer:1900zz}. Their masses allow strong decays only into
$\Lambda_c^+\pi$. In 2001, the CLEO collaboration published mass and
width measurements for the $\Sigma_c^+$ and $\Sigma_c^{\ast +}$~states 
based on their decays into $\Lambda_c^+\pi^0$~\cite{Ammar:2000uh}. 
The CLEO results for the $\Sigma_c^{\ast ++}$ and $\Sigma_c^{\ast 0}$~states
decaying into $\Lambda_c^+\pi^\pm$~\cite{Athar:2004ni} have been much
improved upon statistically by recent Tevatron results~\cite{Aaltonen:2011sf}. 
The CDF~collaboration used a data sample corresponding to 5.2~fb$^{-1}$ 
of integrated luminosity from $p\bar{p}$ collisions at $\sqrt{s}=1.96$~TeV 
and selected $\Lambda_c^+\to pK^-\pi^+$~decays. Figure~\ref{Figure:Sigma_c} 
(left) shows the $m(pK^-\pi^+\pi^-)-m(pK^-\pi^+)$ and Figure~\ref{Figure:Sigma_c} 
(right) the $m(pK^-\pi^+\pi^+)-m(pK^-\pi^+)$~mass difference spectrum. 
The signals are described by a convolution of non-relativistic Breit-Wigner 
and Gaussian resolution functions. Two peaks corresponding to $\Sigma_c^{\ast ++}$ 
and $\Sigma_c^{\ast 0}$ are clearly visible in both distributions.

BABAR observed the resonant decay $B^-\to\Sigma_c(2455)^0\,\bar{p}$ but 
reported a negative result for $B^-\to\Sigma_c(2520)^0\,\bar{p}$~\cite{:2008if}. 
The absence of the $\Sigma_c(2520)^0$ in $B$~decays appears to be in
conflict with a $2.9\sigma$~signal ($13^{+6}_{-5}$~events) based on
$152\times 10^6$ $B\bar{B}$~events reported by Belle~\cite{Abe:2004sr}. 
Figure~\ref{Figure:Spin} (right) shows the helicity-angle distribution from 
BABAR for $\Sigma_c(2455)^0$ candidates~\cite{:2008if}. The data points 
correspond to efficiency-corrected $B^-\to\Sigma_c^0(2455)\,\bar{p}$~events 
and the curves denote the spin-$\frac{1}{2}$ (solid line) and spin-$\frac{3}{2}$ (dashed line) 
hypotheses. From this angular distribution, the spin of the $\Sigma_c(2455)^0$
was determined to be $\frac{1}{2}$ as predicted by quark model calculations.
Neither $J$ nor $P$ has been measured for the $\Sigma_c(2520)$ but 
$J^P = \frac{3}{2}^+$ appears natural rendering this resonance the charm
counterpart of the $\Sigma(1385)$~\cite{Beringer:1900zz}.

A third triplet of $\Sigma_c$~resonances, $\Sigma_c(2800)$, was 
seen in the $\Lambda_c^+\pi$ mass spectra and first reported by Belle
in 2005~\cite{Mizuk:2004yu}. The results for the mass difference and width
measurements are summarized in Table~\ref{Table:Sigma_c2800}. 
BABAR tentatively confirmed the neutral state in the resonant decay 
$B^-\to\Sigma_c(2800)^0\,\bar{p}\to\Lambda_c^+\pi^-\bar{p}$ 
and determined its mass and width to be $(2846\pm 8\pm 10)$~MeV 
and ($86^{+33}_{-22}$)~MeV, respectively~\cite{:2008if}. The widths
are consistent in both experiments, whereas the masses are $3\sigma$~apart. 
This discrepancy may indicate that the two observed signals are not
the same state. Based on the mass of the $\Sigma_c(2800)^+$, Belle
suggested a spin-$\frac{3}{2}$ assignment, while BABAR observes weak evidence
for spin $J=\frac{1}{2}$. The authors conclude that $B$~decays to higher-spin
baryons might be suppressed~\cite{:2008if}.

\begin{table}
\begin{center}
\caption{\label{Table:Sigma_c2800}Mass differences and widths of the
  $\Sigma_c(2800)$ reported by Belle~\cite{Mizuk:2004yu}.}
\begin{tabular}{c|c|c}
\br
 & Mass [\,MeV\,] & Width [\,MeV\,]\\
\mr
$m(\Sigma_c(2800)^{++}) - m(\Lambda_c^+)$ & $514.5^{+3.4+2.8}_{-3.1-4.9}$ & $75^{+18+12}_{-13-11}$\\
$m(\Sigma_c(2800)^{+}) - m(\Lambda_c^+)$ & $505.4^{+5.8+12.4}_{-4.6-2.0}$ & $62^{+37+52}_{-23-38}$\\
$m(\Sigma_c(2800)^{0}) - m(\Lambda_c^+)$ & $515.4^{+3.2+2.1}_{-3.1-6.0}$ & $61^{+18+22}_{-13-13}$\\
\br
\end{tabular}
\end{center}
\end{table}

\subsubsection{The $\Xi_c$ States}
The two states, $\Xi_c^+$ and $\Xi_c^0$, form an isospin doublet with
quark content $usc$ and $dsc$, respectively. The first
evidence for the $\Xi_c^+$~ground state was reported in 1983 by experiments 
studying $\Sigma^-$~nucleon collisions at the CERN SPS hyperon 
beam~\cite{Biagi:1983en}. The $\Lambda K^-\pi^+\pi^+$~mass
distribution showed an excess of 82 events and the signal had a 
$6\sigma$~statistical significance. CLEO discovered the isospin partner
a few years later in its decay to $\Xi^-\pi^+$ in the reaction
$e^+e^-\to\Xi^-\pi^+ + X$~\cite{Avery:1988uh}. According to the PDG,
none of the $I(J^P)$~quantum numbers has been measured, but the
spin-parity assignment $\frac{1}{2}^+$ appears likely from quark-model
calculations.

In 1999, CLEO published evidence for a further $\Xi_c$~doublet, $\Xi_c^{\,\prime +}$ 
and $\Xi_c^{\,\prime 0}$, which was observed to decay via low-energetic photon 
emission to $\Xi_c^{+ (0)}$~\cite{Jessop:1998wt}. The mass differences were 
determined to be
\begin{eqnarray}
 m(\Xi_c^{\,\prime +}) - m(\Xi_c^+) & = 107.8\pm 1.7\pm 2.5~{\rm MeV}\\
 m(\Xi_c^{\,\prime 0}) - m(\Xi_c^0) & = 107.0\pm 1.4\pm 2.5~{\rm MeV}\,,
\end{eqnarray}
which kinematically do not permit the transitions $\Xi^{\,\prime}_c\to\Xi_c\pi$; radiative 
decays are the only allowed decay modes. The two $\Xi_c^{\,\prime}$ states have been 
interpreted as the partners of $\Xi_c$, where the light-quark pairs $sd$ and $su$ are 
symmetric with respect to interchange of the light quarks, i.e. the light-quark pair 
forms mostly a spin-triplet, $S=1$. The masses and widths of the resonances quoted 
by the PDG are summarized in Table~\ref{Table:Xi_c_1}.

\begin{table}
\begin{center}
\caption{\label{Table:Xi_c_1}Masses and widths of the
  $\Xi_c$~states with $J^P = \frac{1}{2}^+$ quoted by the PDG~\cite{Beringer:1900zz}.}
\begin{tabular}{c|c|c||c|c|c}
\br
 & Mass [\,MeV\,] & Mean Life [\,fs\,] & & Mass [\,MeV\,]
 & Decay Mode\\
\mr
$\Xi_c^0$ & $2471.09^{+0.35}_{-1.00}$ & $112^{+13}_{-10}$ &
$\Xi_c^{\,\prime 0}$ & $2577.9\pm 2.9$ & $\Xi_c^0\,\gamma$\\
$\Xi_c^+$ & $2467.6^{+0.4}_{-1.0}$ & $442\pm 26$ & 
$\Xi_c^{\,\prime +}$ & $2575.6\pm 3.1$ & $\Xi_c^+\,\gamma$\\
\br
\end{tabular}
\end{center}
\end{table}

Several additional excited $\Xi_c$~states were discovered by CLEO. A narrow 
state decaying into $\Xi_c^+\,\pi^-$, $\Xi_c(2645)^0$, was observed with a mass 
difference $m(\Xi_c^+\,\pi^-)-m(\Xi_c^+)$ of $178.2\pm 0.5\pm 1.0$~MeV, 
and a width of $< 5.5$~MeV~\cite{Avery:1995ps}. The quantum numbers 
of this state have not been measured but a $J^P = \frac{3}{2}^+$~assignment seems
natural. The isospin partner, $\Xi_c(2645)^+$, was also discovered by CLEO 
with a mass difference of $m(\Xi_c^0\,\pi^+)-m(\Xi_c^0)$ of $174.3\pm 0.5\pm 
1.0$~MeV, and a width of $< 3.1$~MeV~\cite{Gibbons:1996yv}.
A confirmation of these states and the most precise mass measurements were 
reported by Belle~\cite{:2008wz} and are almost identical to the numbers quoted 
by the PDG.
In the same publication~\cite{:2008wz}, Belle reported on a further $\Xi_c$~state, 
$\Xi_c(2815)$, decaying into $\Xi_c(2645)^{+ (0)}\,\pi^\mp$~\cite{:2008wz}. This 
state was originally discovered by CLEO~\cite{Alexander:1999ud} and both experiments 
observed it in the decay into $\Xi_c\,\pi\pi$ via an intermediate $\Xi_c(2645)$. The
quantum numbers were not measured, but the states were interpreted as $J^P = \frac{3}{2}^-$~resonances, 
the charmed-strange analogs of the $\Lambda_c(2625)^+$. Finally about a decade ago, 
CLEO reported on a doublet of states, $\Xi_c(2790)$, decaying to $\Xi_c^{\,\prime}\,\pi$~\cite{Csorna:2000hw}.
The Belle collaboration confirmed the state~\cite{:2008wz}, but did not determine its parameters.
Based on the properties (mass, width, and decay mode), the simplest interpretation of these 
states is as a pair of $J^P = \frac{1}{2}^-$~resonances, the charmed-strange partners of the
$\Lambda_c(2595)^+$. Mass and width measurements of the three states,
$\Xi_c(2645)$, $\Xi_c(2790)$, and $\Xi_c(2815)$ are summarized in Table~\ref{Table:Xi_c_2}.

\begin{table}
\begin{center}
\caption{\label{Table:Xi_c_2}Masses and widths of the $\Xi_c(2645)$,
  $\Xi_c(2790)$, and $\Xi_c(2815)$ resonances.}
\begin{tabular}{c|c|c|c}
\br
 & $\Xi_c(2645)^{+,\,0} $ & $\Xi_c(2790) ^{+,\,0}$ & $\Xi_c(2815) ^{+,\,0}$\\
\mr
Mass [\,MeV\,] & $2645.6\pm 0.2^{+0.6}_{-0.8}$~\cite{:2008wz} &
$2789.1\pm 3.2$~\cite{Csorna:2000hw} & $2817.0\pm 1.2^{+0.7}_{-0.8}$~\cite{:2008wz}\\
& $2645.7\pm 0.2^{+0.6}_{-0.7}$~\cite{:2008wz} & $2791.8\pm
3.3$~\cite{Csorna:2000hw} & $2820.4\pm 1.4^{+0.9}_{-1.0}$~\cite{:2008wz}\\
\mr
Width [\,MeV\,] & $< 3.1$~\cite{Gibbons:1996yv} & $< 15$~\cite{Csorna:2000hw} & 
$< 3.5$~\cite{Alexander:1999ud}\\
& $< 5.5$~\cite{Avery:1995ps} & $< 12$~\cite{Csorna:2000hw} & $< 6.5$~\cite{Alexander:1999ud}\\ 
\br
\end{tabular}
\end{center}
\end{table}

\begin{figure}
\caption{\label{Figure:Xi_c}(Colour online) Signals for charmed-strange $\Xi_c$~states 
  in the invariant $\Lambda_c^+\bar{K}\pi$~mass: (a) $\Lambda_c^+K_S^0 \pi^-$~mass 
  and (b) $\Lambda_c^+ K^-\pi^+$~mass published by the Belle collaboration~\cite{Chistov:2006zj}; 
  (c) and (d) show the corresponding $\Lambda_c^+ K^-\pi^+$~mass from BABAR~\cite{Aubert:2007dt} 
  with the $\Lambda_c^+\pi^+$ mass consistent with $\Sigma_c(2455)$ and the $\Sigma_c(2520)$, 
  respectively.}
\vspace{2mm}
\includegraphics[scale=0.7]{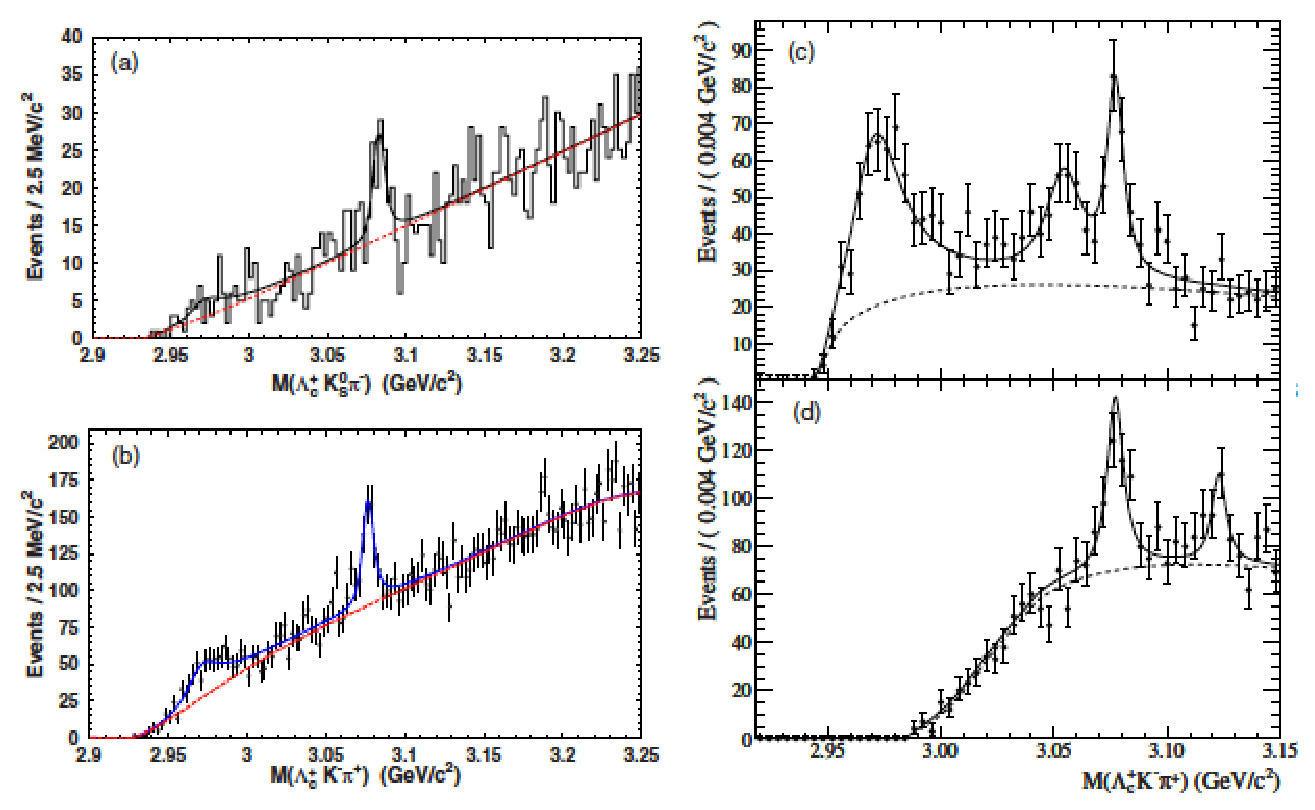}
\end{figure}

In the last few years, even more $\Xi_c$~states have been observed at and above 3~GeV.
Belle observed two new states, $\Xi_c(2980)$ and $\Xi_c(3080)$, decaying into
$\Lambda_c^+ K^-\pi^+$ and $\Lambda_c^+ K_S^0\pi^-$~\cite{Chistov:2006zj}.
The $\Xi_c(2980)$~resonance was later confirmed by Belle in its decay into 
$\Xi_c(2645)\pi$~\cite{:2008wz}, though with a somewhat lower mass. It is 
important to note that, in contrast to all previous $\Xi_c$~states, the masses of 
these states are above the threshold for decays into $\Lambda_c^+ K\pi$. In these 
processes, the charm and strangeness of the initial state are carried away by two 
different final state particles, a charmed baryon and a strange meson. The SELEX 
collaboration has reported a doubly-charmed baryon decaying into $\Lambda_c^+ 
K^-\pi^+$ inspiring further investigation of this promising decay mode~\cite{Mattson:2002vu}. 
However, the claim has still not been confirmed by other experiments. A similar flavour 
separation occurs in the decay mode $\Lambda D^+$, which is allowed above 
3~GeV, but no observation of such a decay has yet been reported.

Figure~\ref{Figure:Xi_c} shows various $\Lambda_c^+ K\pi$~mass distributions.
Results from Belle are given in~(a) for the invariant $\Lambda_c^+ K_S^0\pi^-$~mass
and in (b) for the invariant $\Lambda_c^+ K^-\pi^+$~mass with the overlaid fitting 
curve~\cite{Chistov:2006zj}. The dashed line is the background component of the 
fitting function, and the solid curve is the sum of the background and signal 
(Breit-Wigner function convolved with a Gaussian detector resolution function). 
In~(a), a clear signal around 3080~MeV is observed corresponding to the 
$\Xi_c(3080)^0$. In addition, a broad enhancement near the threshold of the mass 
distribution is visible, which reveals weak evidence for the $\Xi_c(2980)$. The 
distribution in~(b) provides more information on the origin of the states and 
confirms both signals. The narrow peak at 3080~MeV and the broad 
enhancement at threshold are clearly visible. 

The latest confirmation of both states and weak evidence for additional 
$\Xi_c$~states above 3~GeV were reported by BABAR~\cite{Aubert:2007dt}. 
Figure~\ref{Figure:Xi_c}~(c) and~(d) show invariant $\Lambda_c^+ K^-\pi^+$ 
mass distributions. In (c), the corresponding invariant $\Lambda_c^+\,\pi^+$ 
mass is consistent with the $\Sigma_c(2455)^{++}$~mass (within 3.0 natural 
widths) and in (d), the $\Lambda_c^+\,\pi^+$~mass is consistent with the 
$\Sigma_c(2520)^{++}$~mass (within 2.0 natural widths). The very broad and the 
narrow signal for the two states observed by Belle, $\Xi_c(2980)$ and $\Xi_c(3080)^0$, 
are clearly visible. Mass and width measurements of these states are summarized 
in Table~\ref{Table:Xi_c_3}. Moreover, two further signals are observed in the BABAR 
mass distributions. One signal sits at $(3054.2\pm 1.2\pm 0.5)$~MeV with a 
significance of $6.4\sigma$ in the $\Sigma_c(2455)^{++} K^-\to\Lambda_c^+ K^-
\pi^+$~mass spectrum (Figure~\ref{Figure:Xi_c}~(c)). The other signal sits at $(3122.9
\pm 1.3\pm 0.3)$~MeV with a significance of $3.8\sigma$ in the 
$\Sigma_c(2520)^{++} K^-\to\Lambda_c^+ K^-\pi^+$~mass spectrum 
(Figure~\ref{Figure:Xi_c}~(d)). Neither state, $\Xi_c(3055)^+$ and $\Xi_c(3123)^+$, 
has been confirmed by any other experiment. The PDG has omitted these
two states from its summary table.

\begin{table}
\begin{center}
\caption{\label{Table:Xi_c_3}Mass and width measurements of the 
  charmed-strange resonances $\Xi_c(2980)$ and $\Xi_c(3080)$ in
  MeV from Belle and BABAR.}
\begin{tabular}{c|l|c|l}
\br
 & Mass & Width & Experiment\\
\mr
$\Xi_c(2980)^+$ & $2969.3\pm 2.2\pm 1.7$ & $27\pm 8\pm 2$ & BABAR~\cite{Aubert:2007dt}\\
                          & $2967.7\pm 2.3^{+1.1}_{-1.2}$ & $18\pm 6\pm 3$ & Belle~\cite{:2008wz}\\
                          & $2978.5\pm 2.1\pm 2.0$ & $43.5\pm 7.5\pm 7.0$ & Belle~\cite{Chistov:2006zj}\\
$\Xi_c(2980)^0$ & $2972.9\pm 4.4\pm 1.6$ & $31\pm 7\pm 8$ & BABAR~\cite{Aubert:2007dt}\\
                          & $2965.7\pm 2.4^{+1.1}_{-1.2}$ & $15\pm 6\pm 3$ & Belle~\cite{:2008wz}\\
\mr
$\Xi_c(3080)^+$ & $3077.0\pm 0.4\pm 0.2$ & $5.5\pm 1.3\pm 0.6$ & BABAR~\cite{Aubert:2007dt}\\
                          & $3076.7\pm 0.9\pm 0.5$ & $6.2\pm 1.2\pm 0.8$ & Belle~\cite{Chistov:2006zj}\\
$\Xi_c(3080)^0$ & $3079.3\pm 1.1\pm 0.2$ & $5.9\pm 2.3\pm 1.5$ & BABAR~\cite{Aubert:2007dt}\\
                          & $3082.8\pm 1.8\pm 1.5$ & $5.2\pm 3.1\pm 1.8$ & Belle~\cite{Chistov:2006zj}\\
\br
\end{tabular}
\end{center}
\end{table}

\subsubsection{The $\Omega_c$ States}
The $\Omega_c^0$~baryon is the heaviest known singly-charmed hadron
that decays weakly. The quark content is $ssc$ where the $ss$~pair is in 
a symmetric state. In addition to the ground-state $\Omega_c$ with 
$J^P = \frac{1}{2}^+$, a further excited state is fairly well established. The 
observation of the ground state was first reported in 1985 by the WA62 collaboration based
on a cluster of three events in one out of four modes studied in $\Sigma^-$-nucleus 
interactions using the SPS charged hyperon beam at CERN~\cite{Biagi:1984mu}.
Since then, the resonance has been seen and some of its properties measured by a large 
variety of experiments, e.g. at CERN by the WA89 collaboration in its decay into 
$\Xi^- K^-\pi^+\pi^+$ and $\Omega^-\pi^+\pi^-\pi^+$~\cite{Adamovich:1995pf}, 
at Fermilab by the E687 collaboration~\cite{Frabetti:1994dp}, and CLEO~\cite{CroninHennessy:2000bz}. 
The most precise and consistent study was reported in 2009 by Belle via the decay into 
$\Omega^-\pi^-$~\cite{Solovieva:2008fw}. The average mass value given by the PDG 
is~\cite{Beringer:1900zz}
\begin{eqnarray}
M_{\Omega_c^0} \,=\, 2695.2^{+1.8}_{-1.6}~{\rm MeV}\,.
\end{eqnarray}

\begin{figure}
\caption{\label{Figure:Belle-Omega_c}Results on the $\Omega_c^0$ 
  and $\Omega_c^{\ast 0}$~charmed baryons from Belle~\cite{Solovieva:2008fw}. 
  Left: Invariant $M(\Omega^-\pi^+) - M(\Omega^-)+m_{\Omega^-}$ mass, where  
  the latter refers to the 2008 PDG mass~\cite{Amsler:2008zzb} of the $\Omega^-$~state.
  Right: $M(\Omega_c^0\gamma) - M(\Omega_c^0)$ mass difference. The solid line
  represents a fit using a Crystal Ball function for the signal and a second-order 
  polynomial for the background. Pictures from~\cite{Solovieva:2008fw}.}\vspace{-4mm}
\includegraphics[width=0.49\textwidth,height=0.4\textwidth]{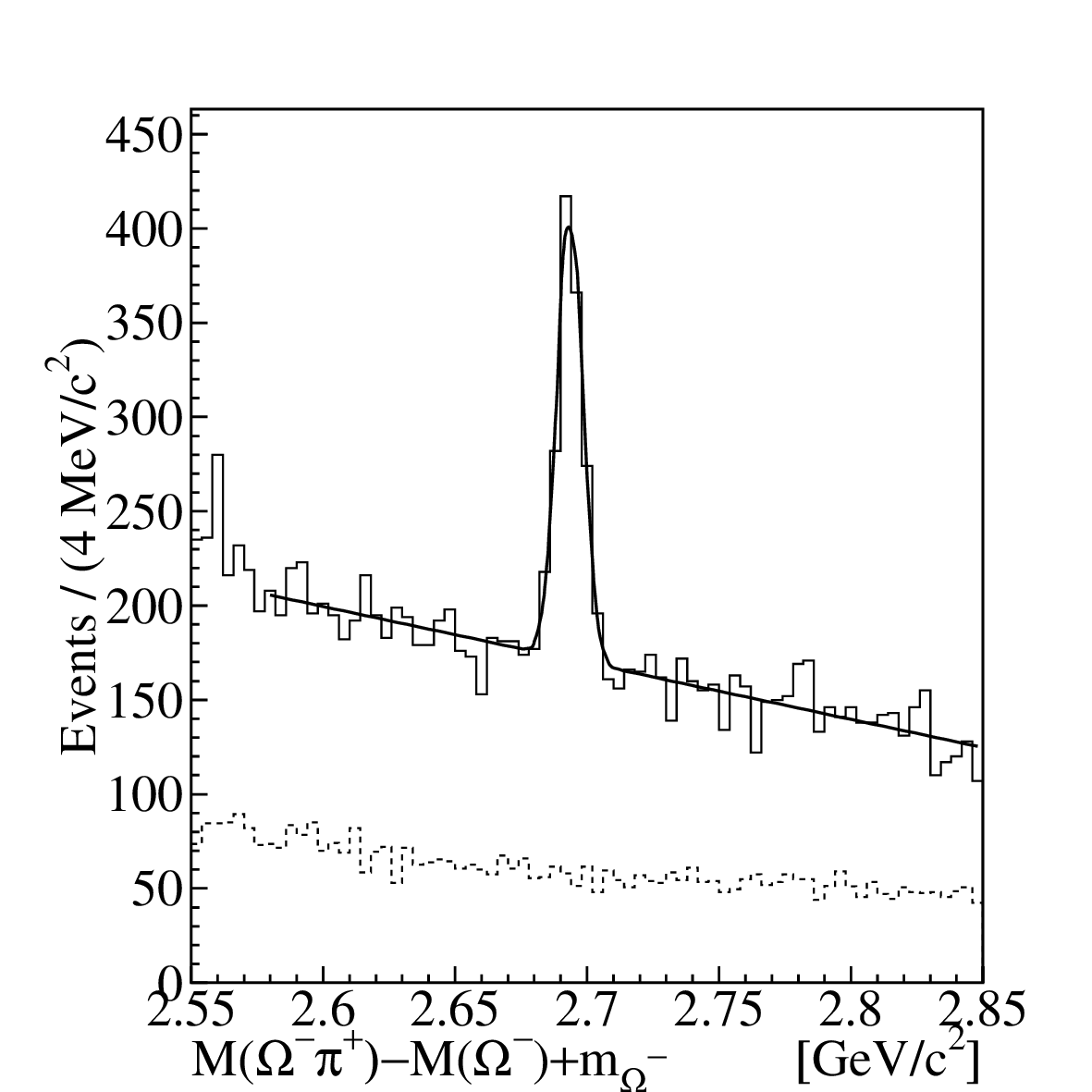}
\includegraphics[width=0.49\textwidth,height=0.4\textwidth]{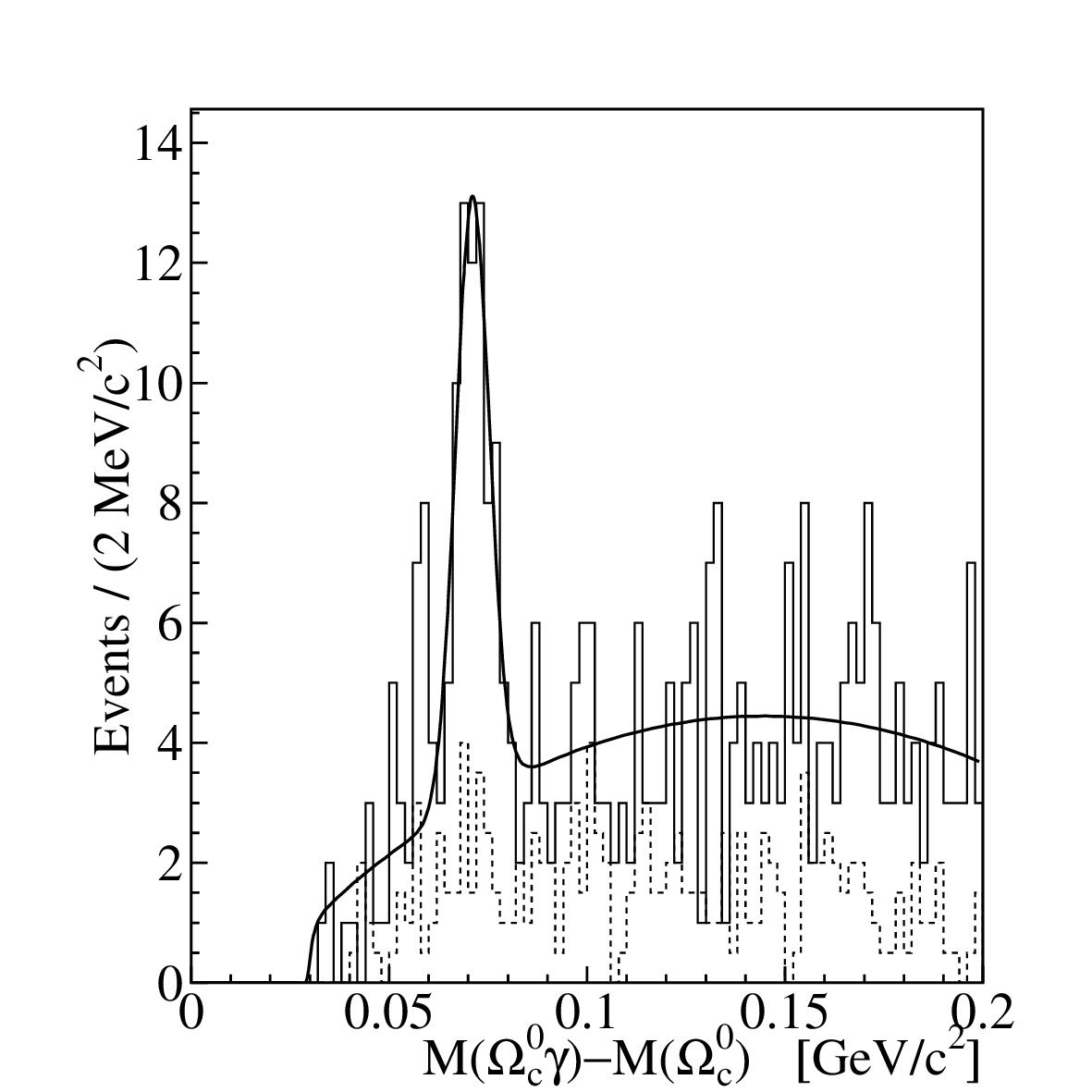}
\end{figure}

An excited $\Omega_c$~baryon was first proposed by BABAR~\cite{Aubert:2006je} 
in the radiative decay $\Omega_c^{\ast 0}\to\Omega_c\gamma$ and in 2009 
confirmed by Belle in the same decay with a significance of $6.4\sigma$~\cite{Solovieva:2008fw}.
The mass difference relative to the ground state, $M_{\Omega_c^{\ast 0}} - 
M_{\Omega_c^0}$, was measured to be $70.7\pm 0.9\,({\rm stat.})^{+0.1}_{-0.9}\,
({\rm syst.})$~MeV, which is in excellent agreement with the BABAR 
result. Figure~\ref{Figure:Belle-Omega_c} presents the results from Belle. The
left side shows the $\Omega_c^0$~signal in the invariant $\Omega^-\pi^+$
mass and the right side shows the signal associated with the $\Omega_c^{\ast 0}$
in the $M(\Omega_c^0\gamma) - M(\Omega_c^0)$ mass difference. The average
mass of the $\Omega_c^{\ast 0}$ quoted by the PDG in the most recent edition
of the RPP is~\cite{Beringer:1900zz}
\begin{eqnarray}
M_{\Omega_c^{\ast 0}} \,=\, 2765.9\pm 2.0~{\rm MeV}\,.
\end{eqnarray}
The $J^P$~quantum numbers have not been measured, but the natural
assignment is $J^P = \frac{3}{2}^+$, which makes the $\Omega_c^{\ast 0}$ a
partner of the $\Sigma_c(2520)$ and $\Xi_c(2645)$~baryons.

\subsubsection{The Status of Doubly-Charmed Baryons}
The lightest doubly-charmed baryons can exist with either quark content 
$ccu$, $\Xi_{cc}^{++}$, or $ccd$, $\Xi_{cc}^+$. Models generally 
predict a mass range of 3.516-3.66~GeV for the $J^P = \frac{1}{2}^+$
ground state and 3.636-3.81~GeV for the $J^P = \frac{3}{2}^+$~excited 
state. In 2002, the SELEX collaboration at Fermilab reported weak evidence 
for a $\Xi_{cc}^+$ candidate decaying into $\Lambda_c^+ K^-\pi^+$ with a
statistical significance of $6.3\sigma$ using a 600~MeV beam of 
$\Sigma^-$~hyperons~\cite{Mattson:2002vu}. The state was later confirmed 
by SELEX in the decay into $p D^+ K^-$~\cite{Ocherashvili:2004hi} with 
an averaged mass from both results of $3518.7\pm 1.7$~MeV. The
observed lifetime of less than 33~fs is significantly shorter than that of the
$\Lambda_c^+$, in contradiction with model calculations based on
heavy quark effective theory (see~\cite{Ocherashvili:2004hi} and further 
references therein).

The observation of doubly-charmed baryons is not undisputed. The two
$B$~factories, BABAR and Belle, have found no evidence for a $\Xi_{cc}^+$~state in 
the $\Lambda_c^+ K^-\pi^+$ decay mode~\cite{Aubert:2006qw,Chistov:2006zj}.
Other experiments with large samples of charmed baryons, e.g. FOCUS with 
about $12\times$ more $\Lambda_c^+$~events in photoproduction or E791 at 
Fermilab with about $5\times$ as many $\Lambda_c^+$~events as SELEX, have 
not confirmed the $\Xi_{cc}^+$~signal. It is interesting to note though that SELEX 
reported the observation only in baryon-induced reactions (using $\Sigma^-$ 
and proton beams), but not in $\pi^-$-induced reactions, and none of the other 
experiments used a baryon beam. Moreover, only SELEX covers the forward 
hemisphere with baryon beams where the doubly-charmed events are observed. 

Some evidence for the doubly-charged partner of the $\Xi_{cc}^+$ as well as for 
a higher-mass state was presented by members of the SELEX collaboration at 
conferences, but BABAR again could not confirm these claims. No evidence for a 
$\Xi^{++}_{cc}$~state was found in the $\Lambda_c^+ K^-\pi^+\pi^+$ and
$\Xi_c^0\pi^+\pi^+$ modes~\cite{Aubert:2006qw} studied by SELEX. Given
the weak experimental evidence for doubly-charm baryons, we will not discuss
these observations further.

\subsection{Recent Results in the Spectroscopy of Beautiful Baryons}
Beautiful or bottom baryons are composed of two light quarks and one heavy
$b$ quark, $qqb~(q = u,d,s)$.  Prior to 2007,
only one bottom baryon, the $\Lambda_b$ with quark content $udb$, 
had been observed directly. It was first discovered at the CERN ISR~\cite{323127,326169} 
and later studied in more details by the LEP collaborations, ALEPH, 
DELPHI, and OPAL. Only indirect observations 
were reported for the other known bottom baryon at that time, 
the $\Xi_b$ baryon, by ALEPH~\cite{CERN-PPE-96-081} and
DELPHI~\cite{CERN-PPE-95-029} via an excess of same-sign $\Xi^\mp
l^\mp$ events as compared to those with an opposite-sign $\Xi^\mp l^\pm$ pair. Many new resonances
have been discovered in recent years, but none of the quantum numbers, $I$, $J$, 
or $P$ has been experimentally determined. All assignments are based on quark 
model predictions.

\subsubsection{The $\Lambda_b$ States}
\begin{figure}
\caption{\label{Figure:LHCb-Lambda_b}(Colour online) Mass spectra from
  LHCb~\cite{Aaij:2012da}. Left: Invariant mass spectrum of $\Lambda_c^+\pi^-$ 
  combinations showing a prominent $\Lambda_b^0$ signal. Right: Discovery 
  of two excited $\Lambda_b^0$~states, $\Lambda_b^{\ast0}$, in the invariant 
  $\Lambda_b^0\pi^+\pi^-$~mass. The fit yields $17.6\pm 4.8$~events for the 
  lower-mass peak, $\Lambda_b(5912)^{\ast0}$, and $52.5\pm 8.1$ events 
  for the higher-mass peak, $\Lambda_b(5920)^{\ast0}$. 
  Pictures from~\cite{Aaij:2012da}.}\vspace{0.5mm}
\includegraphics[scale=0.34]{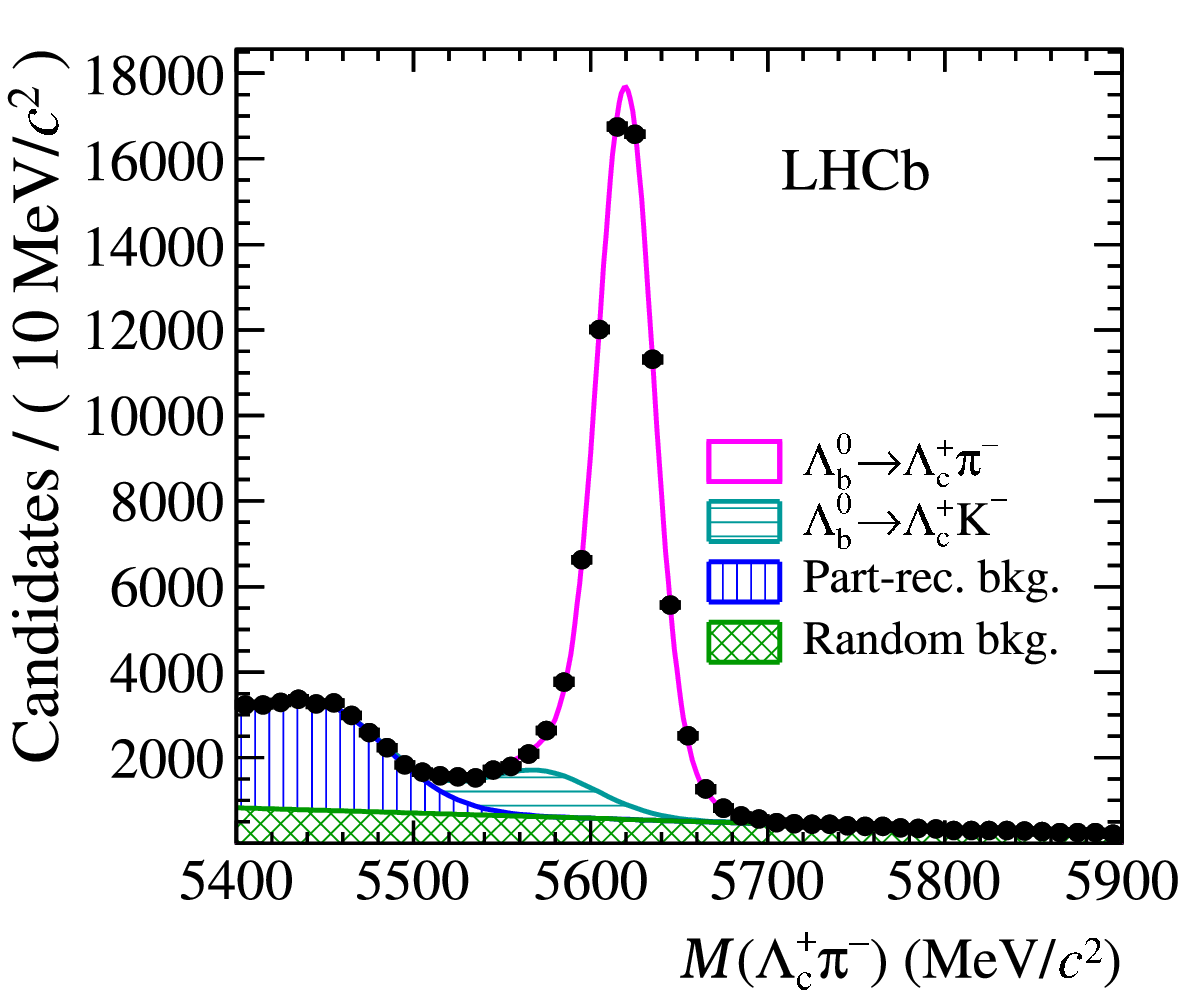}
\includegraphics[scale=0.435]{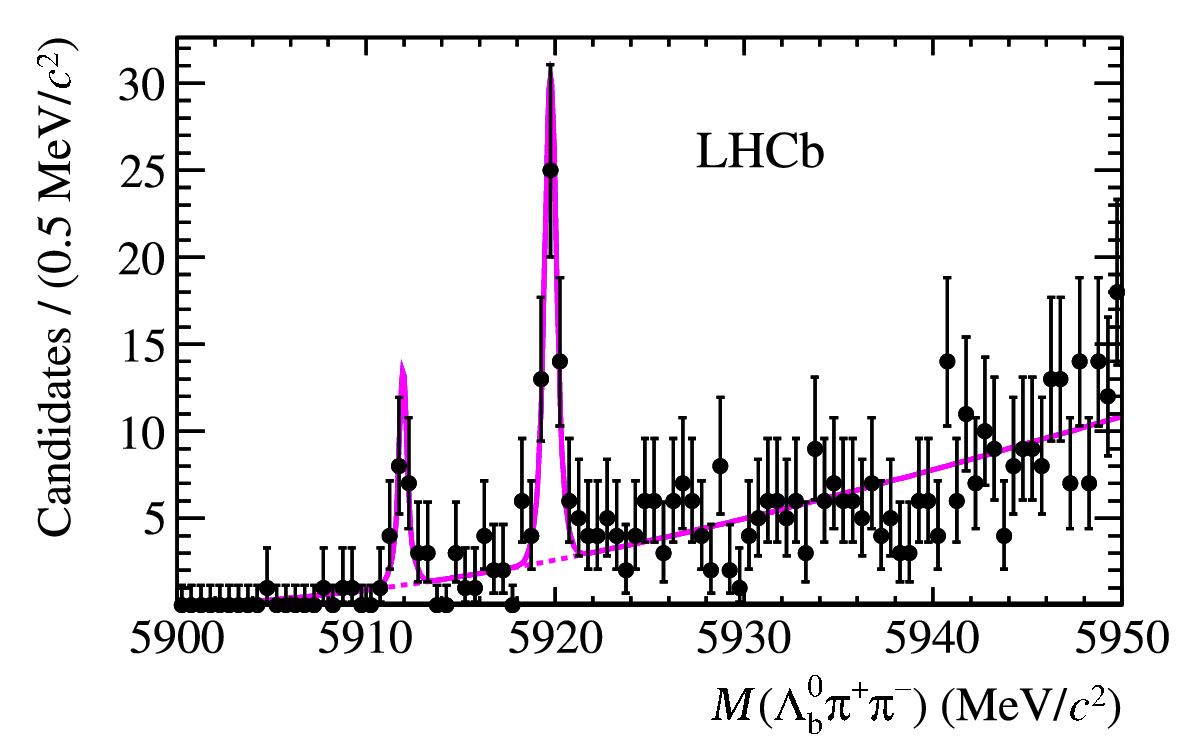}
\end{figure}

The isosinglet $\Lambda_b^0$ baryon has been known for about 20 years, 
but the observation of two new excited $\Lambda_b^{\ast0}$~baryons has 
been the most recent discovery of a new heavy baryon. While the original 
discovery at CERN was based on just 90 events of the reaction $\Lambda_b^0
\to\Lambda_c^+\pi^+\pi^-\pi^-$ in $pp$~collisions at $\sqrt{s} = 62$
GeV~\cite{326169}, LHCb has used $pp$~collision data collected in 2011 
at a center-of-mass energy $\sqrt{s} = 7$~TeV. The left side of 
Figure~\ref{Figure:LHCb-Lambda_b} shows $(70\,540\pm 330)$ 
$\Lambda_b^0$~signal events reconstructed from the $\Lambda_c^+
\pi^-$~decay mode~\cite{Aaij:2012da}. The most precise mass
measurement has been reported by LHCb to be~\cite{Aaij:2013ky}
\begin{eqnarray}
   M_{\Lambda_b^0\,\to\, J/\psi\Lambda} = 5619.53\pm 0.13\,({\rm
     stat})\pm 0.45\,({\rm syst})~{\rm MeV}\,.
\end{eqnarray}

The right side of Figure~\ref{Figure:LHCb-Lambda_b} shows the invariant 
$\Lambda_b^0\pi^+\pi^-$ mass. A higher-mass peak with a significance 
of 10.2 standard deviations is visible around 5920~MeV and a less
significant lower-mass peak with a significance of 5.2 standard deviations
is claimed around 5912~MeV~\cite{Aaij:2012da}. The masses have been 
determined to be
\begin{eqnarray}
   M_{\Lambda_b^{\ast 0}(5912)} \,=\, 5911.97\pm 0.12\pm 0.02\pm 0.66~{\rm MeV}\\
   M_{\Lambda_b^{\ast 0}(5920)} \,=\, 5919.77\pm 0.08\pm 0.02\pm 0.66~{\rm MeV}\,,
\end{eqnarray}
where the first uncertainty is statistical, the second is systematic, and 
the third is the uncertainty due to the the knowledge of the $\Lambda_b^0$ 
mass. The two peaks are associated with two orbitally excited $\Lambda_b^0$ 
states with the quantum numbers $J^P = \frac{1}{2}^-$ and $\frac{3}{2}^-$.

\subsubsection{The $\Sigma_b$ States}
\begin{figure}
\caption{\label{Figure:CDF-Sigma_b}(Colour online) Observation of
  $\Sigma_b^{(\ast)}$~states in $Q$-value spectra from CDF where $Q =
  M_{\Lambda_b^0\pi^\pm} - M_{\Lambda_b^0} - M_{\pi^\pm}$~\cite{CDF:2011ac}. 
  Left: $Q$-value spectrum for $\Sigma_b^{(\ast)-}$~candidates. Right:
  $Q$-value spectrum for $\Sigma_b^{(\ast)+}$~candidates. The solid
  and dashed lines represent the projections of an unbinned
  likelihood fit. The bottom of each plot shows the pull distributions
  of each fit. Picture from~\cite{CDF:2011ac}.}
\includegraphics[width=0.49\textwidth,height=0.4\textwidth]{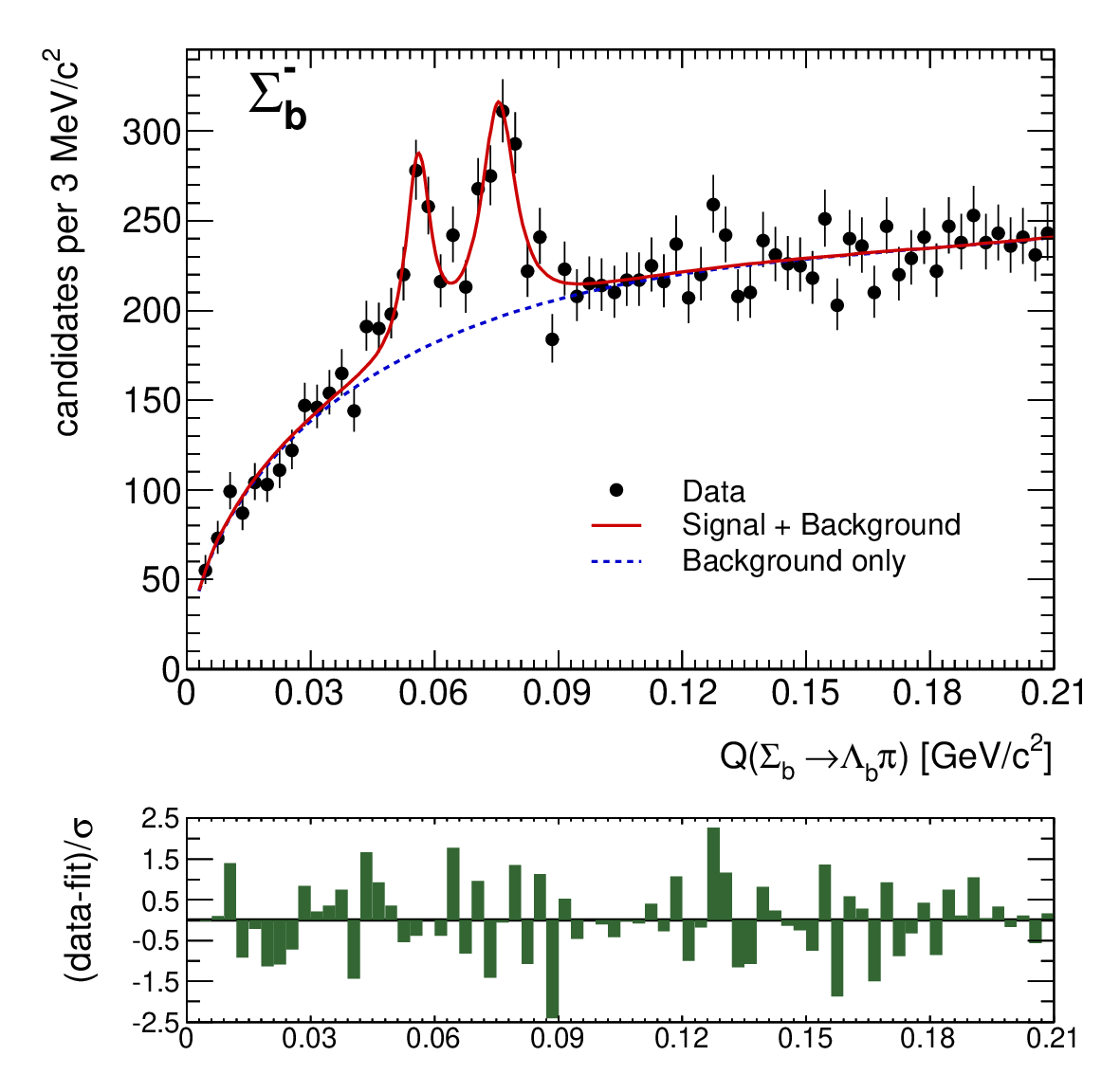}
\includegraphics[width=0.49\textwidth,height=0.4\textwidth]{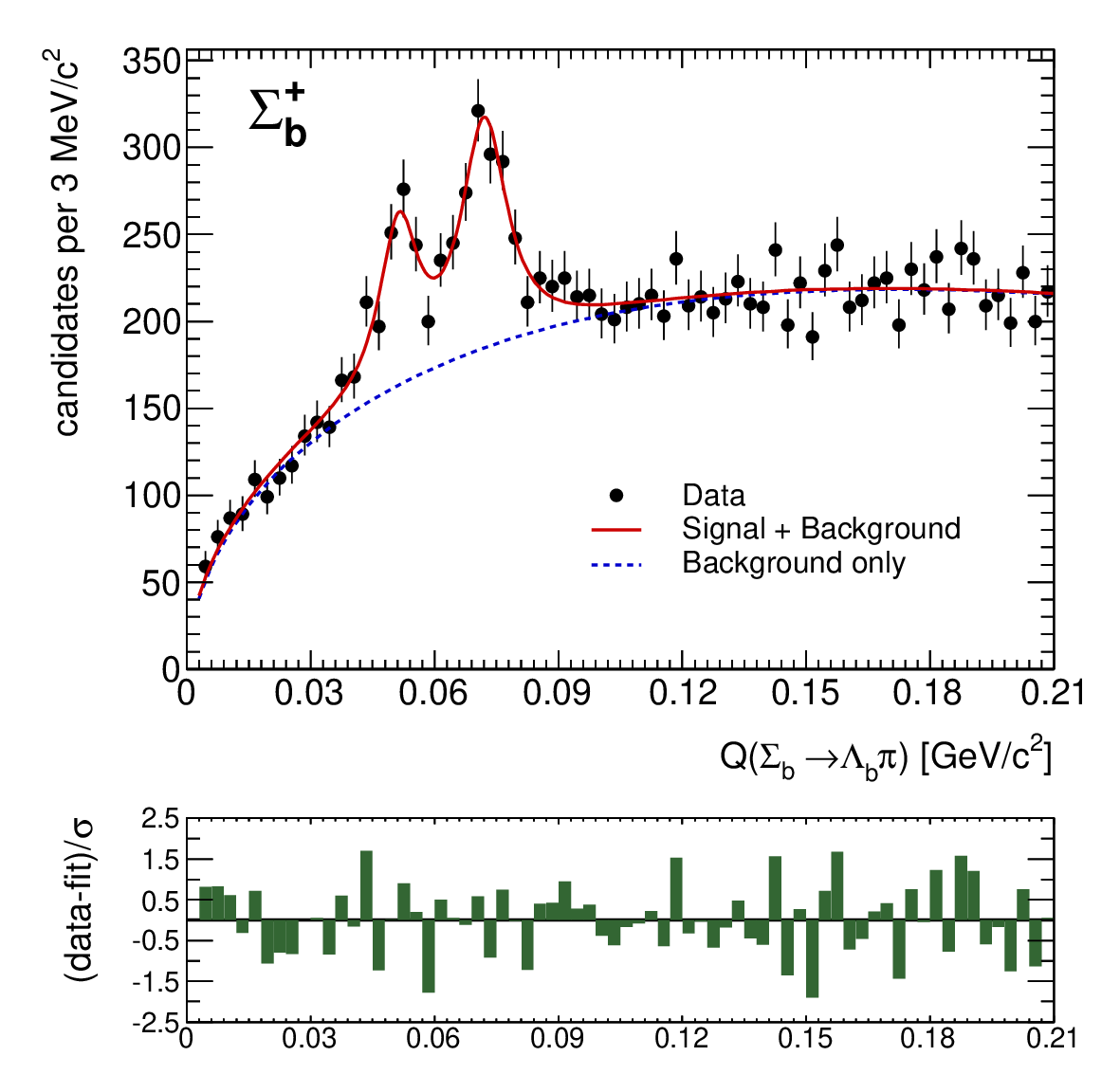}
\end{figure}
 
Only the CDF collaboration has studied baryons with quark content
$qqb~(q = u,d)$ other than the $\Lambda_b^0$. In the quark model,
the three states, $\Sigma_b^+$, $\Sigma_b^0$, and $\Sigma_b^-$, form an
$(uub, udb, ddb)$ isotriplet. Of the two observed $\Sigma_b^{\pm,0}$~triplets 
-- presumably the ground-states with $J^P = \frac{1}{2}^+$ and the lowest-mass 
excitations with $J^P = \frac{3}{2}^+$ -- only the charged states, $\Sigma_b^{(\ast)\pm}$, 
have been observed so far via their decays to $\Lambda_b^0\pi^\pm$. The 
resonances were discovered in 2007, but the most precise measurements
of their properties were reported by CDF in 2012 using a data sample corrsponding
to an integrated luminosity of 6.0~fb$^{-1}$~\cite{CDF:2011ac}. 
Figure~\ref{Figure:CDF-Sigma_b} shows the signals for the $\Sigma_b^{(\ast)}$
states dependent on $Q = M_{\Lambda_b^0\pi^\pm} - M_{\Lambda_b^0} - M_{\pi^\pm}$.
Masses and widths were determined to be
\begin{eqnarray}
   M_{\Sigma_b^+} \,=\, 5811.3^{+0.9}_{-0.8}\pm 1.7~{\rm MeV},\qquad &\Gamma \,=\, 9.7^{+3.8+1.2}_{-2.8-1.1}~{\rm MeV},\\
   M_{\Sigma_b^-} \,=\, 5815.5^{+0.6}_{-0.5}\pm 1.7~{\rm MeV},\qquad &\Gamma \,=\, 4.9^{+3.1}_{-2.1}\pm 1.1~{\rm MeV},\\
   M_{\Sigma_b^{\ast +}} \,=\, 5832.1\pm 0.7^{+1.7}_{-1.8}~{\rm MeV},\qquad &\Gamma \,=\, 11.5^{+2.7+1.0}_{-2.2-1.5}~{\rm MeV},\\
   M_{\Sigma_b^{\ast -}} \,=\, 5835.1^{+1.7}_{-1.8}\pm0.6~{\rm MeV},\qquad &\Gamma \,=\, 7.5^{+2.2+0.9}_{-1.8-1.4}~{\rm MeV}\,.
\end{eqnarray}
Moreover, the authors report on the first-time extraction of the
isospin mass splittings within the $I=1$~triplets, with
\begin{eqnarray}
   M_{\Sigma_b^+} - M_{\Sigma_b^-} & \,=\, -4.2^{+1.1}_{-1.0}\pm 0.1~{\rm MeV},\\
   M_{\Sigma_b^{\ast +}} - M_{\Sigma_b^{\ast -}} & \,=\, -3.0^{+1.0}_{-0.9}\pm 0.1~{\rm MeV}\,.
\end{eqnarray}
The observed widths are in good agreement with theoretical expectations 
and the isospin splittings agree with the predictions of~\cite{Rosner:2006yk}.

\subsubsection{The $\Xi_b$ States}
\begin{figure}
\caption{\label{Figure:CDF-Xib0}(Colour online) Left: First-time observation 
  of the $\Xi_b^-$~baryon at D\O. Picture from~\cite{arXiv:0706.1690}. Right: The 
  discovery of the $\Xi_b^0$ heavy baryon at CDF~\cite{Aaltonen:2011wd}. 
  Both the invariant $\Xi_b^-\to\Xi_c^0\pi^-$ mass distribution in (b) and 
  the $\Xi_b^0\to\Xi_c^+\pi^-$ mass distribution in (c) show evidence for 
  a signal around 5.8~GeV. The dashed lines show the projection of 
  the likelihood fits used to extract the masses and yields. Picture
  from~\cite{Aaltonen:2011wd}.}
\includegraphics[scale=0.24]{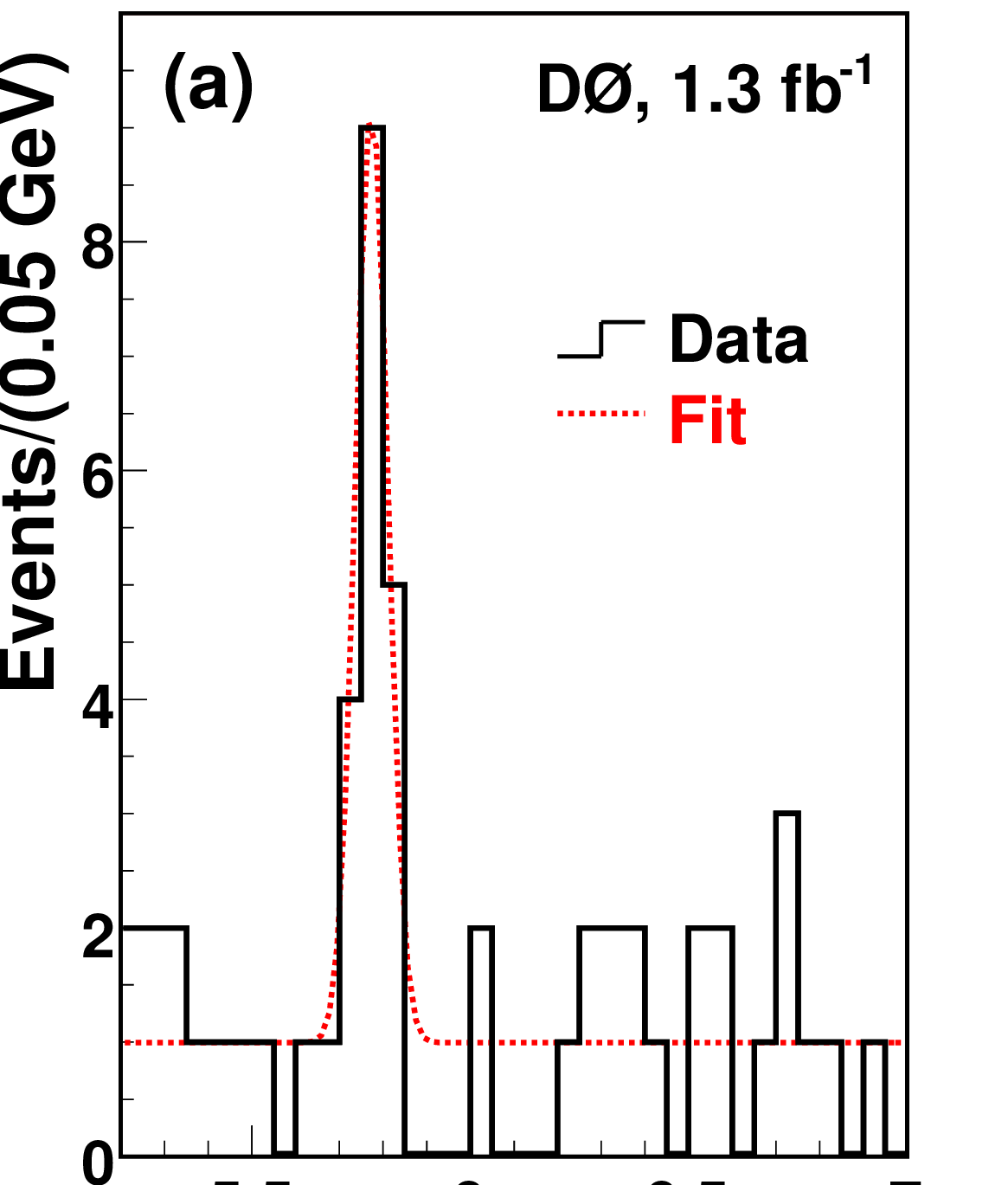}
\includegraphics[scale=0.475]{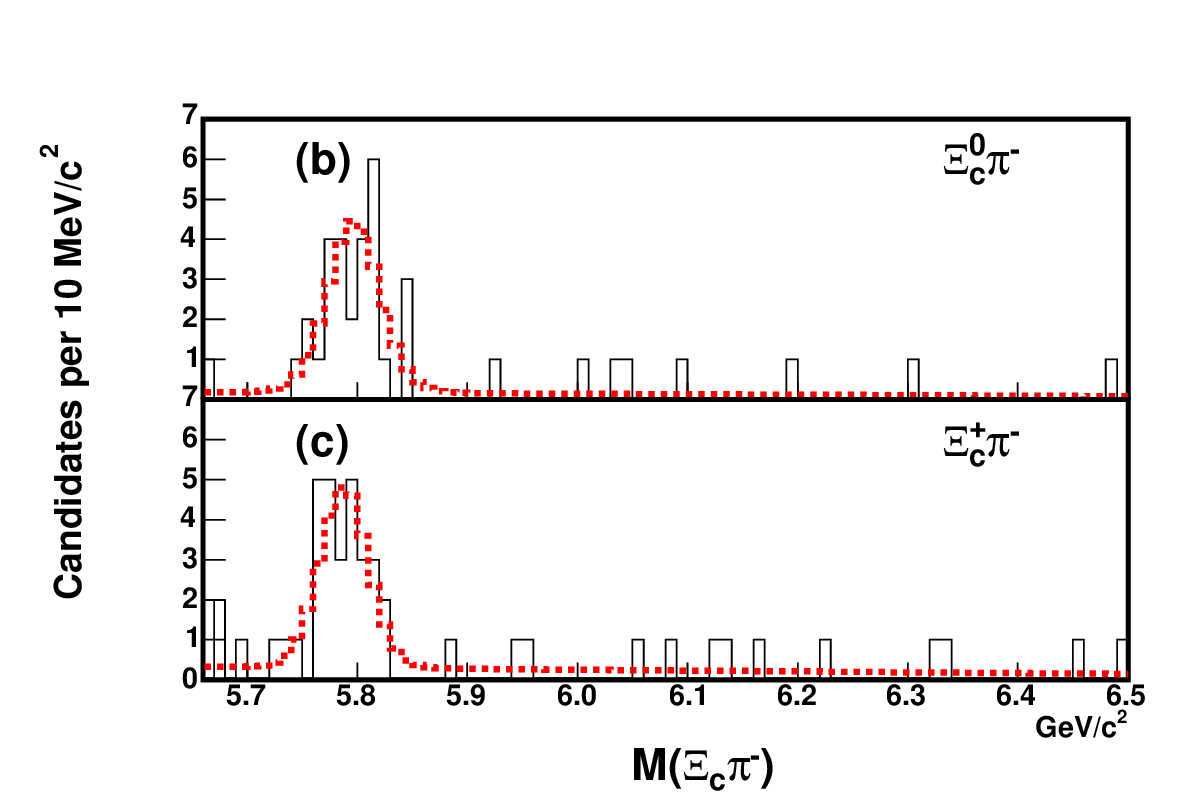}\\
\end{figure}

The first direct observation of a $\Xi_b$~baryon was reported in 2007 by 
the D\O~collaboration at Fermilab in $p\bar{p}$~collisions at $\sqrt{s}
= 1.96$~TeV~\cite{arXiv:0706.1690} and later confirmed by
CDF~\cite{arXiv:0707.0589,arXiv:0905.3123,Aaltonen:2011wd}. 
Earlier experiments at the CERN LEP $e^+e^-$~collider had only observed 
indirect evidence of the $\Xi_b^-$~baryon based on an excess of same-sign 
$\Xi^-l^-$~events in jets~\cite{CERN-PPE-96-081,CERN-PPE-95-029,
Abdallah:2005cw}. The left side of Figure~\ref{Figure:CDF-Xib0} shows 
the $M(\Xi_b^-)$~distribution of the $\Xi_b^-$ candidates from the original 
D\O~analysis. The dotted curve represents an unbinned likelihood fit to 
the model of a constant background plus a Gaussian signal. The fit yields 
$15.2\pm 4.4\,({\rm stat})^{+1.9}_{-0.4}\,({\rm syst})$ signal events with a 
significance of $5.5\sigma$. A higher-statistics CDF~analysis observed 
$66^{+14}_{-~9}$~candidates in $\Xi_b^-\to J/\psi\,\Xi^-$~decays~\cite{arXiv:0905.3123}.
The mass was found to be $5790.9\pm 2.6\,({\rm stat})\pm 0.8\,({\rm 
syst})$~MeV. Figure~\ref{Figure:LHCb_Omega}~(a) shows the 
$J/\psi\,\Xi^-$~mass distribution from CDF. The most precise mass
measurement of the $\Xi_b^-$ was recently reported by LHCb to be 
$5795.8\pm 0.9\,({\rm stat})\pm 0.4\,({\rm syst})$~MeV~\cite{Aaij:2013ky}.

The observation of the $\Xi_b^0$ has 
almost completed the discovery of the ground-state baryons with beauty 
content. The $\Sigma_b^{(\ast) 0}$~states remain hitherto unobserved.
In 2011, CDF reported a signal of $25.3^{+5.6}_{-5.4}$ candidates through 
the decay chain $\Xi_b^0\to\Xi_c^+\pi^-$, where $\Xi_c^+\to\Xi^-\pi^+\pi^+$,
using data from $p\bar{p}$~collisions at $\sqrt{s}=1.96$~TeV~\cite{Aaltonen:2011wd}. 
Figure~\ref{Figure:CDF-Xib0}~(c) shows a clear signal at about 5.8~GeV 
in the invariant $\Xi_c^+\pi^-$ mass distribution. Reconstructed from a similar 
decay sequence involving five final-state tracks and four decay vertices, 
Figure~\ref{Figure:CDF-Xib0}~(b) shows the signal for the $\Xi_b^-$ in 
the invariant $\Xi_c^0\pi^-$ mass distribution. The masses and yields
of the signals were obtained in unbinned likelihood fits (dashed line in
the figure). For the $\Xi_b^0$, a mass of $5787.8\pm 5.0_{\rm \,stat.}\pm 
1.3_{\rm \,sys.}$~MeV was determined.


The most recent discovery of a baryon resonance with beauty-strange
content has been reported from the LHC. In 2012, the CMS collaboration
has reported on the observation of an excited $\Xi_b^0$~state, most likely 
$\Xi_b^{\ast0}$ with $J^P = \frac{3}{2}^+$~\cite{Chatrchyan:2012ni}. The signal of 
about 20~events was observed in the difference between the mass of the
$\Xi_b^-\pi^+$~system and the sum of the masses of the $\Xi_b^-$ and 
$\pi^+$, with a significance exceeding 5 standard deviations. Given the
charged-pion and $\Xi_b^-$~masses from the 2010 edition of the 
RPP~\cite{FERMILAB-PUB-10-665-PPD}, the resulting mass of the new 
resonance is $5945.0\pm 0.7\,({\rm stat.})\pm 0.3\,({\rm syst.})\pm 2.7\,({\rm PDG})$~MeV, 
where the last uncertainty reflects the accuracy of the 
PDG~$\Xi_b^-$~mass~\cite{FERMILAB-PUB-10-665-PPD}. The new excited
$\Xi_b^{\ast 0}$~state was observed in its strong decay into $\Xi_b^-\pi^+$.
The signal yield of the $\Xi_b^-$ was determined to be $108\pm 14$~events,
exceeding the earlier best CDF result by almost a factor of two. This shows
again the big potential of the LHC as a machine for the discovery of new hadron resonances 
in the future.

\subsubsection{The $\Omega_b$ States
}
\begin{figure}
\caption{\label{Figure:LHCb_Omega}(Colour online) Observation of the
  $\Omega_b^-$~baryon. Left: The invariant mass distributions of (a)
  $J/\psi\,\Xi^-$ and (b) $J/\psi\,\Omega^-$~candidates from CDF
  showing the $\Xi_b^-$ and the $\Omega_b^-$, respectively. The 
  projections of the unbinned mass fit are indicated by the dashed 
  lines. Picture from~\cite{arXiv:0905.3123}. Right: Invariant mass 
  distribution of $\Omega_b^-\to J/\psi\,\Omega^-$ candidates from
  LHCb. Picture from~\cite{Aaij:2013ky}.}
\includegraphics[scale=0.38]{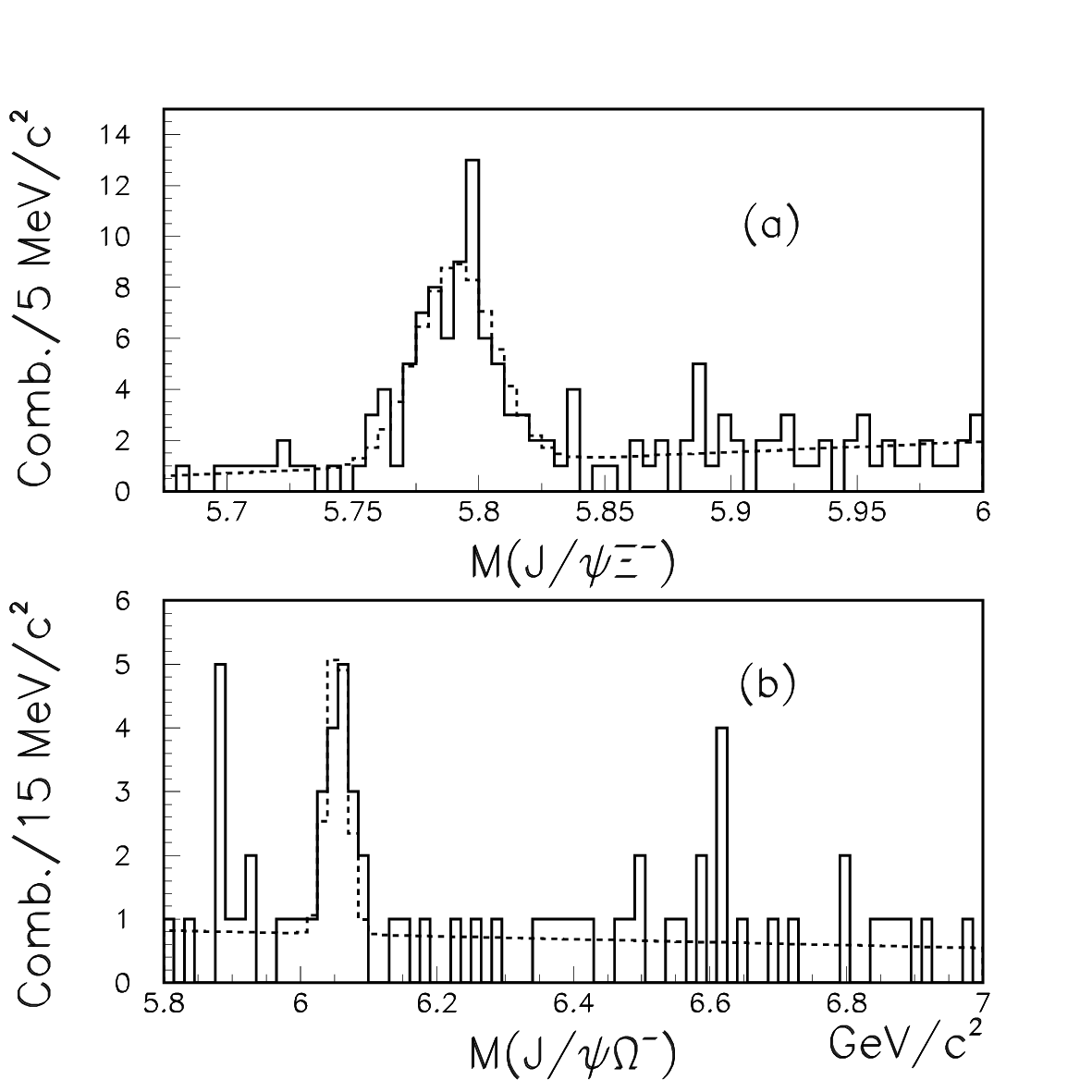}\hspace{-0.6cm}
\includegraphics[width=0.55\textwidth,height=0.4\textwidth]{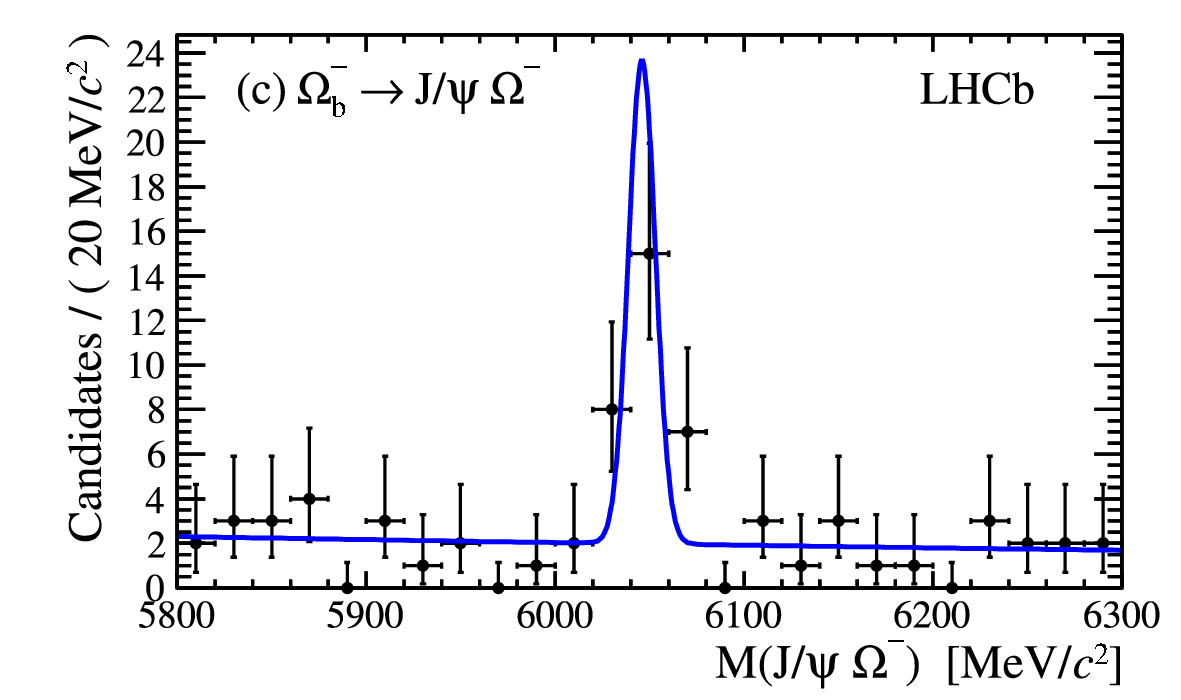}
\end{figure}

The doubly-strange $\Omega_b^-$ baryon (quark content $ssb$),
was just recently discovered at Fermilab. None of its quantum numbers
has been measured, but it is almost certainly the $\frac{1}{2}^+$~ground state.
First reported in 2008 by the D\O~collaboration in $\Omega_b^-\to J/\psi\,
\Omega^-$~decays~\cite{arXiv:0808.4142}, it was confirmed in 2009 by 
the CDF~collaboration in the same decay using 4.2~fb$^{-1}$ of data from 
$p\bar{p}$~collisions at $\sqrt{s} = 1.96$~TeV. Figure~\ref{Figure:LHCb_Omega}~(b)
shows the $ J/\psi\,\Omega^-$~distribution from CDF~\cite{arXiv:0905.3123}.
The signal is observed at a mass of $6054.4\pm 6.8\,({\rm stat.})\pm 0.9\,({\rm 
syst.})$~MeV, in disagreement with the initial mass from D\O, which 
was determined to be $6165\pm 10\pm 13$~MeV.

The mass measurements of the $\Omega_b^-$~baryon is thus particularly
interesting because the CDF and D\O~results differ by more than $6\sigma$. 
The LHCb collaboration has also studied the $\Omega_b^-$ using data 
corresponding to an integrated luminosity of 1.0~fb$^{-1}$ of $pp$~collisions
at a center-of-mass energy of $\sqrt{s} = 7$~TeV~\cite{Aaij:2013ky}. This 
corresponds to the full data sample collected in 2011.
Figure~\ref{Figure:LHCb_Omega}~(right) shows the $ J/\psi\,\Omega^-$
distribution from the LHCb analysis~\cite{Aaij:2013ky}. The mass
of the $\Omega_b^-$~signal has been extracted by performing an unbinned 
extended maximum likelihood fit using a Gaussian function for the signal
and an exponential function for the background. It has been determined 
to~\cite{Aaij:2013ky}
\begin{eqnarray}
M_{\Omega_b^-} = 6046.0\pm 2.2\pm 0.5~{\rm MeV}\,,
\end{eqnarray}
which is in very good agreement with the earlier CDF
result~\cite{arXiv:0905.3123} and closer to most theoretical predictions for
the $\Omega_b^-$ which vary in the range from 5940 to 6120~MeV.
\section{Light-Quark Baryons}
\subsection{From Cross Sections to Baryon Parameters}\label{Section:pwa1}

In contrast with the heavy baryons discussed above, light baryons are not usually found as narrow, isolated peaks in mass distributions or cross section measurements. Indeed, examination of the total or elastic cross section for $\pi N$ scattering shows only three or four structures. Somehow, the parameters of a number of excited nucleons and Deltas have been extracted from the measurements that have been made. How is this done? We illustrate the challenges by discussing the process $\pi N\to \pi N$.

The process $\pi N \to\pi N$ is described in terms of two complex amplitudes. These may be helicity amplitudes, transversity amplitudes, Chew-Goldberger-Low-Nambu (CGLN) amplitudes, or amplitudes defined in some other convention. The fact that these amplitudes are complex means that the process requires four real numbers at every kinematic point for its description. Information about baryon resonances that may be formed as intermediate states in the process is contained in the amplitudes. Thus, the task required is the extraction of the amplitudes from the observables, which are necessarily bilinear in the amplitudes.

The extraction of four real numbers at every kinematic point requires four observables, but since the relationships between the observables and the amplitudes are bilinear, there are necessarily ambiguities in the extracted amplitudes. Furthermore, in order to extract information on different isospin states (nucleons and Deltas), observables from different processes ($\pi^+p \to \pi^+p$, $\pi^-p \to \pi^-p$ and $\pi^-p\to \pi^0 n$) must be combined. In general, these different processes may have been measured in different experiments, with different detectors, having different systematic uncertainties.

Once amplitudes have been extracted, they are expanded in partial waves according to isospin, and the task of searching for resonance signatures begins. Here, one does not look for peaks or bumps, but the phase structure of the amplitudes is used to provide clues about resonant contributions. For a single, narrow, isolated resonance that couples to a single channel, the real part of the scattering amplitude vanishes on resonance, and the imaginary part will go through a maximum. On an Argand plot (real part of the amplitude vs. the imaginary part of the amplitude), the amplitude will describe a circle, or very nearly. When there are nearby resonances in the same channel, the behaviour of the real and imaginary parts of the amplitude become significantly harder to interpret. The real part may no longer have a zero at the resonant mass, or the imaginary part may not go through a maximum. A weakly-coupled resonance may have very little effect on the amplitudes. If the data available do not provide good kinematic coverage, or one or more observables have not been measured, the extraction of resonance information becomes more challenging and more sensitive to model dependences and interpretation. 

Just as it is well nigh impossible to identify resonances from total or differential cross sections alone, so, too, is it extremely difficult to identify resonance signatures from examining a single polarization observable. Such observables may give hints to underlying resonances, but it is unlikely that any sharp structures will be seen. Resonant effects turn on over an energy range that spans the width of the resonance, and this should be the case in any observable examined. Since these resonances are typically 100 to 300 MeV wide, any effect in a polarization observable can be expected to be gradual enough to be almost imperceptible. 

For photoproduction processes, the number of independent amplitudes increases, and so does the number of observables that must be measured in order to extract these amplitudes with the least amount of ambiguity possible. The current status of photoproduction data in {\it any} channel is such that a full partial wave analysis (PWA) cannot be done, as an insufficient number of observables has been measured with sufficient precision. A few of the groups that extract baryon resonance parameters from experimental observables are carrying out {\it bona fide} partial wave analyses of the kind described above. However, many are constructing explicit models of the process, and using these models to obtain information. Whatever method is being used, not enough data yet exist to constrain these extractions. Hence, different groups make different claims about which resonances, old and new, are important in which region and for which process. It must be understood that a claim of a new resonance really means that `inclusion of such a state improves the fit to the data available'. For the analysts who do not see evidence for such a resonance, this really means that `inclusion of such a resonance, at the very best, does not significantly improve our fit to the data available'.

In the following sections, we discuss many `observations' of resonances by various groups. We emphasize that these `observations' really mean extractions, which means that inclusion of the resonance improved the fit to the available data.

\subsection{The Known Baryons}

\begin{table}[b]
\begin{center}
\caption{\label{Table:BaryonSummary}(Colour online) Baryon Summary
  Table for $N^\ast$ and $\Delta$ resonances including recent changes
  from PDG 2010~\cite{FERMILAB-PUB-10-665-PPD} to PDG 2012~\cite{Beringer:1900zz}.}
\begin{tabular}{l|l|l|l||l|l|l|l}
\br
$N^\ast$ & $J^P~(L_{2I,2J})$ & 2010 & 2012 & $\Delta$ & $J^P~(L_{2I,2J})$ & 2010 & 2012\\
\mr
$p$ & $1/2^+\,(P_{11})$ & $\ast\ast\ast\ast$ & $\ast\ast\ast\ast$ & $\Delta(1232)$ & $3/2^+\,(P_{33})$ & $\ast\ast\ast\ast$ & $\ast\ast\ast\ast$\\
$n$ & $1/2^+\,(P_{11})$ & $\ast\ast\ast\ast$ & $\ast\ast\ast\ast$ & $\Delta(1600)$ & $3/2^+\,(P_{33})$ & $\ast\ast\ast$ & $\ast\ast\ast$\\
$N(1440)$ & $1/2^+\,(P_{11})$ & $\ast\ast\ast\ast$ & $\ast\ast\ast\ast$ & $\Delta(1620)$ & $1/2^-\,(S_{31})$ & $\ast\ast\ast\ast$ & $\ast\ast\ast\ast$\\
$N(1520)$ & $3/2^-\,(D_{13})$ & $\ast\ast\ast\ast$ & $\ast\ast\ast\ast$ & $\Delta(1700)$ & $3/2^-\,(D_{33})$ & $\ast\ast\ast\ast$ & $\ast\ast\ast\ast$\\
$N(1535)$ & $1/2^-\,(S_{11})$ & $\ast\ast\ast\ast$ & $\ast\ast\ast\ast$ & $\Delta(1750)$ & $1/2^+\,(P_{31})$ & $\ast$ & $\ast$\\
$N(1650)$ & $1/2^-\,(S_{11})$ & $\ast\ast\ast\ast$ & $\ast\ast\ast\ast$ & $\Delta(1900)$ & $1/2^-\,(S_{31})$ & $\ast\ast$ & $\ast\ast$\\
$N(1675)$ & $5/2^-\,(D_{15})$ & $\ast\ast\ast\ast$ & $\ast\ast\ast\ast$ & $\Delta(1905)$ & $5/2^+\,(F_{35})$ & $\ast\ast\ast\ast$ & $\ast\ast\ast\ast$\\
$N(1680)$ & $5/2^+\,(F_{15})$ & $\ast\ast\ast\ast$ & $\ast\ast\ast\ast$ & $\Delta(1910)$ & $1/2^+\,(P_{31})$ & $\ast\ast\ast\ast$ & $\ast\ast\ast\ast$\\
{\color{red} $N(1685)$} & & & {\color{red} $\ast$} & & & &\\
$N(1700)$ & $3/2^-\,(D_{13})$ & $\ast\ast\ast$ &  $\ast\ast\ast$ & $\Delta(1920)$ & $3/2^+\,(P_{33})$ & $\ast\ast\ast$ & $\ast\ast\ast$\\
$N(1710)$ & $1/2^+\,(P_{11})$ & $\ast\ast\ast$ & $\ast\ast\ast$ & $\Delta(1930)$ & $5/2^-\,(D_{35})$ & $\ast\ast\ast$ & $\ast\ast\ast$\\
$N(1720)$ & $3/2^+\,(P_{13})$ & $\ast\ast\ast\ast$ & $\ast\ast\ast\ast$ & {\color{red} $\Delta(1940)$} & $3/2^-\,(D_{33})$ & $\ast$ & {\color{red} $\ast\ast$}\\
{\color{red} $N(1860)$} & {\color{red} $5/2^+$} & & {\color{red} $\ast\ast$} & & & &\\
{\color{red} $N(1875)$} & {\color{red} $3/2^-$} & & {\color{red} $\ast\ast\ast$} & & & &\\
{\color{red} $N(1880)$} & {\color{red} $1/2^+$} & & {\color{red} $\ast\ast$} & & & &\\
{\color{red} $N(1895)$} & {\color{red} $1/2^-$} & & {\color{red} $\ast\ast$} & & & &\\
{\color{red} $N(1900)$} & $3/2^+\,(P_{13})$ & $\ast\ast$ & {\color{red} $\ast\ast\ast$} & $\Delta(1950)$ & $7/2^+\,(F_{37})$ & $\ast\ast\ast\ast$ & $\ast\ast\ast\ast$\\
$N(1990)$ & $7/2^+\,(F_{17})$ & $\ast\ast$ & $\ast\ast$ & $\Delta(2000)$ & $5/2^+\,(F_{35})$ & $\ast\ast$ & $\ast\ast$\\
$N(2000)$ & $5/2^+\,(F_{15})$ & $\ast\ast$ & $\ast\ast$ & $\Delta(2150)$ & $1/2^-\,(S_{31})$ & $\ast$ & $\ast$\\
\rlap{------------}\,{\color{red} $N(2080)$} & {\color{red} $D_{13}$} & {\color{red} $\ast\ast$} & & $\Delta(2200)$ & $7/2^-\,(G_{37})$ & $\ast$ & $\ast$\\
\rlap{------------}\,{\color{red} $N(2090)$} & {\color{red} $S_{11} $} & {\color{red} $\ast$} & & $\Delta(2300)$ & $9/2^+\,(H_{39})$ & $\ast\ast$ & $\ast\ast$\\
{\color{red} $N(2040)$} & {\color{red} $3/2^+$} & & {\color{red} $\ast$} & & & &\\
{\color{red} $N(2060)$} & {\color{red} $5/2^-$} & & {\color{red} $\ast\ast$} & & & &\\
$N(2100)$ & $1/2^+\,(P_{11})$ & $\ast$ & $\ast$ & $\Delta(2350)$ & $5/2^-\,(D_{35})$ & $\ast$ & $\ast$\\
{\color{red} $N(2120)$} & {\color{red} $3/2^-$} & & {\color{red} $\ast\ast$} & & & &\\
$N(2190)$ & $7/2^-\,(G_{17})$ & $\ast\ast\ast\ast$ & $\ast\ast\ast\ast$ & $\Delta(2390)$ & $7/2^+\,(F_{37})$ & $\ast$ & $\ast$\\
\rlap{------------}\,{\color{red} $N(2200)$} & {\color{red} $D_{15}$} & {\color{red} $\ast\ast$} & & $\Delta(2400)$ & $9/2^-\,(G_{39})$ & $\ast\ast$ & $\ast\ast$\\
$N(2220)$ & $9/2^+\,(H_{19})$ & $\ast\ast\ast\ast$ & $\ast\ast\ast\ast$ & $\Delta(2420)$ & $11/2^+\,(H_{3,11})$ & $\ast\ast\ast\ast$ & $\ast\ast\ast\ast$\\
$N(2250)$ & $9/2^-\,(G_{19})$ & $\ast\ast\ast\ast$ & $\ast\ast\ast\ast$ & $\Delta(2750)$ & $13/2^-\,(I_{3,13})$ & $\ast\ast$ & $\ast\ast$\\
$N(2600)$ & $11/2^-\,(I_{1,11})$ & $\ast\ast\ast$ & $\ast\ast\ast$ & $\Delta(2950)$ & $15/2^+\,(K_{3,15})$ & $\ast\ast$ & $\ast\ast$\\
$N(2700)$ & $13/2^+\,(K_{1,13})$ & $\ast\ast$ &  $\ast\ast$ & & &  &\\
\mr
\br
\end{tabular}
\end{center}
\end{table}

In the 2012 edition of the RPP, the PDG lists 17~$N^\ast$ resonances (including the proton
and neutron) and 10~$\Delta$~resonances in its Summary Table~\cite{Beringer:1900zz} 
which have at least a 3-star assignment. This represents a significant
change from the previous edition. Table~\ref{Table:BaryonSummary}
shows the Baryon Summary Table including recent changes. In the 
$N^\ast$ ~sector, one new 3-star $N^\ast$~resonance, $N(1875)\,\frac{3}{2}^-$,
has been added and one further resonance, $N(1900)\,\frac{3}{2}^+$, has been
upgraded to a 3-star state. Moreover, seven additional resonances have
been proposed and three removed. In the $\Delta$~sector, one negative-parity
resonance, $\Delta(1940)\,\frac{3}{2}^-$, has been upgraded to a slightly
better established state, but the 2-star assignment indicates that the
evidence for its existence is still poor. The list of hyperons remains
unchanged from the 2010 edition. 

\subsection{Experimental Methods and Major Experiments}
Light-flavour baryon resonances can be studied using a large variety of
different production mechanisms. Most laboratories employ fixed-target
experiments and different kinds of either hadronic or electromagnetic 
beams to induce a reaction. Prominent examples are pion, electron, 
and photon beams, but kaon and proton beams have also been used.

Baryons can be produced directly in formation (as opposed to production) 
experiments without a recoil particle, so that they may be identified in 
the cross section, e.g. in the reaction $\pi^+ p\to\Delta^{++}\to p\pi^+$.
All peaks in Figure~\ref{Figure:RPP2012-pip_pid} can be assigned to 
short-lived states, although the broad resonances strongly overlap.
This method allows access only to states with strangeness $S\leq 1$ 
owing to the nature of the available particle beams. A more general 
method is the production of baryons in production experiments. A 
prominent example is the photoproduction of hyperons, e.g. $\gamma 
p\to K Y^\ast$, where the hyperon is produced with a strange partner,
or the production of $\Xi$~resonances in the decay of excited $\Sigma$ 
or $\Lambda$~states, e.g. $Y^\ast\to K\,\Xi$. An alternative approach is 
the production of excited nucleons in the decays of 
heavy mesons like the $J/\psi$ and the $\psi(2S)$.

In this section, we discuss the different production mechanisms
and associated observables which can be determined experimentally. 
A brief survey of major experiments will be given. As has been pointed out 
in Section~\ref{Section:pwa1}, PWA are recognized as the most powerful tools 
for the purpose of extracting resonance contributions. 
For this reason, we will briefly review some approaches in the last part
of this section.

\subsubsection{\label{Section:HadronicProbes}Baryon Production using
  Hadronic Probes}
Most of the information about non-strange baryon resonances listed by
the PDG has been extracted from PWA of $\pi N$~elastic scattering data
in the reaction, $\pi N\to\pi N$, and charge exchange data, e.g. in the
reaction $\pi^- p\to n\pi^0$. The total reaction rate for elastic $\pi
N$~scattering is given as~\cite{Roberts:2004mn}
\begin{eqnarray}
\label{Equation:Rate-piN}
\rho_f I\,=\,I_0\,\biggl[\,1\,+\,\vec{\Lambda}_i\cdot\vec{P}\,+\,\vec{\sigma}\cdot
\vec{P}^{\,\prime}\,+\,\Lambda_i^\alpha\sigma^{\beta^\prime}O_{\alpha\beta^\prime}\,\biggr]\,,
\end{eqnarray}
where $\vec{P}$ represents the polarization asymmetry that arises if
the target nucleon is polarized, $\rho_f = \frac{1}{2} (1+\vec{\sigma}
\cdot\vec{P}^{\,\prime})$ is the density matrix of the recoiling nucleon, 
and $O_{\alpha\beta^\prime}$ represent the double-polarization observables 
if the Cartesian $\alpha$~component of the target polarization is known 
and the Cartesian $\beta^{\,\prime}$~component of the recoil polarization is measured. 
The primes indicate that the recoil observables, defined in terms of helicity 
amplitudes, are measured with respect to the set of axes $x^{\,\prime}$, 
$y^{\,\prime}$,  $z^{\,\prime}$, in which the $z^{\,\prime}$-axis is parallel 
to the momentum of the recoil nucleon. If the observables are defined 
in terms of transversity amplitudes, the $x^{\,\prime}$, $y^{\,\prime}$,  
$z^{\,\prime}$ axes are the same as the $x$, $y$, $z$ axes. 
$\vec{\Lambda}_i$ is the polarization of the initial nucleon. The 
16~quantities in~(\Eref{Equation:Rate-piN}) are not all independent, and many of them vanish. For instance, since the strong
interaction conserves parity, the observables $P_x$, $P_z$, $P_x^{\,\prime}$ and $P_z^{\,\prime}$ must vanish for the process $\pi N \to \pi N$.
In fact, for this process, there are only four linearly-independent non-vanishing observables.
A number of relationships among 
the observables reduce to a single relationship
\begin{eqnarray}
\label{Equation:Relation-piN}
P_y^2\,+\,O^2_{xx^\prime}\,+\,O^2_{xz^\prime}\,=\,1\,.
\end{eqnarray} 
A summary of the relevant formulas including those for the reaction
$\pi N\to\pi\pi N$ and the derivation of~(\Eref{Equation:Relation-piN}) are 
given in~\cite{Roberts:2004mn}. The observables in~(\Eref{Equation:Relation-piN}) 
can also be expressed in terms of the observables, $P$, $R$, and $A$,
more often found in the literature
\begin{eqnarray}
R\,& =\,O_{xx^\prime}\,{\rm cos}\,\theta\,-\,O_{xz^\prime}\,{\rm sin}\,\theta\,,\\[0.5ex]
A\,& =\,O_{xx^\prime}\,{\rm sin}\,\theta\,+\,O_{xz^\prime}\,{\rm cos}\,\theta\,,\qquad{\rm and}\\[0.5ex]
P^2\,& +\,A^2\,+\,R^2\,=\,1\,,
\end{eqnarray}
where $P=P_y$ is the (target or recoil) polarization of the proton perpendicular to the reaction plane, and $R$ and $A$ are the components of the proton polarization transfer along its direction of flight and perpendicular to its direction of flight,
respectively.  The measurement of the two spin-rotation parameters, 
$R$ and $A$, provides experimental information on the relative phase
of the transverse scattering amplitude and is required to unambiguously 
extract the $\pi N$ elastic scattering amplitude. Note that the same observables can be measured in any process
$P B\to P^{\,\prime} B^{\,\prime}$, where $P$ and $P^{\,\prime}$ are pseudoscalar mesons, and $B$ and $B^{\,\prime}$ are spin-$\hlf$ baryons.

The first experiments using strongly-interacting beams were performed between 1957 and
1979 with most of the largest resonance-region experiments done at the
Nimrod accelerator at Rutherford Laboratory. They focused mostly on the 
$\pi N$~system and the production of non-strange baryon resonances. 
Later $\pi N$~experiments extracted cross sections with much smaller 
statistical and systematic uncertainties and also measured polarization 
observables. We refer to the Durham Reaction Database~\cite{durham-database} 
and the Data Analysis Center at George-Washington University~\cite{said-database} 
for the full experimental database. The most recent $\pi N$~elastic and 
charge-exchange results published after 2000 come from experiments at 
the Brookhaven National Laboratory (BNL)~\cite{Starostin:2005pd}, the 
Tri-University Meson Facility (TRIUMF)~\cite{Denz:2005jq}, and the Paul 
Scherrer Institute (PSI)~\cite{Meier:2004kn,Breitschopf:2006gn}, for instance. 
The polarization parameter, $P$, and the spin-rotation parameter, $A$, were 
published recently by the ITEP-PNPI collaboration based on experiments at 
the Institute for Theoretical and Experimental Physics (ITEP)~\cite{Alekseev:2005zr,
Alekseev:2008cw}. Earlier recoil polarization data were accumulated in
the 1980s at Los Alamos, e.g.~\cite{Mokhtari:1987iy,Seftor:1989er}. A 
description of the experimental setups can be found in the references given. 

In addition to $\pi N$~elastic scattering, the Crystal Ball collaboration 
studied inelastic pion and kaon nucleon scattering in experiments at BNL.
At the center of the experimental setup was the Crystal Ball (CB) multiphoton 
spectrometer, which consisted of 672 optically isolated NaI(Tl) crystals
covering 93\,\% of the $4\pi$~solid angle. The typical energy resolution
for electromagnetic showers in the CB spectrometer was $\Delta E/E = 0.020/(E\,[{\rm 
GeV}])^{0.36}$. The detector is described in more detail in~\cite{Starostin:2001zz}.
The collaboration also peformed important measurements of kaon-induced 
reactions. Relevant results can be found in~\cite{Starostin:2001zz,Kozlenko:2003hu,Starostin:2003cc,
Borgh:2003uc,Olmsted:2003is,Craig:2003yd,Sadler:2004yq,Prakhov:2004zv,
Prakhov:2004ri,Prakhov:2004an,Prakhov:2005qb,Prakhov:2008dc,Mekterovic:2009kw}. 
The Crystal Ball detector was later moved to MAMI 
where photoproduction experiments are currently being performed. 

\subsubsection{\label{Section:EM-Baryons}Baryons in Electromagnetically-Induced Reactions}
Many baryon resonances are produced abundantly in electromagnetically-induced 
reactions off the nucleon. While photoproduction experiments serve to establish the 
systematics of the spectrum, electron beams are ideal to measure resonance form 
factors and their corresponding $Q^2$~dependence. The latter provides information 
on the structure of excited nucleons and on the (effective) confining forces of the 
3-quark system.

The electromagnetic transition from the nucleon 
ground state to any resonant excited state can be described in terms of 
photon helicity amplitudes, so that the transverse and longitudinal
virtual photon cross sections at resonance can be written as
\begin{eqnarray}
\sigma_T(\nu_R,\,Q^2)\,=\,\frac{2M}{\Gamma_R M_R}\,[\,|A_{\frac{1}{2}}|^2 \,+\, |A_{\frac{3}{2}}|^2\,]\,,\\[0.5ex]
\sigma_L(\nu_R,\,Q^2)\,=\,\frac{4M}{\Gamma_R M_R}\,[\,|S_{\frac{1}{2}}|^2\,]\,,
\end{eqnarray}
where $M_R$ and $\Gamma_R$ are the invariant mass of the resonance and
its decay width, respectively. The detailed analysis of the $Q^2$~dependence 
of the virtual photon cross sections from electron scattering data can be used
to extract transition form factors for each of the nucleon resonances, see 
e.g.~\cite{Aznauryan:2011qj} for a recent review.

In processes such as $(e,e^{\,\prime}\,\pi)$ and $(e,e^{\,\prime}\,\eta)$, the
initial and final nucleon can each have two possible spin states, and the
photon has one longitudinal~(L) and two transverse~(T) helicity states. Parity considerations reduce 
these 12 initial matrix elements to six independent amplitudes,
which can be expanded in multipoles, i.e. amplitudes ordered according
to angular momentum. The detailed derivation of these expansions in
so-called CGLN amplitudes has been discussed in~\cite{Chew:1957tf}. The
individual contributions to the differential cross section for electroproduction
of a single pseudoscalar meson off the nucleon, $d\sigma_i/d\Omega~(i=
T,\,L,\,TL,\,TT,\,TT,\,TL^{\,\prime},\,TT^{\,\prime}$), are usually parametrised 
in terms of response functions, $R_i^{\beta\alpha}$, which depend on the
independent kinematical variables. The additional indices, $\alpha$ and
$\beta$, describe the target and recoil polarization of the final-state 
baryon, respectively. A nice summary and derivation of formulae relevant to electroproduction is given
in~\cite{Knochlein:1995qz}. A full determination of the six electroproduction 
amplitudes requires a suitable combination of polarization observables, 
such as the beam asymmetry (of the polarized photon) or the target 
asymmetry (of the polarized proton), and the polarization of the recoil 
nucleon. Recent results from electroproduction experiments come from 
MAMI and Jefferson Lab, for instance.

For real photons, the longitudinal component of the photon polarization 
vanishes. The photoproduction of single-pseudoscalar mesons off the 
nucleon is thus described by four complex helicity amplitudes: two for 
the spin states of the photon, two for the target nucleon, and two for
the recoiling baryon; parity considerations reduce these eight amplitudes 
to four. Experimentally, 16 observables can be determined: the unpolarized 
differential cross section, $d\sigma_0$, three single-spin observables, 
$\Sigma$ (beam), $T$ (target), $P$ (recoil), and three sets of four asymmetries: 
beam-target~(BT), beam-recoil~(BR), and target-recoil~(TR). It has been 
shown in~\cite{Chiang:1996em} that angular distributions of at least eight 
carefully chosen observables at each energy are needed for both proton 
and neutron to determine the full scattering amplitude without ambiguities.
The ambitious program of a complete experiment in photoproduction was 
first formulated by Barker {\it et al.} in 1975~\cite{Barker:1975bp}. The 
differential cross section for the three sets of asymmetries can be written
(for BT, BR, and TR, respectively) (colour online)
\begin{eqnarray}\label{Equation:BT}
\eqalign{\frac{d\sigma}{d\Omega}\,=\,\sigma_0\,\{ 1\,-\,P_T\,\bl{\Sigma}\,{\rm
  cos}\,2\phi\, & +\,P_x\,(-P_T\,\bl{H}\,{\rm
  sin}\,2\phi\,+\,P_\odot\,\bl{F})\\ \, & -\,P_y\,(-\bl{T}\,+\,P_T\,\bl{P}\,{\rm
  cos}\,2\phi)\\ \, & -\,P_z\,(-P_T\,\bl{G}\,{\rm sin}\,2\phi\,+\,P_\odot\,\bl{E})\}\,,}
\end{eqnarray}

\begin{eqnarray}
\eqalign{\frac{d\sigma}{d\Omega}\,=\,\sigma_0\,\{ 1\,-\,P_T\,\bl{\Sigma}\,{\rm
  cos}\,2\phi\, & +\,P_{x^{\,\prime}}\,(-P_T\,\bl{O_{x^{\,\prime}}}\,{\rm
  sin}\,2\phi\,-\,P_\odot\,\bl{C_{x^{\,\prime}}})\\ \, & -\,P_{y^{\,\prime}}\,(-\bl{P}\,+\,P_T\,\bl{T}\,{\rm
  cos}\,2\phi)\\ \, & -\,P_{z^{\,\prime}}\,(-P_T\,\bl{O_{z^{\,\prime}}}\,{\rm
  sin}\,2\phi\,+\,P_\odot\,\bl{C_{z^{\,\prime}}})\}\,,}
\end{eqnarray}

\begin{eqnarray}
\eqalign{\frac{d\sigma}{d\Omega}\,=\,\sigma_0\,\{
  1\,+\,P_{y^{\,\prime}}\,\bl{P}\, &
  +\,P_x\,(P_{x^{\,\prime}}\,\bl{T_{x^{\,\prime}}}\,+\,P_{z^{\,\prime}}\,\bl{T_{z^{\,\prime}}})\\
  \, & +\,P_y\,(\bl{T}\,+\,P_{y^{\,\prime}}\,\bl{\Sigma})\\\, &
  -\,P_z\,(P_{x^{\,\prime}}\,\bl{L_{x^{\,\prime}}}\,-\,P_{z^{\,\prime}}\,\bl{L_{z^{\,\prime}}})\}\,,}
\end{eqnarray}
where $\sigma_0$ denotes the unpolarized differential cross section,
$P_T$ and $P_\odot$ denote the linear and circular degree of photon
polarization, respectively, and $\phi$ is the angle between the photon
polarization vector and the reaction plane. The target polarization
vector is represented by $(P_x,\,P_y,\,P_z)$ and the polarization
observables are shown in blue. A total of 16 observables 
exist in photoproduction, whereas in electroproduction, four additional 
observables exist for the exchange of longitudinal photons and further 16
observables arise from longitudinal-transverse
interference~\cite{Knochlein:1995qz}.

\begin{table}
\begin{center}
\caption{\label{Table:PolarizationExperiments} Polarization
  observables in single-pseudoscalar photoproduction using all
  combinations of beam, target, and recoil polarization. The
  observables without hats and tildes can be extracted directly. 
  Moreover, observables with hats are single-polarization observables 
  that can be extracted from double-polarization asymmetries and 
 observables with tildes are double-polarization observables that can
  be extracted from triple-polarization asymmetries~\cite{Aznauryan:2011ub}.}
\begin{tabular}{c|c|ccc|ccc|ccccccccc}
\br
 Beam &  & \multicolumn{3}{c|}{Target} & \multicolumn{3}{c|}{Recoil} & \multicolumn{9}{c}{Target-Recoil}\\
\mr
 & & & & & $x^{\,\prime}$ & $y^{\,\prime}$ & $z^{\,\prime}$ & $x$ & $y$ & $z$ & $x$ & $y$ & $z$ & $x$ & $y$ & $z$\\
 & & $x$ & $y$ & $z$ & & & & $x^{\,\prime}$ & $x^{\,\prime}$ & $x^{\,\prime}$ & $y^{\,\prime}$ & $y^{\,\prime}$ & $y^{\,\prime}$ & $z^{\,\prime}$ & $z^{\,\prime}$ & $z^{\,\prime}$\\
\br
 & & & & & & & & & & & & & & & &\\
 & \rb{$\sigma_0$} & & \rb{$T$} & & & \rb{$P$} & & \rb{$T_{x^\prime}$} & & \rb{$L_{x^\prime}$} & & \rb{$\hat{\Sigma}$} & & \rb{$T_{z^\prime}$} & & \rb{$L_{z^\prime}$}\\
\mr
 & & & & & & & & & & & & & & & &\\
 \rb{$P_T$} & \rb{$\Sigma$} & \rb{$H$} & \rb{$\hat{P}$} & \rb{$G$} &
 \rb{$O_{x^\prime}$} & \rb{$\hat{T}$} & \rb{$O_{z^\prime}$} &
 \rb{$\tilde{L}_{z^\prime}$} & \rb{$\tilde{C}_{z^\prime}$} &
 \rb{$\tilde{T}_{z^\prime}$} & \rb{$\tilde{E}$} & & \rb{$\tilde{F}$} & \rb{$\tilde{L}_{x^\prime}$} & \rb{$\tilde{C}_{x^\prime}$} &
 \rb{$\tilde{T}_{x^\prime}$}\\
\mr
 & & & & & & & & & & & & & & & &\\
 \rb{$P_\odot$} & & \rb{$F$} & & \rb{$E$} & \rb{$C_{x^\prime}$} & &
 \rb{$C_{z^\prime}$} & & \rb{$\tilde{O}_{z^\prime}$} & &
 \rb{$\tilde{G}$} & & \rb{$\tilde{H}$} & & \rb{$\tilde{O}_{x^\prime}$} &\\
\mr
\br
\end{tabular}
\end{center}
\end{table}

The polarization observables that can be extracted from
experiments in single-pseudoscalar photoproduction are
summarised in Table~\ref{Table:PolarizationExperiments}. Since some
observables can be extracted from more than one experiment, this will
allow multiple cross checks of measurements with different
systematics. For example, the beam asymmetry,~$\Sigma$, can be
extracted in a single-polarization experiment with a
linearly-polarized beam and an unpolarized target, but it can also be extracted in a
double-polarization experiment with an unpolarized beam and a
transversely-polarized target and measuring the recoil polarization
$P_{y^{\,\prime}}$.

\subsubsection*{Major Experimental Facilities}
Photo- and electroproduction experiments have been performed for 
decades at many different laboratories. However, high-statistics results and a
large variety of single- and double-polarization observables have
become available only recently.

The CEBAF Large Acceptance Spectrometer (CLAS) at Jefferson Lab can be operated with electron
beams and with energy-tagged photon beams. It utilizes information from 
a set of drift chambers in a toroidal magnetic field and time-of-flight
information to detect and reconstruct charged particles. Each of the six 
drift-chamber sectors is instrumented as an independent spectrometer
with 34~layers of tracking chambers allowing for the full reconstruction 
of the charged particle 3-momentum vectors. A detailed description of 
the spectrometer and its various detector components is given 
in~\cite{Mecking:2003zu}. Since CLAS has only a very limited
photon-detection coverage, an undetected neutral particle is inferred 
through the overdetermined kinematics, making use of the good 
momentum and angle resolution of $\Delta p/p\approx 1\,\%$ and
$\Delta\theta\approx 1$-$2^\circ$, respectively. Recent double-polarization 
measurements in photoproduction have been performed utilizing a frozen-spin 
butanol target for reactions off the proton and a polarized solid HD~target for reactions off the neutron. 
Relevant results from electroproduction experiments are given in~\cite{Thompson:2000by,
Joo:2001tw,Ripani:2002ss,Joo:2003uc,Joo:2004mi,Egiyan:2006ks,Ungaro:2006df,
Denizli:2007tq,Park:2007tn,Biselli:2008aa,Fedotov:2008aa,Carman:2009fi,Park:2012rn,Mokeev:2012vsa}
and results from photoproduction experiments can be found in~\cite{Bradford:2005pt,
Bradford:2006ba,Hleiqawi:2007ad,Dugger:2007bt,Battaglieri:2008ps,Battaglieri:2009aa,
Dugger:2009pn,Williams:2009aa,Williams:2009ab,Williams:2009yj,McCracken:2009ra,Dey:2010hh}.

At ELSA, different
experimental setups were used to extract cross section data and
further polarization observables for several photo-induced reactions using tagged photons. 
In 2001, the CB-ELSA detector recorded events to analyze the reactions 
$\gamma p\to p\pi^0$~\cite{Bartholomy:2004uz,vanPee:2007tw}, 
$\gamma p\to p\eta$~\cite{Crede:2003ax,Bartholomy:2007zz}, 
$\gamma p\to p\pi^0\pi^0$~\cite{Thoma:2007bm}, $\gamma p\to 
p\pi^0\eta$~\cite{Horn:2007pp,Horn:2008qv}, and $\gamma p\to
p\pi^0\omega$~\cite{Junkersfeld:2007yr}. The original experiment
consisted of the Crystal Barrel (CB) calorimeter with its 1380 CsI(Tl)
crystals covering 97.8\,\% of~$4\pi$~\cite{Aker:1992ny}. The CB
calorimeter had an excellent photon-detection efficiency and covered
polar angles from $12^\circ$ to $168^\circ$. Charged reaction products
were detected in a three-layer scintillating fiber detector which
surrounded the 5~cm long liquid hydrogen target. More information about
the experiment can be found in~\cite{vanPee:2007tw}. 

The experimental setup was later modified and in a series of
measurements in 2002/2003, utilized a combination of the CB
calorimeter and the BaF$_2$~TAPS detector in the forward
direction. The experiment covered almost the full $4\pi$ solid 
angle. The CB~detector in its CBELSA/TAPS
configuration consisted of 1290~CsI(Tl) crystals, which had
a trapezoidal shape and were oriented toward the target, with a
thickness of $16~X_R$, and covered the polar angles from $30^\circ$ 
to $168^\circ$. TAPS consisted of 528 hexagonal BaF$_2$~crystals
with a length of approximately $12~X_R$, and covered the polar angles
from $5^\circ$ to $30^\circ$. Both calorimeters covered the full azimuthal
circle. TAPS was configured as a hexagonal wall and served as the
forward endcap of the CB. Forward-going protons were detected by
plastic scintillators (5~mm thick) located in front of each TAPS
module; the other protons were detected by the three-layer
scintillating fiber detector. Results of the CBELSA/TAPS setup on
measurements off the proton can be found in~\cite{Elsner:2007hm,Castelijns:2007qt,
Klein:2008aa,Nanova:2008kr,Elsner:2008sn,Gutz:2009zh,Crede:2009zzb,Sparks:2010vb,Crede:2011dc,
Ewald:2011gw} and off the neutron in~\cite{Jaegle:2008ux,Jaegle:2010jg,Jaegle:2011sw}. 
Results from a modified CBELSA/TAPS setup include a polarized
frozen-spin butanol target~\cite{Thiel:2012yj}.

The SAPHIR detector~\cite{Schwille:1994vg} was a multi-purpose
magnetic spectrometer at ELSA with a large angular acceptance
consisting of a tagging facility, drift chambers and a scintillator
wall for triggering and time-of-flight measurements. The detector
covered the full polar angular range from $0^\circ$ to $180^\circ$,
but the accepted solid angle was limited to approximately $0.6\times
4\pi$~sr due to the magnet pole pieces. The photon flux was measured
using the tagging system and a photon-veto counter downstream of the
detector. Relevant results can be found in~\cite{Tran:1998qw,
Glander:2003jw,Barth:2003bq,Barth:2003kv,Lawall:2005np,Wu:2005wf}.

At the recently upgraded Mainz Microtron (MAMI-C), an experimental
setup utilizing a combination of the NaI(Tl) Crystal Ball and the BaF$_2$
TAPS multiphoton spectrometers has recorded high-quality data 
for several years. While the NaI(Tl)~crystals are arranged in two hemispheres
which cover 93\,\% of the $4\pi$~solid angle, TAPS subtends the full
azimuthal range for polar angles from $1^\circ$ to $20^\circ$. Since
the TAPS calorimeter was installed 1.5~m downstream of the Crystal
Ball center, the resolution of TAPS in the polar angle~$\theta$ was
better than~$1^\circ$. For an electron beam energy of 1508~MeV, a
tagger channel in this experiment has a width of about 2~MeV at
1402~MeV and about 4~MeV at 707~MeV (the $\eta$-production 
threshold). Relevant results from MAMI including the Crystal Ball 
experiment can be found in~\cite{Krusche:1995nv,Beck:2006ye,Weis:2007kf,Ahrens:2007zzj,
Krambrich:2009te,McNicoll:2010qk,Kashevarov:2010gk,Zehr:2012tj,Kashevarov:2012wy}.

The GRAAL facility was located at
the European Synchrotron Radiation Facility (ESRF) in Grenoble, France. 
For a detailed description of the facility, we refer to~\cite{Bartalini:2005wx}. 
The tagged and polarized $\gamma$-ray beam was produced by Compton 
scattering of laser photons off the 6~GeV electrons circulating in the 
storage ring. The photon energy was provided by an internal tagging 
system consisting of silicon microstrips for the detection of the
scattered electron and a set of plastic scintillators for Time-of-Flight 
(ToF) measurements~\cite{Bartalini:2007fg}. At larger angles, the photons
from the decay of a neutral meson were detected in a BGO calorimeter
made of 480~crystals. At forward angles, the photons could be detected
in a lead-scintillator sandwich ToF wall, but the detector was not
used in all analyses, thus limiting somewhat the forward acceptance. 
Results on cross section measurements were recently reported for 
$p\pi^0$~\cite{Bartalini:2005wx}, $p\eta$~\cite{Bartalini:2007fg,
Renard:2000iv}, $p\omega$~\cite{Ajaka:2006bn},
$p\pi^0\pi^0$~\cite{Assafiri:2003mv}, and $n\pi^0\pi^0$~\cite{Ajaka:2007zz}. 
Results on beam asymmetries for a large variety of reactions can be found 
in~\cite{Bartalini:2005wx,Ajaka:2006bn,Assafiri:2003mv,Ajaka:2007zz,
Bartalini:2002cj,Hourany:2005wh,Lleres:2007tx, Lleres:2008em,Ajaka:2008zz,Fantini:2008zz,DiSalvo:2009zz}.

At the SPring-8/LEPS facility, the $\gamma$~beam was produced by
backward-Compton scattering of laser photons off electrons with an
energy of 8~GeV. The liquid hydrogen target had a length of 16.5~cm 
and data were accumulated with about $1.0\times 10^{12}$~photons 
at the target. Charged particles were detected by using the LEPS magnetic
spectrometer with an angular coverage of about $\pm 20^\circ$ and $\pm
10^\circ$ in the horizontal and vertical directions, respectively. Relevant results 
are published in~\cite{Mibe:2005er,Sumihama:2005er,Sumihama:2009gf}.

\subsubsection{Baryon Production in the Decay of Heavy Mesons}

\begin{table}[b]
\begin{center}
\caption{\label{Table:BranchingRatios} Selected branching ratios of heavy
  mesons for decays into a baryon, an anti-baryon, and mesons
  (Particle Data Group~\cite{Beringer:1900zz}).}
\vspace{1mm}
\begin{tabular}{l|c|c|c|c}
\br
 & $J/\psi$ & $\psi(2S)$ & $\psi(3770)$ & $B^0$\\
\mr
$p\bar{p}\pi^0$ & $(1.19\pm 0.08)\times 10^{-3}$ & $(1.50\pm 0.08)\times 10^{-4}$ & $< 1.2\times 10^{-3}$ &\\
$p\bar{p}\pi^+\pi^-$ & $(6.0\pm 0.5)\times 10^{-3}$ & $(6.0\pm 0.4)\times 10^{-4}$ & $< 5.8\times 10^{-4}$ & $< 2.5\times 10^{-4}$\\
$p\bar{p}\eta$ & $(2.00\pm 0.12)\times 10^{-3}$ & $(5.7\pm 0.6)\times 10^{-5}$ & $< 5.4\times 10^{-4}$ &\\
$p\bar{p}\omega$ & $(1.10\pm 0.15)\times 10^{-3}$ & $(6.9\pm 2.1)\times 10^{-5}$ & $< 2.9\times 10^{-4}$ &\\
$p\bar{n}\pi^-$ & $(2.12\pm 0.09)\times 10^{-3}$ & $(2.48\pm 0.17)\times 10^{-4}$ & &\\
\mr
$\Lambda\bar{\Lambda}\pi^0$ & $< 6.4\times 10^{-5}$ & $< 1.2\times 10^{-4}$ & &\\
$\Lambda\bar{\Lambda}\eta$ & $(2.6\pm 0.7)\times 10^{-4}$ & $< 4.9\times 10^{-5}$ & &\\
$pK^-\bar{\Lambda}$ & $(8.9\pm 1.6)\times 10^{-4}$ & & &\\
$pK^-\bar{\Sigma}^0$ & $(2.9\pm 0.8)\times 10^{-4}$ & & &\\
\br
\end{tabular}
\end{center}
\end{table}

The decays of charmonium $(c\bar{c})$ mesons have proven to be a good
laboratory in recent years for studying not only excited nucleon states, 
but also excited hyperons, such as $\Lambda^\ast$, $\Sigma^\ast$, and 
$\Xi^\ast$~states. The production of excited baryons in $J/\psi$~decays 
has a particular advantage since it serves as an isospin filter. The $\pi 
N$~system in decays such as $J/\psi\to\bar{N}N\pi$ is limited to 
isospin~$\frac{1}{2}$. Excited $\Delta$~states are thus excluded, but all excited 
nucleons which are accessible in photo- and electroproduction experiments 
can also be studied in charmonium decays. Table~\ref{Table:BranchingRatios}
shows a selection of branching ratios of heavy mesons into a baryon,
an anti-baryon and mesons. In addition to the $J/\psi$, the $\psi(2S)$
can be used to study excited baryons. Further heavy mesons, e.g. the
$\psi(3770)$ and $B$~mesons, have been observed to decay into
$p\bar{p}$ + mesons, but the branching ratios are small and the
available statistics not yet sufficient to allow partial wave analyses.

Excited nucleon states in $J/\psi$ and $\psi(2S)$~decays have mostly
been studied at the BEijing Spectrometer (BES). The upgraded BES\,III
spectrometer is a general purpose solenoidal detector at the Beijing
Electron Positron Collider~II operating in the energy range 2-4.6~GeV
with a reported peak luminosity of $6\times 10^{32}$~cm$^{-2}$\,s$^{-1}$ 
at a beam energy of 1.88~GeV~\cite{Zhao:2012zz}. Charged particles are 
observed in a small-cell, helium-based drift chamber with a momentum
resolution of 0.5\,\% at $p =1$~GeV. An electromagnetic CsI\,(Tl)
calorimeter allows the detection and reconstruction of photons with an
energy resolution of 2.5\,\% in the barrel and 5\,\% in the endcaps. 
Additional particle identification is based on a Time-of-Flight (ToF) 
system which provides a $2\sigma$~$K/\pi$ separation for momenta
up to about 1~GeV. The BES\,III detector is described in more details
in~\cite{Ablikim:2009aa}. Results relevant to this review are given
in~\cite{Bai:2001ua,Ablikim:2004ug,Ablikim:2005ir,Ablikim:2006aha,
Ablikim:2007ah,Ablikim:2007ac,Ablikim:2009iw,Liang:2012zz,Ablikim:2012,Ablikim:2012qn}.

The reaction $\psi(2S)\to p\bar{p}\pi^0$ was also studied by the Mark-II
collaboration~\cite{Franklin:1983ve} and more recently at CLEO~\cite{Alexander:2010vd}. 
The CLEO detector has been discussed in Section~\ref{Section:HeavyBaryonExperiments}. 
The decay of the $\psi(2S)$ provides a larger phase space than the 
decay of the $J/\psi$, which limits the $N^\ast$~search in the 
$p\bar{p}\pi^0$~channel to states with $M_{N^\ast} \approx 2.1$~GeV.

\subsection{Partial-Wave Analyses}
The recently accumulated high-statistics data sets of
electromagnetically-induced meson production reactions provide an
unprecedented opportunity to obtain a better understanding of the
properties of excited baryons. Owing to the broad and overlapping
nature of these resonances, amplitude or partial-wave analyses
need to be performed to extract $N^\ast$~parameters from the data. 
While such approaches have been well developed in the $\Delta$~region, 
the situation is very complicated at higher energies. Many open channels 
need to be considered and any reliable extraction of resonance properties 
must be based on a coupled-channel approach. Several groups have
significantly contributed to our understanding of baryon resonances,
but a comprehensive PWA based on a larger database of observables has
been performed only at very few institutions. These approaches are
briefly discussed in this section.

PWA formalisms have been developed at several places using different
techniques to extract nucleon resonance parameters. A detailed discussion 
of the theoretical approaches goes beyond the scope of this review, but 
a few general remarks are appropriate. In most coupled-channel analyses, instead of using partial wave amplitudes, 
the database of scattering observables, such as differential and total
cross sections as well as polarization observables, is fitted directly. Masses and widths of resonances
can be determined from Breit-Wigner parametrizations, but more than
one possible Breit-Wigner pole parametrization exists. The $K$-matrix
formalism is used for overlapping resonances in the same partial wave
and provides an elegant way of ensuring that the amplitudes preserve unitarity.
Based on properties of the trace of the $T$-matrix and the associated $K$-matrix, Ceci {\it et al.}~\cite{Ceci:2006jj} proposed an alternative method for extracting resonance parameters from a multi-channel fit to scattering data, without relying on Breit-Wigner parametrizations. However, a recent study
did not find a simple connection between $T$-matrix poles and the
associated $K$-matrix poles~\cite{Workman:2008iv}. The authors also 
note that $K$-matrix poles are not generally independent of background
contributions. The PDG~\cite{Beringer:1900zz} typically gives the $T$-matrix 
poles in the particle listings.

\subsubsection{SAID and MAID}
The SAID (Scattering Analysis Interactive Dial-in) group maintains an
extensive database of (elastic) $\pi N$ (including $\pi d$), $K N$,
and $N N$~scattering data as well as of data on the electromagnetic
production of a single pseudoscalar meson~\cite{SAID}. In addition to
the data, the SAID website also provides access to partial-wave
amplitudes extracted, and to predictions of a large variety of
energy-dependent observables. The most recent solution for
$\pi$~photoproduction including the full SAID database for single
$\pi$~production is discussed in~\cite{Workman:2011vb}.
Baryon masses and widths are determined from a Breit-Wigner
parametrization of the resonance contributions using $\pi
N$~scattering data. The helicity amplitudes $A_{1/2}$ and $A_{3/2}$
are extracted based on the masses, widths, and $\pi N$ branching
fractions determined in earlier analyses of $\pi N$~elastic
scattering data~\cite{Arndt:2006bf}. Progress was reported on the
extraction of photo-decay couplings, but no new resonances have been
proposed. The most recent solution for $\eta$~photoproduction 
was discussed in~\cite{McNicoll:2010qk}.

The MAID partial wave analysis group maintains a similar website,
which gives predictions for multipoles, amplitudes, cross sections,
and polarization observables for photo- and electroproduction in the 
energy range from the pion threshold up to $W=2$~GeV~\cite{MAID}. 
A recent comprehensive study of the world data of pion photo- and
electroproduction was reported in~\cite{Drechsel:2007if} using the
unitary isobar model MAID2007. In addition to resonance properties, 
the longitudinal and transverse helicity amplitudes of nucleon
resonance excitations for all the 4-star resonances below $W = 2$~GeV
were presented. The group has developed similar models for $\eta$ and
$\eta^{\,\prime}$~production. The most recent solution for
$\eta$~photoproduction was discussed in~\cite{McNicoll:2010qk}.
Amplitudes for $K$~photoproduction in a multipole approach were
presented in~\cite{Mart:2006dk}.

Neither MAID nor SAID treats the important
$\pi\pi$~production channels explicitly and their results need to be
re-examined. Nevertheless, the databases and model predictions represent an
invaluable tool for the field.

\subsubsection{The Excited Baryon Analysis Center at Jefferson Lab}
The Excited Baryon Analysis Center (EBAC) at Jefferson Lab was established 
in 2006 with the goal of performing a dynamical coupled-channels analysis of 
the world $\pi N$ and $\gamma^\ast N$~scattering data in order to determine 
meson-baryon partial-wave amplitudes and to extract $N^\ast$~parameters 
from these amplitudes. The developed dynamical coupled-channel (DCC) reaction 
model is based on an energy-independent Hamiltonian formulation~\cite{Matsuyama:2006rp} 
and satisfies the essential two- and three-body unitarity conditions. Since the 
bare $N^\ast$~states are defined as the eigenstates of the Hamiltonian in which 
the couplings to the meson-baryon continuum states are turned off, the extracted
bare states can be related to the hadron states from constituent quark models 
and those from Dyson-Schwinger approaches (static hadron models). Within the 
EBAC model, the bare $N^\ast$~states become resonance states through the 
reaction processes. Thus, the model explicitly allows one to distinguish between the couplings 
of the bare $N^\ast$~states from the meson-cloud (meson-baryon dressing) 
effects. This important and distinct feature will be discussed further in 
Section~\ref{Section:Electroproduction}. A brief overview of EBAC results is 
given in~\cite{Kamano:2012it}.

First results from the DCC model were reported on the pion-induced
reactions $\pi N\to\pi N$~\cite{JuliaDiaz:2007kz}, $\pi N\to\eta
N$~\cite{Durand:2008es}, and $\pi N\to\pi\pi N$~\cite{Kamano:2008gr}. 
The application of the model was also discussed for the analysis of
$\pi$~photoproduction~\cite{JuliaDiaz:2007fa} as well as
electroproduction~\cite{JuliaDiaz:2009ww}, which included the
channels~$\gamma^\ast N,~\pi N,~\eta N$, and $\pi\pi N$. A
simultaneous analysis of both single- and double-pion photoproduction 
is published in~\cite{Kamano:2009im}. The authors note that the analysis 
of the single-pion production reactions alone is not enough to determine
the amplitudes associated with the electromagnetic interactions. Both
channels are required for the extraction of reliable information on
$N^\ast$~states below $W = 2$~GeV.

Several other groups have developed and used dynamical reaction
models. We list here the approaches mentioned in the minireview on
$N^\star$~and $\Delta$~resonances from the latest edition of the RPP:
Argonne-Osaka~\cite{Sato:2009de}, Bonn-J\"ulich~\cite{Doring:2009bi,Doring:2009yv},
Dubna-Mainz-Taipeh~\cite{Chen:2007cy}, and Valencia~\cite{Sarkar:2004jh}.

\subsubsection{The Gie\ss en Coupled-Channel Analysis}
The group at the university in Gie\ss en, Germany, has studied $\pi$- and
$\gamma$-induced reactions for the final states $\gamma N,~\pi N,~\pi\pi
N,~\eta N,$ and $\omega N$ within a coupled-channel phenomenological Lagrangian 
approach in the energy region from the pion threshold up to 2~GeV. In
the model, the Bethe-Salpeter equation is solved in the $K$-matrix
approximation in order to obtain the multi-channel scattering $T$-matrix. 
More details on the model are discussed in~\cite{Penner:2002ma,
Penner:2002md,Shklyar:2004dy}, for instance. Updates of the PWA have
been reported regularly for several meson-production channels,
e.g. more recently for $\omega N$~\cite{Shklyar:2004ba}, 
$K\Lambda$~\cite{Shklyar:2005xg}, $K\Sigma$~\cite{Shyam:2009za}, 
and $\eta N$~\cite{Shklyar:2012js}.

\subsubsection{The Bonn-Gatchina Partial-Wave Analysis}
Among all PWA groups, the Bonn-Gatchina group uses the largest
experimental database in their multichannel approach including results
from multiparticle final states like $p\pi^0\pi^0$ and $p\pi^0\eta$. 
To retain all correlations among the five independent kinematic variables for
the latter reactions, these data are included in event-based likelihood 
fits. The analysis uses a $K$-matrix parametrization deriving the
background terms from phenomenology rather than from a chiral
Lagrangian as in the above Gie\ss en model. The PWA method is
described in more details in~\cite{Anisovich:2004zz,Anisovich:2006bc}
and the main findings from the different analyses are discussed 
in~\cite{Anisovich:2009zy,Anisovich:2010an,Anisovich:2011ye}
for a large variety of reactions. The group also maintains a website 
which provides the Bonn-Gatchina $\pi N$~partial wave amplitudes and
photoproduction multipoles, as well as predictions for many yet to be
determined observables~\cite{Bonn-Gatchina}.

The Bonn-Gatchina group has reported regularly on systematic searches
for new baryon resonances. The latest summary of their results is
presented in~\cite{Anisovich:2011fc} including a list of all data used
in the analysis. The group has proposed several new states and
strongly inspired the recent modifications in the RPP~\cite{Beringer:1900zz}.
\section{Recent Results in the Spectroscopy of Light-Quark Baryons}
\subsection{\label{Section:PionNucleon}Pion-Nucleon Scattering Experiments}
\begin{figure}
\caption{\label{Figure:RPP2012-pip_pid}(Colour online)  Total and elastic
  cross sections for $\pi^\pm p$ and $\pi^\pm d$ (total only) collisions from \url{http://pdg.lbl.gov/current/xsect/}. 
  (Courtesy of the COMPAS Group, IHEP, Protvino, August 2005)}\vspace{0.2cm}
\includegraphics[width=0.88\textwidth]{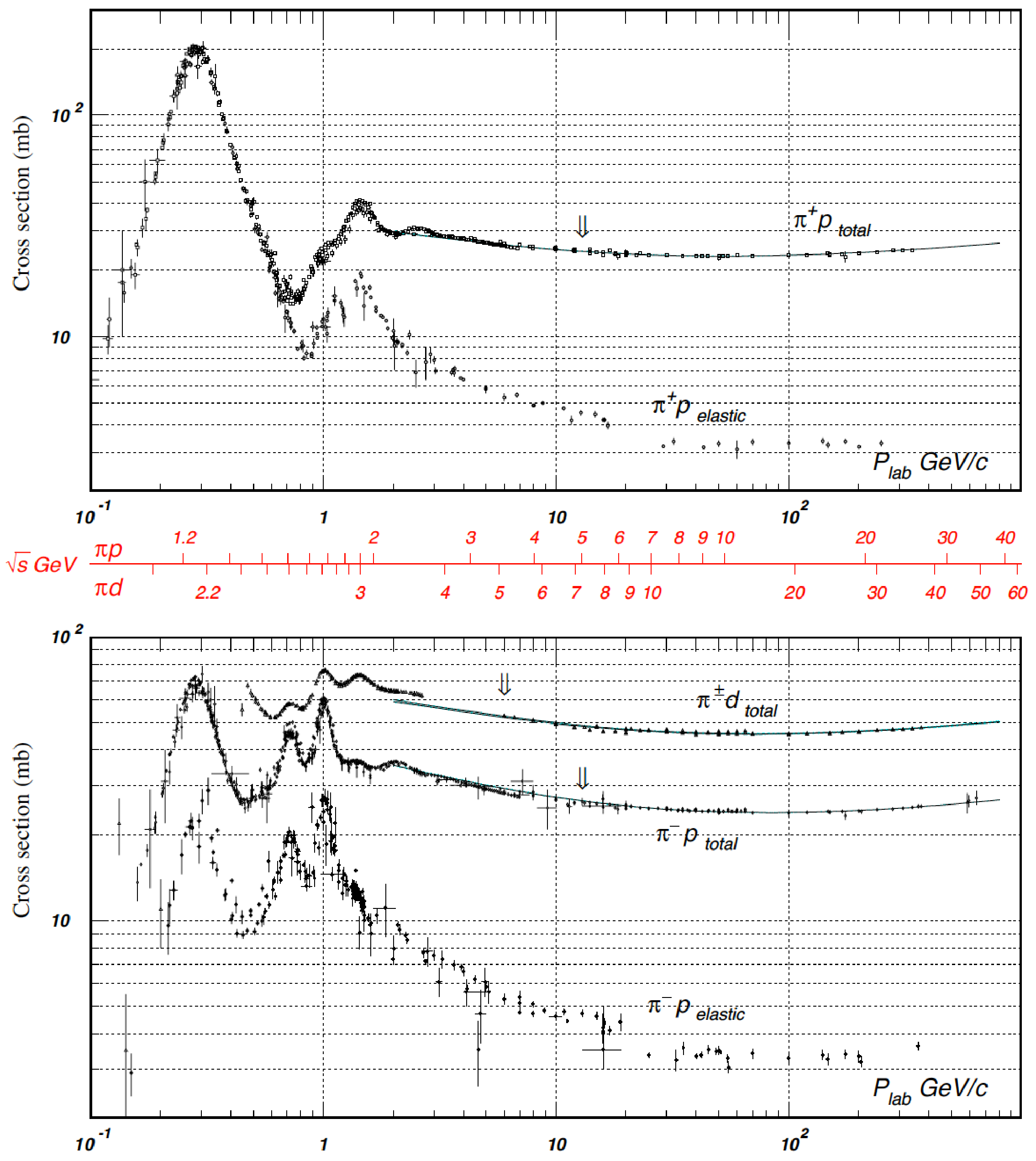}
\end{figure}

The most prominent signal in the $N\pi$~total cross section is that of the $\Delta(1232)$ 
which can be seen in Figure~\ref{Figure:RPP2012-pip_pid} at $p_{\pi^\pm}
\approx 300$~MeV. The higher-mass structures consist of more than one 
resonance, and using a simple Breit-Wigner formula which describes a resonance 
enhancement with spin~$J$ and mass~$M$ in the total cross section for 
meson-nucleon scattering~\cite{Hey:1982aj}
\begin{eqnarray}
\Delta\sigma\,=\,4\pi \biggl(\frac{\hbar}{p_{\rm c.m.}}\biggr)^2 \,(J \,+\,\frac{1}{2})\,x\,/\,(1\,+\,\epsilon^2) 
\end{eqnarray}
does not yield reliable information on resonance properties. Here, $\epsilon=
2(M - E_{\rm c.m.})/\Gamma_{\rm tot}$ and $x=\Gamma_{\rm el}/\Gamma_{\rm tot}$ 
denotes the resonance elasticity, which decreases strongly with mass. For this 
reason, the  total cross section becomes almost featureless, as can 
be seen in Figure~\ref{Figure:RPP2012-pip_pid}. An increasing number of inelastic 
channels becomes accessible with increasing resonance mass. The structures seen 
in the graph have been named first, second, third, etc. resonance
region. 

A large amount of data including polarization observables has been accumulated 
over the decades on the elastic $\pi N$~channel and on the charge-exchange 
reaction, $\pi^- p\to n\pi^0$ (Section~\ref{Section:HadronicProbes}). Much fewer 
statistics are available for many other $\pi$-induced reactions. After 1980, studies 
of several inelastic channels were performed by the Crystal Ball collaboration at BNL 
in reactions such as $\pi^- p\to\eta n$~\cite{Kozlenko:2003hu,Prakhov:2005qb}, 
$\pi^- p\to\pi^0\pi^0 n$~\cite{Craig:2003yd,Prakhov:2004zv}, 
and $\pi^- p\to\pi^0\pi^0\pi^0 n$~\cite{Starostin:2003cc}. A review of the 
experimental situation on $\pi^- p\to\eta n$ before the recent BNL measurements
is given in~\cite{Clajus:1992dh}. In 2009, the EPECUR collaboration started 
data-taking at the ITEP proton synchrotron with the goal of searching for narrow 
states in the second and third resonance regions via $\pi p$~elastic scattering 
and $K\Lambda$~production. The first high-statistics results on the differential 
cross section for $\pi^- p$~elastic scattering have been published for the mass 
range (1650 - 1735)~MeV with an excellent resolution of about 
1~MeV~\cite{Alekseev:2012zu}.

As discussed in Section~\ref{Section:pwa1}, 
all information on the $N^\ast$ and $\Delta$
states listed by the PDG was determined from PWAs of $\pi N$ total, elastic, 
and charge-exchange scattering data. Usually listed as the most comprehensive 
are the $\pi N$~analyses performed by the Karlsruhe-Helsinki~\cite{Hohler:1983},
the Carnegie-Mellon-Berkeley~\cite{Cutkosky:1979fy}, and the George 
Washington~\cite{Arndt:2006bf} groups. The first two analyses used only
data from the experiments performed before 1980, whereas the latter analysis 
represents the most recent study of $\pi N$~scattering data (published in 2006). 
It includes not only an extended database with the most recent polarization 
data, but also data from inelastic channels like $\pi N\to \eta N$, needed for a better 
description of the $\pi N$~$S$-wave.

All 4-star resonances were confirmed in the most recent analysis~\cite{Arndt:2006bf}. 
Surprisingly, however, all but two 3-star resonances and several with a
fewer-star assignment were either not observed or observed with different properties. 
For example, the $N(2000)\,F_{15}$~nucleon state was found, but with a 
mass closer to 1800~MeV. In the latest edition of the
RPP~\cite{Beringer:1900zz}, a further 2-star state with $J = \frac{5}{2}$, 
$N(1860)\,\frac{5}{2}^+$, has been added~(Table~\ref{Table:BaryonSummary}). 
No evidence was found for the two 3-star states, $N(1700)\,\frac{3}{2}^-$ and 
$N(1710)\,\frac{1}{2}^+$. However, a new structure was observed in the 
$H_{1,11}$~wave with a mass around 2200~MeV. In the 
$\Delta$~sector, only a single $P_{31}$~state was found with a pole 
position more in line with the $(\ast)~\Delta(1750)\,\frac{1}{2}^+$ than with
the $(\ast\,\ast\,\ast\,\,\ast)~\Delta(1910)\,\frac{1}{2}^+$. Below 2~GeV, three additional states, 
$(\ast\,\ast)~\Delta(1900)\,\frac{1}{2}^-$, $(\ast\,\ast\,\ast)~\Delta(1920)\,\frac{3}{2}^+$, and 
$(\ast)~\Delta(1940)\,\frac{3}{2}^-$, were absent.

As a step toward performing a complete coupled-channels analysis of 
the world database on the $\pi N$, $\gamma^\ast \to\pi N,~\eta N,~\pi
\pi N$~channels, EBAC has reported first results on the study of the 
$\pi N\to\pi N$~\cite{JuliaDiaz:2007kz} and $\pi N\to \pi\pi N$
reactions~\cite{Kamano:2008gr}. The group observes all well-established 
4-star resonances below 2~GeV. However, several of the resonances 
first announced in the late 1970s and listed by the PDG are also absent in
the EBAC results. No evidence is found for the 3-star state, $N(1700)\,\frac{3}{2}^-$,
consistent with the findings of the George-Washington group~\cite{Arndt:2006bf}.
Moreover, only one $\Delta\,\frac{1}{2}^+$~state is reported with a mass
consistent with the PDG $\Delta(1910)\,\frac{1}{2}^+$~state.

The differential cross section for the reaction $\pi^- p\to\eta n$ was
measured recently by the Crystal Ball collaboration over the full angular 
range at seven incident $\pi^-$~beam momenta from threshold to 
$p_{\pi^-} = 747$~MeV~\cite{Prakhov:2005qb}. The angular
distributions of this reaction are reported to be $S$-wave dominated.
The total cross section was obtained by fully integrating over the
differential cross sections and shows a rapid rise at threshold, as
expected for $S$-wave-dominated production. A coupled-channel 
analysis of $\eta$~production by the Gie\ss en group shows that the 
main effects at $\sqrt{s} < 2$~GeV come from the three resonances,
$N(1535)\,\frac{1}{2}^-$, $N(1650)\,\frac{1}{2}^-$, and $N(1710)\,\frac{1}{2}^+$~\cite{Shklyar:2012js}. 
The well-established higher-spin resonance, $N(1520)\,\frac{3}{2}^-$, is 
required to reproduce the correct shape of the cross section, but the 
second $\frac{3}{2}^-$~state, $N(1900)$, is not consistent with the 3-star 
state, $N(1700)\,\frac{3}{2}^-$, currently listed by the PDG~\cite{Beringer:1900zz}.
Results of the EBAC group confirm the dominance of the $N(1535)\,\frac{1}{2}^-$
resonance in the reaction $\pi^- p\to n\eta$~\cite{Durand:2008es}. 
The authors also discuss the non-negligible roles of the two states, 
$N(1440)\,\frac{1}{2}^+$ and $N(1720)\,\frac{3}{2}^+$. It is worth noting though 
that the scarce experimental situation for most of the $\pi$-induced 
reactions does not provide enough information to reliably constrain 
resonance parameters.

An analysis of the reaction $\pi^- p\to n\pi^0\pi^0$ reveals no 
evidence for the strong production of a low-mass scalar meson, but 
double-pion production appears to become dominated by sequential 
$\pi^0$~decays through the $\Delta(1232)$~resonance as the beam 
momentum increases~\cite{Craig:2003yd,Prakhov:2004zv}. The results
are consistent with a recent PWA which observed large contributions
from $N(1440)\,\frac{1}{2}^+$ interfering with $N(1535)\frac{1}{2}^-$ and 
$N(1520)\,\frac{3}{2}^-$~\cite{Sarantsev:2007aa}. We will discuss resonance
contributions in double-pion production in more details in 
Section~\ref{Section:MultiMeson}. 

\subsection{\label{Section:KaonNucleon}Kaon-Nucleon Scattering Experiments}
The observation of the relatively long-lived $\Sigma$~baryon which can
be produced by the strong interaction in $\pi^- p\to K^+\Sigma^-$, but
decays only weakly via $\Sigma^-\to n\pi^-$ inspired the introduction
of a new quantum number: Strangeness, $S$. The $\Sigma$~hyperon with 
mass $M_{\Sigma^-}=(1197.449\pm 0.030)$~MeV could also decay via
strong interaction, but the decay into $\Lambda\pi^-$ is kinematically
forbidden because $M_\Lambda = (1115.683 \pm 0.006)$~MeV. The
``strangeness'' scheme was confirmed by the subsequent discovery of a
large number of strange particles. The nucleon isospin doublet has
strangeness~$S=0$, the isospin singlet $\Lambda$ and triplet~$\Sigma$ 
have strangeness~$S=-1$, and the isospin doublet~$\Xi$ has strangeness
$S=-2$. The new strangeness quantum number suggested an internal 
SU(3)~symmetry of the hadron spectrum and the simple quark model of
the nucleon was generalized to the flavour~SU(3) group by adding the 
strange quark,~$s$, with strangeness~$S=-1$. By 1964, ``flavour SU(3)'' 
based on the three lightest flavours of quark was established as a new 
symmetry group which naturally fits the hadrons with similar properties 
into its multiplet representations.

The ``strange'' $\Lambda^{(\ast)}$ and $\Sigma^{(\ast)}$~resonances
can of course be more easily produced and studied in $K$-induced
reactions, rather than in reactions with a non-strange beam particle. 
The most recent measurements were presented by the Crystal Ball
collaboration using the low-momentum $K^-$~beam at BNL's Alternating Gradient Synchrotron (AGS) on the reactions
$K^-\, p\to\eta\Lambda$~\cite{Starostin:2001zz}, $K^-\,p\to\pi^0
\Lambda$~\cite{Prakhov:2008dc}, $K^-\, p\to\pi^0\pi^0\Lambda$~\cite{Prakhov:2004ri}, 
$K^-\, p\to\pi^0\pi^0\pi^0\Lambda$~\cite{Borgh:2003uc}, $K^-\, p\to
\pi^0\Sigma^0$~\cite{Olmsted:2003is,Prakhov:2008dc}, $K^-\, p\to\pi^0
\pi^0\Sigma^0$~\cite{Prakhov:2004an}, and $K^-\, p\to \bar{K}^0 
n$~\cite{Prakhov:2008dc}. At SLAC, the Large Aperture Superconducting
Solenoid (LASS) spectrometer performed the last such experiments using an
intense kaon beam of 11~GeV/$c$. Evidence was reported for an
$\Omega^\ast$~resonance with a mass of $2474\pm 12$~MeV and a 
width of $72\pm 33$~MeV~\cite{Aston:1988yn}.

The threshold region in the reaction $K^-\, p\to\eta\Lambda$ is dominated 
by formation of the intermediate $(\ast\,\ast\,\ast\,\,\ast)~\Lambda(1670)\,\frac{1}{2}^-$~state, which
was discovered in 1965 at BNL as a peak in the same reaction~\cite{Berley:1965zz}.
A unitary, multi-channel analysis~\cite{Manley:2002ue} determined the mass
and width of the resonance to be $(1673\pm 2)$~MeV and $(23\pm
6)$~MeV, respectively, which are in excellent agreement with quark-model 
predictions~\cite{Koniuk:1979vy}. Isospin-$1$ states can be observed in
the reaction $K^-\, p\to\pi^0\Lambda$~\cite{Olmsted:2003is,Prakhov:2008dc}. 
It was found that the resonance, $(\ast\,\ast\,\ast\,\,\ast)~\Sigma(1385)\,\frac{3}{2}^+$ contributes,
but no evidence is observed for any of the remaining $\Sigma$~states below
1600~MeV listed by the PDG. In particular, contributions from the
state, $(\ast)~\Sigma(1580)\,\frac{3}{2}^-$, were shown to be inconsistent with 
the observed angular distributions~\cite{Olmsted:2003is}. A recent
effective Lagrangian approach~\cite{Gao:2012zh}, which includes the data 
on $K^-\,p\to\pi^0\Lambda$ from the Crystal Ball experiment~\cite{Prakhov:2008dc},
supports the existence of $(\ast\,\ast\,\ast)~\Sigma(1660)\,\frac{1}{2}^+$
but finds no evidence for $(\ast\,\ast)~\Sigma(1620)\,\frac{1}{2}^-$.

Production of two $\pi^0$~mesons in the reaction $K^-\,p\to\pi^0\pi^0
\Lambda$ was reported in~\cite{Prakhov:2004ri} to proceed dominantly
via $\Lambda^\ast\to\pi^0\Sigma^0(1385)\to\pi^0\pi^0\Lambda$. 
Contributions from the scalar meson, $f_0(600)$, were found to be 
insignificant (a broad uniform $\pi\pi$~band for the $f_0(600)$ is 
absent in the Dalitz plot) and no other $\Sigma^\ast$~intermediate
state was observed. The isospin-related reaction $K^-\,p\to\pi^0\pi^0
\Sigma^0$ was found to be dominated by the decay chain $\Sigma^\ast
\to\pi^0\Lambda(1405)\to\pi^0\pi^0\Sigma^0$ with increasing contributions 
from the $\Lambda(1520)$ toward the highest beam momentum of
750~MeV at which data were obtained~\cite{Prakhov:2004an}. The
authors determine a total cross section for $K^-\,p\to\pi^0\pi^0\Sigma^0$,
which gradually rises from about 35 to 180~$\mu$b. Such behaviour is
consistent with dominant contributions from either $\Sigma(1660)\,\frac{1}{2}^+$ 
or $\Sigma(1670)\,\frac{3}{2}^-$~\cite{Prakhov:2004an}, a 3-star and a 4-star 
state, respectively.

In the limit of massless quarks, the flavour-blindness of the QCD gluon
interactions would result in a perfect SU(3)-flavour symmetry. Neglecting 
additional Coulomb forces, the states corresponding to the $J=\frac{1}{2}$
ground-state octet would be degenerate and, so would the states of the $J=\frac{3}{2}$
ground-state decuplet. For the octet (Figure~\ref{Figure:Multiplets}~(b)), 
the mass differences between the single-strange resonances and the
nucleon are
\begin{eqnarray}
M_\Lambda\,-\,M_N\,=\,177~{\rm MeV},\qquad M_\Sigma\,-\,M_N\,=\,254~{\rm MeV}\,,
\end{eqnarray}
and similarly, the mass difference for the double-strange $\Xi$ to the
$\Lambda$ is
\begin{eqnarray}
M_\Xi\,-\,M_\Lambda\,=\,203~{\rm MeV}\,.
\end{eqnarray}
The mass differences are $\approx 200$~MeV and are much smaller
than the resonance masses themselves. The corresponding mass splittings  
in the decuplet (Figure~\ref{Figure:Multiplets}~(a)) are similar, having the values
\begin{eqnarray}
M_{\Sigma(1385)}\,-\,M_\Delta\,=\,153~{\rm MeV},\\[0.5ex]
M_{\Xi(1530)}\,-\,M_{\Sigma(1385)}\,=\,145~{\rm MeV},\\[0.5ex]
M_{\Omega^-}\,-\,M_{\Xi(1385)}\,=\,142~{\rm MeV}.
\end{eqnarray}
In reality, SU(3)-flavour symmetry is broken by the finite mass of the
strange quark. Although the multiplets exhibit an excellent isospin
symmetry because $m_u\simeq m_d\simeq 0$, the states containing
strange quarks have masses which correspond approximately to integral 
multiples of $m_s$. The mass difference between the quarks is still
much smaller than $\Lambda_{\rm QCD}$ and therefore, an approximate
SU(3)-flavour symmetry can be observed in the baryon spectrum.

\begin{figure}
\caption{\label{Figure:FlavorSymmetry} Total cross sections. Left:
  Results for single-$\eta$ production: $\pi^- p\to n\eta$~\cite{Prakhov:2005qb}, 
  $K^-\,p\to\Lambda\eta$~\cite{Starostin:2001zz}, $\gamma p\to p\eta$~\cite{Krusche:1995nv}.
  The solid line represents the SAID-FA02 solution~\cite{Arndt:2003if}. Picture taken
  from~\cite{Prakhov:2005qb}. Right: Results for $\pi^0\pi^0$~production 
  compared at incident beam momenta such that the reactions have the same 
  $m_{\rm max}(\pi^0\pi^0)$: $\pi^- p\to\pi^0\pi^0 n$~({\tiny $\bigcirc$}),
  $K^-\, p\to\pi^0\pi^0\Lambda$~({\scriptsize $\triangle$}), $K\, p\to\pi^0\pi^0\Sigma^0$~(crosses).
  Picture from~\cite{Nefkens:2002rz}.
}\vspace{0.2cm}
\includegraphics[width=0.48\textwidth,height=0.357\textwidth]{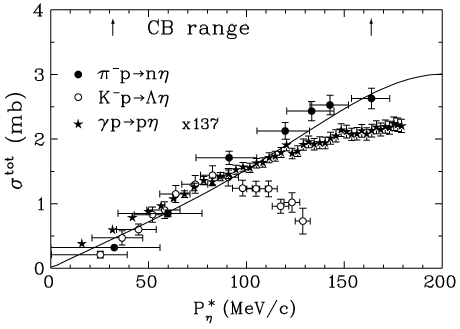}
\includegraphics[width=0.48\textwidth,height=0.35\textwidth]{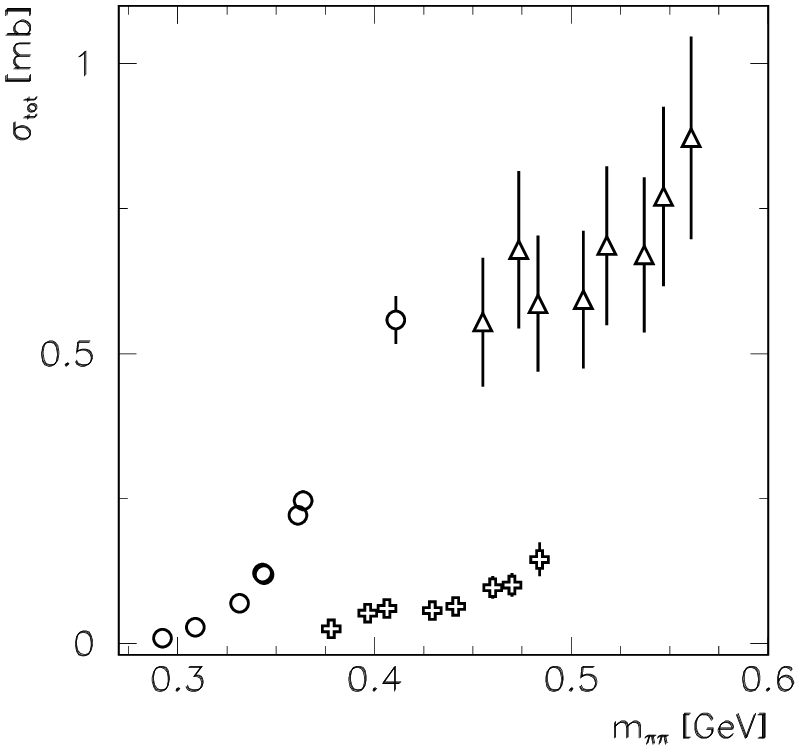}
\end{figure}

It is now generally accepted that the previously mentioned 
$\Lambda(1670)\,\frac{1}{2}^-$ is the partner of the $N(1535)\,\frac{1}{2}^-$ in the 
$S=\frac{1}{2}$ octet of the $({\bf 70},\,1_1^-)$~supermultiplet that contains
the lowest-lying baryon excitations with negative parity. Both
resonances show a sizable branching fraction for decays into $\eta$
and the corresponding baryon ground state. Interestingly, the shapes
of the cross sections of the two reactions $K^-\, p\to\Lambda\eta$ 
and $\pi^- p\to\eta n$ exhibit a remarkable similarity in the threshold
region~(Figure~\ref{Figure:FlavorSymmetry}, left) -- which may be an example of
SU(3)-flavour symmetry. The threshold of the reaction $\gamma p\to 
p\eta$ is also dominated by the $N(1535)\,\frac{1}{2}^-$~resonance, but flavour
symmetry does not apply in electromagnetic processes. However, 
$\eta$~production at threshold via $\gamma p$ and $\pi^- p$ agree
fairly well once the photoproduction cross section is multiplied by a
factor of~137, i.e. $\eta$~photoproduction is about $\alpha$ times
$\eta$~production in $\pi^- p$~reactions.

The authors of~\cite{Nefkens:2002rz} suggest that there are relationships among the processes
\begin{eqnarray}
\pi^- p\,\to\,\eta n \quad\qquad & \pi^- p\,\to\, N^\ast\,\to\,\pi^0\Delta(1232)~3/2^+\,\to\,\pi^0\pi^0 n\,,\\[0.5ex]
K^- p\,\to\,\Lambda\eta \quad\qquad & K^- p\,\to\,\Lambda^\ast\,\to\,\pi^0\Sigma^0(1385)~3/2^+\,\to\,\pi^0\pi^0\Lambda\,,\\[0.5ex]
 & K^- p\,\to\,\Sigma^\ast\,\to\,\pi^0\Lambda(1405)~1/2^-\,\to\,\pi^0\pi^0\Sigma^0\,,\\[0.5ex]
\gamma p\,\to\,\eta p \quad\qquad & \gamma p\,\to\, N^\ast\,\to\,\pi^0\Delta(1232)~3/2^+\,\to\,\pi^0\pi^0 n\,.
\end{eqnarray}
They argue that the first two $\pi^0\pi^0$~reactions have the same SU(3)-flavour structure, 
similar to single-$\eta$ production. The two resonances, $\Delta(1232)$ 
and $\Sigma(1385)^0$, are members of the same ground-state decuplet.
The two $K^-$-induced reactions, however, differ in the flavour structure
of their dominant intermediate states and thus, the cross sections should 
be quite different. Figure~\ref{Figure:FlavorSymmetry} (right) shows the
$\pi^0\pi^0$~cross section data plotted versus $m_{\rm max}(\pi^0\pi^0)$
rather than beam momentum. The total cross section for $K^-\, p\to\pi^0
\pi^0\Sigma^0$ is about a factor of five smaller than the total cross section
for $K^-\,p\to\pi^0\pi^0\Lambda$. This big difference indicates that the
$\pi^0\pi^0$~production in $K^-\, p$~reactions occurs likely via sequential 
baryon decays with very different flavour structure. They go on to argue that if the SU(2) and SU(3) Clebsch-Gordan coefficients are ignored, and these reactions proceed through the same mechanism, then their cross sections should be comparable. The mechanism they assume is that of $ f_0(600)$~production in the intermediate state.  The total cross section for $\pi^- p\to\pi^0\pi^0 n$ is similar 
to the total cross section for $K^-\, p\to\pi^0\pi^0\Lambda$. However, this similarity may be accidental, as the argument provided is not very rigorous.

The concept of flavour-symmetry may provide some guidance, but its
application is certainly limited. The symmetry is broken by the strange-light
quark mass difference and corrections in quark models can be substantial.
The PDG lists four $\Sigma$~states in the mass range 1300-1600~MeV: 
$(\ast\,\ast\,\ast\,\,\ast)~\Sigma(1385)\,\frac{3}{2}^+$, $(\ast)~\Sigma(1480)$, 
$(\ast\,\ast)~\Sigma(1560)$, and $(\ast)~\Sigma(1580)\,\frac{3}{2}^-$.  Without
doubt, the $\Sigma(1385)\,\frac{3}{2}^+$ is a member of the ground-state decuplet. 
Only two $\Lambda$~states are listed in the same mass range:
$(\ast\,\ast\,\ast\,\,\ast)~\Lambda(1405)\,\frac{1}{2}^-$ 
and $(\ast\,\ast\,\ast\,\,\ast)~\Lambda(1520)\,\frac{3}{2}^-$. 
For the first-excitation band, $({\bf 70},\,1_1^-)$, it has been well accepted that 
the two $\Lambda$~states are the members of the $^2{\bf 1}$-plet 
(\Eref{Equation:70plet}). All $\Lambda$~octet states then have 
masses close to or above 1700~MeV. In the framework of  SU(3)-flavour 
symmetry, the low masses of the remaining $\Sigma$~states rule them
out as possible partners and raise doubts about their existence. Similar
arguments have been made about the $\Sigma(1580)\,\frac{3}{2}^-$~state by the
Crystal Ball collaboration~\cite{Olmsted:2003is}. All three of the
$\Sigma$~states mentioned have also been omitted from the recent classification by Klempt 
and Richard~\cite{Klempt:2009pi}.

The spectroscopy of $\Lambda$ and $\Sigma$ resonances remains at a
standstill due to the lack of suitable $K$~beams. A minireview in the
latest version of the RPP~\cite{Beringer:1900zz} refers to the note in the 1990 edition. 
A few measurements on branching fractions have been added. Recent
classifications of the hyperon spectrum can be found in~\cite{Beringer:1900zz,
Klempt:2009pi,Melde:2008yr}.

\subsection{\label{Section:Electroproduction}Electroproduction Experiments}

\begin{figure}[t]
\caption{\label{Figure:CLAS_electroproduction}(Colour online) Left:
  CLAS differential cross sections for $\gamma^\ast p\to n\pi^+$~\cite{Park:2007tn}. 
  Upper, middle, and lower rows are for $Q^2$ = 1.72, 2.44, and 3.48~GeV$^2$, 
  respectively. The solid (dashed) curves denote results using a fixed-$t$ 
  dispersion relations (effective Lagrangian) approach~\cite{Aznauryan:2009mx}. 
  Right: Data on fully integrated $p\pi^+\pi^-$ electroproduction cross
  sections from CLAS~\cite{Ripani:2002ss}. The dashed curves represent
  a description within the framework of the reaction model in~\cite{Mokeev:2008iw}
  considering only conventional $N^\ast$ states, the solid curves 
  include an additional $(1720)\,\frac{3}{2}^+$ candidate. Pictures
  from~\cite{Aznauryan:2011ub}.}
\includegraphics[scale=0.49]{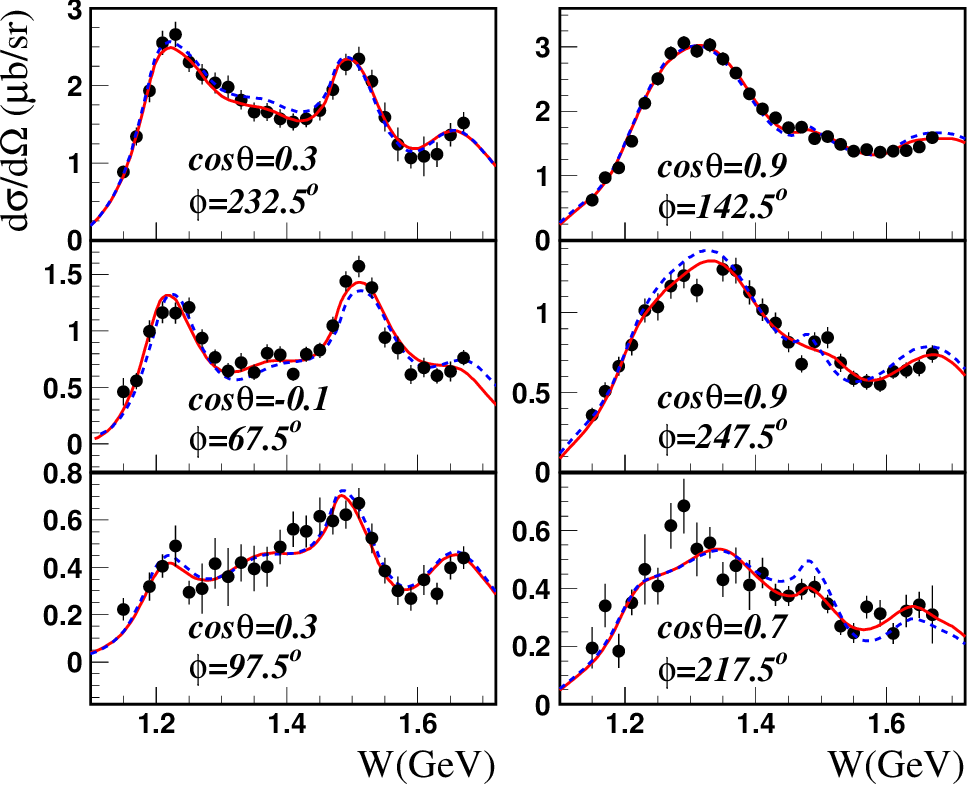}\qquad
\includegraphics[scale=0.36]{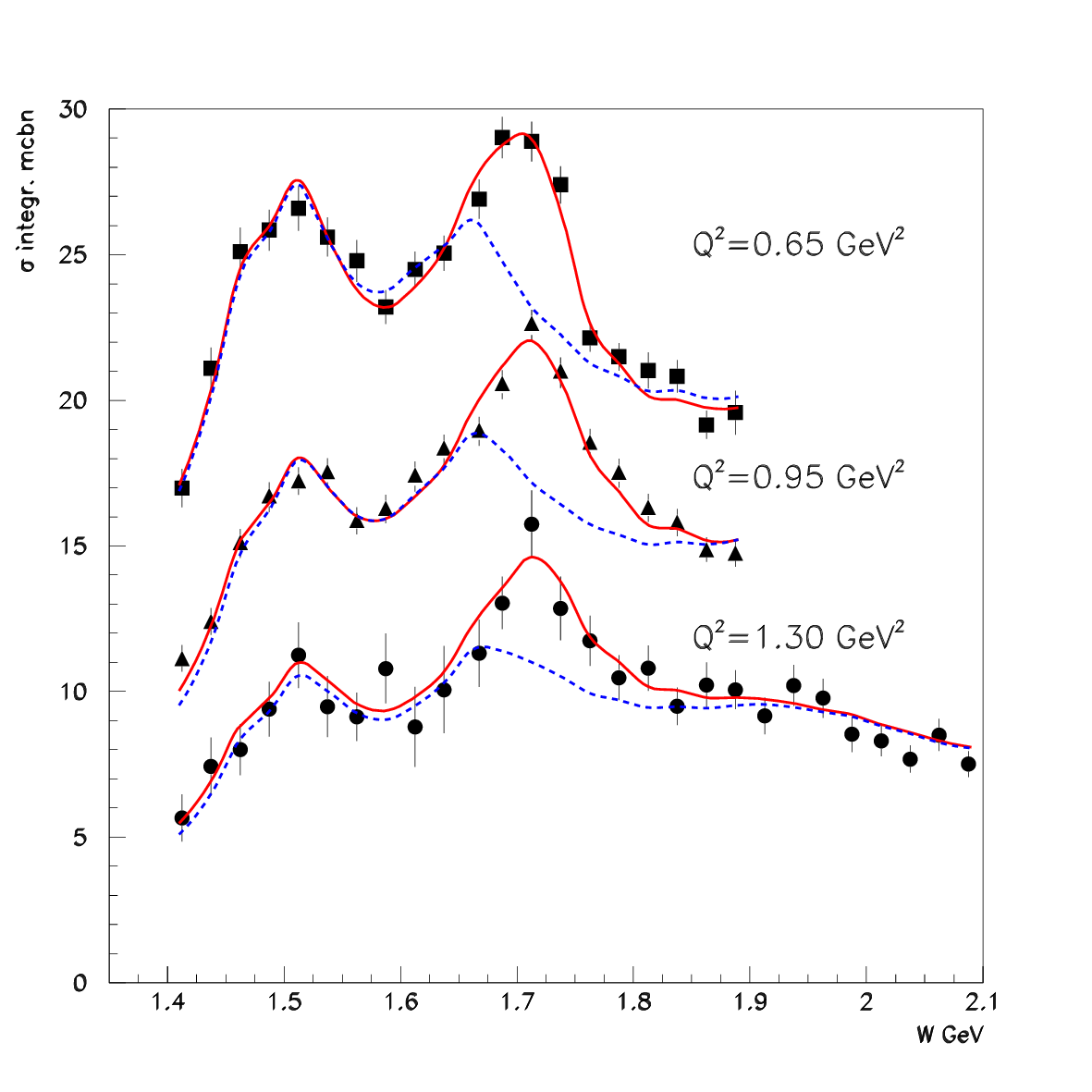}
\end{figure}

An important component of the $N^\ast$~program is the probing of resonance
transitions at different distance scales in order to study the structure of
excited nucleons as well as the (effective) confining forces of the
three-quark system. Electron beams are the ideal tool to study such
resonance form factors. A variety of measurements of single-pion
electroproduction on a proton target have been performed in recent years.
Electroproduction of $\pi$~mesons in the threshold region was studied
at MAMI at very low values of $Q^2$~\cite{Weis:2007kf} and for 
$0.16 < Q^2 < 6$~GeV$^2$ at Jefferson Lab using the CLAS spectrometer. 
The experimental data include differential cross sections for the reactions
$\gamma^\ast p\to p\pi^0$~\cite{Joo:2001tw,Ungaro:2006df} and
$\gamma^\ast p\to n\pi^+$~\cite{Egiyan:2006ks,Park:2007tn,Park:2012rn}, 
beam~\cite{Joo:2003uc,Joo:2004mi,Park:2007tn} and target~\cite{Biselli:2008aa} 
asymmetries using longitudinal polarization, and beam-target 
asymmetries~\cite{Biselli:2008aa}. Differential cross sections from
CLAS were also published for the reactions $\gamma^\ast p\to p\pi^+
\pi^-$~\cite{Ripani:2002ss,Fedotov:2008aa} and $\gamma^\ast p\to p
\eta$~\cite{Thompson:2000by,Denizli:2007tq}.

Figure~\ref{Figure:CLAS_electroproduction} (left) shows examples of
the cross section data on $\gamma^\ast p\to n\pi^+$ for different polar and
azimuthal angles~\cite{Park:2007tn}. The upper, middle, and lower rows
correspond to $Q^2 = 1.72,~2.44$, and 3.48~GeV$^2$, respectively. The
curves denote descriptions of the data using two different approaches. 
The solid curve~\cite{Aznauryan:2009mx} is based on fixed-$t$
dispersion relations (DR), an approach which was developed in the
1950s for pion- and electroproduction on nucleons~\cite{Chew:1957tf,
Fubini:1958zz}. Recent extensions of the model into kinematical
regions covered by the new data, e.g.~\cite{Aznauryan:2009mx}, allowed
a more reliable extraction of the amplitudes for the important
$\gamma^\ast p\to\Delta(1232)\,\frac{3}{2}^+$ transition. The dashed
curve~\cite{Aznauryan:2009mx} is based on an effective Lagrangian
approach -- Unitary Isobar Model~(UIM), which goes back to the late
1990s~\cite{Drechsel:1998hk}.

\begin{figure}[t]
\caption{\label{Figure:CLAS_Roper}(Colour online) Helicity
  amplitudes for the $\gamma^\ast p\to N(1440)\,\frac{1}{2}^+$
  transition in comparison with quark-model predictions. The
  photocouplings ($Q^2 = 0$) are taken from the RPP~\cite{FERMILAB-PUB-10-665-PPD}
  (open square) and an analysis of $\gamma p\to n\pi^+$~\cite{Dugger:2009pn}
  (open triangle). The other data points are from a CLAS analysis of $p\pi^+\pi^-$
  \cite{Fedotov:2008aa} (black triangles) and $N\pi$~electroproduction
  data~\cite{Aznauryan:2009mx} (circles). The solid and dashed
  curves are predictions of light-front relativistic quark models~\cite{Capstick:1994ne,Aznauryan:2007ja}
  considering the $N(1440)\,\frac{1}{2}^+$ a radial excitation of the $3q$ 
  ground state. Results of a covariant valence quark spectator diquark 
  model~\cite{Ramalho:2010js} are shown by the dashed dotted line.
  Pictures from~\cite{Mokeev:2012vsa}.}
\includegraphics[width=0.49\textwidth,height=0.42\textwidth]{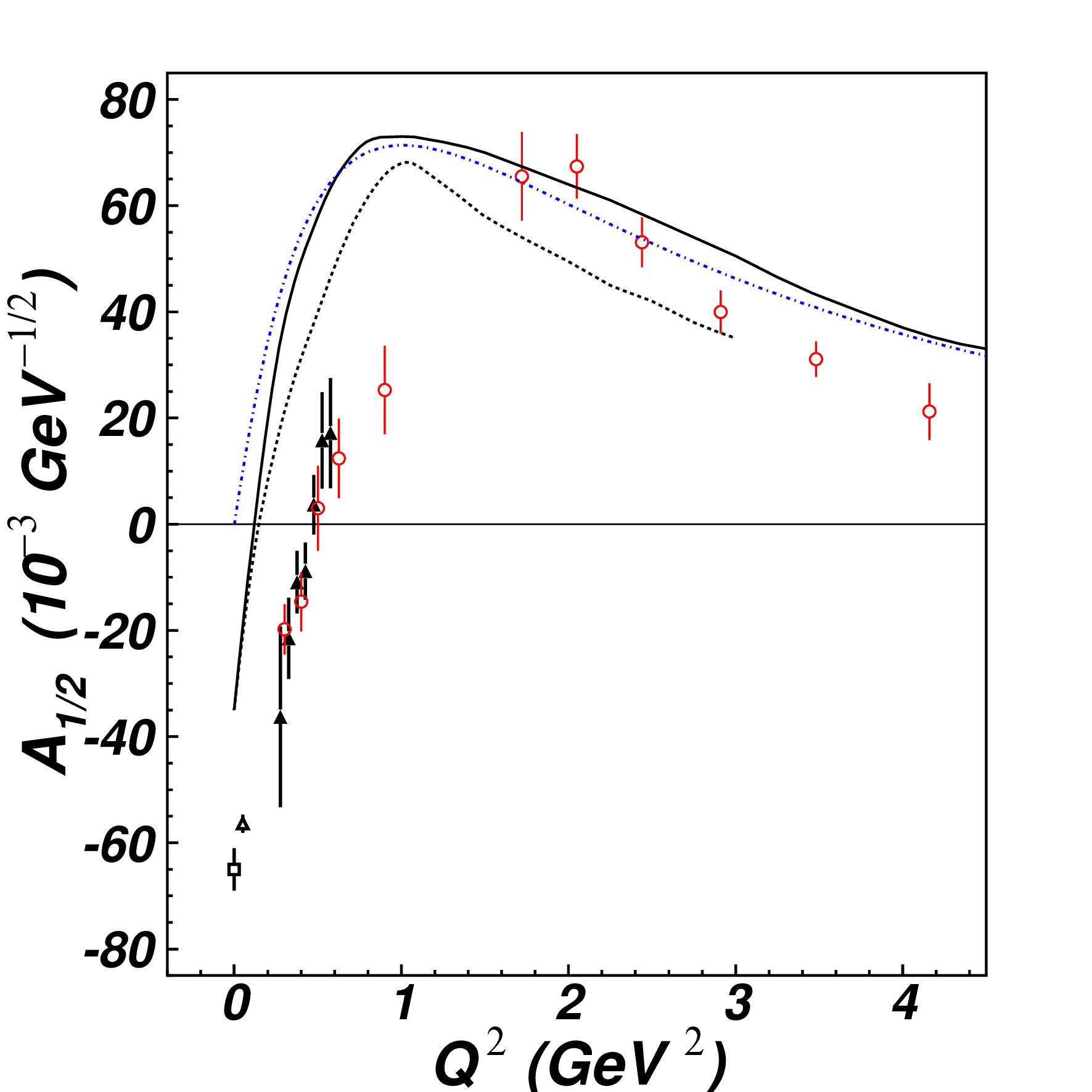}
\includegraphics[width=0.49\textwidth,height=0.42\textwidth]{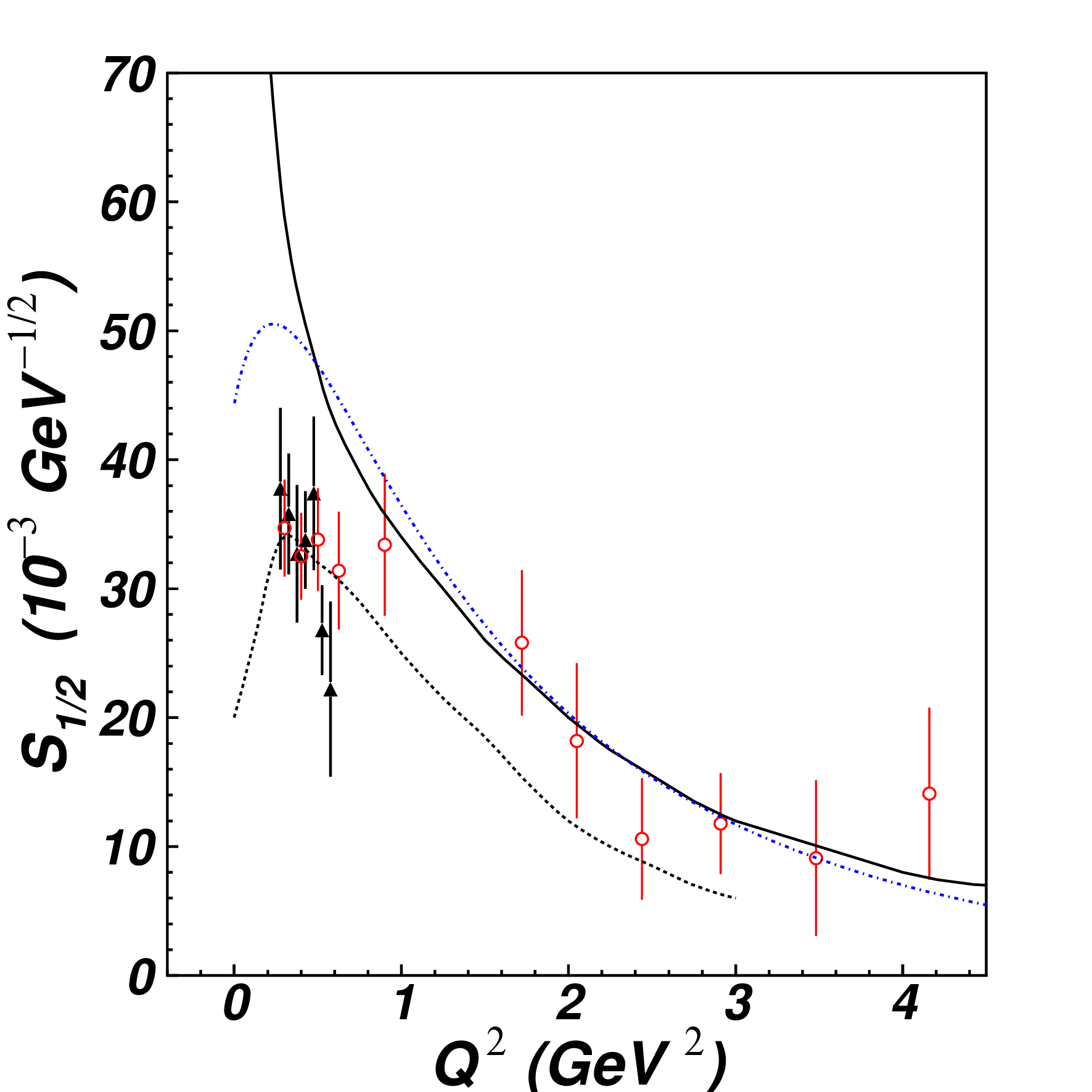}
\end{figure}

Important results of this analysis~\cite{Aznauryan:2009mx} include a
better understanding of the amplitudes for the $\Delta(1232)\,\frac{3}{2}^+$
indicating the importance of a meson-cloud contribution to quantitatively 
explain the magnetic dipole strength~\cite{Matsuyama:2006rp,Sato:2000jf,
Kamalov:2000en}, as well as the electric and scalar quadrupole transitions. 
Recent lattice QCD calculations also identified contributions from the
pion-cloud important for the $\gamma^\ast p\to\Delta(1232)\,\frac{3}{2}^+$
transition~\cite{Alexandrou:2007dt}. Furthermore, the helicity amplitudes 
for the electroexcitation of the low-mass resonances $\Delta(1232)\,\frac{3}{2}^+$, 
$N(1440)\,\frac{1}{2}^+$, $N(1520)\,\frac{3}{2}^-$, and $N(1535)\,\frac{1}{2}^-$ were 
extracted in~\cite{Aznauryan:2009mx}. More recently, very consistent results on
the electrocouplings of the $N(1440)\,\frac{1}{2}^+$ and the $N(1520)\,\frac{3}{2}^-$ 
resonances have become available from the independent analysis of the single-
and double-pion channel~\cite{Mokeev:2012vsa} within the framework of the
phenomenological reaction model in~\cite{Mokeev:2008iw}. As an example, 
Figure~\ref{Figure:CLAS_Roper} shows the helicity amplitudes $A_{\frac{1}{2}}$ (left) 
and $S_{\frac{1}{2}}$ (right) for the $N(1440)\,\frac{1}{2}^+$ ``Roper'' resonance. 
These data provide strong evidence for the state to be a predominantly radial 
excitation of a 3-quark ground state. In particular, the non-zero $S_{\frac{1}{2}}$~amplitude 
rules out an interpretation of the resonance as a gluonic baryon excitation, which 
was predicted in~\cite{Li:1991sh,Li:1991yba}, for instance. In contrast, descriptions 
by light-front relativistic quark models considering the state a radial
excitation of the $3q$ ground state, e.g.~\cite{Capstick:1994ne,
Aznauryan:2007ja}, are in reasonable agreement with the data at large
values of $Q^2$ (short distances). Stronger contributions from the
meson-baryon cloud can explain the discrepancies below $Q^2 <
1$~GeV$^2$ (large distances)~\cite{Aznauryan:2009mx}.

For the $N(1520)\,\frac{3}{2}^-$~resonance, the results show a rapid helicity
switch from the dominant $A_{\frac{3}{2}}$~amplitude at $Q^2 = 0$~GeV$^2$
(photon point) to $A_{\frac{1}{2}}$ at $Q^2 > 1$~GeV$^2$, which confirms
predictions of constituent quark models. A better understanding of the 
$N(1535)\,\frac{1}{2}^-$~resonance remains a challenge, though. Quark models
cannot describe correctly the sign of the $S_{\frac{1}{2}}$ amplitude below $Q^2 
< 3$~GeV$^2$ and also have difficulties in describing the substantial 
$N(1535)\,\frac{1}{2}^-$~coupling to the $N\eta$~channel. This  may
again be indicative of large meson-cloud contributions or of representations 
of this state different from a $3q$~excitation. Alternative interpretations 
range from a quasi-bound meson-baryon molecule~\cite{Kaiser:1995cy} 
to a dynamically-generated resonance coming from the interaction of the
octet of pseudoscalar mesons with the ground-state octet
of baryons~\cite{Jido:2003cb,Jido:2007sm}.

A new resonance, $(1720)\,\frac{3}{2}^+$, was proposed on the basis of
recent CLAS data on $\gamma^\ast p\to p\pi^+\pi^-$~\cite{Ripani:2002ss}.
Figure~\ref{Figure:CLAS_electroproduction} (right) shows the fully integrated
cross sections at three different $Q^2$~values. The data are well described 
by the (red) solid curve, which is based on calculations within the framework 
of the reaction model in~\cite{Mokeev:2008iw}. The difference between the
solid and dashed (blue) curves represents the signal from the proposed new 
state.

\subsection{Photoproduction Experiments}
Most of the information on light-flavour baryon resonances reported by the
PDG is still based on $\pi N$~scattering experiments, but the plethora of 
results coming from the recent photoproduction experiments will likely
have a big influence on the particle listings. Before we discuss the wealth 
of information obtained from individual reactions, it is worth discussing 
the total photoabsorption cross section in comparison with the total
$\pi N$ cross section. 

\begin{figure}
\caption{\label{Figure:GDH}(Colour online) The total cross sections for
  $\gamma p$ and $\pi p$ reactions. Left: Total photoabsorption 
  cross sections, $\sigma_{\frac{3}{2}}$ and $\sigma_{\frac{1}{2}}$ measured by the
  GDH collaboration. Picture from~\cite{Helbing:2006zp} (and data
  references therein). Right: Total cross section for pion-induced
  reactions; courtesy M.~Williams, Ph.D. thesis, Carnegie-Mellon university.}
\includegraphics[width=0.49\textwidth,height=0.32\textwidth]{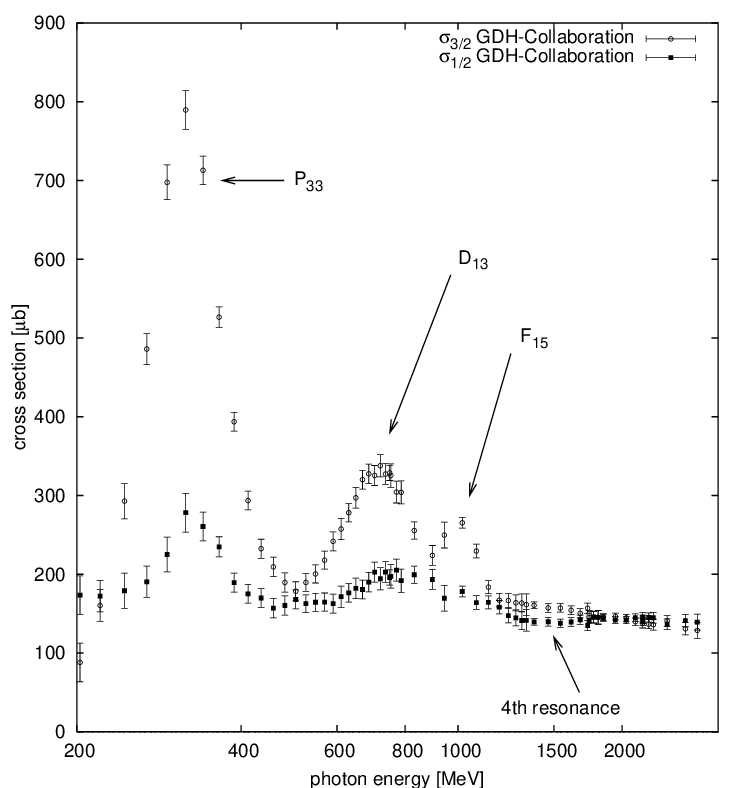}
\includegraphics[width=0.49\textwidth,height=0.35\textwidth]{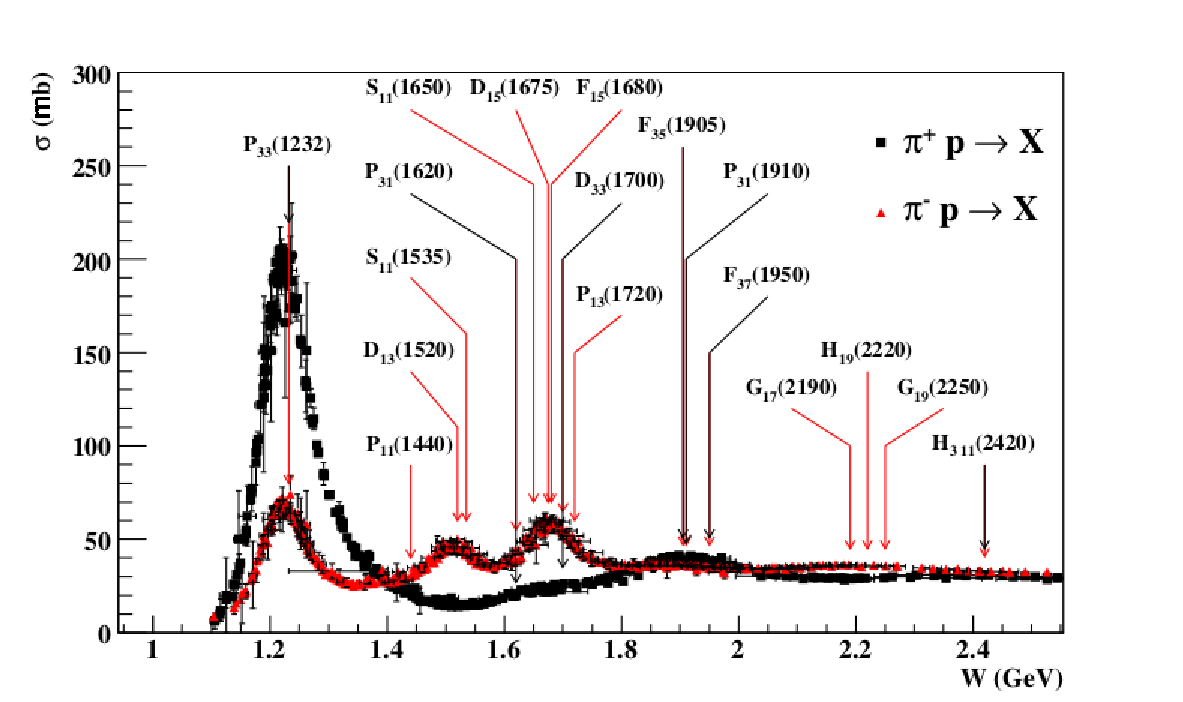}
\end{figure}

The left side of Figure~\ref{Figure:GDH} shows the separate helicity
contributions to the total photoabsorption cross section measured by
the GDH collaboration~\cite{Helbing:2006zp}. The right side of Figure~\ref{Figure:GDH} shows
the corresponding cross sections for $\pi^+ p$ and $\pi^- p$ reactions.
The total $\pi^- p$ cross section exhibits many more structures since 
only $\Delta^{++}$ resonances can contribute to the total $\pi^+ p$~cross 
section owing to the conservation of electric charge. The overall shapes of 
the total $\gamma p$ and $\pi p$ cross sections are similar and the four
resonance regions are clearly observed in both reactions. The large peak at threshold in
all distributions originates from the production of the $\Delta(1232)\,\frac{3}{2}^+$
resonance. In the total photoabsorption cross section, most of the resonance 
strength in the first three resonance regions comes clearly from the 
helicity~$\frac{3}{2}$, which indicates that resonances with 
$J\geq \frac{3}{2}$ provide the dominant contributions. The situation appears to be different for the 
fourth resonance region around $W\approx 1920$~MeV where the $\pi^+p$
cross section shows a broad structure and the $\pi^-p$ cross section
is featureless. 

The photoproduction cross section clearly depends on the helicity of
the proton and photon and therefore, the helicity difference defined as
\begin{eqnarray}
    E\,=\,\frac{\sigma_{\frac{3}{2}} - \sigma_{\frac{1}{2}}}{\sigma_{\frac{3}{2}} + \sigma_{\frac{1}{2}}}
\end{eqnarray}
is a very important quantity, not only for the total cross section, but
also for individual meson-production channels (\Eref{Equation:BT}). At long distance scales
in the confinement regime, the integrated total helicity-difference cross 
section can be related to the static properties of the nucleon. One such 
sum rule was derived by Gerasimov, Drell, and Hearn (GDH)~\cite{Gerasimov:1969zz,Drell:1966jv} in the mid-1960s 
and provides an elegant connection between the nucleon structure functions 
obtained in high-energy lepton-scattering experiments and the anomalous 
magnetic moment, $\kappa$, and the nucleon mass, $m$. At $Q^2=0$ (for 
real photons), the GDH integral is given as
\begin{eqnarray}
\int_{\nu_{\rm th}}^\infty\,\frac{d\nu}{\nu}\,[\,\sigma_{\frac{3}{2}}(\nu)\,-\,\sigma_{\frac{1}{2}}(\nu)\,]
\,=\,\frac{2\pi^2\alpha}{m^2}\,\kappa^2\,,
\end{eqnarray}
where $\nu = E_\gamma$ is the incident photon energy in the lab frame
and $\nu_{\rm th}$ is the photo-absorption threshold.
Consequently, it provides a bridge between perturbative and non-pertubative 
QCD. 

The GDH sum rule was confirmed experimentally by the GDH collaboration in 
measurements from a proton target at MAMI~\cite{Ahrens:2000bc,Ahrens:2001qt, 
Ahrens:2006yx} and ELSA~\cite{Dutz:2003mm}. The uncertainties for the neutron
measurements are larger, though. A nice review on the Gerasimov-Drell-Hearn
sum rule is given in~\cite{Helbing:2006zp}. Including some extrapolation to 
high energies, the experimental value of the GDH integral for the proton was 
determined to be $212\pm 6_{\rm \,stat.}\pm 16_{\rm \,syst.}~\mu$b~\cite{Helbing:2006zp}, 
which agrees with the prediction of $205~\mu$b.

The total photo-absorption cross section in Figure~\ref{Figure:GDH} clearly exhibits 
features that suggest resonance production, but such inclusive measurements do not allow
a detailed investigation of the closely-spaced baryon resonances contributing 
to the different resonance regions. The dominant decay of any excited nucleon 
state is via the emission of mesons in the strong interaction, e.g. $N^\ast\to 
N\pi,~N\eta,~N\omega,~N\pi\pi,~KY$, etc. Electromagnetic decays via photon 
emission have typical branching ratios below the 1\,\%~level and are usually 
difficult to identify in the presence of large background contributions to the 
final states. Although meson production has been studied mostly in pion-induced 
reactions, many nucleon resonances have been predicted to couple only very 
weakly to the $\pi N$~channel. This has inspired new programs in baryon 
spectroscopy where the focus has shifted to studying resonances using 
electromagnetic probes.

In the following sections, recent photoproduction results obtained in a 
large variety of meson-production channels are presented. The current focus at the laboratories
worldwide is on the so-called complete experiments which involve the 
determination of all 16~spin-dependent observables in single- and double-polarization 
experiments, discussed in Section~\ref{Section:EM-Baryons}. While 
only eight well-chosen observables are needed in the mathematical problem to
extract the scattering amplitude without ambiguities, an experimental approach 
will require a significantly larger number of observables because of unavoidable 
experimental uncertainties. Observables for complete experiments will also come
from different facilities which use detectors optimized for different reactions and 
kinematical regions. Many cross section results are not statistically limited 
anymore, but the various theoretical approaches to better understand the
underlying production mechanisms face large systematic discrepancies 
among results from different experiments. In some cases, a combined analysis
of the available data is even impossible at the moment.

\subsubsection{Photoproduction of a Single $\pi$ Meson}
The single-meson reactions $\gamma p\to p\pi^0$ and $\gamma p\to 
n\pi^+$ are among the best studied photoproduction channels, but a
comprehensive study of polarization observables has only begun recently. 
For the full database of differential cross section results, we refer to the 
GWU Data Analysis Center~\cite{said-database}. A good summary of 
references on $\pi^0$~production is also given in~\cite{vanPee:2007tw}.
The most recent cross section results for $\pi^0$~production, which cover
a large angular and energy range, come from JLab~\cite{Dugger:2007bt} 
and ELSA~\cite{Crede:2011dc}. The latter data cover in particular the very 
forward and backward angles of the $\pi^0$ in the c.m. system, which are 
sensitive to higher-spin resonances. Very precise differential cross sections 
for $\gamma p\to n\pi^+$ were measured at CLAS for energies from 0.725 
to 2.875~GeV~\cite{Dugger:2009pn}.

\begin{figure}
\caption{\label{Figure:pi0FixedAngle}(Colour online) Differential cross
  sections for the reaction $\gamma p\to p\pi^0$ for $E_\gamma\in
  [1.0,\,2.6]$~GeV at two different polar angles of the $\pi^0$ in
  the c.m. system: $\theta \approx 148^\circ$ (left) and
  $\theta \approx 41^\circ$ (right). Data are from CLAS~{\color{magenta}
  $\blacksquare$}~\cite{Dugger:2007bt}, CB-ELSA~{\color{green}
  $\blacktriangledown$}~\cite{Bartholomy:2004uz,vanPee:2007tw},
  CBELSA/TAPS~{\large\color{blue} $\bullet$}~\cite{Crede:2011dc},
  Becks {\it et al.}~$\vartriangle$~\cite{Becks:1973ws}, and Brefeld
  {\it et al.}~$\footnotesize\bigstar$~\cite{Brefeld:1975dv}. Statistical
  and systematic errors have been added in quadrature.}
\includegraphics[width=0.51\textwidth]{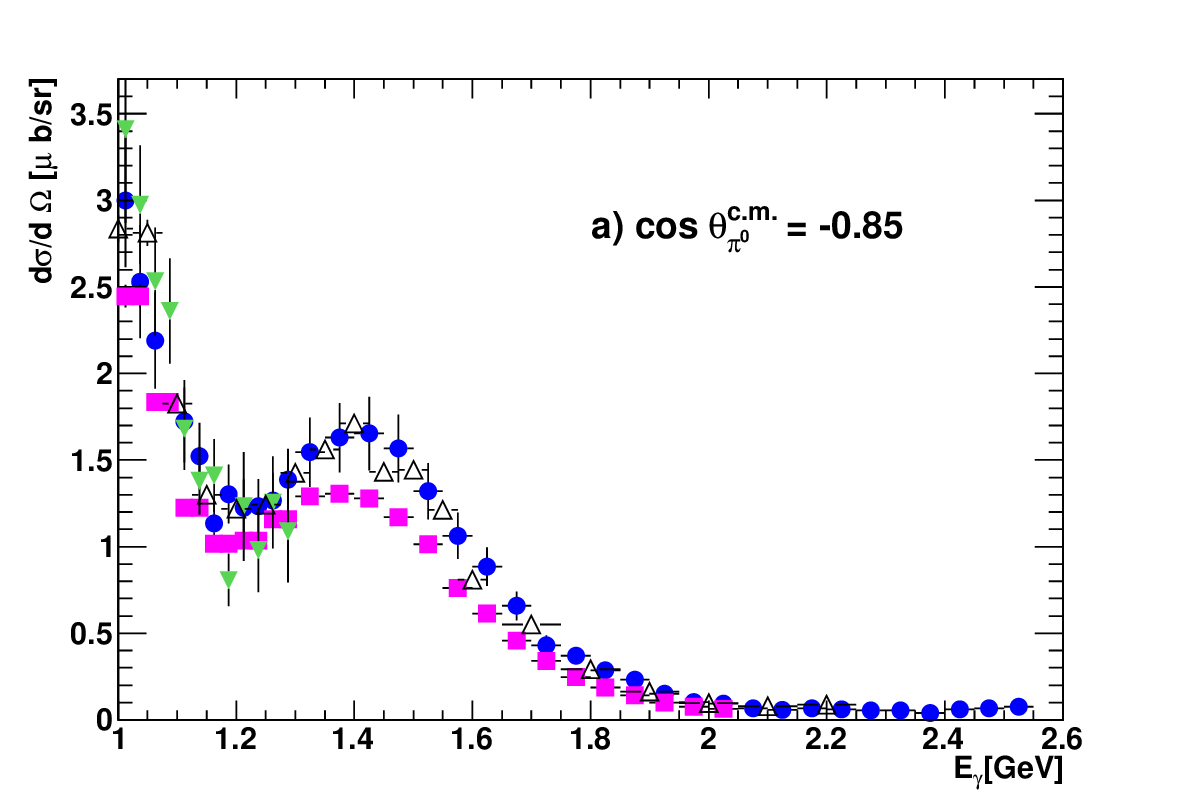}
\includegraphics[width=0.51\textwidth]{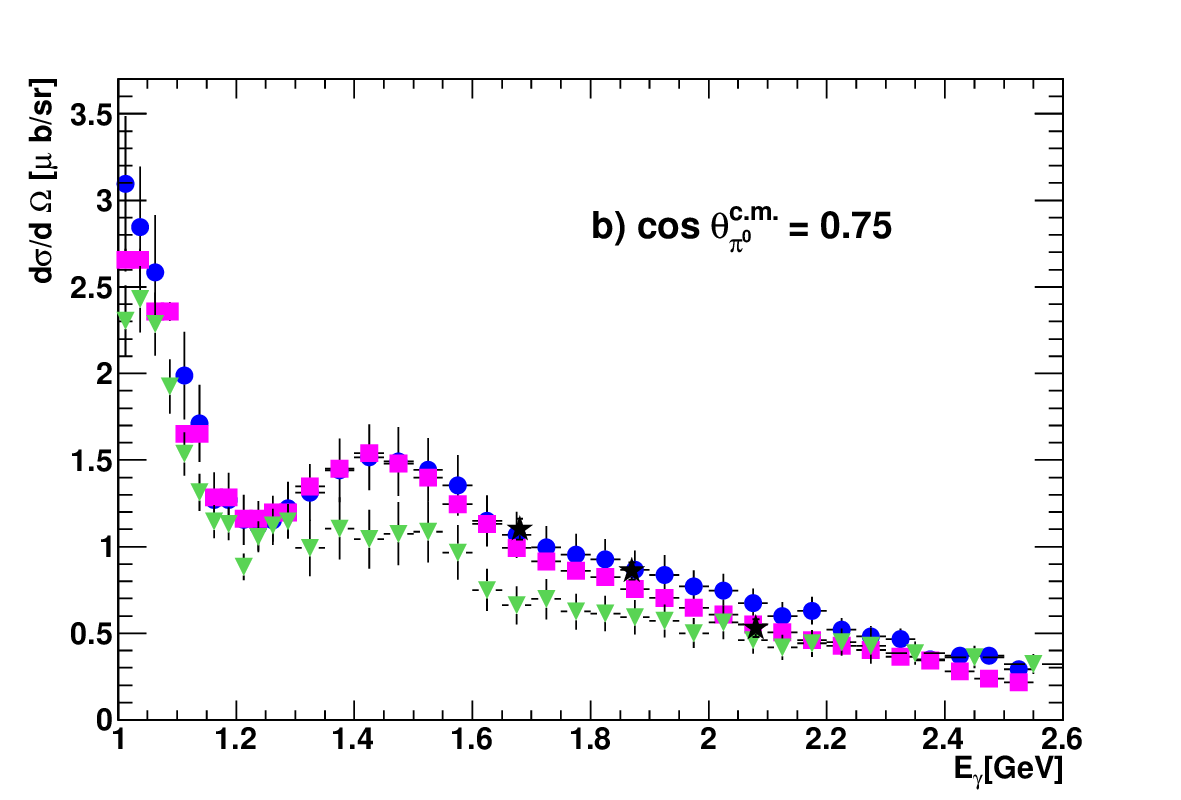}
\end{figure}

The overall agreement of the available $\pi^0$~cross section results is 
very good, but some discrepancies have been observed and need to be
resolved in order to reliably extract resonance properties. As an example,
Figure~\ref{Figure:pi0FixedAngle} shows some of the available data for a 
large energy range from 1.0 to 2.6 GeV for fixed angles of cos$\,\theta_{\rm 
\,c.m.} = -0.85$~(left) and cos$\,\theta_{\rm \,c.m.} = 0.75$~(right). At the 
backward angle, most of the experiments show a very consistent behaviour. 
However, the CLAS (g1c) data~({\footnotesize\color{magenta} $\blacksquare$})
are clearly lower. At the forward angle, the CB-ELSA data~({\small\color{green} 
$\blacktriangledown$}) are somewhat lower than the other data sets. An overall 
large normalization problem is generally not observed for $\pi^0$~production 
and we conclude that the photon flux normalization appears to be reasonably 
well understood, but likely some detector acceptance effects are not correctly 
accounted for. 

Some data on target and proton recoil polarization are available in the 
reaction databases~\cite{durham-database,said-database} in addition to 
a few data on double-polarization. The best studied polarization observable 
is the beam asymmetry, $\Sigma$, which has been recently measured for
$\vec{\gamma} p\to p\pi^0$ at GRAAL~\cite{Bartalini:2005wx} in the energy 
range from 550 to 1500~MeV, at MAMI at low energies in the $\Delta$~resonance 
region~\cite{Beck:2006ye}, and at ELSA~\cite{Elsner:2008sn,Sparks:2010vb}.
High-statistics data are expected from CLAS in the very near future. GRAAL 
also reported measurements for $\vec{\gamma} p\to n\pi^+$~\cite{Bartalini:2002cj}
as well as $\vec{\gamma} n\to n\pi^0$~\cite{DiSalvo:2009zz} and found that 
the asymmetries from the quasi-free proton and the quasi-free neutron were 
equal up to 0.82 GeV, but substantially different at higher energies. The beam 
asymmetry, $\Sigma$, arises from a linearly-polarized photon beam and
addresses the non-spin-flip terms in the transition current (e.g. convection 
currents and double spin-flip contributions), whereas spin-flip contributions
are projected out by a circularly-polarized photon beam. At low photon energies,
between the production threshold and the $\Delta$~region, the $\pi^0$~beam 
asymmetry is important in addition to the differential cross section to measure 
the $S$-wave amplitude, $E_{0+}~(l_\pi = 0)$, and the $P$-wave amplitudes, 
$M_{1+},~M_{1-}$, and $E_{1+}~(l_\pi = 1)$ separately~\cite{Beck:2006ye}. In the 
second resonance region, detailed analysis of the data in $\pi$~photoproduction yields clear signals 
for the $N(1520)\,\frac{3}{2}^-$ and $N(1535)\,\frac{1}{2}^-$~states. The unpolarized 
cross section and the $\Sigma$~observable are somewhat less sensitive to the 
$N(1440)\,\frac{1}{2}^+$~resonance. Additional observables are needed to understand 
the reaction at higher energies.

\begin{figure}
\caption{\label{Figure:pi0-G}(Colour online) Recent results for the double-polarization
  observable $G$ from CBELSA/TAPS~\cite{Thiel:2012yj} for the reaction 
  $\gamma p\to p\pi^0$ as a function of cos\,$\theta_\pi$ (left) and as 
  a function of energy for two selected cos\,$\theta_\pi$ bins (right). 
  Systematic errors are shown as gray bars. The curves represent predictions 
  from the SAID (red, dashed)~\cite{Workman:2011vb}, MAID (blue, 
  dotted)~\cite{Drechsel:1998hk}, and Bonn-Gatchina (black, 
  solid)~\cite{Anisovich:2011fc} PWAs. The black, long-dashed and dashed-dotted
  lines denote Bonn-Gatchina with the $E_{0+}$ and $E_{2-}$ amplitudes
  from SAID and MAID, respectively.   
  Pictures from~\cite{Thiel:2012yj}.}
\includegraphics[width=0.46\textwidth,height=0.437\textwidth]{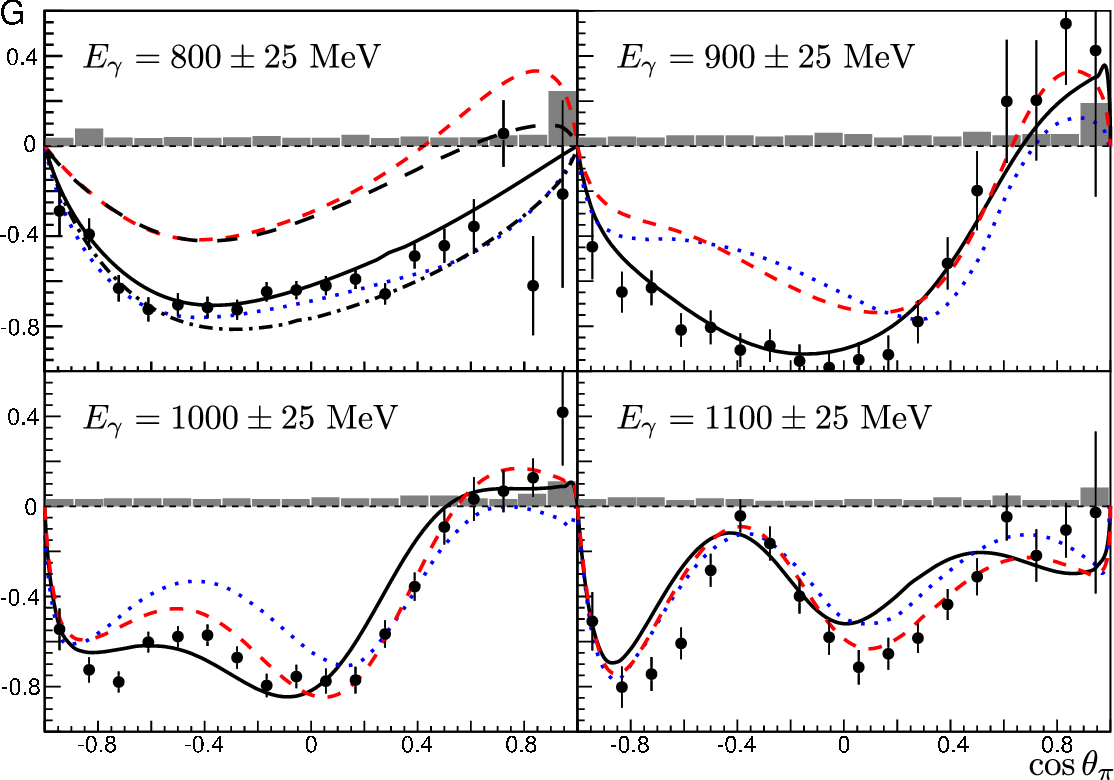}\qquad
\includegraphics[width=0.46\textwidth]{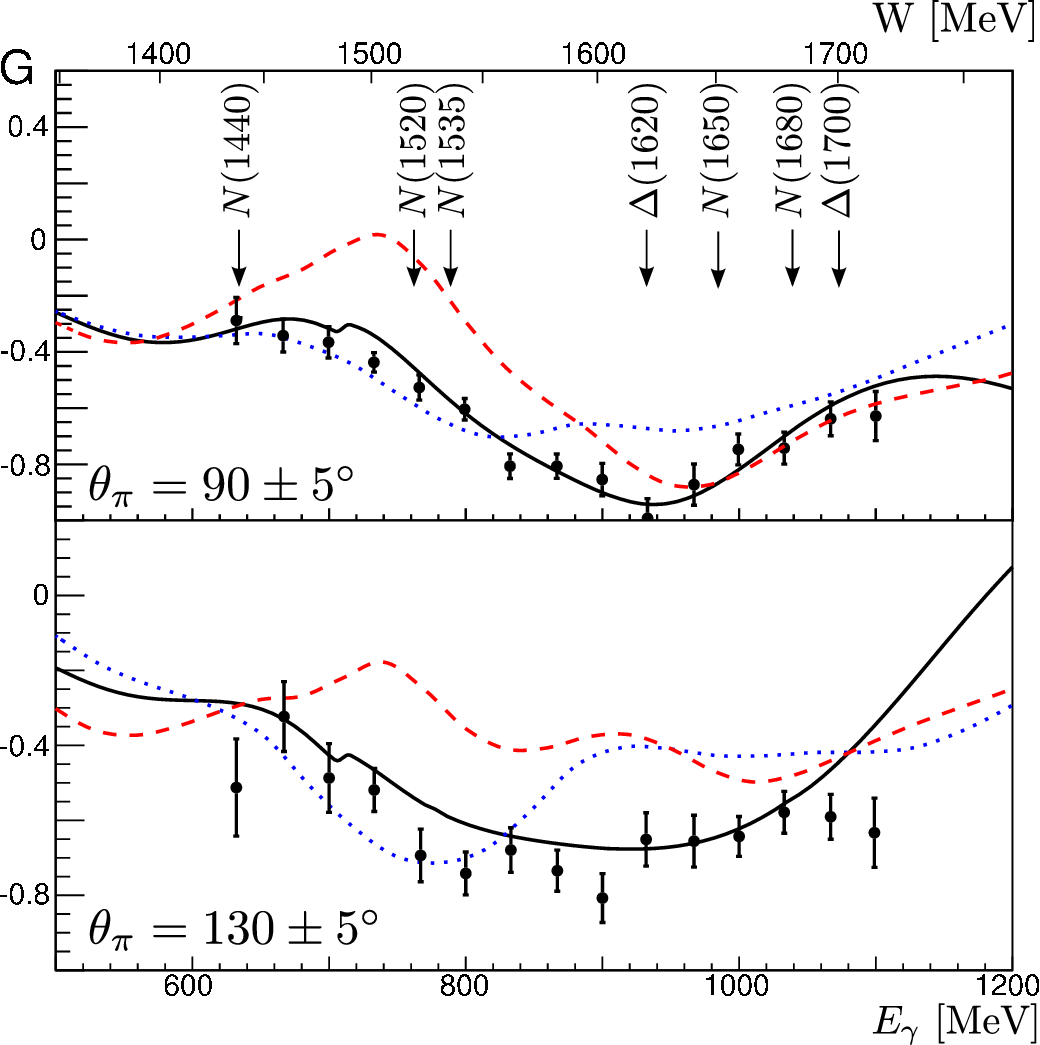}
\end{figure}

New high-statistics data on (single- and double-) polarization observables for 
$\gamma p\to p\pi^0$, e.g. the $E,~P,~H,~T$~observables (\Eref{Equation:BT}), 
have been presented at recent conferences. Surprisingly, big discrepancies have 
been observed in single-$\pi$~photoproduction, even at lower energies, between 
the latest predictions and the new polarization data. In 2012, CBELSA/TAPS 
reported on the first ever measurement of the double-polarization observable, 
$G$, for the photon energy range from 620 to 1120~MeV and the full solid 
angle~\cite{Thiel:2012yj}. The $G$~observable describes the correlation between the photon 
polarization plane and the scattering plane for protons polarized along the 
direction of the incoming photon. Figure~\ref{Figure:pi0-G} shows the observable 
for fixed energies (left) and fixed angles (right). The deviations of the current
model predictions from the data are clearly observed even for photon energies 
below 1~GeV. In particular, SAID shows large deviations for the two shown 
angles. It has been discussed~\cite{Thiel:2012yj} that the discrepancies among 
the models arise from the two multipoles, $E_{0+}$ and $E_{2-}$, which receive 
significant contributions from the $\frac{1}{2}^-$ resonances, $N(1535)$, 
$\Delta(1620)$, $N(1650)$, and the $\frac{3}{2}^-$~resonances, 
$N(1520)$, $\Delta(1700)$, respectively.

Several new resonances which are also observed to decay into $\pi N$ have been proposed in the latest edition of the RPP. 
While most of the new entries
have been inspired by recent results of the Bonn-Gatchina PWA group~\cite{Anisovich:2011fc},
some (or all) of them were already reported earlier by other groups, e.g. by 
Manley {\it et al.}~\cite{Manley:1992yb}. However, most of the new states
have not been observed in the recent analysis of $\pi N$~data by 
Arndt~{\it et al.}~\cite{Arndt:2006bf} including the $N(1700)\,\frac{3}{2}^-$~resonance. 
The new states including more recent references are listed in 
Table~\ref{Table:Baryons-piN}.

\begin{table}
\begin{center}
\caption{\label{Table:Baryons-piN} Summary of newly proposed
  resonances observed in $\pi N$. We list all the resonances for the
  partial waves with at least a 2-star assignment by the PDG (without 
  the nucleon ground state). New states are given in red.}
\begin{tabular}{@{}|l|c|c|l|l|}
\br
 $J^P$ & \multicolumn{4}{c|}{Resonance Region} \\
\mr
$1/2^+$ & $N(1440)$ & $N(1710)$ & \re{$N(1880)$}~\cite{Anisovich:2011fc,Manley:1992yb} &\\
$1/2^-$ & $N(1535)$ & $N(1650)$ & \re{$N(1895)$}~\cite{Anisovich:2011fc,Manley:1992yb} &\\
$3/2^-$ & $N(1520)$ & $N(1700)$ & \re{$N(1875)$}~\cite{Anisovich:2011fc,Manley:1992yb} & \re{$N(2120)$}~\cite{Anisovich:2011fc}\\
$5/2^+$ & & $N(1680)$ & \re{$N(1860)$}~\cite{Arndt:2006bf,Anisovich:2011fc,Manley:1992yb} & $N(2000)$\\
$5/2^-$ & & $N(1675)$ & & \re{$N(2060)$}~\cite{Anisovich:2011fc}\\
\br
\end{tabular}
\end{center}
\end{table}

\subsubsection{\label{Section:photoeta}Photoproduction of Single $\eta$ and $\eta^{\,\prime}$ Mesons}
\begin{figure}
\caption{\label{Figure:eta}(Colour online) Cross sections for the
  reaction $\gamma p\to p\eta$. Left: Total cross section. Picture
  from~\cite{McNicoll:2010qk}. Right: Differential cross sections for 
  $E_\gamma \in [0.8,\,1.4]$~GeV and different angle bins in the forward 
  direction. Data are taken from CLAS g11a~{\footnotesize\color{green}
  $\blacksquare$}~\cite{Williams:2009yj}, CBELSA/TAPS~{\large\color{magenta} 
  $\circ$}~\cite{Crede:2009zzb}, MAMI~{\large\color{red} $\bullet$}~\cite{McNicoll:2010qk}, 
  and GRAAL~{\small\color{yellow} $\blacktriangledown$}~\cite{Bartalini:2007fg}. The error 
  bars comprise statistical and systematic errors added in quadrature.}\vspace{2mm}
\includegraphics[width=0.335\textwidth,angle=90]{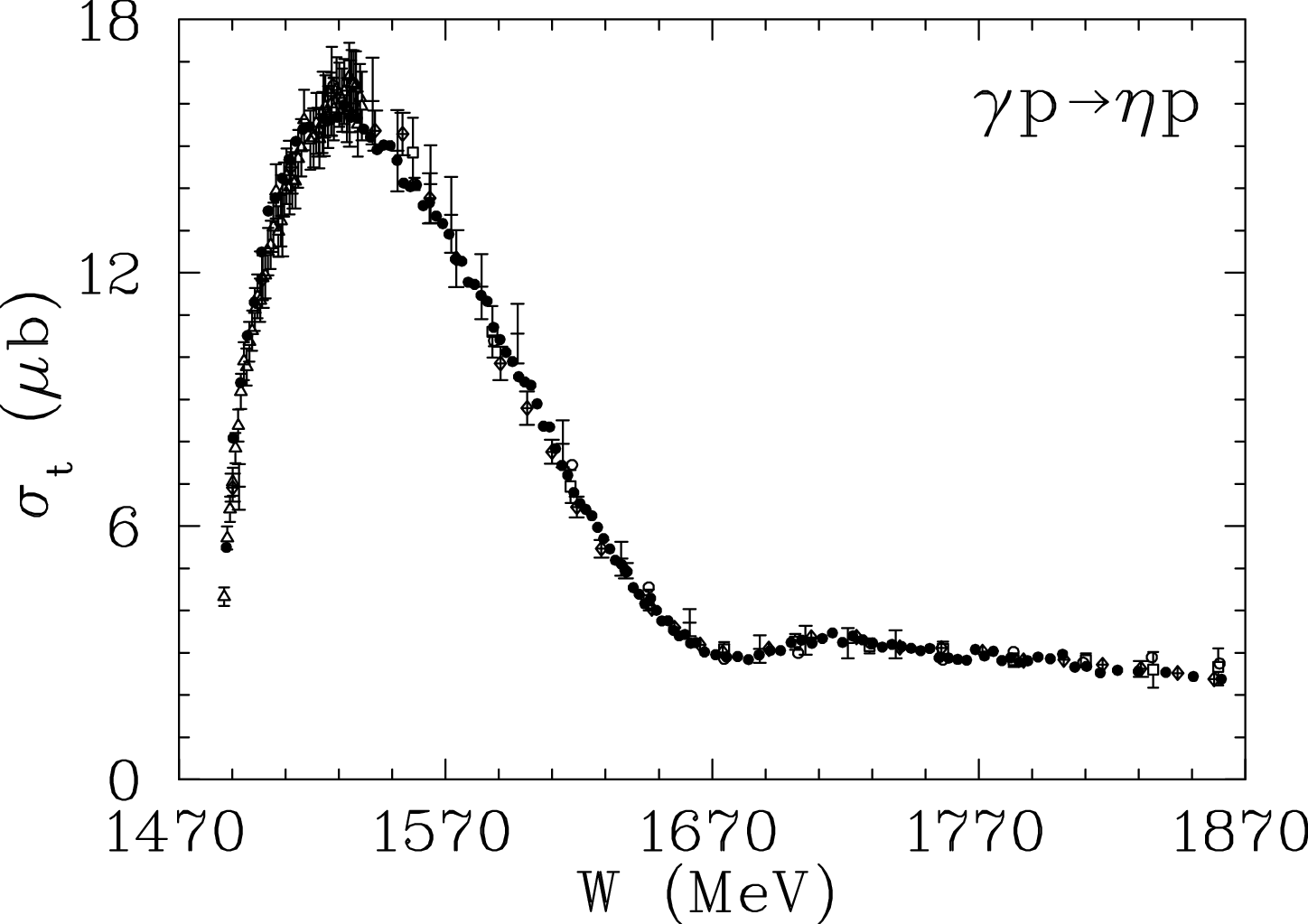}
\includegraphics[width=0.51\textwidth]{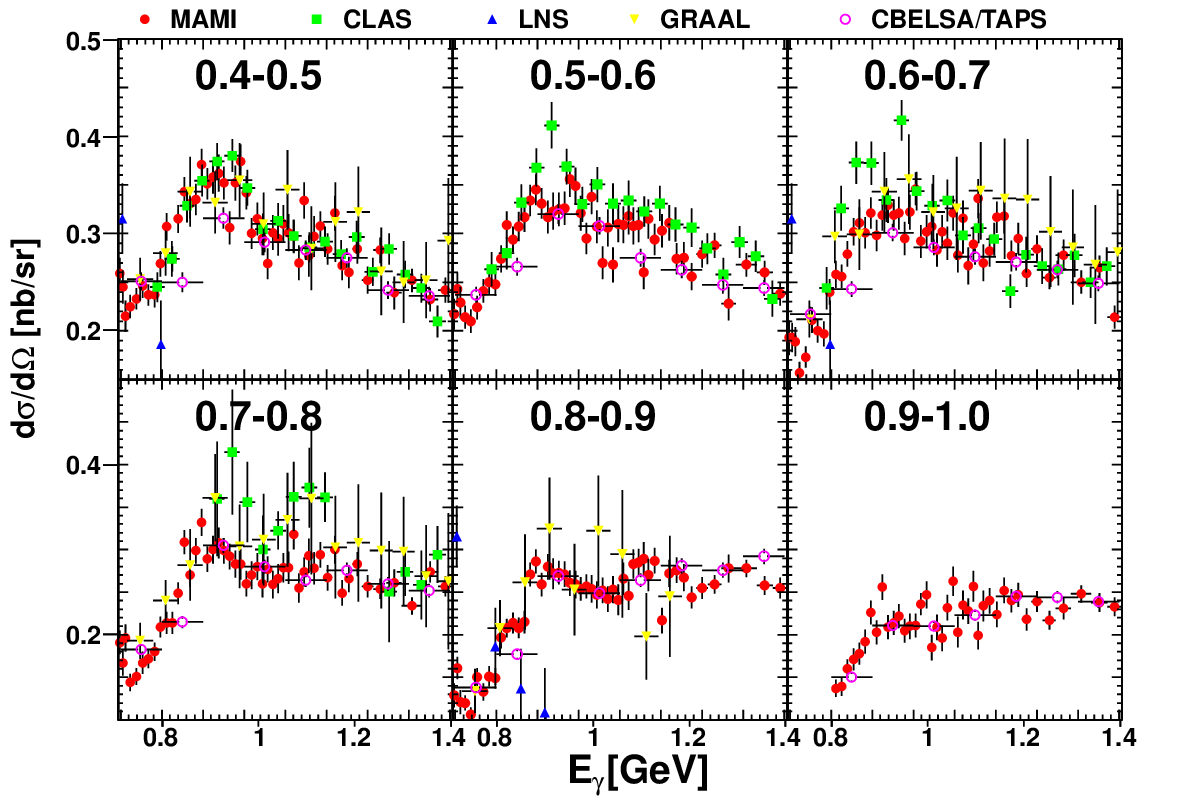}
\end{figure}

The total cross section for the reaction $\gamma p\to p\eta$ is shown
in Figure~\ref{Figure:eta}~(left). It exhibits a steep rise at the reaction 
threshold of $E_{\rm th} \approx 706$~MeV, reaches a maximum of about 
16~$\mu$b very close to the threshold, and is dominated by the broad peak 
around $W\approx 1540$~MeV. Apart from a small dip at $W = 1680$~MeV, 
the total cross section for this reaction shows no further structures.

The latest data come from MAMI with unprecedented statistical quality, but 
are limited in energy to 1.4~GeV~\cite{McNicoll:2010qk}. In recent years, the 
differential cross section has also been measured at CLAS~\cite{Williams:2009yj}, 
at ELSA using the CB-ELSA~\cite{Crede:2003ax,Bartholomy:2007zz} and CBELSA/TAPS~\cite{Crede:2009zzb} 
setups, and at GRAAL~\cite{Bartalini:2007fg}. Overall, great experimental progress can
be reported for $\eta$~photoproducton. In addition to the improved statistical quality, 
the differential cross section covers the full angular range for energies below
$E_\gamma = 2.5$~GeV. As an example, Figure~\ref{Figure:eta}~(right) shows results
for $E_\gamma \in [0.7,\,1.4]$~GeV and $0.4 < {\rm cos}\,\theta_{\rm c.m.}^{\,\eta} < 1.0$.
However, a fairly large normalization discrepancy of almost 60\,\% at higher energies (above 2~GeV)
between the results from ELSA and CLAS render a combined analysis of all available data 
sets currently almost impossible. The effect of this discrepancy on PWA results was discussed 
in~\cite{Sibirtsev:2010yj}, for instance. The authors point out that all recent experimental 
results on differential cross sections are in good agreement with each other, except for 
the recent CLAS data. In the analysis, overall consistency can be achieved by introducing
an energy-dependent renormalization factor for the CLAS data~\cite{Sibirtsev:2010yj}. 
The situation is unsatisfactory and needs to be resolved soon. It is worth mentioning 
that results from LEPS~\cite{Sumihama:2009gf} and older results from the Daresbury 
facility~\cite{Bussey:1976si} are in excellent agreement with CBELSA/TAPS at energies 
above 2~GeV.

Precise measurements of polarization observables in $\eta$~photoproduction are 
currently only available for the beam asymmetry, $\Sigma$. Very few older data can be 
found in the reaction databases, e.g.~\cite{said-database}. In 2007, results on 
$\Sigma$ were reported from CBELSA/TAPS~\cite{Elsner:2007hm} and 
GRAAL~\cite{Bartalini:2007fg}. GRAAL also reported the beam asymmetry 
for the photoproduction reaction off the neutron~\cite{Fantini:2008zz}. Many other observables including 
double-polarization observables from ELSA and CLAS have been presented at recent 
conferences and will be available soon, e.g.~\cite{Beck:HADRON2011}.

\begin{figure}
\caption{\label{Figure:eta-n}(Colour online) Left: Total cross sections
  for $\eta$~photoproduction as a function of the final-state invariant mass,
  $W$; for $\gamma d$~reactions, no cut on the spectator momentum was applied. The $n$~cross section
  was scaled up 3/2 to match the $p$~results. The insert shows the
  ratio of quasi-free neutron to proton data. Data and picture from~\cite{Jaegle:2011sw}. 
  Right: 
  Total cross sections for $\eta^{\,\prime}$ photoproduction. Data 
  and picture from~\cite{Jaegle:2010jg}. Other data are free proton
  results from~\cite{Crede:2009zzb} (open circles) and~\cite{Williams:2009yj} (magenta stars).
  With kind permission of The European Physical Journal (EPJ).}\vspace{4mm}
\includegraphics[height=0.36\textwidth,width=0.47\textwidth]{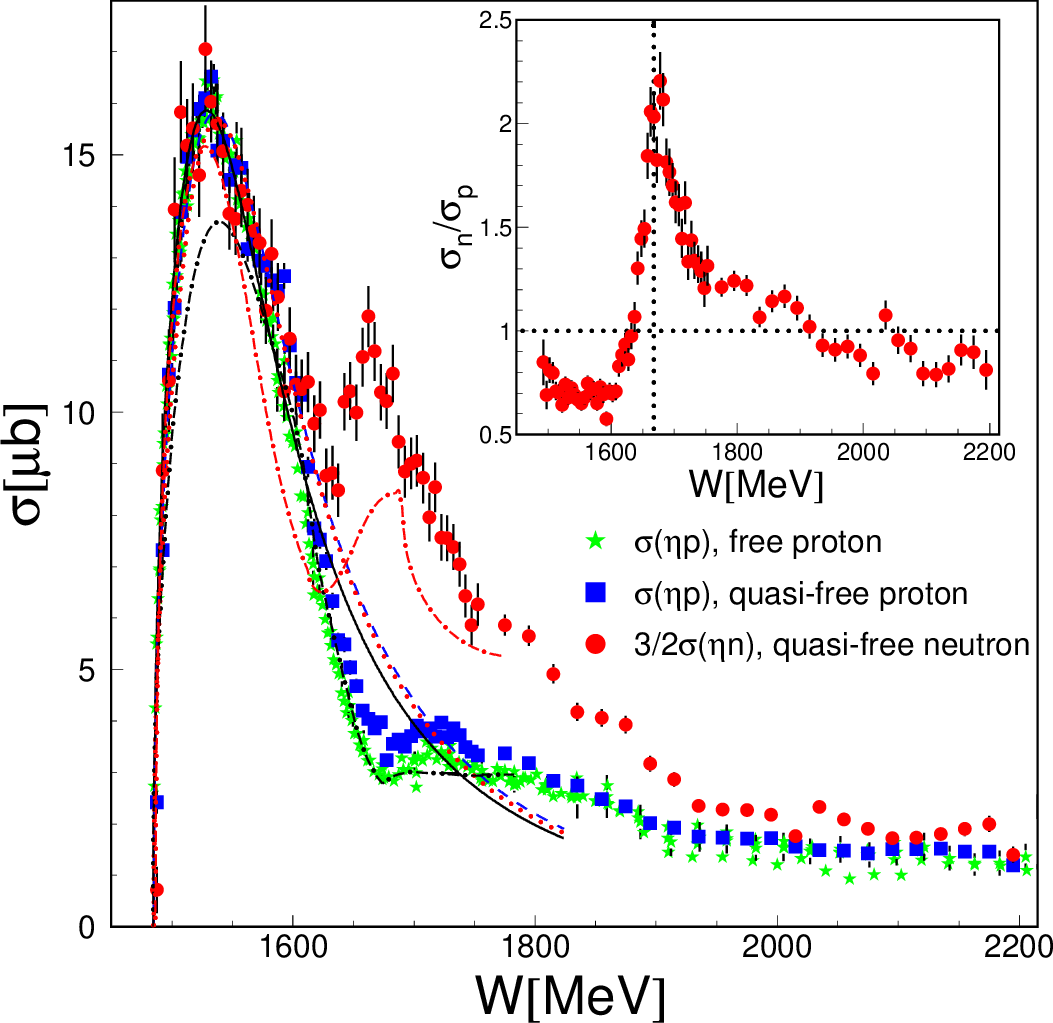}\qquad
\includegraphics[height=0.36\textwidth,width=0.47\textwidth]{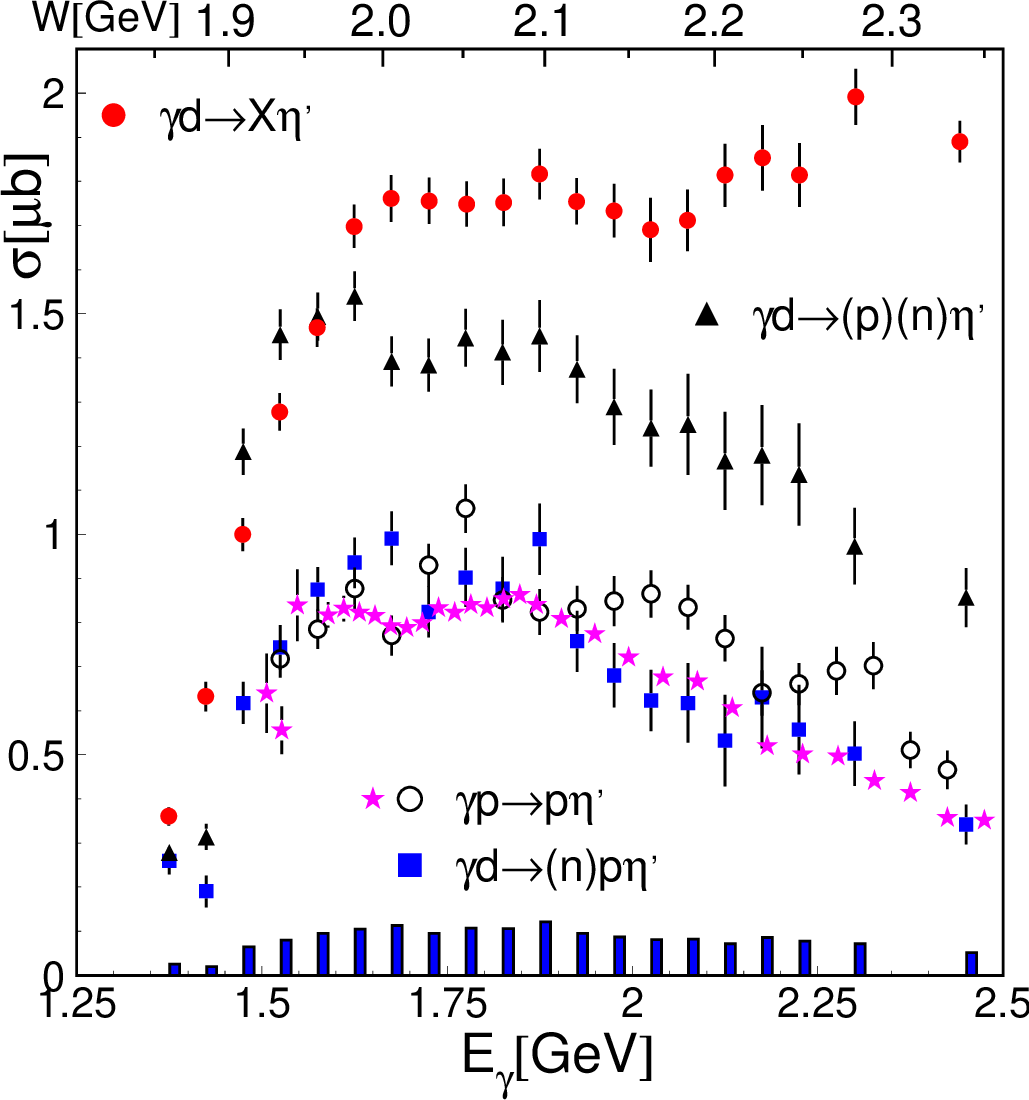}
\end{figure}

A striking observation has been made recently in the reaction $\gamma n\to 
n\eta$~\cite{Jaegle:2008ux,Jaegle:2011sw,Kuznetsov:2007gr,Kuznetsov:2010as}
where a narrow peak at $W\approx 1660$~MeV is observed, absent in the reaction 
$\gamma p\to p\eta$. The peak position coincides with the dip observed in the
cross section off the proton. Figure~\ref{Figure:eta-n} (left) shows
total cross sections for $\eta$~photoproduction as a function of the final
state invariant mass~\cite{Jaegle:2011sw}. The quasi-free neutron cross section was scaled
by a factor of~3/2 to match the broad peak at the threshold. The inset
shows the ratio of quasi-free neutron to proton data and the narrow peak 
is clearly visible. The observation has given rise to speculations about a 
narrow resonance, but no direct evidence that supports this
conjecture exists. Alternative explanations range from an interference effect
in the $\frac{1}{2}^-$-waves~\cite{Anisovich:2008wd} to coupled channel effects 
related to the $N(1650)\,\frac{1}{2}^-$ and the $N(1710)\,\frac{1}{2}^+$
resonances~\cite{Shklyar:2006xw} or a cusp effect from the $K\Sigma$ and $K\Lambda$
rescattering channels~\cite{Doring:2009qr}. The PDG lists the state as $N(1685)$ with a
1-star assignment and unknown quantum numbers. Polarization 
observables should shed more light on the nature of this structure.

There is no doubt about the dominant contribution of the $N(1535)\,\frac{1}{2}^-$
state to $\eta$~photoproduction at threshold. Other $\frac{1}{2}^-$~resonances 
also couple strongly to the $p\eta$~channel. The dip in the total cross 
section off the proton is likely due to destructive interference 
between the contributions from $N(1535)\,\frac{1}{2}^-$ and $N(1650)\,\frac{1}{2}^-$~\cite{Shklyar:2012js}.  
The large positive values of the $\eta$~beam asymmetry in this mass
range have been interpreted as being due 
to $N(1520)\,\frac{3}{2}^-$, although the effect of this resonance on
the differential cross section is small, see e.g. the discussion 
in~\cite{Krusche:2003ik}. More recently, the $N(1710)\,\frac{1}{2}^+$~\cite{Shklyar:2012js,
Anisovich:2011fc} and $N(1720)\,\frac{3}{2}^+$~\cite{Arndt:2006bf,Anisovich:2011fc}~resonances 
have been found to play an important role in $\eta$~production. Indications 
for contributions from additional $\frac{1}{2}^+$ and $\frac{3}{2}^-$~resonances are very
weak. Further evidence for resonances in $\eta$~production comes from the 
Bonn-Gatchina group~\cite{Anisovich:2011fc}. They observe the new 
resonances listed by the PDG, $N(1895)\,\frac{1}{2}^-$ and $N(2060)\,\frac{5}{2}^-$,
as well as the known $N(1900)\,\frac{3}{2}^+$ and $N(2000)\,\frac{5}{2}^+$ decaying 
to $p\eta$ in their multichannel analysis.

We briefly mention results for $\eta^{\,\prime}$~production, which has been 
measured with a fair amount of statistics at CBELSA/TAPS~\cite{Crede:2009zzb}. High-statistics 
results on the reaction $\gamma p\to p\eta^{\,\prime}$ are available from CLAS, 
e.g.~\cite{Williams:2009yj}. Moreover, $\gamma n\to n\eta^{\,\prime}$ has
been studied recently at CBELSA/TAPS and the total cross section found to be very
small~\cite{Jaegle:2010jg}. Figure~\ref{Figure:eta-n} (right) shows the corresponding 
total quasi-free proton cross section from the same analysis. The data are in good 
agreement with the results for the free proton, which indicates that nuclear effects 
have no significant impact. The total cross section off the proton reaches a maximum
of about $1~\mu$b in the photon energy range 1.8-1.9~MeV ($E_{\rm th}\approx 1448$~MeV) 
and then slowly declines without any structures.

A recent combined analysis of $\eta^{\,\prime}$~production in photon- and 
hadron-induced reactions by Huang {\it et al.}~\cite{Huang:2012xj}, based on an 
effective Lagrangian approach, finds the importance of three high-mass
resonances which the authors tentatively identify with $(\ast\,\ast)~N(1895)\,\frac{1}{2}^-$, 
$(\ast)~N(2100)\,\frac{1}{2}^+$, and $(\ast)~N(2040)\,\frac{3}{2}^+$. Additional contributions 
from the sub-threshold resonance, $N(1720)\,\frac{3}{2}^+$, are required for a good fit 
in their analysis. The authors stress that the normalization discrepancy between 
the results from CLAS~\cite{Williams:2009yj} and CBELSA/TAPS~\cite{Crede:2009zzb}, 
similar to the discrepancy observed in $\eta$~photoproduction, has an impact on the 
extracted resonance parameters but do not favour one data set over the other.

\subsubsection{Photoproduction of $K$~Mesons}
Production of open strangeness, particularly in the $\gamma p\to 
K^+Y~(Y = \Sigma^0,\,\Lambda)$ channels, has received much attention recently since a true 
complete experiment appears feasible. None of the current experimental facilities 
employs a recoil polarimeter, but the parity-violating weak decay of the 
hyperon (self-analyzing particle) still provides access to the polarization of the recoiling
baryon.

The differential cross sections for the reactions $\gamma p\to K^+\Lambda$ 
and $\gamma p\to K^+\Sigma^0$ were measured with good statistics at JLab
using the CLAS detector, e.g.~\cite{Bradford:2005pt,McCracken:2009ra,Dey:2010hh},
at ELSA using the SAPHIR detector~\cite{Glander:2003jw}, at GRAAL~\cite{Lleres:2007tx},
and, in the forward region, at LEPS~\cite{Sumihama:2005er}. The reactions $\gamma p\to 
K^0\Sigma^+$ and $\gamma p\to K^{\ast 0}\Sigma^+$ were studied at CLAS~\cite{Hleiqawi:2007ad}, 
CBELSA/TAPS~\cite{Castelijns:2007qt,Nanova:2008kr,Ewald:2011gw}, and SAPHIR~\cite{Lawall:2005np}.
Recent measurements of the beam-recoil observables, $C_{x,\,z}$ at CLAS~\cite{Bradford:2006ba} and 
$O_{x,\,z}$ at GRAAL~\cite{Lleres:2008em}, in $\gamma p\to K^+Y$ mark an important step toward 
complete experiments for these reactions. Despite the use of an unpolarized target at GRAAL, 
values for the target asymmetry $T$ could also be extracted for energies 
below 1.5~GeV~\cite{Lleres:2008em}.

\begin{figure}
\caption{\label{Figure:Lambda-cz}(Colour online) Beam-recoil observables in the reaction 
  $\gamma p\to K^+\Lambda$. Left: The observable $C_z$ from CLAS using circularly-polarized 
  photons. The data and picture are from~\cite{Bradford:2006ba}. We refer to~\cite{Bradford:2006ba} 
  for a full description of the curves denoting a large variety of different models. Right: The 
  observable $O_z$ from GRAAL using linearly-polarized photons. The data and picture are
  from~\cite{Lleres:2008em}. The curves denote the Bonn-Gatchina (solid line) and Ghent 
  RPR (Regge-plus-resonance) isobar~\cite{Corthals:2007kc} (dashed line) models.
  With kind permission of The European Physical Journal (EPJ).}\vspace{-8mm}
\includegraphics[width=0.6\textwidth,height=0.55\textwidth]{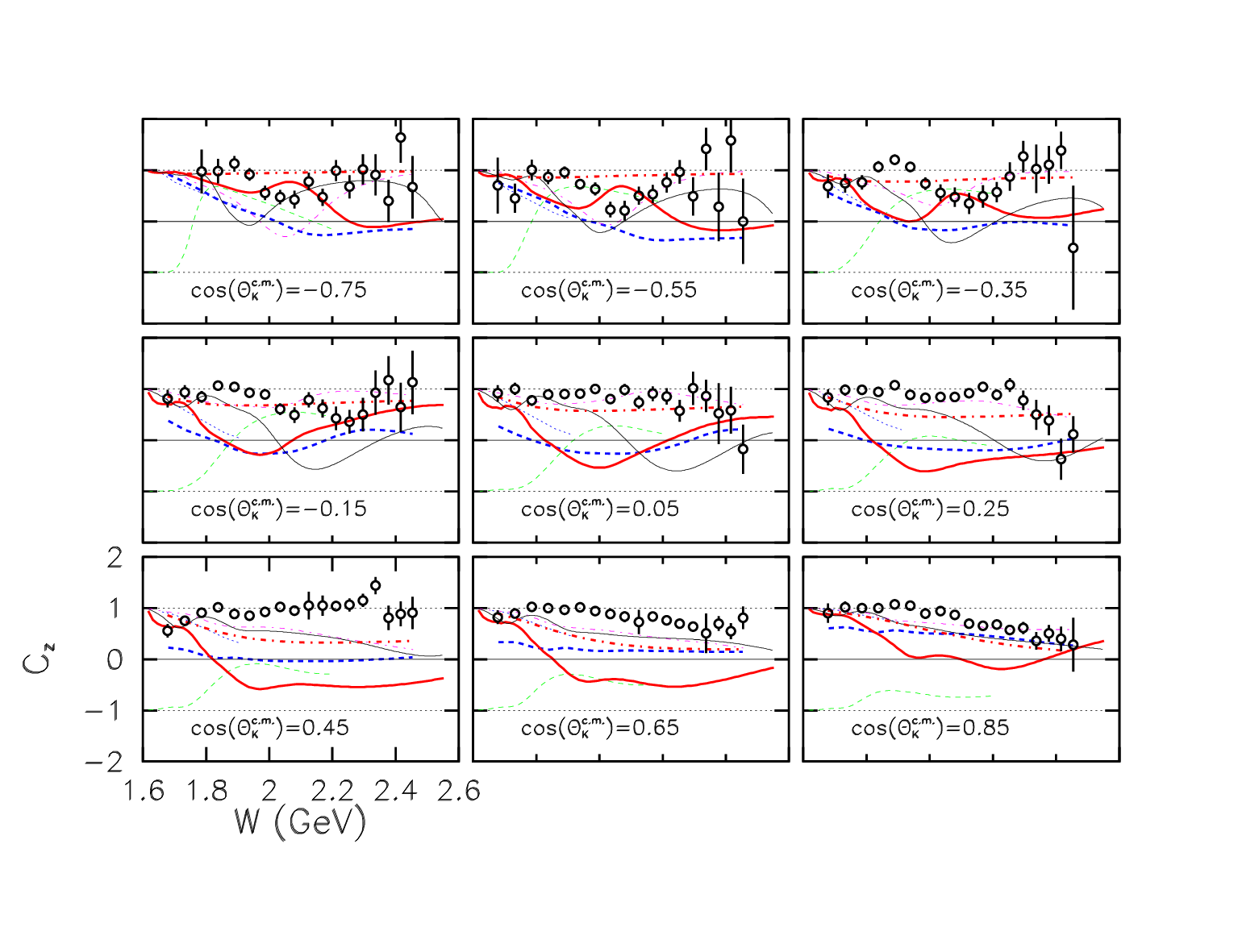}
\includegraphics[width=0.4\textwidth,height=0.53\textwidth]{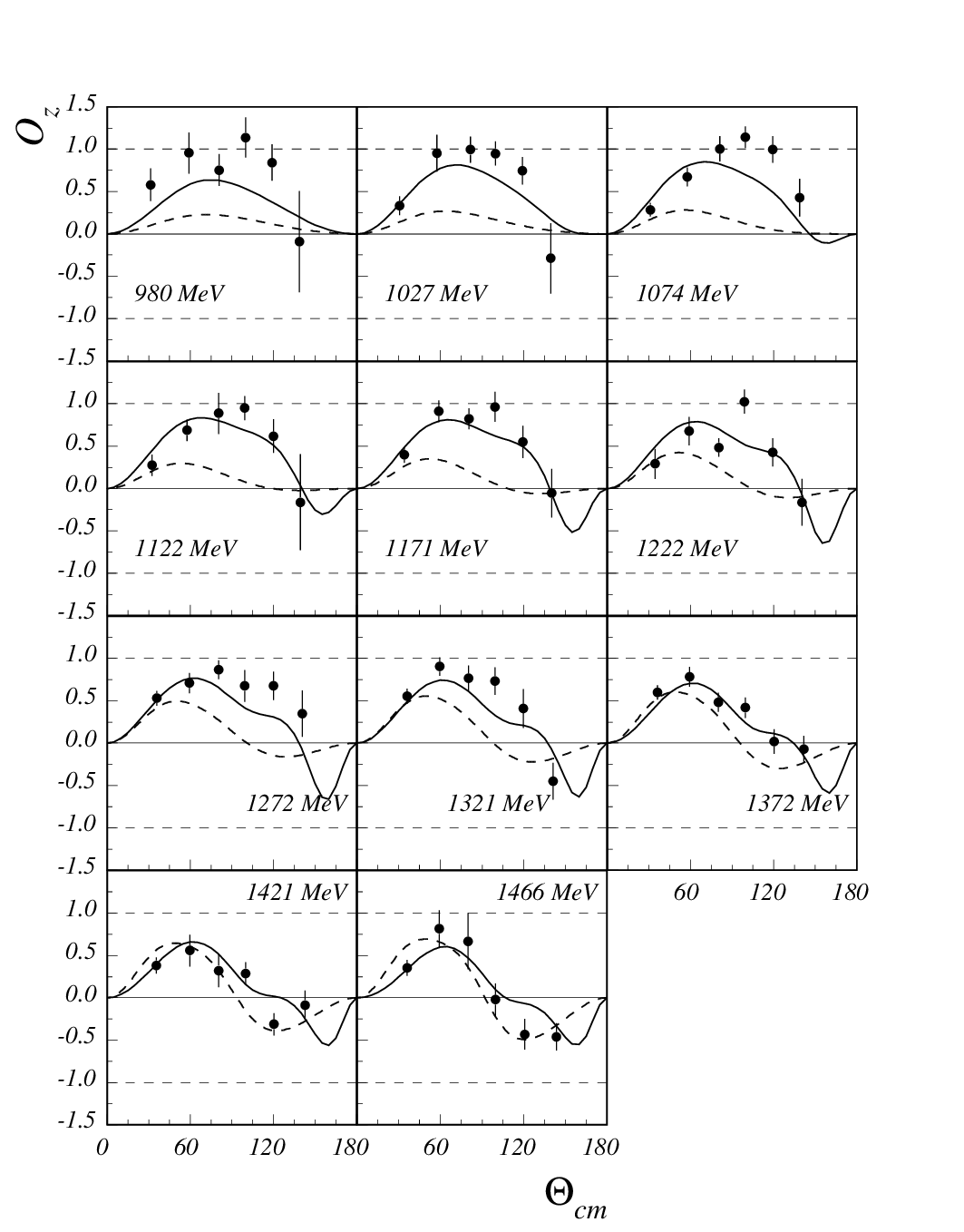}\\[-5ex]
\end{figure}

The observables, $C_{x,\,z}$, describe the spin transfer from the circularly-polarized 
photon to the recoiling hyperon along and perpendicular to the beam axis in the c.m. 
system, respectively. Most remarkable, the CLAS collaboration reported that the total 
$\Lambda$~polarization vector, $R_\Lambda = \sqrt{P^2 + C_x^2 + C_z^2}$, is consistent 
with unity and that the results also suggest $C_z \simeq C_x + 1$ over a large range of 
$W$~values and $K^+$~angles~\cite{Bradford:2006ba}. This observation implies that the 
$\Lambda$~ hyperons produced in the reaction $\vec{\gamma} p\to K^+\vec{\Lambda}$ 
using circularly-polarized photons are almost 100\,\% spin polarized, which is not required by 
the reaction kinematics. Figure~\ref{Figure:Lambda-cz} (left) shows the CLAS results for 
$C_z$~\cite{Bradford:2006ba}. None of the models shown in the figure reproduces 
the experimental results well. The phenomenon is still not properly understood. The right 
side of Figure~\ref{Figure:Lambda-cz} shows results from GRAAL for the observables, 
$O_z$~\cite{Lleres:2008em}. The authors of \cite{Bradford:2006ba}  used a number of inequalities and the two 
equations
\begin{eqnarray}
  C_x^2 \,+\, C_z^2 \,+\, O_x^2 \,+\, O_z^2\,=\, 1 \,+\, T^2 \,+\, P^2 \,+\, \Sigma^2\\[0.5ex]
  C_zO_x \,-\, C_xO_z \,=\, T \,-\, P\Sigma
\end{eqnarray}
to show the overall consistency of the CLAS and GRAAL results. In principle, one 
additional double-polarization observable (either beam-target or target recoil) would 
be sufficient in the energy range below 1.5~GeV to extract the helicity amplitudes for 
$K^+\Lambda$~photoproduction (see also the discussion in Section~\ref{Section:EM-Baryons}). 
The large polarization transfer in $K^+\Lambda$~production was confirmed in CLAS
electroproduction experiments, $\vec{e}p\to e^{\,\prime} K^+\vec{\Lambda}$, which 
span a wide range of momentum transfer $Q^2$ from 0.7 to 5.4~GeV$^2$~\cite{Carman:2009fi}. 
The $\Lambda$~polarization was found to be largest along the direction of the virtual photon.
The authors of~\cite{Carman:2009fi} also discuss the reaction dynamics in the framework of a partonic model.
It appears to be important to shed light on the relevance of the quark-gluon dynamics 
when the $s\bar{s}$~pair is created in a domain thought to be dominated by meson/baryon 
degrees of freedom. The $\Sigma^0$~hyperon differs substantially from the simple 
behaviour of the $\Lambda$~hyperon.

A full discussion of the relevant nucleon resonance contributions in hyperon 
photoproduction goes beyond the scope of this review. We only give a short 
summary in this section. Since the $\Lambda$~hyperon is an isoscalar, only 
intermediate $I = \frac{1}{2}$ nucleon resonances can couple to $K^+\Lambda$
and the reaction serves thus as an isospin filter. The $\Sigma^0$~hyperon is an 
isovector, which allows it to couple to both $I = \frac{1}{2}$~$N^\ast$ and 
$I=\frac{3}{2}~\Delta$~states. The total cross section for $K^+ Y~(Y=\Lambda,\,\Sigma^0$) 
reaches about $2.5~\mu$b and a structure around 1900~MeV is observed 
in both reactions. However, its interpretation in terms of $s$-channel baryon 
resonances has been highly controversial. The latest double-polarization 
observables have revealed indications for baryon resonances and additional
observables will further our understanding of this peak structure. The cross 
section for $K^+\Sigma^0$ photoproduction is about a factor of four bigger 
than for $K^0\Sigma^+$, partly owing to forbidden background contributions 
from $K$~exchange in the latter reaction.

Strong contributions from the new resonances listed by the PDG are observed 
in the Bonn-Gatchina multichannel approach~\cite{Anisovich:2011fc}.  
The $N(1880)\,\frac{1}{2}^+$~state was first proposed on the basis of data on 
$\gamma p\to\Sigma^+ K_S$~\cite{Castelijns:2007qt}, but was also seen earlier 
in an analysis of $\pi N\to\pi N$ and $\pi N \to\pi\pi N$~\cite{Manley:1992yb}. Overall 
evidence for the existence of the $N(1895)\,\frac{1}{2}^-$~resonance appears to be 
stronger. It was recently observed in $K^+\Lambda$ and $K^+\Sigma^0$
production~\cite{Anisovich:2011fc}, but also found earlier in several analyses
of $\pi N\to\pi N$ data~\cite{Beringer:1900zz}. The latest analysis of the SAID 
group~\cite{Arndt:2006bf} finds no indication for this resonance, though. Evidence for 
$N(1875)\,\frac{3}{2}^-$ was claimed in~\cite{Mart:1999ed} based on $\gamma p
\to K^+\Lambda$ from SAPHIR~\cite{Glander:2003jw}. It was confirmed in 
several more recent multichannel analyses, e.g.~\cite{Penner:2002ma,
Anisovich:2011fc}. The recently published beam-recoil observables provide 
important evidence for the existence of the $N(1900)\,\frac{3}{2}^+$~state and 
inspired an upgrade of the resonance's star assignment from two to three. 
The right side of Figure~\ref{Figure:Lambda-cz} shows two curves in addition 
to the GRAAL data on the $O_z$~observable denoting the Bonn-Gatchina solution (solid line) and a
description within the Ghent RPR (Regge-plus-resonance) isobar model 
(dashed line)~\cite{Corthals:2007kc}. Both models require the $N(1900)\,\frac{3}{2}^+$ 
to describe the data at the highest energies. Initial evidence for this resonance 
in the Bonn-Gatchina model came from the CLAS data on the 
$C_{x,\,z}$~observables~\cite{Nikonov:2007br}.

\subsubsection{Photoproduction of Vector Mesons}
The vector mesons, $\rho,~\omega,$ and $\phi$, carry spin-parity quantum
numbers $J^{PC} = 1^{--}$ and thus have the same quantum numbers as
the photon. For this reason, when a photon probes a nucleon, one should 
expect vector mesons to play an important role in the spectrum of intermediate 
hadronic states. The total cross sections for photoproduction of these three mesons reach about $24~\mu$b, $8.5~\mu$b,
and $0.2~\mu$b, respectively. The total $\omega$~cross section exhibits
a pronounced peak structure at about $E_\gamma = 1.3$~GeV in addition to 
a broader peak similar in shape to that seen in $\rho$ and $\phi$ production. The
differential cross sections show an exponential fall-off at small values of
the squared recoil momentum, $t$. Further strong contributions that cannot be explained by $t$-channel 
pion (kaon)~exchange are 
observed at larger values of $t$.

In contrast to pseudoscalar mesons, vector-meson decays give rise to 
additional observables. The decays provide a measure of their spin-density 
matrices, which is equivalent to determining the intensity, polarization, 
and tensor polarization of these vector mesons. Without polarization
observables, a more detailed analysis of resonance contributions as
well as the study of the relative strength of the diffractive production
and $\pi^0$~exchange is impossible. At CLAS, a determination of a
large variety of such observables in $\omega$~production including
spin-density matrix elements (SDMEs) has become feasible and has consequently
received renewed interest in recent years. The full decay angular distributions
were initially discussed in~\cite{Schilling:1969um} for decays into pseudoscalar
mesons and more recently also for the radiative decay of the $\omega$~meson 
into $\pi^0\gamma$~\cite{Zhao:2005vh}.

Differential cross sections for $\rho$ and $\phi$~production were 
reported by SAPHIR~\cite{Barth:2003bq,Wu:2005wf}, LEPS~\cite{Mibe:2005er} 
and CLAS~\cite{Dey:2011zv}. Production of $\omega$~mesons 
was studied at SAPHIR~\cite{Barth:2003kv}, GRAAL~\cite{Ajaka:2006bn}, and 
recently, with large statistics, at CLAS~\cite{Williams:2009aa} for energies from 
threshold to $W = 2.4$~GeV. The experimental situation for polarization observables
is still scarce. Data on the $\omega$~beam asymmetry, $\Sigma$, come from 
CBELSA/TAPS~\cite{Klein:2008aa}  and high-statistics results on the SDMEs,  
$\rho^0_{00},~\rho^0_{10},~\rho^0_{1-1}$, using an unpolarized photon beam 
from CLAS~\cite{Williams:2009ab}. The first measurements of $\omega$
double-polarization observables from the CLAS-FROST program were presented 
recently at conferences~\cite{Collins:CIPANP2012}.

At lower energies, $W$~$<$~2~GeV, an earlier analysis included the beam asymmtries 
measured by the GRAAL collaboration and reported significant sensitivity
to $N(1720)\,\frac{3}{2}^+$ and the off-shell states $N(1520)\,\frac{3}{2}^-$ and $N(1680)\,\frac{5}{2}^+$ 
in the energy ranges $E_\gamma\in [1.108,\,1.218]$~GeV and $E_\gamma\in
[1.327,\,1.423]$~GeV~\cite{Hourany:2005wh}. A PWA of the Bonn-Gatchina group included the SAPHIR 
and GRAAL data and also found the $\frac{3}{2}^+$~wave as the dominant contribution, 
which was identified with the $N(1720)\,\frac{3}{2}^+$. The CLAS PWA found strong 
contributions from a $\frac{3}{2}^-$~wave at threshold, but did not include other data in the
analysis. The Gie\ss en group studied $\omega$~production within a coupled-channel 
effective Lagrangian approach in the energy range from the pion threshold up to 2~GeV 
and included data on $\pi$- and $\gamma$-induced reactions available at that time 
for the final states $\gamma N$, $\pi N$, $2\pi N$, $\eta N$, and $\omega N$~\cite{Shklyar:2004ba}. 
The analysis indicates that the initial resonance peak around 1700~MeV is primarily 
a result of the two sub-threshold resonances, $N(1675)\,\frac{5}{2}^-$ and $N(1680)\,\frac{5}{2}^+$. 
The latter hardly influences the reaction $\pi N\to\omega N$, but is significant in 
$\omega$~photoproduction due to its large $A_{\frac{3}{2}}$~helicity
amplitude. 

At higher energies, the $\gamma p\to p\omega$ differential cross section is
dominated by $t$-channel exchange, but resonance production is still observed. 
At energies  above $W = 2$~GeV, a PWA based on the CLAS unpolarized cross section 
alone~\cite{Williams:2009aa} required contributions from at least two spin-$\frac{5}{2}$ 
resonances and from a heavier spin-$\frac{7}{2}$ resonance to correctly describe the cross 
section and the corresponding phase motion. The higher-spin state was identified 
as the $N(2190)\,\frac{7}{2}^-$~resonance listed as a 4-star state in the RPP~\cite{Beringer:1900zz}. 
The first-time observation of this resonance in photoproduction was later confirmed in 
$\gamma p\to p\pi^0$~\cite{Crede:2011dc}. The two $\frac{5}{2}$-states were identified as 
the well-known $N(1680)\,\frac{5}{2}^+$ and the $N(2000)\,\frac{5}{2}^+$. The latter is a 2-star 
resonance that was also observed in the latest SAID analysis~\cite{Arndt:2006bf}.

\subsubsection{\label{Section:MultiMeson}Photoproduction of Multi-Meson Final States}
\begin{figure}
\caption{\label{Figure:pi0pi0}(Colour online) Left: Total cross section for
  the reaction $\gamma p\to p\pi^0\pi^0$ as a function of the
  incident-photon energy. Picture from~\cite{Kashevarov:2012wy}. Right: 
  Beam-helicity asymmetry, $I^\odot$, for $\vec{\gamma} p\to n\pi^+\pi^0$ 
  from MAMI~\cite{Krambrich:2009te}. The different curves denote model descriptions
  from~\cite{Fix:2005if} (red) and~\cite{Roca:2004vs} (blue, black). Picture 
  from~\cite{Krambrich:2009te}.}\vspace{2mm}
\includegraphics[width=0.46\textwidth,height=0.405\textwidth]{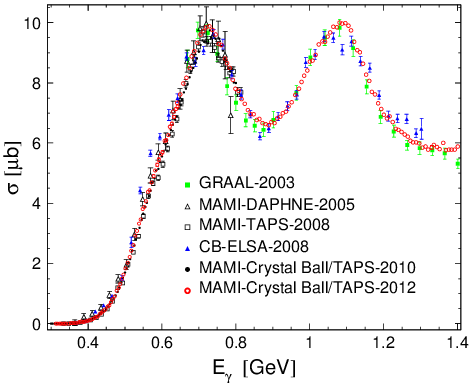}\qquad
\includegraphics[width=0.46\textwidth,height=0.4\textwidth]{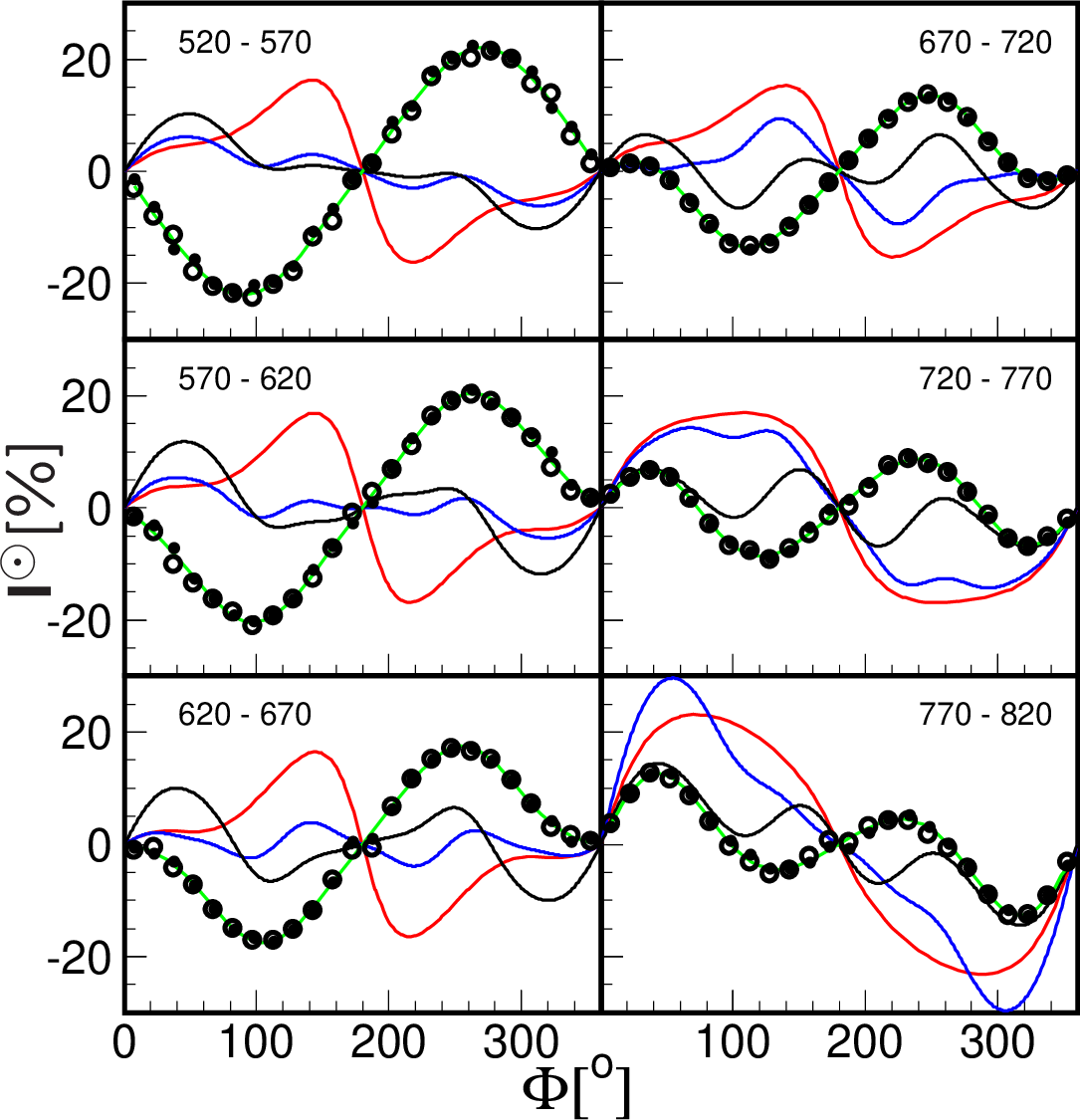}
\end{figure}

The photoproduction of two mesons, in particular the production of two pions, 
plays an increasing role toward higher energies. Above $E_\gamma = 2$~GeV, the 
double-pion cross section is the biggest contributor to the total photoabsorption 
cross section. The two-pion cross section exhibits a double-peak structure, which 
is most pronounced in the reaction $\gamma p\to p\pi^0\pi^0$. The total cross 
section for this reaction is shown in Figure~\ref{Figure:pi0pi0} (left). The first peak
is in the second resonance region around $W\approx 1500$~GeV, the second peak 
around $W\approx 1700$~GeV. For the $p\pi^0\pi^0$ final state, it reaches a maximum 
of about $10~\mu$b in the first peak, which is about a factor of eight smaller than 
the strength in the $p\pi^+\pi^-$ cross section. A complete description of the
production of two pseudoscaler mesons requires five independent kinematic variables.
Since the reaction plane and the decay plane spanned by the final state particles
form an angle, the reaction does not proceed {\it in plane} as the production of 
a single meson does. The differential cross section is thus five-fold differential and 
using a polarized beam and an unpolarized target, can be written as
\begin{eqnarray}
\frac{d\sigma}{dx_i}\,=\,\biggl(\frac{d\sigma}{dx_i}\biggr)_{\rm
  unpol.}\,(1\,+\,\delta_\odot\,I^\odot\,+\,\delta_l\,(I^s\,{\rm
  sin}\,2\beta\,+\,I^c\,{\rm cos}\,2\beta))\,,
\end{eqnarray}
where $\delta_\odot$ and $\delta_l$ denote the degrees of circular and linear 
polarization, $\beta$ the angle between the reaction plane and the direction of 
the linear beam polarization vector, and $I^\odot$, $I^s$, $I^c$ are the circular and linear 
polarization observables. A full theoretical discussion of the 64 possible polarization observables is given in~\cite{Roberts:2004mn}.

The photoproduction cross sections of two neutral pions in the reaction $\gamma p\to 
p\pi^0\pi^0$ have been studied at ELSA~\cite{Thoma:2007bm}, at GRAAL~\cite{Assafiri:2003mv},
and very recently at MAMI with a fine energy binning from threshold up to 0.8~GeV~\cite{Zehr:2012tj} 
and 1.4~GeV~\cite{Kashevarov:2012wy}. We refer to the references in~\cite{Kashevarov:2012wy} 
for the large variety of earlier MAMI results obtained with the DAPHNE and TAPS detectors. 
In~\cite{Zehr:2012tj}, results are also discussed for the reaction $\gamma p\to n\pi^+\pi^0$.
The reaction $\gamma n\to n\pi^0\pi^0$ was studied at GRAAL~\cite{Ajaka:2007zz}. The
photoproduction of two charged pions was studied at SAPHIR~\cite{Wu:2005wf} and, at 
higher energies ($E_\gamma > 3$~GeV) with a focus on meson spectroscopy, at 
CLAS~\cite{Battaglieri:2008ps,Battaglieri:2009aa}. Polarization results are scarce. At 
low energies ($E_\gamma < 820$~MeV), the beam-helicity asymmetry, $I^\odot$, 
has been recently measured at MAMI~\cite{Krambrich:2009te} for the reactions 
$\vec{\gamma} p\to p\pi^+\pi^-$, $n\pi^+\pi^0$, and $p\pi^0\pi^0$. Earlier, the 
helicity difference, $P_z^\odot$, was determined by the GDH collaboration at MAMI 
for $p\pi^+\pi^-$~\cite{Ahrens:2007zzj}. CLAS published data on $I^\odot$ in 
$p\pi^+\pi^-$ for the $1.4<W<2.4$~GeV range and some data on the linear beam 
asymmetry, $\Sigma$, are available from GRAAL~\cite{Assafiri:2003mv}.

Double-pion photoproduction allows the study of sequential decays of nucleon 
resonances via intermediate excited states, but also the investigation of direct 
decays into $N\rho$ and $N\sigma$. However, large background contributions,
in particular in the reaction $\gamma p\to p\pi^+\pi^-$, render an analysis 
challenging. Resonance contributions are more pronounced in $\pi^0\pi^0$
\cite{Thoma:2007bm,Kashevarov:2012wy,Assafiri:2003mv}, but contributions from $N\rho$ intermediate states are excluded. Even at low energies, in and below 
the second resonance region, double-pion production is still poorly understood. 
While the unpolarized cross sections are fairly well described, models disagree 
substantially on the description of the available polarization data. As an example, 
Figure~\ref{Figure:pi0pi0} (right) shows the unprecedented quality of the data 
for the beam-helicity observable, $I^\odot$, in $\vec{\gamma} p\to n\pi^+\pi^0$, 
but also the discrepancies between the models discussed in~\cite{Fix:2005if,Roca:2004vs}. 

It is widely believed that double-pion production is dominated by intermediate production 
of a $\Delta\pi$~state. In the second resonance region, the $N(1520)\,\frac{3}{2}^-$~state 
plays an important role in $\gamma p\to p\pi^0\pi^0$. A large contribution from the 
$J=\frac{3}{2}$ wave (as a reflection of the $J^P=\frac{3}{2}^+$ fraction of $\pi^+\pi^-
\to\pi^0\pi^0$ rescattering) appears also crucial at energies below
the second resonance region~\cite{Kashevarov:2012wy}. 
In~\cite{Sarantsev:2007aa}, the same low-energy behaviour and the large peak around 
$E_\gamma = 700$~MeV (Figure~\ref{Figure:pi0pi0}, left) is explained by interference of 
the $N(1520)\,\frac{3}{2}^-$ and $\Delta(1700)\,\frac{3}{2}^-\to(\Delta\pi)_{L=0}$~resonances. 
However, the strong contribution of the $\Delta(1700)\,\frac{3}{2}^-$~state is in conflict 
with the model results discussed in the GRAAL publication~\cite{Assafiri:2003mv} and the 
Valencia model, e.g.~\cite{Nacher:2000eq}. The $N(1720)\,\frac{3}{2}^+$ was found
in~\cite{Thoma:2007bm} to decay strongly into $\Delta\pi$, not reported by the 
PDG at that time. A similar discrepancy in the width was observed in $\pi^+\pi^-$
electroproduction data from JLab but interpreted as a second (rather narrow) 
$N\,\frac{3}{2}^+$~resonance, e.g.~\cite{Ripani:2002ss}. At higher energies, 
the Bonn-Gatchina group observes a strong coupling of the (newly-listed) 
$N(1875)\,\frac{3}{2}^-$~state to $N\sigma$~\cite{Anisovich:2011fc}. In 
conclusion, double-pion production is far from understood and the full 
spectrum of polarization observables is likely required for progress in
this exciting field. 

\begin{figure}
\caption{\label{Figure:pi0eta}(Colour online) Beam asymmetries in
  $\vec{\gamma}p\to p\pi^0\eta$. Left: The observable~$I^\odot$ measured at 
  MAMI~\cite{Kashevarov:2010gk}. The dotted curve denotes a Fourier expansion, the
  remaining curves represent predictions of a full isobar model with six resonances 
  (solid line) and with only the $\Delta\,\frac{3}{2}^-$ amplitude (dashed line)~\cite{Fix:2010bd}.
  Data and picture from~\cite{Kashevarov:2010gk}. Right: The observable~$I^s$ from 
  CBELSA/TAPS~\cite{Gutz:2009zh}. The rows (top to bottom) show the proton, 
  $\pi^0$, and $\eta$ as recoiling particle in the c.m. system, respectively (open 
  circles: $-I^s\,(2\pi - \phi^\ast$)). The curves denote the Bonn-Gatchina fit for 
  the full solution (solid line) and without the $\frac{3}{2}^-$ wave. 
  Data and picture from~\cite{Gutz:2009zh}.}\vspace{4mm}
\includegraphics[width=0.5\textwidth,height=0.372\textwidth]{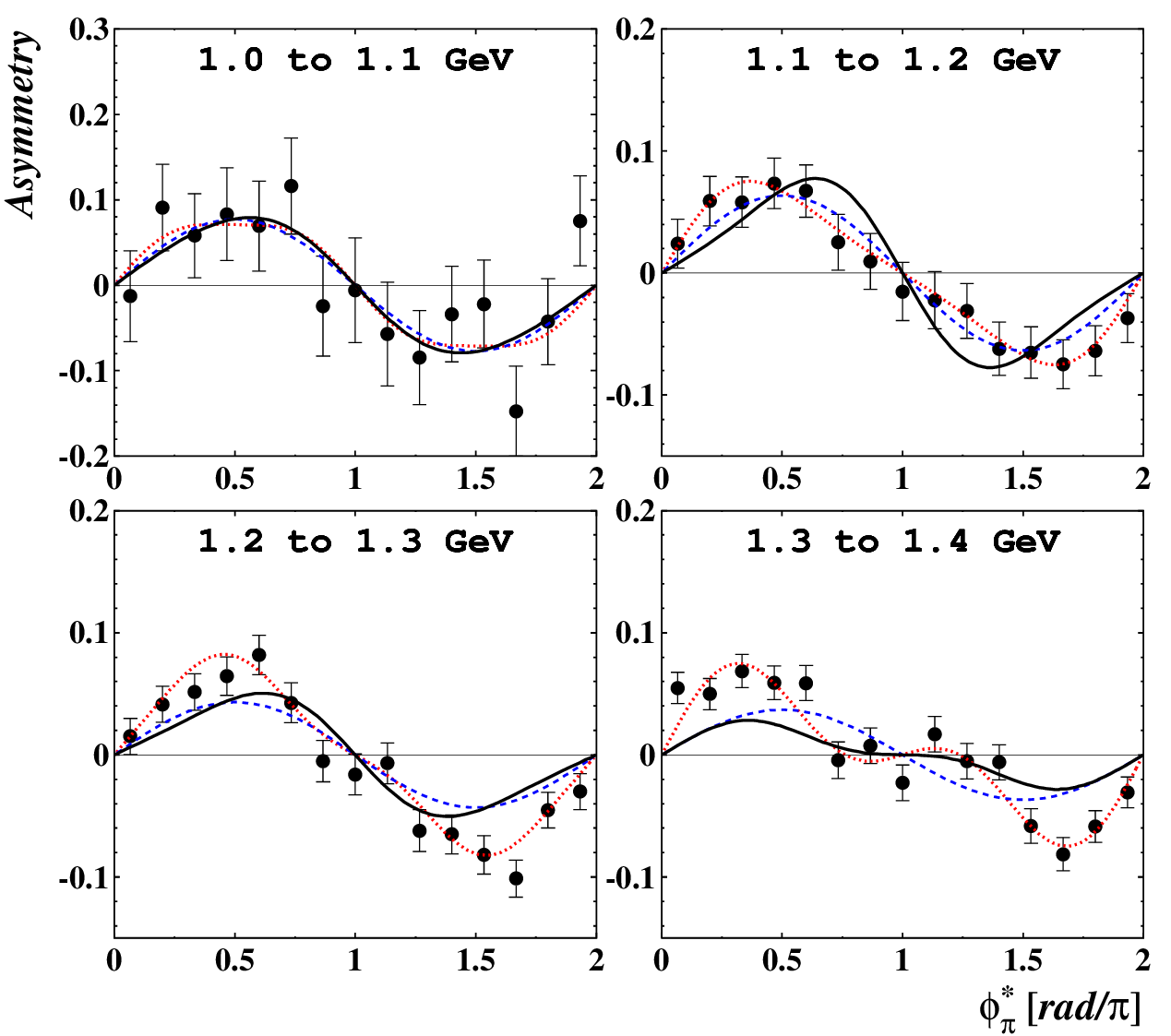}\quad
\includegraphics[width=0.45\textwidth,height=0.38\textwidth]{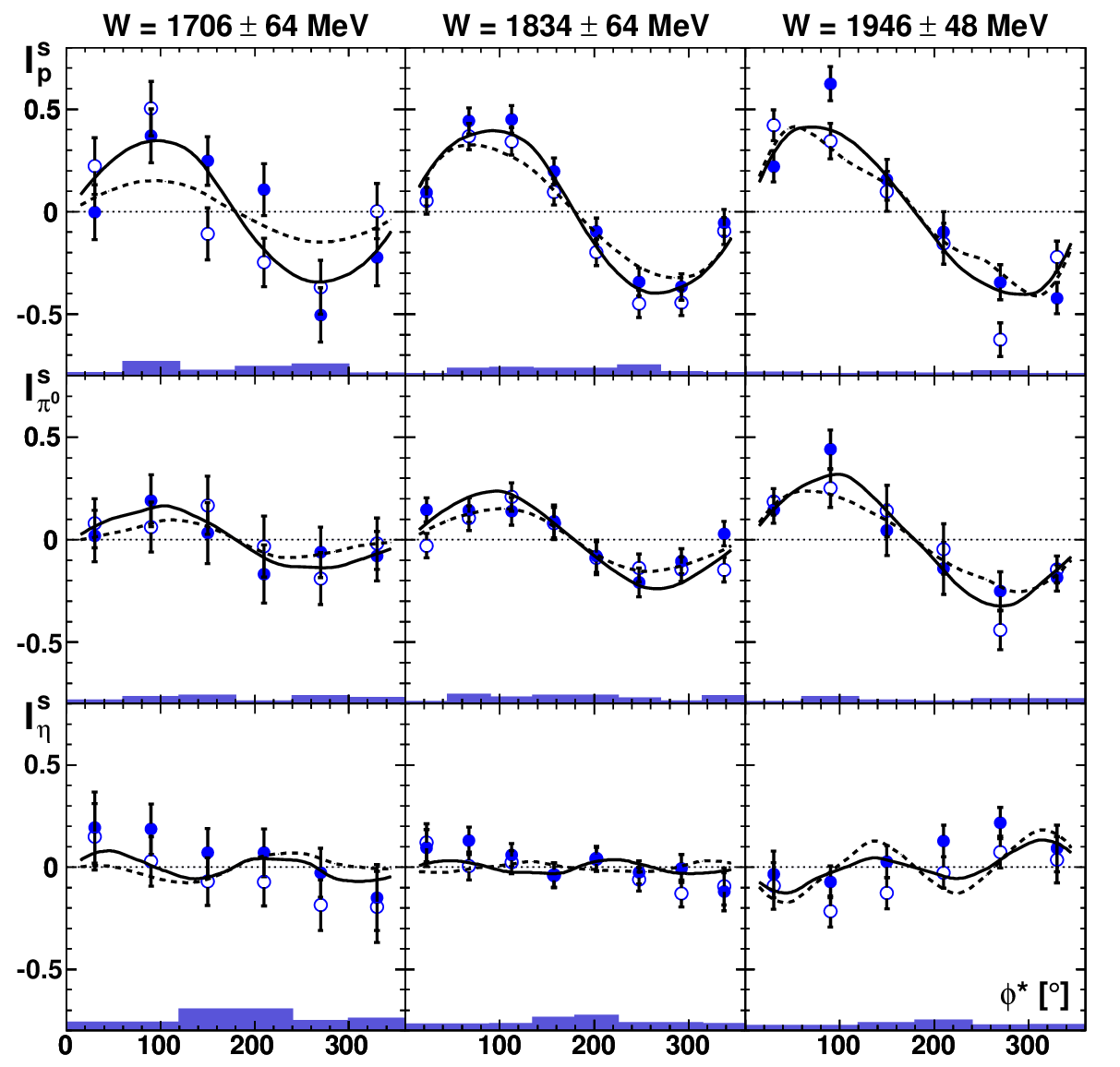}
\end{figure}

The reaction $\gamma p\to p\pi^0\eta$ was measured at GRAAL~\cite{Ajaka:2008zz}, 
MAMI~\cite{Kashevarov:2010gk}, and ELSA~\cite{Horn:2007pp,Horn:2008qv,Gutz:2009zh}.
The published results include cross section measurements and the first-time extraction 
of the beam asymmetries using circularly- (Figure~\ref{Figure:pi0eta}, left) and 
linearly-polarized photons (Figure~\ref{Figure:pi0eta}, right). The total unpolarized cross section 
rises rapidly from the reaction threshold at $E_\gamma\approx 930$~MeV to a maximum 
of about $3.5~\mu$b at 1500~MeV and then slowly declines without any structures. 
Most of the reaction rate proceeds via the $\Delta\eta$ intermediate state, which 
serves as an isospin filter for $\Delta^\ast$~resonances. Other dominant contributions 
include the $p a_0(980)$ and $N(1535)\,\frac{1}{2}^-\,\pi^0$ intermediate states. 

Several analyses have observed strong contributions from the $\Delta\,\frac{3}{2}$
wave to $p\pi^0\eta$. This is not surprising. The $\Delta(1232)$ and the $\eta$ have
$J^P=\frac{3}{2}^+$ and $J^P=0^-$, respectively, and can thus form the total spin-parity 
state $J = \frac{3}{2}^-$ with orbital angular momentum $L = 0$. Little doubt exists 
about the importance of the $(\ast\,\ast\,\ast\,\,\ast)~\Delta(1700)\,\frac{3}{2}^-$~resonance in this 
reaction~\cite{Anisovich:2011fc,Fix:2010bd,Doring:2010fw}, but
contributions from the higher-mass 
$\Delta(1940)\,\frac{3}{2}^-$ resonance seem to be confirmed. Figure~\ref{Figure:pi0eta} 
(left) shows a description of the beam asymmetry, $I^\odot$, using the model in~\cite{Fix:2010bd} 
for the full solution (solid line), and for the $\Delta\,\frac{3}{2}^-$~amplitude (dashed 
line) only. Figure~\ref{Figure:pi0eta} (right) shows a description of the beam asymmtry, $I^s$, 
using the Bonn-Gatchina framework~\cite{Anisovich:2011fc}. The full solution
is again given by the solid line, whereas the dashed line denotes the full solution 
without the $\Delta\,\frac{3}{2}^-$, which demonstrates the importance of this 
contribution. Both models also include the positive-parity $\Delta\,\frac{3}{2}^+$ 
waves, $\Delta(1600)$ and $\Delta(1920)$. Weaker contributions are also observed 
from the higher-spin resonance, $\Delta(1905)\,\frac{5}{2}^+$, and from a corresponding
negative-parity state above $W = 2$~GeV~\cite{Gutz:2009zh}. However, the experimental 
situation clearly becomes worse at these higher energies. It is also important to emphasize 
that despite the observation of the $\Delta(1700)\,\frac{3}{2}^-$~state
in the reaction $\gamma p\to p\pi^0\eta$, no agreement on the nature of this 
resonance has been reached. Interpretations range from a dynamically-generated 
resonance~\cite{Doring:2010fw} to a regular $3q$~quark model state.

The reaction $\gamma p\to p\pi^0\omega$ gives access to the $\Delta\omega$
system, which also serves as an isospin filter for $\Delta^\ast$~resonances. The
reaction is very interesting and promises a rich source of baryon production. Similar 
to the $p\pi^0\eta$ final state, first results indicate that $\gamma p\to p\pi^0\omega$ 
proceeds dominantly via $\Delta\omega$. Since the $\omega$ has $J^P = 1^-$, 
$\Delta\,\frac{3}{2}$ and $\Delta\,\frac{5}{2}$~states should strongly contribute. 
The current experimental situation is weak, though. Cross sections based on about 
2000~events from CB-ELSA were published in 2007~\cite{Junkersfeld:2007yr}, but 
a higher-statistics data set from CBELSA/TAPS with almost an order of magnitude
more events has been presented at conferences~\cite{Crede:2011zz} and will be
available soon. Signatures of the reaction $\gamma p\to\Delta\rho$ have been 
observed in CLAS data at high photon energies~\cite{Bookwalter:PhD}.

\subsection{Observation of Baryons in $J/\psi$~Decays}
An alternative approach to studying nucleon resonances in induced
reactions from a nucleon target is the investigation of baryons which
are observed in $J/\psi$ and $\psi^{\,\prime}$~decays into a baryon, 
an antibaryon, and an additional meson. In reactions such as $\psi^{\,\prime}\to
p\bar{p}\pi^0$, $N^\ast$~resonances coupling tp $p\pi^0$ or $\bar{p}
\pi^0$ can be studied. In these decays, contributions of the $\Delta$
resonance are excluded owing to isospin conservation, which greatly
facilitates this type of analysis. Excited hyperons, such as $\Lambda^\ast,~
\Sigma^\ast,~{\rm and}~\Xi^\ast$, can also be studied. First measurements
of $J/\psi$ and $\psi(2S)$~decays into $\Lambda\bar{\Lambda}\pi^0$
and $\Lambda\bar{\Lambda}\eta$ were published by the BES 
collaboration~\cite{Ablikim:2007ah}, and more recently even
branching fractions for the decay $J/\psi\to\Lambda\bar{\Lambda}\pi^+\pi^-$~\cite{Ablikim:2012qn}.

\begin{figure}[t]
\caption{\label{Figure:NSTAR_from_ccbar}(Colour online) Left: Invariant
$p\pi^-$~distribution for $J/\psi\to p\pi^-\bar{n}$ from
BES~\cite{Ablikim:2004ug}; the phasespace Monte Carlo distribution is
shown as a dotted line. Middle: Invariant $p\pi^0$~mass spectrum for 
$J/\psi\to p\bar{p} \pi^0$ from BES~\cite{Ablikim:2009iw}. The data
are represented by crosses and the histogram denotes the fit. Pictures
from~\cite{Ablikim:2004ug,Ablikim:2009iw}. Right: Invariant $p\pi^0$~spectrum for 
$\psi^{\,\prime} \to p\bar{p}\pi^0$ from CLEO~\cite{Alexander:2010vd}. 
The solid line denotes the sum of the Monte Carlo distributions labeled 
in the figure for $N^\ast(1440)$~(1), $N^\ast(2300)$~(2), and two additional 
$p\bar{p}$~states (3,\,4). Picture from~\cite{Alexander:2010vd}.}
\includegraphics[width=5cm,height=5cm]{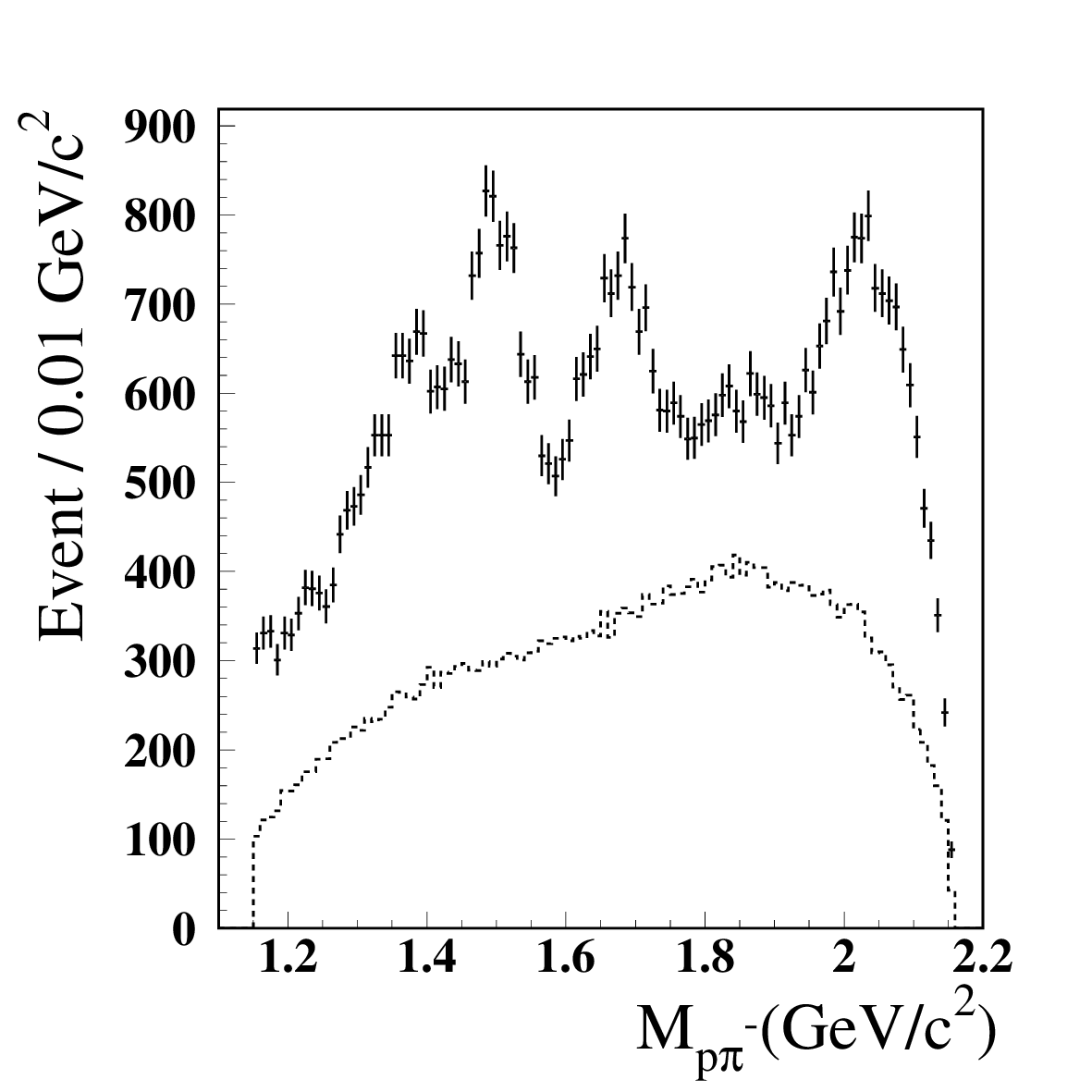} \hspace{-0.2cm}
\includegraphics[width=5cm,height=4.5cm]{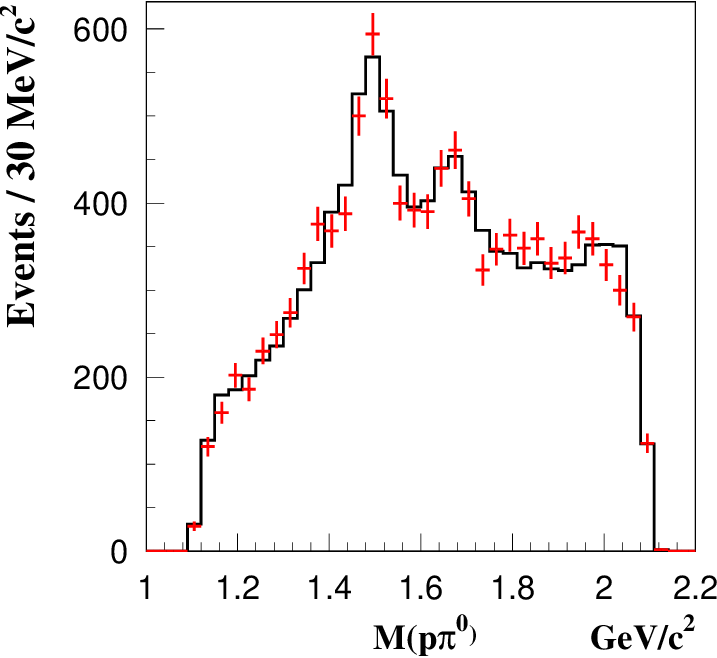} \hspace{0.15cm}
\includegraphics[width=5cm,height=4.75cm]{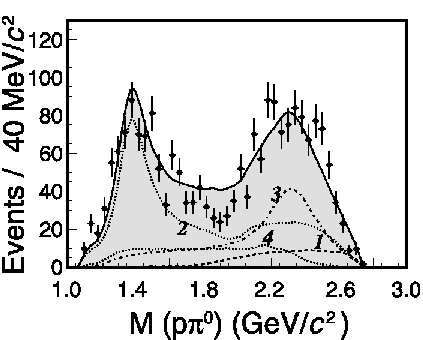}
\end{figure}

$N^\ast$~production was studied at BES in the reactions $J/\psi\to p\bar{p}
\pi^0$~\cite{Ablikim:2009iw}, $J/\psi\to p\bar{p}\eta$~\cite{Bai:2001ua}, 
and $J/\psi\to p\bar{n}\pi^-$~\cite{Ablikim:2004ug}. In 2005, the decays 
$\psi^{\,\prime}\to p\bar{p}\pi^0$ and $\psi^{\,\prime}\to p\bar{p}\eta$ were 
presented in~\cite{Ablikim:2005ir}, but first results of a PWA were reported 
just recently in~\cite{Liang:2012zz,Ablikim:2012}. Branching fractions were 
also published for $J/\psi\to p\bar{p}\omega$~\cite{Ablikim:2007ac}, $\psi^{\,\prime}
\to p\bar{n}\pi^-$~\cite{Ablikim:2006aha}, and $\psi^{\,\prime}\to\bar{p}n
\pi^+$~\cite{Ablikim:2006aha}, but the statistics are not sufficient to perform a 
PWA. Even double-pion production was observed in the reactions $\psi^{\,\prime}
\to\bar{p}n\pi^0\pi^+$~\cite{Ablikim:2006aha}. Moreover, BES has studied the 
interesting $J/\psi$ and $\psi^{\,\prime}$~decays into $B_8\bar{B}_8~(p\bar{p},~n
\bar{n},~\Lambda\bar{\Lambda},~\Sigma^0\bar{\Sigma}^0,~\Xi^-
\bar{\Xi}^+)$~\cite{Bai:2000ye,Ablikim:2005cda,Ablikim:2006aw,Ablikim:2008tj,
Ablikim:2012eu} and $\psi^{\,\prime}\to B_{10}\bar{B}_{10}~(\Delta^{++}\bar{\Delta}^{--},~
\Sigma^+(1385)\bar{\Sigma}^-(1385),~\Xi^0(1530)\bar{\Xi}^0(1530),~
\Omega^-\bar{\Omega}^+)$~\cite{Bai:2000ye}. Studying excited hyperons in
such decays remains thus a strong possibility. CLEO has 
recently presented PWA results for the reactions $\psi^{\,\prime}\to p\bar{p}
\pi^0$ and $\psi^{\,\prime}\to p\bar{p}\eta$~\cite{Alexander:2010vd}.

An early PWA of the $J/\psi\to p\pi^-\bar{n}$~channel at BES, based on
about $5.5\times 10^4$~events, observed four peaks~\cite{Ablikim:2004ug},
which the authors identify with the first direct observation of the $N^\ast(1440)$
peak, the well-known second and third resonance regions, and
a new $N^\ast$~state with a mass around 2030~MeV. The invariant
$p\pi^-$~mass distribution is shown in Figure~\ref{Figure:NSTAR_from_ccbar}
(left). A more recent PWA of the $J/\psi\to p\bar{p}\pi^0$~channel based 
on about $11,000$~events confirms the resonance above 2~GeV and determines 
a mass and width of $2040^{+3}_{-4} \pm 25$~MeV and $230^{+8}_{-8} 
\pm 52$~MeV, respectively~\cite{Ablikim:2009iw}. The analysis used 
the relativistic covariant tensor amplitude formalism described 
in~\cite{Rarita:1941mf,Liang:2002tk}. The observed invariant $p\pi^0$~mass 
distribution is shown in Figure~\ref{Figure:NSTAR_from_ccbar} (middle). 
Since $L=1$ is suppressed in the fits, the spin-parity of the $N^\ast$~state 
is limited to $\frac{1}{2}^+$ and $\frac{3}{2}^+$, which may explain the strong peak
in the mass spectrum (Figure~\ref{Figure:NSTAR_from_ccbar} (left)). The
analysis favours a $\frac{3}{2}^+$~assignment. Baryon production in $\pi N$ and
$\gamma N$~reactions allows all $N^\ast$~quantum numbers including
contributions from their $\Delta$~partners. 

\begin{table}
\begin{center}
\caption{\label{Table:BaryonsCharmonium} Summary of all nucleon
  resonances which have been observed by the BES collaboration in charmonium $c\bar{c}$~decays.}
\begin{tabular}{@{}|l|c|c|}
\br
 $N^\ast$ & 2012~\cite{Beringer:1900zz} & Reference\\
\mr
$N(1440)~1/2^+$ & $\ast\ast\ast\ast$ & $J/\psi\to p\bar{p}\pi^0,~p\pi^-\bar{n} +
c.c.$~\cite{Ablikim:2004ug,Ablikim:2009iw}, $\psi(2S)\to p\bar{p}\pi^0$~\cite{Ablikim:2012}\\
$N(1520)~3/2^-$ & $\ast\ast\ast\ast$ & $J/\psi~\&~\psi(2S)\to p\bar{p}\pi^0$~\cite{Ablikim:2009iw,Ablikim:2012}\\
$N(1535)~1/2^-$ & $\ast\ast\ast\ast$ & $J/\psi~\&~\psi(2S)\to
p\bar{p}\pi^0$~\cite{Ablikim:2009iw,Ablikim:2012}, $J/\psi\to p\bar{p}\eta$~\cite{Bai:2001ua}\\
$N(1650)~1/2^-$ & $\ast\ast\ast\ast$ & $J/\psi~\&~\psi(2S)\to
p\bar{p}\pi^0$~\cite{Ablikim:2009iw,Ablikim:2012}, $J/\psi\to p\bar{p}\eta$~\cite{Bai:2001ua}\\
$N(1710)~1/2^+$ & $\ast\ast\ast$ & $J/\psi\to p\bar{p}\pi^0$~\cite{Ablikim:2009iw}\\
$N(1720)~3/2^+$ & $\ast\ast\ast\ast$ & $\psi(2S)\to p\bar{p}\pi^0$~\cite{Ablikim:2012}\\
$N(2065)~3/2^+$ & $\ast$ & $J/\psi\to p\pi^-\bar{n} +
c.c.$~\cite{Ablikim:2004ug}, $J/\psi\to p\bar{p}\pi^0$~\cite{Ablikim:2009iw}\\
$N(2300)~1/2^+$ & & $\psi(2S)\to p\bar{p}\pi^0$~\cite{Ablikim:2012}
(also observed by CLEO~\cite{Alexander:2010vd})\\
$N(2570)~5/2^-$ & & $\psi(2S)\to p\bar{p}\pi^0$~\cite{Ablikim:2012}\\
\br
\end{tabular}
\end{center}
\end{table}

The CLEO collaboration reported results on the analysis of the $\psi(2S)\to
p\bar{p}\pi^0$~channel~\cite{Alexander:2010vd} and observed two clear
enhancements in the $p\pi^0$~mass around 1400 and 2300~MeV 
(Figure~\ref{Figure:NSTAR_from_ccbar} (right)). Without considering
interferences between resonances, the spectrum was described by two
$N^\ast$ (N(1440) and N(2300)) and two further $p\bar{p}$~resonances
(the solid curve in the figure denotes the sum of corresponding Monte Carlo
distributions). Mass and width of the $N^\ast(2300)$ were determined
to be $2300\pm 25$~MeV and $300\pm 30$~MeV, respectively. 
Very recently, the BES collaboration announced the observation of two
new $N^\ast$~resonances in $\psi(2S)\to p\bar{p}\pi^0$~\cite{Ablikim:2012}
based on a $\psi(2S)$~data set, which is about four times larger than the 
CLEO statistics. The earlier reported $N^\ast(2300)$ is confirmed by BES
with mass and width of $2300^{+40+109}_{-30-0}$~MeV and 
$340^{+30+110}_{-30-58}$~MeV, respectively, consistent with the
CLEO results. A $J^P = \frac{1}{2}^+$~assignment for this state is preferred
in the analysis. The second new state has been reported with a mass of
$2570^{+19+34}_{-30-0}$~MeV, a width of $250^{+14+69}
_{-24-21}$~MeV, and a $J^P$~assignment of $\frac{5}{2}^-$. A total of 
seven intermediate $N^\ast$~states has been observed to contribute
significantly in the $\psi(2S)$~decay, but the authors do not claim
evidence for the $N(2065)$~state previously observed by BES in 
$J/\psi$~decays~\cite{Ablikim:2009iw} or any further resonance in the
1900~MeV mass region. A summary of all nucleon resonances 
observed in charmonium decays is given in Table~\ref{Table:BaryonsCharmonium}.
\section{Models and Phenomenology}
\label{Section:Theory}
To get a better understanding of baryon structure, dynamics must be added to the symmetry schemes outlined in the early sections of this article. Theoretical studies of baryons are just about as old as the entire field of particle physics, and the literature is vast. Any attempt at summarizing this quantity of material in a few pages is futile: there are review articles on specfic approaches that are longer than the space allowed for this entire review.  Ruthless selectivity has to be exercised in choosing those approaches discussed, and even for those, only a very brief synopsis can be given. 

The overarching goals of baryon spectroscopy are to understand the relevant degrees of freedom in a baryon, and how the strong interaction in the non-perturbative regime gives rise to the states we observe. Different hypotheses regarding the relevant degrees of freedom, symmetries and dynamics lead to different spectra, and comparison with experiment will serve to validate or repudiate one set of assumptions or another. Unfortunately, no models can yet be completely eliminated.

Perhaps the most prevalent approach for attempting to understand the internal structure and dynamics of a baryon is through some kind of quark model. More recently, the lattice has been brought to bear with increasing success, and the large $N_c$ limit of QCD has also been used. Relativistic bound state equations have also been applied to the study of baryon spectra and properties, as have holographic models. In the next few subsections, the salient points of some of these approaches are discussed. Later in the manuscript,  the numerical predictions for the masses of baryons are shown and compared among the approaches discussed, as well as with the masses extracted from experiments. 

\subsection{Quark Models}

\subsubsection{Non- and Semi-Relativistic Models}
The constituent quark model provides perhaps the simplest, most intuitive insight into the structure of a baryon. Originally constructed as a classification scheme that predates QCD,  to explain the regularities among the new particles being discovered in the 1960s, this model has been through a number of incarnations, with varying degrees of sophistication and refinement, as well as varying degrees of success. The main appeal of such models is that they provide a simple framework within which a vast array of phenomenology may be integrated and understood.

\begin{figure}[h]
\caption{Quark model depiction of a baryon.}
\vspace{2mm}
\centerline{\includegraphics[width=2.1in]{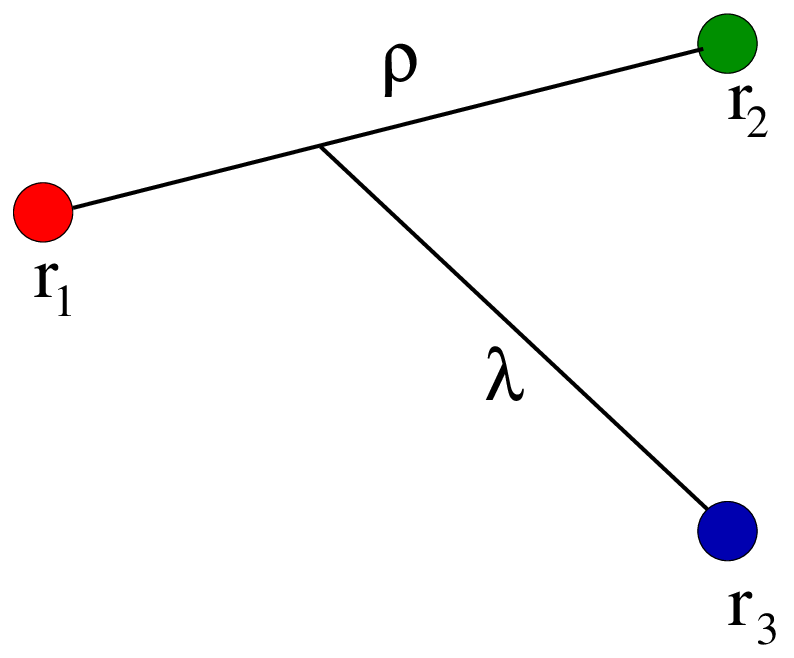}}
\label{fig:baryon}
\end{figure}

In its simplest version the quark model posits that a baryon consists of three valence quarks, as depicted in Figure \ref{fig:baryon}. All of the nonperturbative dynamics of the gluons being exchanged between pairs of quarks are assumed to be mimicked by a potential that at least provides the expected confinement when the separation between the quarks is large. The baryon spectrum is thus obtained by solving the wave equation for three particles moving in a potential. As far as possible, the symmetries discussed are built into the model.

The full wave function of the baryon is written as
\beq
\psi=\psi_{\rm colour}\psi_{\rm spin}\psi_{\rm space}\psi_{\rm flavour}.
\eeq
Since baryons are colour singlets, $\psi_{\rm colour}$ is antisymmetric under exchange of any pair of quarks. The product $\psi_{\rm spin}\psi_{\rm space}\psi_{\rm flavour}$ must therefore be symmetric under exchange of any pair of identical quarks. For baryons constructed of only $u$, $d$ and $s$ quarks, the flavour part of the wave function, $\psi_{\rm flavour}$,  is constructed in accordance with the representations discussed in Section \ref{sec:flavours}.

The spins of the quarks are coupled to either spin $\thlf$ with a totally symmetric wave function, or spin $\hlf$  with mixed symmetry in the wave function. The maximally stretched spin wave functions are
\beqy
\left|\frac{3}{2},\frac{3}{2}\right\rangle_S&=&\left|\uparrow\uparrow\uparrow\right\rangle,\nonumber\\
\left|\frac{1}{2},\frac{1}{2}\right\rangle_\lambda&=&-\frac{1}{\sqrt{6}}\left|\uparrow\downarrow\uparrow+ \downarrow\uparrow\uparrow-2\uparrow\uparrow\downarrow\right\rangle,\nonumber\\
\left|\frac{1}{2},\frac{1}{2}\right\rangle_\rho&=&\frac{1}{\sqrt{2}}\left|\uparrow\downarrow\uparrow- \downarrow\uparrow\uparrow\right\rangle,
\eeqy
where the subscripts indicate fully symmetric ($S$), symmetric under exchage of the first two quarks ($\lambda$) or antisymmetric under exchange of the first two quarks ($\rho$).

The spatial part of the wave function is obtained by solving a Hamiltonian equation that treats the baryon as consisting of three valence quarks that interact through a potential that includes spin-independent confining terms, as well as any number of spin-dependent contributions. The dynamics of the quarks in the baryon are described in terms of the Jacobi coordinates, $\vec{\rho}$, $\vec{\lambda}$ and $\vec{R}$ which are related to the quark positions by
\beqy
\vec{\rho}&=&\frac{1}{\sqrt{2}}\left(\vec{r}_1-\vec{r}_2\right),\nonumber\\ 
\vec{\lambda}&=&\sqrt{\frac{2}{3}}\left(\frac{m_1\vec{r}_1+m_2\vec{r}_2}{m_1+m_2}-\vec{r}_3\right),\nonumber\\
\vec{R}&=&\frac{1}{M}\left(m_1\vec{r}_1+m_2\vec{r}_2+m_3\vec{r}_3\right),
\eeqy
where
\beq
M=m_1+m_2+m_3.
\eeq
Here $\vec{\rho}$ is proportional to the separation between quarks 1 and 2, and $\lambda$ is proportional to the separation between quark 3 and the centre of mass of quarks 1 and 2. The choice of normalization of the coordinates is arbitrary. This particular choice leads to 
\beq
\vec{\lambda}=\frac{1}{\sqrt{6}}\left(\vec{r}_1+\vec{r}_2-2\vec{r}_3\right)
\eeq
when all quarks have the same mass.

In the nonrelativistic version of the quark model, the kinetic part of the Hamiltonian is
\beq
T=\sum_{i=1}^3\left(\frac{p_i^2}{2m_i}+m_i\right).
\eeq
This can be rewritten in terms of $p_\rho$, $p_\lambda$ and $p_R$, the momenta conjugate to the Jacobi coordinates defined above. This part of the Hamiltonian then becomes
\beq
T=-\frac{\nabla_\rho^2}{2m_\rho}-\frac{\nabla_\lambda^2}{2m_\lambda}-\frac{\nabla_R^2}{2M}+M,
\eeq
where $m_\rho=\frac{2m_1m_2}{m_1+m_2}$ and $m_\lambda=\frac{2}{3}\frac{m_3(m_1+m_2)}{m_1+m_2+m_3}$. The spatial part of the wave function is then written as a sum of separable components, namely
\beq
\psi_{\rm space}=\sum \psi_\rho(\vec{\rho})\psi_\lambda(\vec{\lambda}),
\eeq
and the dependence on the centre-of-mass coordinate $R$ is trivial. In some models, the kinetic energy term is treated relativistically, and is written
\beq
K_i= \sqrt{m_i^2+p_i^2}.
\eeq
Such versions of the model are referred to as `semi-relativistic', since the kinetic energies of the quarks in the baryon are treated relativistically, but the model lacks Lorentz covariance.

In this picture of a baryon, each Jacobi coordinate may be independently excited, either radially, orbitally, or both.  The total orbital angular momentum of the baryon is 
\beq
\vec{L}=\vec{\ell}_\rho+\vec{\ell}_\lambda,
\eeq
and the total angular momentum is
\beq
\vec{J}=\vec{L}+\vec{S},
\eeq
where $\vec{S}$ is the total spin of the quarks. The parity, $P$, of the baryon constructed in this way is
\beq
P=(-1)^{\ell_\rho+\ell_\lambda}.
\eeq

A variety of forms for the potentials have been used in the literature. Many models treat the inter-quark potential as arising from exchange of gluons. In such models, the spin-independent confining potentials are approximated by linear and Coulomb
forms, 
\beq
V^{ij}_{\rm conf}= \left({br_{ij}\over 2}-
{2\alpha_{\rm Coul}\over3r_{ij}}\right),
\eeq
but other forms, including logarithmic forms, have been used. In such models, the spin-dependent parts of the potential take the forms
\beqy\label{hyp}
H^{ij}_{\rm hyp} &=&\sum_{i<j=1}^3\left[V_{\rm contact}(\rmb{r}_{ij})\frac{1}{m_i m_j}\right.\nonumber\\
&+& \left.V_{\rm tensor}(\rmb{r}_{ij})\frac{1}{m_i m_j}\left({3\rmb{S}_i\cdot\rmb {r}_{ij} \rmb{S}_j\cdot\rmb {r}_{ij}\over {r}^2_{ij}} -\rmb{S}_i\cdot\rmb{S}_j\right) \right]
\eeqy
where the functions $V_{\rm contact}(\rmb{r}_{ij})$ and $V_{\rm tensor}(\rmb{r}_{ij})$ are scalar functions of the separation~$r_{ij}$. These functions vary from model to model. For instance, many authors use a Dirac delta function for $V_{\rm contact}(\rmb{r}_{ij})$, while others use a functional form that is not quite so singular at the origin. In addition, there are spin-orbit interactions, which usually have a small effect on the spectrum.

In models like this, as in most models of baryons, the baryons of the ground state octet and decuplet sit in the ({\bf 56},\,$0^+_0$) of Table \ref{Table:Classification}, while excited states belong to some of the other multiplets shown in the same table. If all spin-dependent interactions are neglected, the states of the ground-state octet are degenerate with their flavour analogs in the decuplet. The contact hyperfine interaction removes this degeneracy, making the nucleon lighter and the $\Delta$ heavier, for instance. The tensor and spin-orbit interactions induce mixings with the other supermultiplets of Table \ref{Table:Classification}, leading to components of the nucleon and $\Delta$ wave functions having non-zero orbital angular momentum, for instance. The spin-orbit interaction also removes degeneracies between states with the same non-zero total orbital angular momentum and spin, but having different total angular momentum. One example of this is the $\Lambda(1405)\,\hlf^-$ and the $\Lambda(1520)\,\thlf^-$.

Results from one model of this kind, that of~\cite{Capstick:1986bm}, are shown in the fourth columns of Tables \ref{nucleonspec} to \ref{xiomegaspec}, for baryons consisting of $u$, $d$ and $s$ quarks. It can be seen from these tables that this model, and others of similar ilk, provide a moderately successful description of both light and heavy baryon spectra, but they have a few notable shortcomings. One example is that it is customary to identify the lightest excited $\hlf^+$ nucleon, $N(1440)\,\hlf^+$, as the first radial excitation of the nucleon. With potentials motivated by gluon exchange, most models fail to get this state as light as it is.

A number of authors have treated the inter-quark potential as arising from exchange of mesons, starting with pseudoscalar mesons as Goldstone bosons, but expanding the list of exchanged particles to include vector and scalar mesons. A significant part of the motivation in such models is the failure of gluon-exchange models to reproduce the relatively light mass of the $N(1440)\,\hlf^+$. In one such model~\cite{Melde:2008yr}, the interaction potential takes the form
\beqy
V(ij)_{\rm conf}&=& V(ij)^{ps}+V(ij)^v+V(ij)^s\nonumber\\
&=& \sum_{a=1}^3\left[V_\pi(ij)+V_\rho(ij)\right]\lambda^a_i\lambda^a_j+\sum_{a=4}^7\left[V_K(ij) +
V_{K^*}(ij)\right]\lambda^a_i\lambda^a_j\nonumber\\
&+&\left[V_\eta(ij)+V_{\omega_8}(ij)\right]\lambda^8_i\lambda^8_j+\frac{2}{3}\left[V_{\eta^\prime}(ij)+V_{\omega_1}(ij)\right]+V_\sigma(ij).
\eeqy
Exchange of mesons in the pseudoscalar nonet provides spin-spin and tensor forces, the vectors provide central, spin-spin, tensor and spin-orbit forces, and the scalar singlet leads to central and spin-orbit forces.

Results from this model (reference \cite{Melde:2008yr}) are shown in column five of  Tables \ref{nucleonspec} to \ref{xiomegaspec}. Although the mass of the radially excited nucleon is improved, the decription of the spectra of baryons is not significantly better than that obtained with gluon-exchange forces.

\begin{center}
\begin{table}[h]
\caption{Predictions for excited nucleons from a number of theoretical treatments. The first column shows the $J^P$ of the state, while the second shows the PDG designation, including the star rating. Column three shows the PDG Breit-Wigner mass range. Numbers with no parentheses are PDG averages. Numbers in this column in parentheses show the range of measurements reported, with no average given by the PDG. Hyphens in columns two and three indicate that there is no experimental state that matches model states shown in the later columns. Column four shows the numbers from a relativized quark model (QM) with one-gluon-exchange (OGE) forces, that of Capstick and Isgur~\cite{Capstick:1986bm}. The numbers in column five come from a model in which the inter-quark interactions arise from meson exchanges (GBE)~\cite{Melde:2008yr}. Column six shows the results from a diquark model~\cite{Ferretti:2011zz}. Column seven shows the results obtained in the relativistic quark model (RQM) of L\"oring {\it et al.}~\cite{Loring:2001kv,Loring:2001kx,Loring:2001ky}.  The numbers in column eight are obtained in the large $N_c$ treatment~\cite{Goity:2002pu,Goity:2003ab,Goity:2004pw,Goity:2008zz,Schat:2001xr,Carlson:1998vx,Matagne:2004pm,Matagne:2011fr,Matagne:2012tm}, and the numbers in column nine are the results of the lattice simulations of Edwards {\it et al.}~\cite{Edwards:2011jj,Edwards:2012fx,Dudek:2012ag}. For the diquark model (column six), a hyphen indicates that such a state cannot exist in the model. For other models, a blank line indicates that the model does predict such a state, but the prediction has not been published.
\label{nucleonspec}}
\scriptsize
\vspace{5mm}
\begin{tabular}{|c|l|c||l|l|l|l|l|l|}
\hline
$J^P$& State&PDG Mass &QM&QM&Di-&RQM&Large&Lattice\\ 
&& Range&OGE&GBE&quark&& $N_c$&\\ \hline

\multirow{6}{*}{$\hlf^+$}&938$^{\starfour}$&938&960&939&939&939&&1196$\pm $11\\\cline{2-9}
&1440$^{\starfour}$&1420-1470&1540&1459&1513&1518&1450&2187$\pm$45\\\cline{2-9}
&1710$^{\starthree}$&1680-1740&1770&1776&1768&1729&1712&2255$\pm$28\\\cline{2-9}
&1880$^{\startwo}$&(1835-1915)&1880&&1893&1950&&2351$\pm$37\\\cline{2-9}
&2100$^{\starone}$&(2030-2200)&1975&&---&1996&1983&2244$\pm$28\\\cline{2-9}
&---&---&2065&&---&2009&&2544$\pm$51$^*$\\\cline{1-9}

\multirow{4}{*}{$\thlf^+$}&1720$^{\starfour}$&1700-1750&1795&&1768&1688&1674&2146$\pm$16\\ \cline{2-9}
&1900$^{\starthree}$&(1862-1975)&1870&&1808&1809&1885&2314$\pm$20\\ \cline{2-9}
&2040$^{\starone}$&2031-2065&1910&&--- &1936&&2334$\pm$22\\ \cline{2-9}
&---&---&1950&&---&1969&&2401$\pm$17\\ \cline{1-9}

\multirow{4}{*}{$\fhlf^+$}&1680$^{\starfour}$&1680-1690&1770&&1808&1723&1689&2143$\pm$17\\ \cline{2-9}
&1860$^{\startwo}$&1820-1960&1980&&---&1934&&2352$\pm$22\\ \cline{2-9}
&2000$^{\startwo}$&1950-2150&1995&&---&1959&1850&2415$\pm$18\\ \cline{2-9}
&---&---&1995&&---&2120&&2943$\pm$68$^*$\\ \cline{1-9}

\multirow{3}{*}{$\shlf^+$}&1990$^{\startwo}$&(1920-2155)&2000&&---&1989&1872&2481$\pm$20\\ [+5pt]\cline{2-9}
&---&---&1995&&---&2190&&2900$\pm$57\\ \cline{2-9}
&---&---&&&---&2365&2240&\\ \cline{1-9}

$\nhlf^+$&2220$^{\starfour}$&2200-2300&2345&&&2221&2245&\\[+5pt]\cline{1-9}

$\thhlf^+$&2700$^{\startwo}$&(2570-3100)&2820&&&2616&&\\[+5pt]\cline{1-9}

\multirow{4}{*}{$\hlf^-$}&1535$^{\starfour}$&1525-1545&1460&1519&1527&1435&1541&1707$\pm$21\\\cline{2-9}
&1650$^{\starfour}$&1645-1670&1535&1647&1671&1660&1660&1860$\pm$27\\ \cline{2-9}
&1895$^{\startwo}$&(1860-2260)&\hspace*{-5pt}$\begin{array}{l}1945\\2030\end{array}$&&1882&\hspace*{-5pt}$\begin{array}{l}1901\\1918\end{array}$&&2357$\pm$113\\ \cline{1-9}

\multirow{5}{*}{$\thlf^-$}&1520$^{\starfour}$&1515-1525&1495&1519&1527&1476&1532&1811$\pm$22\\\cline{2-9}
&1700$^{\starthree}$&1650-1750&1625&1647&1671&1606&1699&1889$\pm$21\\ \cline{2-9}
&1875$^{\starthree}$&1820-1920&1960&&1882&1926&&2513$\pm$54\\ \cline{2-9}
&2120$^{\startwo}$&(1980-2210)&\hspace*{-5pt}$\begin{array}{l}2055\\2095\end{array}$&& &\hspace*{-5pt}$\begin{array}{l}1959\\2070\end{array}$&&2673$\pm$20\\ \cline{1-9}

\multirow{3}{*}{$\fhlf^-$}&1675$^{\starfour}$&1670-1680&1630&1647&1671&1655&1671&1987$\pm$17\\\cline{2-9}
&2060$^{\startwo}$&(1900-2260)&\hspace*{-5pt}$\begin{array}{l}2080\\2095\end{array}$&&&\hspace*{-5pt}$\begin{array}{l}1970\\2104\end{array}$&&2486$\pm$24\\ \cline{1-9}

$\shlf^-$&2190$^{\starfour}$&2100-2200&2090&&&2015&&2635$\pm$22\\ [+5pt]\cline{1-9}

$\nhlf^-$&2250$^{\starfour}$&2200-2350&2215&&&2212&&\\ [+5pt]\cline{1-9}

$\elhlf^-$&2600$^{\starthree}$&2550-2750&2600&&&\hspace*{-5pt}$\begin{array}{l}2425\\2600\end{array}$&&\\[+5pt]\hline
\end{tabular}

\end{table}
\end{center}

\begin{center}
\begin{table}[h]
\caption{Predictions for excited $\Delta$ baryons from a number of theoretical approaches. The key is as in Table \ref{nucleonspec}.
\label{deltaspec}}

\vspace{5mm}
\begin{tabular}{|c|l|c||l|l|l|l|l|l|}
\hline
$J^P$& State & PDG Mass &QM&QM&Di-&RQM&Large&Lattice\\ 
& &Range &OGE&GBE&quark&& $N_c$&\\ \hline

\multirow{2}{*}{$\hlf^+$}&1750$^{\starone}$&(1700-1780)&1835&&1858&1866&1746&2193$\pm$34\\\cline{2-9}
&1910$^{\starfour}$&1860-1910&1875&&1952&1906&1897&2343$\pm$20\\\cline{1-9}

\multirow{4}{*}{$\thlf^+$}&1232$^{\starfour}$&1230-1234&1230&1240&1233&1260&&1505$\pm$13\\\cline{2-9}
&1600$^{\starthree}$&1500-1700&1795&1718&1602&1810&1625&2300$\pm$28\\\cline{2-9}
&1920$^{\starthree}$&1900-1970&1915&&1952&1871&1906&2380$\pm$40\\\cline{2-9}
&---&---&1985&&---&1950&&2436$\pm$57\\\cline{1-9}

\multirow{3}{*}{$\fhlf^+$}&1905$^{\starfour}$&1855-1910&1910&&1952&1897&1921&2334$\pm$18\\ \cline{2-9}
&2000$^{\startwo}$&(1600-2325)&1990&&---&1985&1756&2422$\pm$17\\ \cline{2-9}
&&&&&---&&2368&2672$\pm$56\\ \cline{1-9}

\multirow{2}{*}{$\shlf^+$}&1950$^{\starfour}$&1915-1950&1940&&1952&1956&1942&2320$\pm$34\\ \cline{2-9}
&2390$^{\starone}$&(2250-2485)&2370&&&2339&2372&\\ \cline{1-9}
$\nhlf^+$&2300$^{\startwo}$&(2240-2550)&2420&&&2393&2378&\\ \cline{1-9}
$\elhlf^+$&2420$^{\starfour}$&2300-2500&2450&&&2442&2385&\\ \cline{1-9}
$\fihlf^+$&2950$^{\startwo}$&(2750-3090)&2920&&&2824&&\\ [+5pt]\cline{1-9}

\multirow{3}{*}{$\hlf^-$}&1620$^{\starfour}$&1600-1660&1555&1642&1554&1654&1645&1897$\pm$18\\\cline{2-9}
&1900$^{\startwo}$&1840-1920&2035&&1986&2100&&2572$\pm$53\\ \cline{2-9}
&2150$^{\starone}$&(2050-2250)&2140&&&2141&&2656$\pm$30\\ \cline{1-9}

\multirow{4}{*}{$\thlf^-$}&1700$^{\starfour}$&1670-1750&1620&1642&1554&1628&1720&1945$\pm$19\\\cline{2-9}
&1940$^{\startwo}$&1940-2060&\hspace*{-5pt}$\begin{array}{l}2080\\2145\\2155\end{array}$&&1986&\hspace*{-5pt}$\begin{array}{l}2089\\2156\\2170\end{array}$&&2751$\pm$24\\ \cline{1-9}

\multirow{2}{*}{$\fhlf^-$}&1930$^{\starthree}$&1900-2000&2155&&2005&2170&&2748$\pm$21\\\cline{2-9}
&2350$^{\starone}$&(2160-2525)&2165&&&2187&&\\ \cline{1-9}

$\shlf^-$&2200$^{\starone}$&(2120-2360)&2230&&&2181&&2677$\pm$26\\ [+5pt]\cline{1-9}

$\nhlf^-$&2400$^{\startwo}$&(2100-2780)&2295&&&2280&&\\ [+5pt]\cline{1-9}

$\thhlf^-$&2750$^{\startwo}$&(2550-2870)&2750&&&2685&&\\[+5pt]\hline
\end{tabular}
\end{table}
\end{center}

One feature apparent in the results of columns four and five of these tables is the prediction of many more states than have been observed (this is less apparent for the results shown in column five: that model predicts about the same number of states as the model whose results are shown in column four, but the results are not published). This is the so-called missing or undiscovered baryon problem, one that is key to understanding the effective degrees of freedom in a baryon. A number of authors have suggested that the primary reason for the mismatch between the numbers of predicted and observed states is that models such as those discussed have the wrong degrees of freedom. Instead of being a system of three quarks, in which any of the three can be excited, the baryon should be treated as a quark-diquark system, with excitations in the diquark frozen out, or occuring at significantly higher energies. With fewer excitations possible, the numbers of predicted and observed states are closer to each other. Such models still predict as-yet undiscovered states, but they are far fewer. 

Before exploring one of these diquark models further, we point out that there is an alternative explanation that has been proposed for the number of missing states. This is the suggestion, first made by Koniuk and Isgur~\cite{Koniuk:1979vy}, and later confirmed by Capstick and Roberts~\cite{capstickroberts}, that the missing states couple weakly to the channel that has been used predominantly for production of excited baryons, namely the $N\pi$ channel. This suggestion has provided the motivation for a significant part of the baryon spectroscopy program at many labs around the world.

In diquark models, the wave function is no longer symmetrized for all three quarks, but only for the two quarks in the diquark. In SU(6), the diquark representations are
\beq
{\bf 6}\,\otimes\,{\bf 6} = {\bf 21}\,\oplus\,{\bf 15},
\eeq
and the symmetry of the wave function of the diquark requires it to be the {\bf 21}, since the color wave function is antisymmetric. Thus the baryon multiplets possible are~\cite{Lichtenberg:1981pp}
\beq
{\bf 6}\,\otimes\,{\bf 21} = {\bf 56}\,\oplus\,{\bf 70}.
\eeq
In this variant, there is no {\bf 20} of baryons, and only one of the {\bf 70} multiplets exists. Since this multiplet has mixed-symmetry, it means that many states in the diquark model do not belong to irreducible representations of SU(6).

In the diquark model of Ferretti {\it et al.}~\cite{Ferretti:2011zz}, the mass operator for $N^*$ and $\Delta$ resonances is chosen to be
\beq
M=E_0+\sqrt{q^2+m_1^2}+\sqrt{q^2+m_2^2}+M_{\rm dir}(r)+M_{\rm cont}(r)+M_{\rm ex}(r)
\eeq
where $E_0$~is a constant, $m_1$~is the mass of the diquark, $m_2$~is the mass of the quark, $r$~is their separation and $q$~is the magnitude of the momentum conjugate to~$r$. In this operator, 
\beq
V_{\rm dir}(r)=-\frac{\tau}{r}\left(1-e^{-\mu r}\right)+\beta r,
\eeq
is the so-called direct term and
\beq
V_{\rm ex}(r)=(-1)^{l+1}e^{-\sigma r}\left[A_S\vec{s}_1\cdot\vec{s}_2+A_I\vec{t}_1\cdot\vec{t}_2+
A_{SI}\left(\vec{s}_1\cdot\vec{s}_2\right)\left(\vec{t}_1\cdot\vec{t}_2\right) \right]
\eeq
is the crucial exchange term first proposed by Lichtenberg~\cite{Lichtenberg:1981pp}. $\vec{s}_i$ and $\vec{t}_i$ are spin and isospin operators, respectively. $V_{\rm cont}$ is the contact term. There is no spin-orbit or tensor interaction in this model. This exchange term was introduced by Lichtenberg as a way to ensure that the ({\bf 70},\,$1^-_1$) would emerge to be lighter than the ({\bf 70},\,$0^+$) supermultiplet. 

The results obtained in the model of Ferretti {\it et al.}~\cite{Ferretti:2011zz}  are shown in column six of Tables \ref{nucleonspec} and~\ref{deltaspec}. 
In this column,  a `---' indicates that such a state cannot exist in the diquark model, whereas a blank indicates that such a state could exist (from a higher oscillator band, for instance), but a prediction has not been made. Comparison of these results with those of columns five and four shows clearly that the diquark model predicts fewer states, as expected, and that there is a very good correspondance between the states predicted and those observed. There are a number of states that have been observed, but which do not exist in the diquark model. However, in all cases, evidence for such states is weak at present.

\begin{center}
\begin{table}[h]
\caption{Predictions for excited $\Lambda$ baryons from a number of theoretical approaches. Columns one, two and three have the same meaning as in Tables \ref{nucleonspec} and \ref{deltaspec}. For column headings for columns four, five, six and seven have the same meaning and the same sources as columns four, five, seven and eight, respectively, of Tables \ref{nucleonspec} and \ref{deltaspec}. 
\label{lambdaspec}}

\vspace{5mm}
\begin{tabular}{|c|l|c||l|l|l|l|l|}
\hline
$J^P$& State & PDG Mass &QM&QM&RQM&Large&Lattice\\ 
& &Range &OGE&GBE& &$N_c$&\\ \hline

\multirow{4}{*}{$\hlf^+$}&1116$^{\starfour}$&1114-1116&1115&1136&1108&&1279$\pm$ 11\\\cline{2-8}
&1600$^{\starthree}$&1560-1700&1680&1625&1677&1630&2170$\pm$29\\\cline{2-8}
&1810$^{\starthree}$&1750-1850&1830&1799&1747&1742&2195$\pm$27\\\cline{2-8}
&---&---&1910&&1898&&2198$\pm$32\\\cline{1-8}

\multirow{4}{*}{$\thlf^+$}&1890$^{\starfour}$&1850-1910&1900&&1823&1876&2225$\pm$15\\\cline{2-8}
&---&---&1960&&1952&&2287$\pm$36\\\cline{2-8}
&---&---&1995&&2045&&2318$\pm$36\\\cline{2-8}
&---&---&2050&&2087&&2367$\pm$14\\\cline{1-8}

\multirow{2}{*}{$\fhlf^+$}&1820$^{\starfour}$&1815-1825&1890&&1834&1816&2228$\pm$12\\ \cline{2-8}
&2110$^{\starthree}$&2090-2140&2035&&1999&2104&2390$\pm$18\\ \cline{2-8}
&---&---&2115&&2078&&2419$\pm$15\\ \cline{1-8}

\multirow{2}{*}{$\shlf^+$}&2020$^{\starone}$&(2000-2130)&2120&&2130&2125&2545$\pm$14\\ \cline{2-8}
&---&---&2447&&2331&2350&3033$\pm$24\\ \cline{1-8}

\multirow{2}{*}{$\nhlf^+$}&2350$^{\starthree}$&2340-2370&2423&&2340&2355&\\ \cline{2-8}
&---&---&2518&&2479&&\\ \cline{1-8}

\multirow{3}{*}{$\hlf^-$}&1405$^{\starfour}$&1404-1406&1550&1556&1524&1407&1709$\pm$17\\\cline{2-8}
&1670$^{\starfour}$&1660-1680&1615&1682&1630&1667&1776$\pm$16\\ \cline{2-8}
&1800$^{\starthree}$&1720-1850&1675&1778&1816&1806&1847$\pm$20\\ \cline{1-8}

\multirow{4}{*}{$\thlf^-$}&1520$^{\starfour}$&1518-1521&1545&1556&1508&1520&1816$\pm$12\\\cline{2-8}
&1690$^{\starfour}$&1685-1695&1645&1682&1662&1676&1905$\pm$13\\ \cline{2-8}
&---&---&1770&&1775&1864&1936$\pm$17\\ \cline{2-8}
&2325$^{\starone}$&(2305-2375)&2290&&1987&&2626$\pm$31\\ \cline{1-8}

\multirow{2}{*}{$\fhlf^-$}&1830$^{\starfour}$&1810-1830&1775&1778&1828&1836&2059$\pm$11\\\cline{2-8}
&---&---&2180&&2080&&2571$\pm$19\\ \cline{1-8}

$\shlf^-$&2100$^{\starfour}$&2090-2110&2150&&2090&&2694$\pm$16\\ [+5pt]\hline
\end{tabular}

\end{table}
\end{center}

\begin{center}
\begin{table}[h]
\caption{Predictions for excited $\Sigma$ baryons from a number of theoretical approaches. The sources of the numbers are the same as in Table \ref{lambdaspec}.
\label{sigmaspec}}
\vspace{5mm}
\begin{tabular}{|c|l|c||l|l|l|l|l|l|}
\hline
$J^P$& State & PDG Mass &QM&QM&RQM&Large&Lattice\\ 
& &Range &OGE&GBE& &$N_c$&\\ \hline

\multirow{5}{*}{$\hlf^+$}&1193$^{\starfour}$&1190-1197&1190&1180&1190&&1308$\pm$ 7\\\cline{2-8}
&1660$^{\starthree}$&1630-1690&1720&1616&1760&1660&2270$\pm$14\\\cline{2-8}
&1770$^{\starone}$&(1730-1790)&1915&1911&1947&1776&2251$\pm$24\\\cline{2-8}
&1880$^{\startwo}$&(1800-2035)&1970&&2009&1810&2258$\pm$16\\\cline{2-8}
&---&---&2005&&2052&2068&2326$\pm$22\\\cline{1-8}

\multirow{4}{*}{$\thlf^+$}&1385$^{\starfour}$&1382-1388&1370&1389&1411&&1579$\pm$ 9\\\cline{2-8}
&1840$^{\starone}$&(1725-2125)&1920&1865&1896&1790&2243$\pm$14\\\cline{2-8}
&2080$^{\startwo}$&(2040-2120)&1970&&1961&2061&2317$\pm$16\\\cline{2-8}
&---&---&2010&&2011&&2366$\pm$13\\\cline{1-8}

\multirow{3}{*}{$\fhlf^+$}&1915$^{\starfour}$&1900-1935&1955&&1956&1920&2228$\pm$14\\ \cline{2-8}
&2070$^{\starone}$&(2025-2080)&2030&&2027&2051&2367$\pm$13\\ \cline{2-8}
&&&&&2071&2478&2437$\pm$14\\ \cline{1-8}

\multirow{2}{*}{$\shlf^+$}&2030$^{\starfour}$&2025-2040&2060&&2070&2036&2427$\pm$24\\ \cline{2-8}
&---&---&2390&&2161&2350&2546$\pm$15\\ \cline{2-8}
&---&---&&&&2482&3021$\pm$25\\ \cline{1-8}

\multirow{2}{*}{$\nhlf^+$}&&&&&&2355&\\ \cline{2-8}
&---&---&2390&&&2488&\\ \cline{1-8}

\multirow{2}{*}{$\elhlf^+$}&&&&&&2495&\\ \cline{2-8}
&---&---&2390&&&&\\ \cline{1-8}

\multirow{3}{*}{$\hlf^-$}&1620$^{\startwo}$&(1600-1645)&1630&1677&1628&1637&1780$\pm$16\\\cline{2-8}
&1750$^{\starthree}$&1730-1800&1675&1736&1771&1755&1837$\pm$20\\ \cline{2-8}
&2000$^{\starone}$&(1755-2040)&2110&1759&1798&&1951$\pm$14\\ \cline{2-8}
&---&---&1695&&2111&1784&2545$\pm$49\\ \cline{1-8}

\multirow{4}{*}{$\thlf^-$}&1580$^{\starone}$&(1578-1587)&1655&1677&1669&&1903$\pm$15\\\cline{2-8}
&1670$^{\starfour}$&1665-1685&1750&1736&1771&1667&1948$\pm$14\\ \cline{2-8}
&---&---&1755&1759&2139&1769&1956$\pm$18\\ \cline{2-8}
&1940$^{\starthree}$&1900-1950&2120&&1798&1847&2667$\pm$25\\ \cline{1-8}

\multirow{2}{*}{$\fhlf^-$}&1775$^{\starfour}$&1770-1780&1755&1736&1770&1784&2057$\pm$12\\\cline{2-8}
&---&---&2205&&2174&&2521$\pm$24\\ \cline{1-8}

$\shlf^-$&2100$^{\starone}$&(2040-2150)&2245&&2236&&2720$\pm$15\\ [+5pt]\hline
\end{tabular}
\end{table}
\end{center}

\begin{center}
\begin{table}[h]
\caption{Predictions for $\Xi$ and $\Omega$ baryons in a number of theoretical approaches. The sources of the numbers are the same as in Table \ref{lambdaspec}.
\label{xiomegaspec}}

\vspace{5mm}
\begin{tabular}{|c|c|l|c||l|l|l|l|l|l|}
\hline
Flavor&$J^P$&State& PDG Mass &QM&QM&RQM&Large&Lattice\\ 
& &&Range &OGE&GBE& &$N_c$&\\ \hline
\multirow{10}{*}{$\Xi$}&\multirow{2}{*}{$\hlf^+$}&1318$^{\starfour}$&1314-1322&1305&1348&1310&&1351$\pm $9\\\cline{3-9}
&&---&---&1840&1805&1876&1825&2281$\pm$17\\\cline{2-9}

&\multirow{2}{*}{$\thlf^+$}&1530$^{\starfour}$&1530-1532&1505&1528&1539&&1635$\pm $8\\ \cline{3-9}
&&---&---&2045&&1988&1955&2262$\pm$18\\ \cline{2-9}

&\multirow{2}{*}{$\fhlf^+$}&---&---&2045&&2013&1997&2296$\pm$13\\ \cline{3-9}
&&---&---&2165&&2141&2181&2428$\pm$14\\ \cline{2-9}

&\multirow{2}{*}{$\hlf^-$}&1690$^{\starthree}$&1680-1700&1755&&1770&1779&1845$\pm$17\\\cline{3-9}
&&---&---&1810&&1922&1927&1875$\pm$15\\\cline{2-9}

&\multirow{2}{*}{$\thlf^-$}&1820$^{\starthree}$&1818-1828&1785&1792&1780&1815&1973$\pm$12\\\cline{3-9}
&&---&---&1880&&1870&1980&1998$\pm$17\\\cline{2-9}

&$\fhlf^-$&---&---&1900&1881&1955&1974&2127$\pm$11\\\hline

\multirow{10}{*}{$\Omega$}&\multirow{2}{*}{$\hlf^+$}&---&---&2220&&2232&&2379$\pm$22\\\cline{3-9}
&&&&2255&&2256&&2449$\pm$18\\\cline{2-9}

&\multirow{2}{*}{$\thlf^+$}&1672$^{\starfour}$&1671-1673&1635&&1636&&1692$\pm $7\\ \cline{3-9}
&&---&---&2165&&2177&2120&2419$\pm$18\\ \cline{2-9}

&$\fhlf^+$&&&2280&&2253&&2486$\pm$13\\ \cline{2-9}

&\multirow{2}{*}{$\hlf^-$}&&&1950&&1992&2061&2011$\pm$23\\\cline{3-9}
&&---&---&2410&&2456&&2764$\pm$19\\\cline{2-9}

&\multirow{2}{*}{$\thlf^-$}&&&2000&&1976&2100&2104$\pm$14\\\cline{3-9}
&&---&---&2440&&2446&&2771$\pm$27\\\cline{2-9}

&$\fhlf^-$&&&2490&&2528&&2807$\pm$19\\[+5pt]\hline
\end{tabular}
\end{table}
\end{center}

Extension of this kind of model to baryons containing at least one heavy quark is relatively straightforward, as it simply requires changing the masses of the quarks. If anything, such models, especially the non-relativistic versions, might be expected to work better when at least one of the quarks in the baryon is heavy. In addition, the symmetries expected from effective theories like the heavy quark effective theory may either be used to constrain the model {\it a priori}, or to check the predictions of the model {\it a posteriori.} Results from extensions of such quark models to heavy baryons are shown in columns four \cite{pr}  and five \cite{Capstick:1986bm} of Tables \ref{baryonspec3} and~\ref{baryonspec4}, as well as column four \cite{pr} of Tables \ref{baryonspec5} and~\ref{baryonspec6}. The model of Roberts and Pervin~\cite{pr} has also been extended to make predictions for the masses and properties of baryons containing two and three heavy quarks. In addition, there are a few subclasses of models that warrant a few words of discussion.

The fundamental tenet of the heavy quark effective theory (HQET) is that a single heavy quark in a hadron acts as a static source of colour. Consequently, the dynamics of the light quarks in the presence of this static colour source are independent of the flavour, spin and other properties of the heavy quark. In the context of quark models, this suggests a departure from the Jacobi coordinates used earlier, in favour of a set of coordinates that highlight the special role of the heavy quark. With the heavy quark chosen as quark three, Albertus and collaborators~\cite{Albertus:2003sx} define the set of coordinates
\beq
\vec{\rho}_1=\vec{r}_1-\vec{r_3},\,\,\,\,\vec{\rho}_2=\vec{r}_2-\vec{r_3}.
\eeq
In terms of these, the nonrelativistic kinetic energy of the quarks becomes
\beq
T=-\frac{\nabla_{\rho_1}^2}{2\mu_1}-\frac{\nabla_{\rho_2}^2}{2\mu_2}-\frac{\nabla_R^2}{2M}-\frac{\nabla_{\rho_1}\cdot\nabla_{\rho_2}}{m_3}+M,
\eeq
where $\nabla_{\rho_i}=\frac{\partial}{\partial\rho_i}$ and $\mu_i=(1/m_i+1/m_3)^{-1}$. 

The quark-quark interactions used by these authors are similar in their general structure to those used by other authors, in that there is a confining term (chosen to be logarithmic) and a Coulomb-like term. There is also a `contact' hyperfine term, but no tensor or spin-orbit interactions. These authors also explore the effects of interactions that arise from exchange of pseudoscalar and scalar mesons. They confine their discussion to the lowest-lying, positive parity states. In Tables \ref{baryonspec3} to \ref{baryonspec6}, the results obtained by Albertus {\it et al.}~\cite{Albertus:2003sx} are shown in columns 10, nine, eight and seven respectively.

\begin{center}
\begin{table}[h]
\caption{Predictions for $\Lambda_c$, $\Sigma_c$ and $\Omega_c$ baryons from a number of theoretical approaches. The numbers in columns four (QM1) and five (QM2) result from `conventional' quark models. Column six is the result of a relativistic model (RQM1) in which there are no excitations between the two light quarks. Column seven results from a relativistic model with instanton forces (RQM2). The numbers in column eight are from a Fadeev study (FD), while those in column nine result from relativistic three-quark equations that include dispersion techniques (RQM3). The numbers in column ten are from a nonrelatvistic quark model that include heavy quark symmetry constraints (QM3), and column eleven results from a lattice simulation with 2+1+1 flavours of dynamical quarks.
\label{baryonspec3}}
\scriptsize
\vspace{5mm}
\begin{tabular}{|c|l|c||l|l|l|l|l|l|l|l|}
\hline
Flavor&$J^P$& Expt. &QM1& QM2&RQM1&RQM2 &FD& RQM3 \cite{Gerasyuta:1999pc}&QM3&Lattice \\ 
&& Mass& \cite{pr}& \cite {Capstick:1986bm}& \cite{Ebert1}& \cite{Migura:2006ep} & \cite{Garcilazo:2007eh} &\cite{Gerasyuta:2007un,Gerasyuta:2008zy}& \cite{Albertus:2003sx}&\cite{Briceno:2012wt}\\ \hline
\multirow{10}{*}{$\Lambda_c$}&\multirow{2}{*}{$\hlf^+$}&2286&2268&2265&2297&2272&2292&2284&2296&2291$\pm$44\\\cline{3-11}
&&&2791&2775&2772&2769&2669&&&\\\cline{2-11}
&\multirow{2}{*}{$\thlf^+$}&&2887&2910&2874&2848&2906&&&\\ \cline{3-11}
&&&3073&3035&3262&3100&3061&&&\\ \cline{2-11}
&$\fhlf^+$&&2887&2910&2883&&&&&\\ \cline{2-11}
&\multirow{2}{*}{$\hlf^-$}&2592&2625&2630&2598&2594&2559&2400&&\\\cline{3-11}
&&&2816&2780&3017&2853&2779&2635&&\\\cline{2-11}
&\multirow{2}{*}{$\thlf^-$}&2628&2636&2640&2628&2586&2559&2625&&\\\cline{3-11}
&&&2830&2840&3034&2874&2779&2630&&\\\cline{2-11}
&$\fhlf^-$&&2872&2900&3061&&&2765&&\\\hline
\multirow{10}{*}{$\Sigma_c$}&\multirow{2}{*}{$\hlf^+$}&2453&2455&2440&2439&2459&2448&2458&2466&2481$\pm$29\\\cline{3-11}
&&&2958&2890&2864&2947&2793&&&\\\cline{2-11}
&\multirow{2}{*}{$\thlf^+$}&2518&2519&2495&2518&2539&2505&2516&2548&2559$\pm$34\\ \cline{3-11}
&&&2995&2985&2912&3010&2825&&&\\ \cline{2-11}
&$\fhlf^+$&&3003&3065&3001&&&&&\\ \cline{2-11}
&\multirow{2}{*}{$\hlf^-$}&&2748&2765&2795&2769&2706&2700&&\\\cline{3-11}
&&&2768&2770&2805&2817&2791&2915&&\\\cline{2-11}
&\multirow{2}{*}{$\thlf^-$}&&2763&2770&2761&2799&2706&2570&&\\\cline{3-11}
&&&2776&2805&2799&2815&2791&2570&&\\\cline{2-11}
&$\fhlf^-$&&2790&2815&2790&&&2740&&\\\hline
\multirow{10}{*}{$\Omega_c$}&\multirow{2}{*}{$\hlf^+$}&2695&2718&&2698&2688&2701&2806&2681&2681$\pm$34\\\cline{3-11}
&&&3152&&3065&3169&3044&&&\\\cline{2-11}
&\multirow{2}{*}{$\thlf^+$}&2766&2776&&2768&2721&2759&3108&2755&2764$\pm$33\\ \cline{3-11}
&&&3190&&3119&&3080&&&\\ \cline{2-11}
&$\fhlf^+$&&3196&&3218&&&&&\\ \cline{2-11}
&\multirow{2}{*}{$\hlf^-$}&&2977&&3020&&2959&&&\\\cline{3-11}
&&&2990&&3025&&3029&&&\\\cline{2-11}
&\multirow{2}{*}{$\thlf^-$}&&2986&&2998&&2959&&&\\\cline{3-11}
&&&2994&&3026&&3029&&&\\\cline{2-11}
&$\fhlf^-$&&3014&&3022&&&&&\\\hline
\end{tabular}

\end{table}
\end{center}

\begin{center}
\begin{table}[h]
\caption{Predictions for $\Lambda_b$, $\Sigma_b$ and $\Omega_b$ baryons from a number of theoretical approaches. The lattice numbers are from a quenched simulation, and only the statistical uncertainties are shown. The numbers in columns four (QM1) and five (QM2) result from `conventional' quark models. Column six is the result of a relativistic model (RQM1) in which there are no excitations between the two light quarks. Column seven results from a Fadeev study (FD),  while column eight results from relativistic three-quark equations that include dispersion techniques (RQM2). The numbers  in column nine are from a nonrelativistic quark model that include heavy quark symmetry constraints (QM3). Column ten results from a quenched lattice simulation.
\label{baryonspec4}}
\vspace{5mm}
\begin{tabular}{|c|l|c||l|l|l|l|l|l|l|}
\hline
Flavor&$J^P$& Expt.&QM1 &QM2 &RQM1 &FD &RQM2 \cite{Gerasyuta:1999pc}&QM3&Lattice \\
&&Mass&\cite{pr}&\cite {Capstick:1986bm}&\cite{Ebert1} &\cite{Garcilazo:2007eh}&\cite{Gerasyuta:2007un,Gerasyuta:2008zy}& \cite{Albertus:2003sx}&\cite{AliKhan:1999yb}\\\hline
\multirow{10}{*}{$\Lambda_b$}&$\hlf^+$&5619&5612&5585&5622&5624&5624&5643&5679$\pm$71\\\cline{2-10}
&\multirow{2}{*}{$\thlf^+$}&&6181&6145&6189&6246&&&\\\cline{3-10}
&&&6401&&6540&&&&\\ \cline{2-10}
&$\fhlf^+$&&6183&6165&6548&&&&\\ \cline{2-10}
&\multirow{2}{*}{$\hlf^-$}&5912&5939&5912&5930&5890&&&\\\cline{3-10}
&&&6180&5780&6328&5853&&&\\\cline{2-10}
&\multirow{2}{*}{$\thlf^-$}&5920&5941&5920&5947&5890&&&\\\cline{3-10}
&&&6191&5840&6337&5874&&&\\\cline{2-10}
&$\fhlf^-$&&6206&6205&6421&&&&\\\hline
\multirow{4}{*}{$\Sigma_b$}
&$\hlf^+$&5811&5833&5795&5805&5789&5808&5851&5887$\pm$49\\\cline{2-10}
&$\thlf^+$&5832&5858&5805&5834&5844&5829&5882&5909$\pm$47\\\cline{2-10}
&$\hlf^-$&&6099&6070&6108&6039&&&\\\cline{2-10}
&$\thlf^-$&&6101&6070&6076&6039&&&\\\hline
\multirow{4}{*}{$\Omega_b$}
&$\hlf^+$&6071&6081&&6065&6037&6120&6033&6048$\pm$33\\\cline{2-10}
&$\thlf^+$&&6102&&6088&6090&6220&6063&6069$\pm$34\\\cline{2-10}
&$\hlf^-$&&6301&&6352&6278&&&\\\cline{2-10}
&$\thlf^-$&&6304&&6330&6278&&&\\\hline
\end{tabular}
\end{table}
\end{center}

\begin{center}
\begin{table}[h]
\caption{Predictions for $\Xi_c$ baryons from a number of theoretical approaches. The numbers in column three are from a conventional quark model (QM1). Those in column four result from a relativistic model with no excitations between the light quarks (RQM1), while those in column five are from a relativistic model in which all excitations are allowed (RQM2). Column six results from a Fadeev study (FD),  while column seven results from relativistic three-quark equations that include dispersion techniques (RQM3). Column eight is from a nonrelativistic model with heavy quark symmetry constraints (QM2), and column nine is from a lattice simulation with 2+1+1 dynamical fermions.
\label{baryonspec5}}
\vspace{5mm}
\begin{tabular}{|l|c||l|l|l|l|l|l|l|}
\hline
$J^P$& Expt. &QM1&RQM1 &RQM2 &FD &RQM3 \cite{Gerasyuta:1999pc}&QM2 &Lattice \\
& mass& \cite{pr}&  \cite{Ebert1}&\cite {Migura:2006ep}&\cite{Garcilazo:2007eh}&\cite{Gerasyuta:2007un,Gerasyuta:2008zy}&\cite{Albertus:2003sx}&\cite{Briceno:2012wt}\\ \hline

\multirow{2}{*}{$\hlf^+$}&2467&2466&2481&2469&2496&2467&2474&2439$\pm$39\\\cline{2-9}
&2576&2594&2578&2595&2574&2565&2578&2568$\pm$28\\\hline
\multirow{2}{*}{$\thlf^+$}&2646&2649&2654&2651&2633&2725&2655&2655$\pm$28\\ \cline{2-9}
&&3012&3030&&2951&&&\\ \hline
$\fhlf^+$&&3004&3042&&&&&\\ \hline
\multirow{2}{*}{$\hlf^-$}&2789&2773&2801&2769&2749&&&\\\cline{2-9}
&2816?&2855&2928&&2829&&&\\\hline
\multirow{2}{*}{$\thlf^-$}&2816?&2783&2820&2771&2749&&&\\\cline{2-9}
&&2866&2900&&2829&&&\\\hline
$\fhlf^-$&&2989&2921&&&&&\\\hline
\end{tabular}
\end{table}
\end{center}

\begin{center}
\begin{table}[h]
\caption{Predictions for $\Xi_b$ baryons from a number of theoretical approaches.  The numbers in column three are from a conventional quark model (QM1). Those in column four result from a relativistic model with no excitations between the light quarks (RQM1). Column five results from a Fadeev study (FD),  while column six  results from relativistic three-quark equations that include dispersion techniques (RQM2). Column seven is from a nonrelativistic model with heavy quark symmetry constraints (QM2), and column eight is from  a quenched lattice simulation, and only the statistical uncertainties are shown.
\label{baryonspec6}}
\vspace{5mm}
\begin{tabular}{|l|c||l|l|l|l|l|l|}
\hline
$J^P$&Expt. &QM1 &RQM1 &FD &RQM2 \cite{Gerasyuta:1999pc}&QM2&Lattice \\
&mass&\cite{pr}& \cite{Ebert1}&\cite{Garcilazo:2007eh}&\cite{Gerasyuta:2007un,Gerasyuta:2008zy}& \cite{Albertus:2003sx}&\cite{AliKhan:1999yb}\\\hline
\multirow{2}{*}{$\hlf^+$}&5788&5806&5812&5825&5761&5808&5795$\pm$53\\\cline{2-8}
&5945?&5970&5806&5937&6007&5825&5968$\pm$39\\\hline
$\thlf^+$&5945?&5980&5963&5967&6066&5975&5989$\pm$39\\\hline
$\hlf^-$&&6090&6119&6076&&&\\\hline
$\thlf^-$&&6093&6130&6076&&&\\\hline
\end{tabular}
\end{table}
\end{center}

\subsubsection{Relativistic Equations}
A number of relativistic treatments of baryons,  exist in the literature. Among these are bound-state equations like the Bethe-Salpeter (BS) equation, Dyson-Schwinger (DS) treatments \cite{Chen:2012qr}, and QCD sum rules. All of these treatments are very technical, so that a full description is well beyond the scope of this review article. However, the salient points of one of these approaches are presented here. 

\begin{figure}[h]
\caption{Diagrammatic representation of the integral equation for the three-fermion propagator.}
\centerline{\includegraphics[height=0.7in]{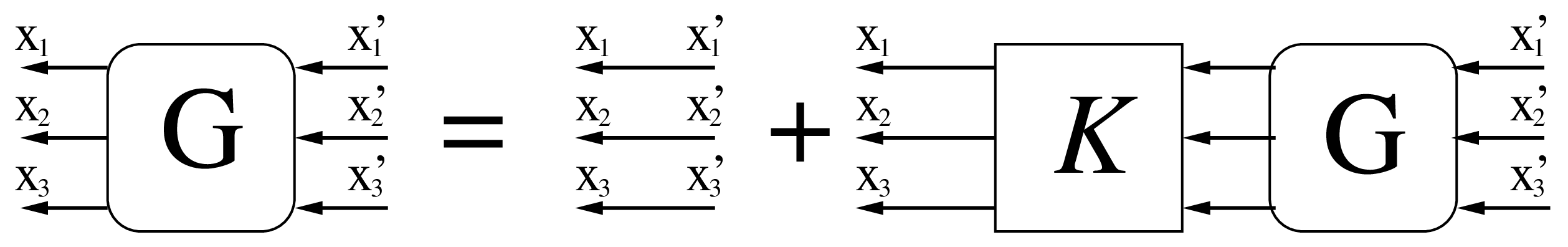}}
\label{fig:bseqn}

\end{figure}

The starting point in the BS approach is the six-point Green's function (or three-fermion propagator) for three interacting fermions.This is the vacuum expectation value of the time-ordered product of three fermion field operators $\Psi^i$ and their adjoints $\overline{\Psi}^{\,i}$:
\beqy
&&G_{a_1a_2a_3;a^\prime_1a^\prime_2a^\prime_3}(x_1,x_2,x_3;x_1^\prime,x_2^\prime,x_3^\prime):=\nonumber\\
&&-\langle0|T\Psi^1_{a_1}(x_1)\Psi^2_{a_2}(x_2)\Psi^3_{a_3}(x_3)\overline{\Psi}^1_{a^\prime_1}(x^\prime_1)\overline{\Psi}^2_{a^\prime_2}(x^\prime_2)\overline{\Psi}^3_{a^\prime_3}(x^\prime_3)|0\rangle.
\eeqy
Here $|0\rangle$ is the physical vacuum, $T$ is the time ordering operator and the $a_i$ denote combined Dirac, colour and flavour indices. The integral equation that is obtained for the full three-fermion propagator is shown diagramatically in Figure \ref{fig:bseqn}, and is written symbolically as
\beq
G=G_0-iG_0KG.
\eeq
In this equation, $G$ is the full propagator, $G_0$ is the propagator with no interactions between the quarks, and $K$ is the sum of the full two-particle and three-particle irreducible kernels. There is a similar integral equation for the full propagator of each quark, as well as for the kernel itself. Baryon resonances are extracted as poles in the full three-particle propagator. We note here, however, that not only is the formalism very technical, but obtaining solutions is also very challenging. More details of this formalism can be found, for instance, in the work by L\"oring and collaborators~\cite{Loring:2001kv,Loring:2001kx,Loring:2001ky}. These authors used a linear confining potential with instanton-induced forces.  A non-relativistic reduction of this force has the form
\beq 
\hspace{-1.2cm}V^{\rm NR}_{\rm instanton}(\vec{x}_i-\vec{x}_j)\approx -4 {\cal P}^s_{S_{ij}=0}\otimes\left(g_{nn} {\cal P}^{\cal F}_A(nn) +
g_{nn}{\cal P}^{\cal F}_A(ns)\right)\delta^{(3)}(\vec{x}_i-\vec{x}_j),
\eeq
where
\beq
{\cal P}^s_{S_{ij}=0}=\frac{1}{4}\left( \mathds{1}\otimes \mathds{1}-\vec{\sigma}\cdot\vec{\sigma}\right)
\eeq
is the projector on quark pairs with $S_{ij}=0$, and ${\cal P}^{\cal F}_A(nn)$ and ${\cal P}^{\cal F}_A(ns)$ denote the flavour projectors on antisymmetric nonstrange ($nn$) and nonstrange-strange ($ns$) quark pairs, respectively. $g_{nn}$ and $g_{ns}$ are coupling constants. Since this force acts only on flavour-antisymmetric quark pairs, it does not contribute to the spectrum of $\Delta$~resonances. However, it plays a role in the spectrum of nucleons and other states that are not fully symmetric in flavour. This is in marked contrast to the one-gluon or one-Goldstone-boson induced spin-dependent forces, which affect the spectra of both nucleons and~$\Delta$s. Further details of this calculation and others like it are beyond the scope of this manuscript.

The results obtained in the relativistic model described above are shown in column seven of Tables \ref{nucleonspec}, \ref{deltaspec} and \ref{xiomegaspec}, and column six of  Tables \ref{lambdaspec} and~\ref{sigmaspec}. This model has also been extended to treat charmed baryons~\cite{Migura:2006ep} and the results are shown in column seven of Table \ref{baryonspec3} and column five of Table~\ref{baryonspec5}.

The BS formalism can be simplified somewhat when one or more of the quarks in a baryon is heavy. In the approach used by Ebert and collaborators~\cite{Ebert1}, the relativistic three-body problem is reduced to a sequence of relativistic two-body problems. This is accomplished in a quasipotential approach by approximating the heavy baryon as a light diquark composed of the two light quarks that interact with each other, with a heavy quark interacting with the diquark. The diquark wave function, $\Psi_d$, and the baryon wave function, $\Psi_B$, each satisfies a quasipotential equation of the Schr\"odinger type
\beq
\left(\frac{b^2(M)}{2\mu_R}-\frac{p^2}{2\mu_R}\right)\Psi_{d,B}(\vec{p})=\int\frac{d^3q}{(2\pi)^3}V(\vec{p},\vec{q};M)\Psi_{d,B}(\vec{q}),
\eeq
where the relativistic reduced mass is
\beq
\mu_R=\frac{M^4-(m_1^2-m_2^2)^2}{4M^3},
\eeq
and $E_1,\,\,E_2$ are  given by
\beq
E_1=\frac{M^2-m_2^2+m_2^2}{2M},\,\,\,\,E_2=\frac{M^2-m_1^2+m_2^2}{2M}\,.
\eeq
The bound state mass $M$ (of the diquark or the baryon) is the sum of $E_1$ and $E_2$. $m_1$ and $m_2$ are the masses of the two quarks in the diquark, or the mass of the diquark and the heavy quark in the baryon, and $\vec{p}$ is their relative momentum. In the centre of mass, the relative momentum squared on shell is
\beq
b^2(M)=\frac{\lambda(M^2,m_1^2,m_2^2)}{4M^2},
\eeq
where $\lambda(a,b,c)$ is the K\"allen function.

The kernel $V(\vec{p},\vec{q};M)$ is the quasipotential operator of the quark-quark or quark-diquark interaction. Both interactions include scalar and vector confinement components, as well as an interaction arising from gluon exchange. In addition, the quarks have an anomalous chromomagnetic dipole moment. The diquarks are either scalar diquarks (with total quark spin zero) or axial-vector diquarks. In the work of Ebert {\it et al.}~\cite{Ebert1}, no radial or orbital excitations are allowed in the light diquarks, and this has a significant effect on the spectrum they predict.

This quasipotential treatment has also been applied to baryons with two heavy quarks. In this case, the diquark is composed of the two heavy quarks that interact with each other, and the light quark interacts with the diquark. Unlike the light diquark, radial and orbital excitations are allowed in the heavy diquark, in this particular model. The interactions between the heavy quarks have the same character, namely, vector and scalar confinement and spin-dependent interactions arising from gluon exchange, but the spatial dependences of the interactions are different from those between light quarks.

The results obtained in this model for baryons containing a single heavy quark are shown in column six of Tables~\ref{baryonspec3} and~\ref{baryonspec4}, and column four of Tables~\ref{baryonspec5} and~\ref{baryonspec6}. The effects of the diquark approximation employed by these authors is seen in the masses of some of the $\Lambda_c$ and $\Lambda_b$~states shown in Tables~\ref{baryonspec3} and~\ref{baryonspec4}. The second $\thlf^+$, $\hlf^-$ and $\thlf^-$ states, as well as the only $\fhlf^-$ state shown, are significantly heavier than the corresponding states obtained in other approaches. This is because in the other approaches, the states shown include excitations of the diquark: the Ebert model assumes that states with such excitations lie much higher in mass.

\subsubsection{Large $N_c$}
It was first noted by 't Hooft~\cite{'tHooft:1973jz} that many features of QCD become easier to understand if the gauge group is changed from SU(3) to SU($N_c$), in the limit $N_c\to \infty$. With this gauge group, colour-singlet mesons are still made from a quark and an antiquark, but a baryon requires $N_c$~quarks. Thus, the mass of the baryon grows with~$N_c$. Although $N_c\to\infty$ would appear to have little to do with the physical world, this limit provides an organization scheme in which operators and processes can be treated with `$1/N_c$~counting rules'. The physical limit of $N_c=3$ is then taken, and this counting scheme works remarkably well in that limit.

Large-$N_c$~baryon representations contain all spin and isospin representations~$(J, I)$ that are consistent with a given value of $K$~\cite{Jenkins:1998wy}, where
\beq
\vec{K}=\vec{J}+\vec{I}.
\eeq
For the physical case of $N_c$ odd, the lowest allowed baryon angular momentum is $J=1/2$. The irreducible representation with $K=0$ corresponds to the infinite tower of $(J,I)$~states
\beq
\left(\hlf,\hlf\right), \,\left(\thlf,\thlf\right), \,\left(\fhlf,\fhlf\right),\dots.
\eeq
The irreducible representation with $K=1/2$ corresponds to the tower
\beq
\left(\hlf,0\right), \,\,\,\left(\hlf,1\right), \,\,\,\left(\thlf,1\right), \,\,\,\,\,\left(\thlf,2\right),\,\,\left(\fhlf,2\right), \,\,\,\,\,\left(\fhlf,3\right),\dots.
\eeq
The irreducible representation with $K=1$ corresponds to the tower
\beqy
&&\left(\hlf,\hlf\right), \,\left(\hlf,\thlf\right), \,\left(\thlf,\hlf\right), \,\left(\thlf,\thlf\right),\,\left(\thlf,\fhlf\right), \,\left(\fhlf,\thlf\right),\nonumber\\
&&\left(\fhlf,\fhlf\right),\,\left(\fhlf,\shlf\right),\dots.
\eeqy
The irreducible representation with $K=3/2$ corresponds to the tower
\beq
\left(\hlf,1\right), \,\left(\hlf,2\right), \,\left(\thlf,0\right), \,\left(\thlf,1\right),\,\left(\thlf,2\right), \,\left(\thlf,3\right),\dots.
\eeq

If the quantum number $K$ is related to the number of strange quarks in a baryon by $K=N_s/2$, then the low-spin states in each tower have the correct spin, isospin and strangeness to be identified with the spin-$\hlf$ octet and spin-$\thlf$ decuplet. The  $K=0$~tower contains the strangeness-0 baryons, the nucleon, $\left(\hlf,\hlf\right)$ and $\Delta$, $\left(\thlf,\thlf\right)$. The $K=\hlf$ tower contains the strangeness $-1$ baryons $\Lambda$ $\left(\hlf,0\right)$, $\Sigma$ $\left(\hlf, 1\right)$ and $\Sigma^*$ $\left(\thlf,1\right)$. Similarly, the $K=1$ tower contains the strangess $-2$ baryons $\Xi$ $\left(\hlf,\hlf\right)$ and $\Xi^*$ $\left(\thlf,\hlf\right)$, while the $K=\thlf$ tower contains the $\Omega^-$ $\left(\thlf,0\right)$.

Implementation of  an analysis of baryons in the framework of the $1/N_c$ expansion requires the establishment of the counting rules associated with the different operators needed in the theory.  This is most easily done in the limit where the quark masses are large enough that a non-relativistic picture becomes reliable. Since the $1/N_c$ counting is largely unaffected by the quark masses, the counting rules established should also hold when the current masses are small. Using this approach, Witten~\cite{Witten:1979kh,Witten:1978bc,Witten:1978qu} showed that baryon masses are proportional to $N_c$, but that their sizes are only affected by corrections ${\cal O}(1/N_c)$. Thus, baryons are compact systems, allowing for rigorous usage of the effective potential approach \`a la Hartree.

As with the conventional quark model, the colour wave function of a baryon is completely antisymmetric, which means that the rest of the wave function must be completely symmetric in identical quarks. A convenient basis is furnished by using wave functions factorized into a spatial part and a spin-flavour part, summed over permutations. The spin-flavour wave functions belong to the irreducible representations of SU($2N_f$) when $N_f$~flavours with degenerate or nearly-degenerate masses are considered. Such wave functions therefore also belong to irreducible representations of the permutation group of the $N_c$~quark indices. In a Hartree picture, the spatial wave function takes the form of a product of $N_c$ one-quark wave functions. In the large $N_c$ limit, ground state baryons have wave functions of the form~\cite{Goity:2004pw}
\beq
\Psi^{GS}_{\xi_1,\dots,\xi_{N_c}}\left(x_1,\dots,x_{N_c}\right)=\chi^S_{\xi_1,\dots,\xi_{N_c}}\prod_{i=1}^{N_c}\phi\left(x_i\right),
\eeq
where the one-quark spatial wave function $\phi(x)$ is an $S$-wave.

Excited baryons result from exciting one or more quarks, leaving a core of quarks in their ground states. A quark in the core has, up to corrections proportional to $1/N_c$, the same wave function as a quark in the ground state baryons. To date, only excited states with a single excited quark have been discussed in detail, but the generalization to two or more excited quarks can be carried out easily. The wave functions with one excited quark are of two types, namely symmetric (S) and mixed-symmetric (MS) in spin flavour. They take the form
\beqy
\fl\Psi^{S}_{\xi_1,\dots,\xi_{N_c}}\left(x_1,\dots,x_{N_c}\right)&=&\frac{1}{\sqrt{N_c}}\chi^S_{\xi_1,\dots,\xi_{N_c}}\sum_{i=1}^{N_c}\phi\left(x_1\right)\dots\phi^\prime(x_i)\dots\phi(x_{N_c}),\nonumber\\
\fl\Psi^{MS}_{\xi_1,\dots,\xi_{N_c}}\left(x_1,\dots,x_{N_c}\right)&=&\frac{1}{\sqrt{N_c}(N_c-1)!}\sum_\sigma\chi^{MS}_{\xi_{\sigma 1},\dots,\xi_{\sigma N_c}}\phi\left(x_{\sigma 1}\right)\dots\nonumber\\
\fl&&\times\phi\left(x_{\sigma N_c-1}\right)\phi^\prime\left(x_{\sigma N_c}\right),
\eeqy
where $\phi^{\,\prime(x)}$ is the excited quark wave function, taken to be orthonormal to the ground state wave function $\phi(x)$. The label $\sigma$ indicates the permutation: the sum in the second equation is over all possible permutations of the quarks. The generalization to states with more than one excited quark can be easily carried out.

It must be noted that these wave functions do not treat the centre-of-mass motion properly. The effects introduced by this deficiency are generally subleading in $1/N_c$ and should not affect the power countings addressed so far in this approach. Nevertheless there exists the possibility of modification to countings that are suppressed solely on the grounds of orthonormality of the single-quark wave functions used, when the centre-of-mass motion is properly treated.

With three flavours of quark, baryons are classified into SU(6) multiplets. This group has 35 generators, $\left\{S_i,\,T_a,\,G_{ia}\right\}$, with $i=1,2,3$ and $a=1,\dots,8$. The first three are the SU(2) generators of spin, the second eight are the generators of flavour SU(3), and the last 24 are identified as the components of an octet of axial-vector currents in the limit of zero momentum transfer. In a nonrelativistic picture, these generators are expressed in terms of the quark fields as
\beq
S_i=q^\dag \frac{\sigma_i}{2}q,\,\,\,T_a=q^\dag\frac{\lambda_a}{2}q,\,\,\,G_{ia}=q^\dag\frac{\sigma_i\lambda_a}{4}q,
\eeq
with the Gell-Mann matrices normalized to give $Tr\left[\lambda_a\lambda_b\right]=2\delta_{ab}$.

The mass operator in the large $N_c$ limit can be expressed as a linear combination of composite operators sorted according to their order in $1/N_c$.  A basis of such operators can be constructed using the O(3)$\,\otimes\,$SU(6) generators, with a distinction drawn between operators that act on quarks in the core, denoted $\left\{S_i^c,\,T_a^c,\,G_{ia}^c\right\}$, and those that act on the excited quark (in the case of excited baryons), denoted $\left\{s_i,\,t_a,\,g_{ia}\right\}$. Operators are also classified as $n$-body operators. The mass operators must be rotationally invariant, parity and time-reversal even and isospin symmetric. A generic $n$-body mass operator has the general structure
\beq
O^{(n)}=\frac{1}{N_c^{n-1}}O_\ell O_qO_c,
\eeq
where the factors $O_\ell$, $O_q$ and $O_c$ can be expressed in terms of products of generators of orbital angular momentum ($\ell_i$), spin flavour of the excited quark ($s_i$, $t_a$ and $g_{ia}$), and spin-flavour of the core ($S_i^c$, $T_a^c$ and $G_{ia}^c$).  The explicit $1/N_c$ factors originate in the $n-1$ gluon exchanges required to give an $n$-body operator. For a given O(3)$\,\otimes\,$SU(6) multiplet, the coefficients associated with each operator are obtained by fitting to the empirical masses of the states in the multiplet.

To a first approximation, the main features of the spectrum can be described by taking into account a few operators. These are the ${\cal O}(N_c)$ spin-flavour singlet operator that essentially counts the number of quarks in the baryon. The strangeness operator appears at ${\cal O}(N_c^0m_s)$, and counts the number of strange quarks in the baryon. The hyperfine operator appears at ${\cal O}(1/N_c)$. For a few multiplets, the hyperfine SU(3) breaking ${\cal O}(m_s/N_c)$ operator $S_i G_{i8}-\frac{1}{2\sqrt{3}}S_iS_i$ is necessary for achieving a consistent fit. Other operators play smaller roles.

By considering the leading operators as described above, mass formulae can be written for the states of a particular multiplet. For instance, for the ground state $(56,\,0^+_0)$, the mass formula obtained is
\beq
\fl M_{\rm GS}= N_c c_1 \mathds{1}+\frac{1}{N_c} C_{\rm HF}\left(S^2-\frac{3}{4}N_c\right) -c_{\cal S} {\cal S}
+\frac{1}{N_c}c_4\left(I^2-S^2-\frac{1}{4}{\cal S}^2\right),
\eeq
where $S$ is the spin of the baryon, $I$ is its isospin and ${\cal S}$ is the strangeness operator. The expressions for the masses of excited multiplets have similar forms, perhaps with contributions from additional operators.

The values obtained for the coefficient $c_1$ for the various multiplets can be fit by the form~\cite{Goity:2007sc}
\beq
\left(3c_1\right)^2=1.179\pm 0.003+ (1.05\pm 0.01)\ell
\eeq
for the $(56,\ell)$ multiplets, and 
\beq
\left(3c_1\right)^2=1.34\pm 0.02+ (1.18\pm 0.02)\ell
\eeq
for the $(70,\ell)$ multiplets. Thus, the leading contributions to the masses are consistent with linear Regge trajectories.

Operators that contain $T_a$ and/or $G_{ia}$ correspond to changes of flavour in the quarks, and may be interpreted in terms of meson exchanges. 
In most calculations in the large $N_c$ approach,  there is no mixing among the multiplets of SU(6)$\,\times\,$O(3). There is therefore also no mixing of different orbital angular components in the states. For instance, the ground-state nucleon is completely devoid of any $D$-wave component. More recent work has included such contributions to the masses and wave functions. They appear at different orders in the $\frac{1}{N_c}$~expansion, depending on the specific SU(6)$\,\otimes\,$O(3) multiplets being considered.

For baryons containing one or more heavy quarks, simply extending the spin-flavour symmetry group from SU(6) to include the heavier flavours of quark (SU(8) for the inclusion of charm quarks, for instance) is not appropriate, as this symmetry is so badly broken. Instead, the spin-flavour symmetry group is expanded to SU(6)$\,\otimes\,$SU(2)$_c\,\otimes\,$SU(2)$_b$. There is therefore a separate spin symmetry for each heavy flavour. As far as can be determined, such an approach is yet to be applied to excited heavy baryons.

The results from the large $N_c$ treatment of baryons are shown in Tables \ref{nucleonspec} to~\ref{xiomegaspec} from a number of authors. For the nucleons and $\Delta$~resonances, Tables \ref{nucleonspec} and~\ref{deltaspec}, respectively, the numbers are shown in column eight. For the $\Lambda$ and $\Sigma$~states in Tables \ref{lambdaspec} and~\ref{sigmaspec}, respectively, they are shown in column seven. For the $\Xi$ and $\Omega$~states, the number are shown in column eight of Table~\ref{xiomegaspec}. In this approach, fits are carried out for baryons in a particular (SU(6)$\,\otimes\,$O(3)) multiplet. Masses of the (56,\,$2^+$) and the (70,\,$1^-$) are from the work by Goity and collaborators, while those for the 
(70,\,$0^+$), (70,\,$ 2^+$) and (56,\,$4^+$) are from the work by Matagne and Stancu~\cite{Matagne:2004pm,Matagne:2006zf}. The masses for radially excited states of the (56,\,$0^{+\prime}$) are taken from~\cite{Carlson:2000zr}.

\subsection{Lattice}
Lattice QCD is a discretized version of QCD~\cite{Fodor:2012gf,hashimoto}. The path integral over fields at infinitely many Minkowski space-time points is approximated by a finite number of points in a Euclidean space lattice with periodic boundary conditions. The finite lattice spacing provides an automatic ultraviolet cut-off, while the finite box-size provides an infrared cut-off, and it is crucial that any physics extracted from this technique not be affected by these cut-offs. Any results obtained on the lattice must be extrapolated to the physical limit of infinite box size and zero lattice spacing.

At present, lattice simulations are the only {\it ab initio} treatment of QCD in the non-perturbative regime. Advances in computer capabilities and improvements in calculational techniques have made it possible for ever more precise, unquenched  extractions of hadron properties, at more and more realistic quark masses. 
An observable corresponding to the  vacuum expectation value of an operator (${\cal O}$) may be written
\beq
\langle{\cal O}\rangle=\frac{1}{\cal Z}\int\left[dU\right]\left[d\psi\right]\left[d\overline\psi\right] e^{-S_F(U,\psi,\overline\psi)-S_G(U)} {\cal O}(U,\psi, \overline\psi),
\eeq
where
\beq
{\cal Z}=\int\left[dU\right]\left[d\psi\right]\left[d\overline\psi\right] e^{-S_F(U,\psi,\overline\psi)-S_G(U)}.
\eeq
$S_G$ is the gauge action and $S_F=\overline\psi M\psi$ is the fermion action with Dirac operator $M$. A particle of interest is created by constructing an operator with the appropriate quantum numbers. For the nucleon, for exampple, one possible choice is
\beq
\chi_N(x)=\epsilon^{abc}\left(u^{T,a}(x)C\gamma_5d^b(x)\right)u^c(x),
\eeq
where $u,\,\,d$ are the spinors for the $u$ and $d$ quark, respectively. Note that all three quarks reside at the same lattice point.

For a baryon created at a specific space-time point $y=(\vec{y},0)$ and annihilated at point $x=(\vec{x},t)$, the Euclidean two-point function is
\beq
C^(2)(x,y)=\frac{1}{\cal Z}\int\left[dU\right]\left[d\psi\right]\left[d\overline\psi\right] e^{-S(U,\psi,\overline\psi)}\overline{\chi}_N(x)\chi_ N(y).
\eeq
When $t$ is large enough, the lowest-energy state dominates, and its mass can be obtained by fitting the correlator to an exponential as
\beq
C^{(2)}(t)\approx Ae^{-mt}.
\eeq

Obtaining the spectrum of excited states on the lattice requires overcoming a number of challenges. First, excited-state contributions to the correlation functions decay faster than those of the ground state. Thus, at large times, the signals for excited states are swamped by signals for lower-energy states. This can be mitigated by the use of anisotropic lattices that have a finer spacing in the Euclidean time dimension than in the space dimensions.  A larger basis of independent operators that allow maximum overlap with high-spin operators is also needed. Since rotational symmetry is lost on the lattice, irreducible representations of the cubic group must be used. For a chosen parity, there are two two-dimensional irreducible representations $G_1$ and $G_2$, and one four-dimensional irreducible representation $H$. $G_1$ contains $J=\hlf,\shlf,\nhlf,\elhlf,\dots$, $G_2$ contains $J=\thlf,\fhlf,\shlf,\nhlf,\dots$, and $H$ contains $J=\fhlf,\shlf,\elhlf,\dots$. The continuum-limit angular momenta $J$ of lattice states are identified by examining patterns of degeneracies among the different irreducible representations.

In addition to the challenges discussed briefly above, some lattice simulations are carried out with light quark-masses that lead to pions that are significantly heavier than the physical pion. For instance, the results of Edwards {\it et al.}~\cite{Edwards:2011jj,Edwards:2012fx,Dudek:2012ag} are obtained using pion masses of 524 MeV, 444 MeV and 396 MeV. For many collaborations, significant effort is placed on extrapolating the results to the physical pion mass, using techniques taken from chiral perturbation theory. Furthermore, there is the question of quenched vs unquenched similations. For quenched calculations, quark-pair creation is forbidden, which affects the values of properties extracted. More realistic, unquenched calculations, however, are very expensive computationally. Nevertheless there has been some progress in this area. In addition, calculations with lighter quark masses come with another set of complications that stem from the resonant nature of the states being treated and the existence of nearby thresholds. It then becomes necessary to include multi-particle states in lattice simulations in order to extract the properties of the states of interest. For a discussion of this and other challenges inherent in lattice computations, see reference \cite{hashimoto}.

There are a number of latice simulations of light baryons. Some of the more recent studies include chiral extrapolations to the physical mass of the pion, and others start with pion masses that are quite light \cite{Engel:2013rba}. However, to date, we know of a single collaboration that has yet examined states with $J\ge \thlf$, and these are the results that are shown in Tables~\ref{nucleonspec} to \ref{xiomegaspec}. For baryons composed only of $u$, $d$ and $s$ quarks, lattice results for the masses are shown in the last column of Tables~\ref{nucleonspec} to \ref{xiomegaspec}. The results shown in those tables are from the work of Edwards and collaborators~\cite{Edwards:2011jj,Edwards:2012fx,Dudek:2012ag}, who use improved gauge and fermion actions with 2+1 dynamical flavours.  For charmed baryons, Tables~\ref{baryonspec3} and~\ref{baryonspec5}, the lattice results are taken from the work by Brice\~no {\it et al.}~\cite{Briceno:2012wt}, who use 2+1+1 flavours of dynamical quarks.  For $b$-flavoured baryons, the results are taken from the work of Ali Khan {\it et al.}~\cite{AliKhan:1999yb}, whose calculation is a quenched one.

\subsection{Other Approaches}
There are a number of other approaches to baryon spectroscopy that cannot be  treated in any depth in this manuscript, but which warrant a brief mention.

\subsubsection{Effective Field Theories}
Traditionally, effective field theories have little to say about excited hadrons. The quintessential example is chiral perturbation theory, in which an expansion parameter that is small compared with $\Lambda_{\rm QCD}$ is needed. Since baryonic excitation energies are typically of the same order as   $\Lambda_{\rm QCD}$, there is no easy way to include more than one resonance in such a theory, or predict the mass differences between them. In the case of baryons containing a single heavy quark, however, the expansion parameter is the inverse mass of the heavy quark~\cite{hqet}. In the limit of an infinitely massive quark, all baryonic mass differences are small compared with the mass of the heavy quark, and a limited number of predictions can be made.  

When one of the quarks in a baryon is very heavy ($m_Q>>\Lambda_{\rm QCD}$), that quark acts like a static source of colour around which the light quarks and gluons move. The spin of the heavy quark decouples from the light degrees of freedom. Consequently, states in the spectrum are arranged in nearly degenerate doublets: when the light degrees of freedom have total angular momentum and parity $j^P$, the possible baryons have total angular momentum $J=j\pm \hlf$ and parity $P$. Since these two states are almost degenerate, the doublet is denoted $((j-\hlf)^P, (j+\hlf)^P)$. The exception occurs when $j=0$, leading to a singlet with $J=\hlf$. 

Since the mass difference between the states of a doublet arises from the chromomagnetic interaction with the heavy quark, the effective theory that has been developed  predicts that this splitting is inversely proportional to the mass of the heavy quark. If the charm and beauty quarks are treated as heavy, this implies that mass splittings within multiplets of $b$-flavoured baryons can be predicted once the analogous splittings within the corresponding multiplets of $c$-flavoured baryons are known. Furthermore, since the dynamics of the light quarks and gluons are completely insensitive to the flavour of the heavy quark, mass differences between multiplets should also be independent of the flavour of heavy quark. This means, for instance, that the mass difference between the spin-averaged mass of the $(\hlf^-,\thlf^-)$ and the spin averaged mass of the ground-state $(\hlf^+,\thlf^+)$ should be the same for the $\Sigma_c$ and the $\Sigma_b$, modulo $1/m_b$ and $1/m_c$ corrections.

The heavy quark expansion (expansion of the QCD Lagrangian in powers of $1/m_Q$) has been combined with other approaches, such as quark models and QCD sum rules, to simplify the formalisms, or to try to enhance the predictive power. On its own, this expansion can only be used in the ways already described.

\subsubsection{Bag Models}

The bag model was originally designed to describe the confinement of quarks and gluons inside hadrons. In the original formulation, the bag is a spherical cavity with rigid walls, and the quarks move freely within the cavity. Each quark satisfies the Dirac equation for a free particle, subject to the boundary condition of a confining bag. Many variants have been proposed, including deformed bags, cloudy bags, bags with partially restored chiral symmetry, etc. In the early work of Donoghue {\it et al.}~\cite{Donoghue:1975yg}, the mass of the baryon is written
\beq
E(R)=\sum_iN_i(m_i^2+x^2/R^2)^{1/2}-\frac{Z_0}{R}+B\frac{4\pi R^3}{3},
\eeq
where $R$ is the radius of the bag, $N_i$ is the number of quarks of type $i$ and $x$ is the quark momentum in units of $1/R$. The last term is a phenomenological bag pressure added to keep the bag from expanding to infinite radius. Equilibrium arises when $E(R)$ is minimized. Chromoelectric and chromomagnetic corrections can then be included in $E(R)$.

One challenge experienced with the bag model was the existence of a number of spurious excitations connected with the motion of the centre of mass of the system. This has been dealt with for ground state baryons, but becomes increasingly challenging for higher excitations. This is one of the main reasons why bag models have not been applied to highly excited states in the baryon spectrum. When one of the quarks in the baryon is heavy, it is treated as being anchored at the centre of the bag, with the light quarks in motion around it, but the challenge of isolating the centre-of-mass motion remains.

\subsubsection{QCD Sum Rules}

As with lattice simulations and relativistic bound-state equations, the starting point for sum-rule calculations of baryon spectra is a correlation function
\beq
\Pi(q)=i\int d^4x e^{iq\cdot x}\langle 0|T\left[B(x)\overline{B}(0)\right]|0\rangle,
\eeq
where $B$ is the interpolating field for a spin-$\hlf$  baryon. This approach thus has the same starting point as that of the relativistic bound-state equations, and of the lattice simulations. As the angular momentum of the baryon increases, the form of the correlation function becomes more complicated. For $J=-\thlf$, the correlation function is~\cite{Lee:2006bu}
\beq
\Pi_{\mu\nu}(q)=i\int d^4x e^{iq\cdot x}\langle 0|T\left[B_\mu(x)\overline{B_\nu}(0)\right]|0\rangle,
\eeq
where $B_\mu$ is the vector-spinor that represents a particle with spin $\thlf$. 

For the spin-$\hlf$ baryon, the most general form for the correlation function is
\beq
\Pi(q)=\Pi_1(q^2)+\slash{q}\Pi_2(q^2),
\eeq
and a sum rule can be written for each of the $\Pi_i$. For the spin-$\thlf$ case, the correlator is a tensor of rank two that can be constructed from the momentum $p$, the metric tensor and Dirac $\gamma$ matrices. It takes the form
\beqy
\fl\Pi_{\mu\nu}(p)&=&\lambda_{3/2}^2\left\{-g_{\mu\nu}\slash{p}+\frac{1}{3}\gamma_\mu\gamma_\nu\slash{p}-\frac{1}{3}\left(\gamma_\mu p_\nu-\gamma_\nu p_\mu\right)+\frac{2}{3}\frac{p_\mu p_\nu}{M^2_{3/2}}\slash{p}\right.\nonumber\\
\fl&\pm&M_{3/2}\left.\left[g_{\mu\nu}-\frac{1}{3}\gamma_\mu\gamma_\nu+\frac{1}{3M_{3/2}^2}\left(\gamma_\mu p_\nu-\gamma_\nu p_\mu\right)\slash{p}-\frac{2}{3}\frac{p_\mu p_\nu}{M^2_{3/2}}\right]\right\}\nonumber\\
\fl&+&\dots
\eeqy
where the contributions from the spin-$\hlf$ sector of the vector-spinor, as well as of other spin-$\thlf$ resonances, are denoted by `$\dots$' in the equation above. 

The correlation for a particular spin is calculated in two ways. In the hadronic representation, a complete set of baryon states with the appropriate quantum numbers is inserted into the correlation function. For the spin-$\hlf$ correlator, the pole terms of the lowest lying states are obtained as
\beq
\Pi^\pm(p)=\lambda_+^2\frac{\slash{p}+M_+}{M_+^2-p^2}+\lambda_-^2\frac{\slash{p}+M_-}{M_-^2-p^2}+\dots.
\eeq
On the QCD side, the correlator is calculated using a variety of field theory methods, such as the operator product expansion. Matching both sides, and using a few other mathematical tools (such as Borel transforms) allows the extraction of baryon masses.

Because the correlators for higher-spin states become increasingly complicated, this method has not been applied to states with spin beyond $\thlf$, to the best of our knowledge. However, they have been applied to obtain a variety of properties of baryons, none of which are discussed in this manuscript.

\subsubsection{Dynamical Generation}

A number of authors have examined the properties of a few baryons using chiral dynamics. They find that a number of states are `dynamically-generated' states. That is, they arise as result of the strong interaction (chiral dynamics) between baryons and mesons, and manifest themselves as poles in the scattering amplitude~\cite{Oller:2000ma}. This approach has been used to describe properties of scattering amplitudes in the neighbourhood of the dynamically generated states, but it has not been used to predict where such states should lie.

\subsubsection{Parity Doublets}
Chiral symmetry predicts that there should exist approximately degenerate hadronic states of opposite parity. Along with the nucleon, there should be a state with the same mass and opposite parity. There is a very simple interpretation to this prediction. In the chiral limit, in which the pion is massless, a state containing a nucleon and a pion at rest has the same mass as a state with a single nucleon, but the two states have opposite parity. More recently, the interpretation is that there should be approximately degenerate states in the baryon spectrum, for instance, that have the same total angular momentum and opposite parity. It has been argued that the presence of such doublets is consistent with the restoration of chiral symmetry high in the baryon spectrum~\cite{Cohen:2001gb}.

L\"oring {\it et al.}~\cite{Loring:2001kx} have noted that, in their model, approximate parity doublets arise from the dynamics of the 't Hooft force that they use, with no explicit influence from chiral symmetry. In other models, such as that of Capstick and Isgur~\cite{Capstick:1986bm}, pairs of approximately degenerate states with opposite parity abound, even though that model was constructed with no reference to chiral symmetry. In order to more firmly establish the idea of chiral doublets, there need to be predictions of ratios of decay amplitudes, for instance, for the states assigned to a particular doublet.

\subsubsection{Holography} A recent approach to quantum field theory is that has garnered some interest is the so-called AdS/CFT correspondence (Anti de Sitter/Conformal Field Theory), which establishes a duality between string theories defined on the five-dimensional AdS space-time and conformal field theories in physical space-time~\cite{Brodsky:2006uq}. The strong coupling is assumed to be approximately constant in an appropriate range of momentum transfer, and that the quark masses can be neglected. With these assumptions,  QCD becomes a nearly conformal field theory and the AdS/CFT correspondence can be applied. The hadron spectrum and strong interaction dynamics can then be calculated from a holographic dual string theory defined on the five-dimensional AdS space. A very useful summary of the calculation methodology and the results obtained is presented in the review by Klempt and Richard~\cite{Klempt:2009pi}.
\section{\label{Section:Discussion}Discussion and Open Questions}
One of the important questions to be addressed is that of the relevant degrees of freedom in a baryon. Does a baryon consist of three quarks, a quark and a diquark, or does it even have a very different, more complex structure. A better understanding of the properties of the known resonances and an improved mapping of the baryon spectra will help elucidate this question. Recent photoproduction experiments in the light-baryon sector that aim at finding new or undiscovered states have yielded a number of candidates, but none of these is yet without doubt. It is worth pointing out again that, unlike for heavy baryons, the confirmation of a new light-baryon state is more challenging since the discovery is not inferred from a direct observation. 

\subsection{Heavy-Quark Baryons}
A glance at Tables~\ref{nucleonspec} to \ref{baryonspec6} reveals that, among the light baryons, all of the theoretical approaches we have discussed in Section~\ref{Section:Theory} lead to rather similar results. In the heavy sector, the predictions of the diquark model show more striking differences from the three-quark model. In the model of Ebert {\it et al.} \cite{Ebert1}, the lack of excitations in the diquark means that there is a single negative-parity $\hlf^-$~state along with a single $\thlf^-$~$\Lambda_Q$ in the region of low-lying excitations.  These two states form an HQET $(\hlf^-,\thlf^-)$ doublet. The second $\hlf^-$ state in this model is 417~MeV heavier than the first, whereas the mass difference is between 150 and 250~MeV in other models. Similarly, there are two $\hlf^-$ and two $\thlf^-$~$\Sigma_Q$ along with a single $\fhlf^-$~state.  These form a $(\hlf^-,\thlf^-)$ doublet, a $(\thlf^-,\fhlf^-)$ doublet and a $(\hlf^-)$ singlet. In models in which all three quarks can be excited, there are three $\hlf^-$ states, three $\thlf^-$ states and a single $\fhlf^-$, in each isospin sector, all within 250~MeV of each other (in the mass range 2600 to 2890~MeV for the $\Lambda_c$ and 2750 to 2880~MeV for the $\Sigma_c$). In each sector, these seven states constitute a $(\hlf^-)$ singlet, two $(\hlf^-,\thlf^-)$ doublets, and a $(\thlf^-,\fhlf^-)$ doublet, although spin-dependent forces induce some mixing among states with the same $J^P$. In the limit in which the mass of the quark becomes infinitely massive, the multiplet structure should be as described. This structure should therefore be more apparent for the $b$-flavoured baryons than for the charmed ones, and in the model of Roberts and Pervin~\cite{pr}, this is indeed found to be the case.
A number of experimentally observed $\Lambda_c$ states have masses between 2700 and 2900 MeV, but the quantum numbers are unknown. If any of these states are found to have negative parity, it would suggest that the basic hypothesis of the quark-diquark model would not be correct, at least for baryons containing a heavy quark. 

Among the heavy baryons, it can be seen that most approaches get the masses of the ground states roughly correct, mainly because these states are often used as input to the fits to the spectra. The details of the inter-quark potential then determines where the excited states lie, and this is where differences in the predictions of the masses emerge. Nevertheless, for the low-lying excitations, most models are quite consistent, often predicting masses within a range of 50 MeV of each other. The same is true of the lattice studies. It should be noted that the lattice predictions for the $b$-flavoured baryons are from a quenched study, while the results for the $c$-flavoured baryons are from an unquenched calculation.

A number of the known charmed baryons have been omitted from Tables \ref{baryonspec3} and \ref{baryonspec5}. This is because for each of these experimental states, there are a number of model candidates that are sufficiently close in mass to make an assignment ambiguous. Two examples are the $\Lambda_c(2880)^+$ and the $\Lambda_c(2940)^+$. The former is close in mass to a $\thlf^+$~state with a predicted mass of  2887~MeV in the model by Roberts and Pervin~\cite{pr}, but also to a $\fhlf^+$~state with the same mass, and a $\fhlf^-$~state with a predicted mass of 2872~MeV. This is not very different from the other models shown. The state at 2940~MeV could be one of a number of positive-parity excitations. Thus, other information, such as the strong and radiative decay widths of the state, will be needed before a definite assignment can be made. Among the $b$-flavoured baryons, the $\Xi_b^{\,\prime}$ ground-state doublet is still to be identified experimentally, if the state at 5945~MeV is indeed the $\thlf^+$~state. In the Fadeev study of Garcilazo {\it et al.} for instance \cite{Garcilazo:2007eh}, the $\Xi_b^{\,\prime}$ with $J^P=\hlf^+$ is predicted to have a mass of 5937~MeV, in very good agreement with the experimentally known state, and the $\thlf^+$~state is predicted to be 30~MeV heavier at 5967~MeV. In Tables \ref{baryonspec4} and \ref{baryonspec5}, there are experimental states that are included twice, alongside possible model states that they match. These are the $\Xi_c$(2816) that is close in mass to model states with $J^P=\hlf^-$ and $\thlf^-$, and the $\Xi_b$(5945) that could be either the $\Xi_b^{\,\prime}$ or the $\Xi_b^\ast$.

Examination of the multiplet structure of Figure \ref{Figure:Multiplets} provides some guidance regarding the states that can be expected. There should be an $\Omega_c$ state for every triplet of $\Sigma_c$ found. There should be a doublet of $\Xi_c$ for every $\Lambda_c$, and another doublet for every triplet of $\Sigma_c$. Similarly, in the light sector, there should be a doublet of $\Xi$ resonances for each quartet of $\Delta$ states, and another doublet for each doublet of nucleons. There should be an $\Omega$ for each $\Delta$ quartet. There should also be a triplet of $\Sigma$ states for each nucleon and $\Delta$ isospin multiplet, and there should be more $\Lambda$ states than nucleon doublets.

\subsection{Light-Quark Baryons}
In the light baryon sector, we first note that the lattice results of Edwards and collaborators~\cite{Edwards:2011jj,Edwards:2012fx,Dudek:2012ag} are significantly higher than the experimental masses for the nucleons and $\Delta$ resonances. This is primarily because the results shown correspond to a pion mass of 396~MeV. There are simulations that have been carried out with lower pion masses, but they have focused on states with $J\le \thlf$ to date. Among the hyperons, the predictions for the ground-state masses improve steadily as the number of strange quarks in the baryon increases. In this study, although the dependence of the results on the pion mass has been examined, no extrapolations to the physical pion mass have been carried out. At present, chiral extrapolations for excited states are not yet very reliable. In this simulation, the ordering of the states is consistent with the conventional quark model, in that the excited $\hlf^+$~nucleon is more massive than the $\hlf^-$~state. More generally, in all flavour sectors, the lower-lying positive-parity excitations are all more massive than the low-lying negative-parity states.  We note, however, that there has been at least one lattice study in which the lowest-lying $\hlf^+$~excitation has a mass that is lower than the lowest-lying $\hlf^-$~state~\cite{Lee:2002qw}. A second feature that emerges from this lattice simulation is that the number of lower-lying excitations is consistent with the conventional quark model, not the diquark version. Of course, these  results may change significantly as the pion mass decreases toward the physical value. In addition, the present calculation, as well as that of~\cite{Lee:2002qw}, includes only one-particle states. The conclusions about which states are {\it bona fide} resonances may change when multi-particle states are included.

\subsubsection{Multiplet Assignments}
Tables~\ref{Table:BaryonOctets} and~\ref{Table:BaryonDecuplets} show our suggestions for assignments of the known light baryons to the lowest-lying quark model spin-flavour SU(6)\,$\otimes$\,O(3) multiplets. The assignments to the ground-state octet and decuplet of the ({\bf 56},\,$0^+_0$) are straightforward and have been discussed in Section~\ref{sec:flavours}, Table~\ref{Table:Classification}. This is the only supermultiplet that is completely filled.

The first excitation band contains one supermultiplet, ({\bf 70},\,$1^-_1$), and the states have one unit of orbital angular momentum and negative parity. This leads to a total of five flavour octets from coupling the orbital angular momentum, $L=1$, with the two possible spins, $S=\frac{1}{2}$ and $S=\frac{3}{2}$. These are a doublet of total angular momentum and parity $(\frac{1}{2}^-,\,\frac{3}{2}^-)$ ($^2{\bf 8}_{\hlf^-}$ and $^2{\bf 8}_{\thlf^-}$) and a triplet with $J^P=(\frac{1}{2}^-,\,\frac{3}{2}^-,\,\frac{5}{2}^-)$ ($^4{\bf 8}_{\hlf^-}$, $^4{\bf 8}_{\thlf^-}$ and $^4{\bf 8}_{\fhlf^-}$). The notation in parentheses is $^{2S+1}{\bf D}_{J^P}$ to denote the multiplets. Here $S$ is the total spin of the three quarks in the baryon, ${\bf D}$ denotes the SU(3) flavour multiplet to which it belongs, and $J^P$ are the total angular momentum and the parity. There will also be multiplets $^2{\bf 10}_{\hlf^-}$, $^2{\bf 10}_{\thlf^-}$, $^2{\bf 1}_{\hlf^-}$ and $^2{\bf 1}_{\thlf^-}$. The $N(1535)\hlf^-$ and $N(1520)\thlf^-$ are members of the $^2{\bf 8}_{J^-}$ doublet (with $J=\hlf$ or $\thlf$), and $N(1650)\hlf^-$, $N(1700)\thlf^-$ and $N(1675)\fhlf^-$ belong to the $^4{\bf 8}_{J^-}$ triplet. $\Lambda(1670)\hlf^-$ and $\Lambda(1690)\thlf^-$ can be identified as members of the $^2{\bf 8}_{J^-}$ doublet, while two members of the $^4{\bf 8}_{J^-}$ triplet,  $\Lambda(1800)\hlf^-$ and $\Lambda(1830)\fhlf^-$ can be identified, with the third member, the state with $J^P=\thlf^-$, missing or undiscovered. The assignments for the $\Sigma$ and $\Xi$ hyperons are more difficult because they can be members of either flavour octets or flavour decuplets. Moreover, quantum numbers for most of the few $\Xi$ resonances known have not been measured. Only the $\Xi(1820)$ is listed with $J^P = \frac{3}{2}^-$ in the RPP~\cite{Beringer:1900zz}, and there is evidence that the $\Xi(1690)$ has $J^P=\hlf^-$ \cite{Aubert:2008ty}. 
Assignments for excited $\Sigma$~states remain educated guesses. In~\cite{Klempt:2009pi,Klempt:2012fy}, the authors suggest that $\Sigma(1620)\hlf^-$ and $\Sigma(1670)\thlf^-$ belong to the $^2{\bf 8}_{J^-}$ doublet, with $\Sigma(1750)\hlf^-$,  $\Sigma(1775)\fhlf^-$ and a missing the $\Sigma(?)\thlf^-$ belonging to the $^4{\bf 8}_{J^-}$ triplet. We have included these states in Table~\ref{Table:BaryonOctets}. In~\cite{Melde:2008yr}, $\Sigma(1620)\hlf^-$ was assigned to the $^4{\bf 8}_{J^-}$ triplet and the $(\ast\ast)$~state, $\Sigma(1560)?^?$, to the $^2{\bf 8}_{J^-}$ doublet ($?^?$ indicates that the total angular momentum and the parity of the state are not known). However, as discussed in Section~\ref{Section:KaonNucleon}, the low-mass of $\Sigma(1560)?^?$ raises serious doubts about its existence. 

We assign $\Lambda(1405)\hlf^-$ and $\Lambda(1520)\thlf^-$ to the $^2{\bf 1}_{J^-}$ doublet, a classification not questioned in the literature. $\Delta(1620)\hlf^-$ and $\Delta(1700)\thlf^-$ belong to the $^2{\bf 10}_{J^-}$. $\Sigma(1940)\,\frac{3}{2}^-$ could be a member of the $^2{\bf 8}_{J^-}$ with $J = \frac{3}{2}$, but its high mass makes an assignment to the third excitation band more likely. We therefore consider the two $\Sigma$~states to be missing. Quantum numbers for excited $\Omega$~states are unknown and we do not assign them to any of the multiplets.

The second excitation band contains several positive-parity supermultiplets. These are the first radial excitation of the ground-state octet and decuplet, ({\bf 56},\,$0^+_2$), along with multiplets in which either of the oscillators is excited with two units of orbital angular momentum, ({\bf 56},\,$2^+_2$), or where each oscillator carries one unit of orbital angular momentum and these couple to give  $L = 0, 1, 2$. These states sit in the three multiplets ({\bf 70},\,$0^+_2$), ({\bf 20},\,$1^+_2$), and ({\bf 70},\,$2^+_2$). The {\bf 20} does not contain a decuplet~(\Eref{Equation:20plet}) and is therefore not listed in Table~\ref{Table:BaryonDecuplets}. The ({\bf 56},\,$0^+_2$) comprises a radially excited $^2{\bf 8}_{\hlf^+}$ and $^4{\bf 10}_{\thlf^+}$ multiplets. The spin-flavour momentum multiplets of the ({\bf 70},\,$0^+_2$) consist of $^2{\bf 10}_{\hlf^+}$, $^2{\bf 8}_{\hlf^+}$, $^2{\bf 1}_{\hlf^+}$ and $^4{\bf 8}_{\thlf^+}$. The ({\bf 56},\,$2^+_2$) contains $^4{\bf 10}_{J^+}$ , with $J^P=(\hlf^+, \thlf^+,\fhlf^+,\shlf^+)$  and $^2{\bf 8}_{J^+}$ with $J^P=(\thlf^+,\fhlf^+)$. In the ({\bf 70},\,$2^+_2$), the spin-flavour multiplets consist of $^4{\bf 8}_{J^+}$ with $J^P=(\hlf^+, \thlf^+,\fhlf^+,\shlf^+)$, as well as  $^2{\bf 8}_{J^+}$, $^2{\bf 10}_{J^+}$ and $^2{\bf 1}_{J^+}$, each with $J^P=(\thlf^+,\fhlf^+)$.

The $^2{\bf 8}_{\hlf^+}$ multiplet of the ({\bf 56},\,$0^+_2$) includes the states $N(1440)\hlf^+$, $\Lambda(1600)\hlf^+$, and $\Sigma(1660)\hlf^+$, while the $\Delta(1600)\thlf^+$~resonance is a member of the $^4{\bf 10}_{\thlf^+}$.  The $^2{\bf 8}_{\hlf^+}$ spin-flavour multiplet of the ({\bf 70},\,$0^+_2$) contains $N(1710)$, $\Lambda(1810)$, and presumably $(\ast)$ $\Sigma(1770)$. The $\Lambda$~state may also be the $^2{\bf 1}_{\hlf^+}$ state. Evidence for the $^2{\bf 10}_{\hlf^+}$ members of the ({\bf 70},\,$0^+_2$) is poor. These could be $(\ast)$ $\Delta(1750)$ and $(\ast\ast)$ $\Sigma(1880)$, but the latter state could also be a member of the $^2{\bf 8}_{\hlf^+}$. However,  its mass makes a decuplet assignment more likely, and the  $\Sigma(1770)\hlf^+$ has already been assigned as a $^2{\bf 8}_{\hlf^+}$ state in this supermultiplet.

For the members of the ({\bf 56},\,$2^+_2$) we suggest ($N(1720)\,\frac{3}{2}^+$,~$N(1680)\,\frac{5}{2}^+$), ($\Lambda(1890)\,\frac{3}{2}^+$,~$\Lambda(1820)\,\frac{5}{2}^+$), and ($(\ast)~\Sigma(1840)\,\frac{3}{2}^+$,~$\Sigma(1915)\,\frac{5}{2}^+$). These assignment are speculative and all these states could also be assigned to the  $^2{\bf 8}_{J^+}$ in the ({\bf 70},\,$2^+_2$) (Table~\ref{Table:BaryonOctets}). The masses of the $\Sigma$~states are too low to make them likely $^2{\bf 10}_{J^+}$ members. The group of $\Delta$s below 2~GeV, $\Delta(1910)$, $\Delta(1920)$, $\Delta(1905)$, and $\Delta(1910)$, with $J=\frac{1}{2}, \frac{3}{2}, \frac{5}{2}, \frac{7}{2}$, respectively, are ideal candidates for the $^4{\bf 10}_{J^+}$. The corresponding $\Sigma$~resonances are missing.

The ({\bf 70},\,$2^+_2$) multiplet contains $^4{\bf 8}_{J^+}$, $^2{\bf 8}_{J^+}$, $^2{\bf 10}_{J^+}$ and $^2{\bf 1}_{J^+}$. Assuming the existence of the new nucleon resonances listed in the RPP~\cite{Beringer:1900zz}, we assign $(\ast\ast)~N(1860)\,\frac{5}{2}^+$ to $^2{\bf 8}_{J^+}$ with the $J=\frac{3}{2}$~state missing, and the four states, $(\ast\ast)~N(1880)\,\frac{1}{2}^+$, $N(1900)\,\frac{3}{2}^+$, $(\ast\ast)~N(2000)\,\frac{5}{2}^+$, and $(\ast\ast)~N(1990)\,\frac{7}{2}^+$ to the $^4{\bf 8}_{J^+}$. The two $\Lambda$s, $\Lambda(2110)\,\frac{5}{2}^+$ and $(\ast)~\Lambda(2020)\,\frac{7}{2}^+$ are also likely members of this spin-flavour multiplet, with two $\Lambda$~members missing. The assignment of the $\Lambda(2110)$ is not certain, as it could also be the $J=\frac{5}{2}$~ $^2{\bf 1}_{J^+}$ state. The corresponding $\Sigma$~resonances are expected with similar masses and the three states, $(\ast\ast)~\Sigma(2080)\,\frac{3}{2}^+$, $(\ast)~\Sigma(2070)\,\frac{5}{2}^+$, and $\Sigma(2030)\,\frac{7}{2}^+$ are ideal candidates with the $J = \frac{1}{2}$~state missing. However, the assignment of the $\Sigma$~states is also ambiguous since they can be members of the $^4{\bf 10}_{J^+}$  in the ({\bf 56},\,$2^+_2$) supermultiplet (this is the assignment made in the large $N_c$ calculation, and in \cite{Klempt:2009pi}). We finally assign $(\ast)~N(2100)\,\frac{1}{2}^+$ and $(\ast)~N(2040)\,\frac{3}{2}^+$ to the $^2{\bf 8}_{J^+}$ in the ({\bf 20},\,$1^+_2$) supermultiplet but evidence for these states is poor. The $^4{\bf 1}_{J^+}$  of the ({\bf 20},\,$1^+_2$) with $J^P=(\hlf^+,\thlf^+,\fhlf^+)$ are all missing.

In the higher excitation bands the number of multiplets increases substantially and the classification of resonances is very ambiguous. We refer to our assignments in Tables~\ref{Table:BaryonOctets} and~\ref{Table:BaryonDecuplets}. We assign the two new, lower-mass nucleon resonances listed in the RPP~\cite{Beringer:1900zz}, namely $(\ast\ast)~N(1895)\,\frac{1}{2}^-$ and $(\ast\ast)~N(1875)\,\frac{3}{2}^-$,  to the ({\bf 56},\,$1^-_3$) with $S=\frac{1}{2}$ in the third excitation band. The likely partners of these states are the group of three $\Delta$s, $(\ast\ast)~\Delta(1900)$, $(\ast\ast)~\Delta(1940)$, and $\Delta(1930)$ with $J=\frac{1}{2}^-, \frac{3}{2}^-, \frac{5}{2}^-$, respectively. Some of the higher-spin $\Delta$~resonances find clear assignments but all partners are missing.

\begin{table}
\begin{center}
\caption{\label{Table:BaryonOctets}(Colour online) Tentative assignments of the known light baryons to the lowest-lying SU(6)\,$\otimes$\,O(3) singlets and octets. States marked with $^\dagger$ are merely educated guesses because the evidence for their existence is poor or they can be assigned to other multiplets. A hyphen indicates that the state does not exist, an empty box that it is missing. For the higher multiplets, Nx gives the number of expected states.}
\begin{tabular}{l|l|l|c|l|l|l|l||c}
\br
$N$ & $(D,\,L^P_N)$ & $S$ & $J^P$ & \multicolumn{4}{c||}{Octet Members} & Singlets\\
\mr
0 & $(56,\,0_0^+)$ & $\frac{1}{2}$ & $\frac{1}{2}^+$ & $N(939)$ & $\Lambda(1116)$ & $\Sigma(1193)$ & $\Xi(1318)$ & $-$\\
\mr
1 & $(70,\,1_1^-)$ & $\frac{1}{2}$ & $\frac{1}{2}^-$ & $N(1535)$ & $\Lambda(1670)$ & $\Sigma(1620)$ & $\Xi(1690)$& $\Lambda(1405)$\\
   &                        &                      & $\frac{3}{2}^-$ & $N(1520)$ & $\Lambda(1690)$ & $\Sigma(1670)$ & $\Xi(1820)$ & $\Lambda(1520)$\\
   &                        & $\frac{3}{2}$ & $\frac{1}{2}^-$ & $N(1650)$ & $\Lambda(1800)$ & $\Sigma(1750)$ & & $-$\\
   &                        &                      & $\frac{3}{2}^-$ & $N(1700)$ & & & & $-$\\
   &                        &                      & $\frac{5}{2}^-$ & $N(1675)$ & $\Lambda(1830)$ & $\Sigma(1775)$ & & $-$\\
\mr
2 & $(56,\,0_2^+)$ & $\frac{1}{2}$ & $\frac{1}{2}^+$ & $N(1440)$ & $\Lambda(1600)$ & $\Sigma(1660)$ & & $-$\\
   & $(70,\,0_2^+)$ & $\frac{1}{2}$ & $\frac{1}{2}^+$ & $N(1710)$ & $\Lambda(1810)^\dagger$ & $\Sigma(1770)^\dagger$ & &\\
   &                         & $\frac{3}{2}$ & $\frac{3}{2}^+$ &  & & & & $-$\\
   & $(56,\,2_2^+)$ & $\frac{1}{2}$ & $\frac{3}{2}^+$ & $N(1720)^\dagger$ & $\Lambda(1890)^\dagger$ & $\Sigma(1840)^\dagger$ & & $-$\\
   &                         &                      & $\frac{5}{2}^+$ & $N(1680)$ & $\Lambda(1820)^\dagger$ & $\Sigma(1915)^\dagger$ & & $-$\\
   & $(70,\,2_2^+)$ & $\frac{1}{2}$ & $\frac{3}{2}^+$ &  & & & &\\
   &                         &                      & $\frac{5}{2}^+$ & $N(1860)$ & & & &\\
   &                         & $\frac{3}{2}$ & $\frac{1}{2}^+$ & $N(1880)$ & & & & $-$\\
   &                         &                      & $\frac{3}{2}^+$ & $N(1900)^\dagger$ & & $\Sigma(2080)^\dagger$ & & $-$\\
   &                         &                      & $\frac{5}{2}^+$ & $N(2000)$ & $\Lambda(2110)^\dagger$ & $\Sigma(2070)^\dagger$ & & $-$\\
   &                         &                      & $\frac{7}{2}^+$ & $N(1990)$ & $\Lambda(2020)$ & $\Sigma(2030)^\dagger$ & & $-$\\
   & $(20,\,1_2^+)$ & $\frac{1}{2}$ & $\frac{1}{2}^+$ & $N(2100)^\dagger$ & & & &\\
   &                         &                      & $\frac{3}{2}^+$ & $N(2040)^\dagger$ & & & &\\
   &                         &                      & $\frac{5}{2}^+$ & \multicolumn{1}{c|}{$-$} & \multicolumn{1}{c|}{$-$} &
                                                                                       \multicolumn{1}{c|}{$-$} & \multicolumn{1}{c||}{$-$} &\\
\mr
3 & $(56,\,1_3^-)$ & $\frac{1}{2}$ & $\frac{1}{2}^-$ & $N(1895)^\dagger$ & & & & $-$\\
   &                        &                      & $\frac{3}{2}^-$ & $N(1875)^\dagger$ & & $\Sigma(1940)^\dagger$ & & $-$\\
  & $(70,\,1_3^-)$ & & & \multicolumn{1}{c|}{5\,x} & & & &\\
  & $(70,\,1_3^-)$ & & & \multicolumn{1}{c|}{5\,x} & & & &\\
  & $(20,\,1_3^-)$ & $\frac{1}{2}$ & & \multicolumn{1}{c|}{2\,x} & & & &\\
  & $(70,\,2_3^-)$ & & & \multicolumn{1}{c|}{6\,x} & & & &\\
  & $(56,\,3_3^-)$ & $\frac{1}{2}$ & & \multicolumn{1}{c|}{2\,x} & & & & $-$\\
  & $(70,\,3_3^-)$ & $\frac{1}{2}$ & $\frac{7}{2}^-$ & $N(2190)^\dagger$ & $\Lambda(2100)^\dagger$ & & &\\
  &                        & $\frac{3}{2}$ & $\frac{9}{2}^-$ & $N(2250)$ & & & &\\
  & $(20,\,3_3^-)$ & $\frac{1}{2}$ & & \multicolumn{1}{c|}{2\,x} & & & &\\
\mr
4 & & & $\frac{9}{2}^+$ & $N(2220)$ & $\Lambda(2350)$ & & &\\
\mr
5 & & & $\frac{11}{2}^-$ & $N(2600)$ & & & &\\
\mr
\br
\end{tabular}
\end{center}
\end{table}

\begin{table}
\begin{center}
\caption{\label{Table:BaryonDecuplets}(Colour online) Tentative assignments of the known light baryons to the lowest-lying SU(6)\,$\otimes$\,O(3) decuplets. States marked with $^\dagger$ are merely educated guesses because the evidence for their existence is poor or they can be assigned to other multiplets. A hyphen indicates that the state does not exist, an empty box that it is missing. For the higher multiplets, Nx gives the number of expected states.}
\begin{tabular}{l|l|c|c|l|l|l|l}
\br
$N$ & $(D,\,L^P_N)$ & Spin, $S$ & $J^P$ & \multicolumn{4}{c}{Decuplet Members}\\
\mr
0 & $(56,\,0_0^+)$ & $\frac{3}{2}$ & $\frac{3}{2}^+$ & $\Delta(1232)$ & $\Sigma(1385)$ & $\Xi(1530)$ & $\Omega(1672)$\\
\mr
1 & $(70,\,1_1^-)$ & $\frac{1}{2}$ & $\frac{1}{2}^-$ & $\Delta(1620)$ & & & \\
   &                        &                      & $\frac{3}{2}^-$ & $\Delta(1700)$ & &  & \\
\mr
2 & $(56,\,0_2^+)$ & $\frac{3}{2}$ & $\frac{3}{2}^+$ & $\Delta(1600)$ & & & \\
   & $(70,\,0_2^+)$ & $\frac{1}{2}$ & $\frac{1}{2}^+$ & $\Delta(1750)^\dagger$ & $\Sigma(1880)^\dagger$ & & \\
   & $(56,\,2_2^+)$ & $\frac{3}{2}$ & $\frac{1}{2}^+$ & $\Delta(1910)^\dagger$ & & & \\
   &                         &                      & $\frac{3}{2}^+$ & $\Delta(1920)^\dagger$ & & & \\
   &                         &                      & $\frac{5}{2}^+$ & $\Delta(1905)$ & & & \\
   &                         &                      & $\frac{7}{2}^+$ & $\Delta(1950)$ & & & \\
   & $(70,\,2_2^+)$ & $\frac{1}{2}$ & $\frac{3}{2}^+$ &  & & & \\
   &                         &                      & $\frac{5}{2}^+$ & $\Delta(2000)$ & & & \\
\mr
3 & $(56,\,1_3^-)$ & $\frac{3}{2}$ & $\frac{1}{2}^-$ & $\Delta(1900)^\dagger$ & & &\\
   &                        &                      & $\frac{3}{2}^-$ & $\Delta(1940)^\dagger$ & & &\\
   &                        &                      & $\frac{5}{2}^-$ & $\Delta(1930)^\dagger$ & & &\\
   & $(70,\,1_3^-)$ & $\frac{1}{2}$ & & \multicolumn{1}{c|}{2\,x} & & &\\
   & $(70,\,1_3^-)$ & $\frac{1}{2}$ & & \multicolumn{1}{c|}{2\,x} & & &\\
   & $(70,\,2_3^-)$ & $\frac{1}{2}$ & & \multicolumn{1}{c|}{2\,x} & & &\\
   & $(56,\,3_3^-)$ & $\frac{3}{2}$ & & \multicolumn{1}{c|}{4\,x} & & &\\
   & $(70,\,3_3^-)$ & $\frac{1}{2}$ & & \multicolumn{1}{c|}{2\,x} & & &\\
\mr
4 & $(56,\,4_4^+)$ & $\frac{3}{2}$ & $\frac{11}{2}^+$ & $\Delta(2420)$ & & & \\
\mr
5 & & & $\frac{13}{2}^-$ & $\Delta(2750)$ & & & \\
\mr
6 & & & $\frac{15}{2}^+$ & $\Delta(2950)$ & & & \\
\mr
\br
\end{tabular}
\end{center}
\end{table}

The evidence for many states is still weak and the assignments of the known states to multiplets is certainly ambiguous. The ground-state octet and decuplet are completely filled. In the first excitation band with negative parity, the octet multiplet with $S=\frac{3}{2}$ and $J=\frac{3}{2}$ appears to be empty. Some doubt has recently emerged about the existence of the $N(1700)$ as will be discussed briefly in the next section. Looking over Table~\ref{Table:BaryonOctets}, it is interesting to note that in the second excitation band with positive parity, many multiplets with $J=\frac{3}{2}$ are either completely empty or have many missing members, which puts into question the already poor evidence of the assigned states. 

From a theoretical point of view, the only multiplet which is completely inconsistent with the quark-diquark picture discussed in Section~\ref{Section:Theory} is the~{\bf 20}. The evidence for the two $(\ast)$~$N^\ast$~states which we assigned to the ({\bf 20},\,$1^+_2$) supermultiplet is weak. If the evidence for these states grows stronger in the future, the missing states in the table will have to be regarded as simply undiscovered states, not forbidden ones. Therefore, these states spell trouble for diquark models. We also note that the predictions from the diquark model~\cite{Ferretti:2011zz} in Table~\ref{nucleonspec} include only one low-mass $\frac{5}{2}^+$~state and no $\frac{7}{2}^+$~state. However, beyond the well-established $N(1680)\,\frac{5}{2}^+$, two further $N\,\frac{5}{2}^+$~candidates and one $N\,\frac{7}{2}^+$~candidate are currently known at and below 2~GeV~(Table~\ref{Table:BaryonSummary}). Solid evidence for these candidates is still missing, but it appears that the dynamics inside of a baryon go beyond the quark-diquark picture.

A comparison of the results from the different models does not indicate that any of the other models is significantly better than any of the others in predicting the masses of the light (or heavy) excited baryons. Apart from the diquark model, all the models predict roughly the same number of states, in roughly the same ordering with some variation in the masses. From the results shown, it is not glaringly obvious that the relativistic models provide a better description of the baryon spectrum than the nonrelativistic ones (or vice versa), for instance. More significant differences between the models will emerge if they are used to examine other properties of the baryons, such as their strong and electromagnetic decay widths. 

\subsubsection{Does the $N(1700)\,\frac{3}{2}^-$ really exist?}
In recent years, some doubts have been raised about the existence of $N(1700)\,\frac{3}{2}^-$. The various analyses do
not agree very well and support has mostly come from the Bonn-Gatchina
group. The latest analysis by the SAID group~\cite{Arndt:2006bf} finds 
no evidence for this resonance. Moreover, the EBAC group reported negative 
evidence in their latest analysis~\cite{JuliaDiaz:2007kz,Kamano:2008gr}. It is 
also interesting to note that the $N(1700)\,\frac{3}{2}^-$ appears to be
the only low-lying, ``well established'' resonance with $J\leq\frac{3}{2}$ 
which is not observed by the BES collaboration in $J/\psi$~decays. Interestingly, experimental 
evidence for the other strange members of the $J^P\,=\,\frac{3}{2}^-$~multiplet is 
also currently non-existent, although all other octets of ({\bf 70}$,\,1_1^-)$ appear 
to be filled with the exception of the doubly-strange $\Xi$~baryons. All models 
predict a group of $\frac{3}{2}^-$~baryons with masses around 1700~MeV and 
the non-observation of these states would pose an interesting challenge for quark 
models.  

\subsubsection{Has a second $N(1720)\,\frac{3}{2}^+$ been discovered?}
The CLAS collaboration has reported indications for a second nucleon resonance 
with $J^P = \frac{3}{2}^+$ and a mass of about 1720~MeV~\cite{Mokeev:2008iw} 
based on $\gamma^\ast p\to p\pi^+\pi^-$ data~\cite{Ripani:2002ss}. While a 
second low-mass $\frac{3}{2}^+$~resonance is needed to completely fill the 
({\bf 70}$,\,0_2^+)$ and ({\bf 56}$,\,2_2^+)$ supermultiplets of the second excitation band, 
quark models typically predict a further $\frac{3}{2}^+$~state with a mass about 150-200~MeV higher (Table~\ref{nucleonspec}). 
The new CLAS state requires a large $\Delta\pi$ coupling and a suppressed $N\rho$~coupling, very different from 
the RPP listings of recent years. An alternative hypothesis is that, the well known 
$\frac{3}{2}^+$~resonance is much broader than originally thought, with the additional width coming from a large 
$\Delta\pi$~partial width. The $N(1720)$ Breit-Wigner width reported in the RPP 
was changed from $\Gamma = $~150-300~MeV (2010 edition) to $\Gamma =$~150-400~MeV (2012 edition). 

The question of a second $N(1720)\,\frac{3}{2}^+$ resonance with a
mass around 1700~MeV is certainly interesting. If it exists,
such a state could be a candidate for a resonance beyond the quark
model with structure different from the conventional $(qqq)$~picture. One possibility would be a hybrid baryon.

\subsection{Exotic Baryons and the Interpretation of $N(1685)$}
The prediction of exotic mesons -- states beyond the conventional quark model with three quark degrees of freedom -- has 
inspired the search for hybrid mesons and glueballs for many decades, see 
e.g.~\cite{Klempt:2007cp} for a very comprehensive review. These states receive strong 
contributions to their overall quantum numbers from gluonic components and 
some have $J^P$~assignments which are not allowed for a conventional
$q\bar{q}$~system. Candidates have been identified through
overpopulation of some conventional quark-model multiplets~\cite{Crede:2008vw,Meyer:2010ku}. 

Hybrid baryons are also expected to exist, but no candidates for baryonic hybrids
currently exist. Some lattice simulations have examined baryon states with significant gluon components in their correlation functions~\cite{Edwards:2011jj} and additional nucleons with $J^P = \frac{1}{2}^+,~\frac{3}{2}^+,~\frac{5}{2}^+$ 
and $\Delta$ resonances with  $J^P = \frac{1}{2}^+,~\frac{3}{2}^+$, having significant hybrid components in their wave functions, have resulted from such simulations. For the most part, these have been found to lie above the lowest band of excited, positive-parity, three-quark baryons. However, there are a few instances in which the hybrids penetrate into the band of conventional baryons. In the case of the $\Lambda$ resonances, there are $\hlf^+$ hybrids at $2213\pm 19$~MeV and $2452\pm 40$~MeV. Among the $\Sigma$, the lightest hybrid is again $\hlf^+$, and has a mass of $2364\pm 54$~MeV. Generally, though, the hybrids do have masses that are significantly larger than the first band of positive-parity excitations, particularly those with higher total angular momenta. We also note that where the hybrids sit in the spectrum is sensitive to the pion mass used. For a pion mass of 524~MeV, none of the hybrid masses fall within the band of conventional positive-parity excitations.

It will be challenging to identify hybrid baryons experimentally as there are no
obviously exotic signals as is the case with exotic mesons. Since all $J^P$ can be accessed by conventional three-quark baryons, there are no exotic quantum numbers for exotic baryons. The extraction of helicity amplitudes in electroproduction
experiments can help our understanding of the structure of baryon resonances.
Recent results from the CLAS collaboration~\cite{Aznauryan:2009mx} have  ruled 
out the Roper resonance, $N(1440)\,\frac{1}{2}^+$, as a $q^3G$~hybrid
state. Other baryons have received attention owing to their unusual
properties and have been interpreted beyond the regular quark model.
For example, the nucleon resonance, $N(1535)\,\frac{1}{2}^-$, has an 
unusually large coupling to the $N\eta$~decay mode and the hyperon 
resonance, $\Lambda(1405)\,\frac{1}{2}^-$, is very low in mass compared 
to predictions. An interpretation of these states as dynamically-generated 
resonances has been proposed, e.g.~\cite{Jido:2003cb,Jido:2007sm,Hyodo:2011ur,Oset:1997it,Oller:2000fj}.

A different group of exotic baryons are those which are composed of more than three constituent quarks. 
The idea of such multiquark hadrons is intriguing and was further nourished
in 2003 by the announcement of a pentaquark candidate, the $\Theta(1540)^+$, in photoproduction, see e.g. the review in~\cite{Hicks:2005gp}. 
The observed candidate triggered much excitement because its mass appeared to be almost exactly as predicted: the chiral soliton model of \cite{Diakonov:1997mm} predicted an antidecuplet of narrow $qqqq\bar{q}$~baryons. The $\Theta(1540)^+$ was initially confirmed by several
(low-statistics) experiments but the experimental evidence has mostly
died out in recent years. We refer to the literature for a more detailed 
discussion of this state.

In its latest (2012) edition, the RPP lists a nucleon resonance,
$N(1685)$, with unknown quantum numbers and a one-star rating. This
new state is associated with a narrow peak discovered in the reaction
$\gamma n\to n\eta$~\cite{Jaegle:2008ux,Jaegle:2011sw,Kuznetsov:2007gr}. While the 
experimental evidence for this narrow state appears solid, its interpretation 
as a pentaquark is extremely weak. Other explanations appear more likely and 
may even rule out the particle as an excited nucleon state. As discussed in 
Section~\ref{Section:photoeta}, an interference effect was proposed involving the two low-lying 
$S_{11}$~states, $N(1535)\,\frac{1}{2}^-$ and $N(1650)\,\frac{1}{2}^-$.
Alternatively, a threshold effect has been discussed which explains
the observed structure in $n\eta$ by a strong (resonant or non-resonant) 
$\gamma p\to p\omega$ coupling in the $S_{11}$~partial wave. The measurement of polarization observables in 
$\gamma n\to n\eta$ will
help elucidate the origin of this new structure.

\subsection{Future Experiments}
Baryon spectroscopy remains an active and lively field. Many more results 
on heavy-flavour baryons can be expected from the LHC experiments and
new surprises are just a matter of time. In the light-flavour sector, programs 
to extract polarization observables in electromagnetically-induced reactions 
will continue at the ELSA and MAMI facilities. At the time of this writing, the recording of new data at Jefferson Lab has been suspended 
owing to the energy upgrade of the CEBAF accelerator from 6 to 12~GeV. Data 
taking will likely resume in 2014/15 and hadron spectroscopy will continue at 
the GlueX and CLAS12 experiments. The scientific focus of the future excited 
baryon programs is on the spectroscopy of doubly- and triply-strange baryons, 
$\Xi$s and $\Omega$s. The properties of these multi-strange states are poorly 
known as discussed earlier and only the $J^P$ of the $\Xi(1820)\,\frac{3}{2}^-$ 
has been (directly) determined experimentally. Future experiments will 
provide better statistics and improved detector acceptance, which are considered 
the prime limitations of previous experiments.

The $\Xi$~hyperons have the unique feature of double strangeness: the quark content is $ssu$ 
and $ssd$. If the confining potential is independent of quark flavour, the
energy of spatial excitations of a given pair of quarks decreases as their reduced mass is increased in many theoretical approaches. This means that the lightest excitations in each partial wave 
occur between the two strange quarks. In a spectator decay model, such states will 
not decay to the ground state $\Xi$ and a pion because of orthogonality of the 
spatial wave functions of the two strange quarks in the excited state and the
ground state. This removes the decay channel with the largest phase space for 
the lightest states in each partial wave, substantially reducing their widths. 
Typically, $\Gamma_{\Xi^\ast}$ is about 10-20~MeV for the known lower-mass 
resonances, which is 5-30~times narrower than for $N^\ast$, $\Delta^\ast$,
$\Lambda^\ast$, and $\Sigma^\ast$~states. 

These features, combined with their isospin of $\frac{1}{2}$, render possible the 
study of the $\Xi$ and its excited states in reactions such as $\gamma 
p\to KY^\ast\to K(\bar{K} Y)_{\Xi^\ast} K$, complementary to the challenging study of 
broad and overlapping $N^\ast$~states. This reaction will also render possible the study of excited $\Sigma$ and $\Lambda$~states~\cite{Man:2011np}. At GSI, the PANDA collaboration has developed 
plans to study multi-strange hyperons in proton-antiproton annihilations in reactions 
such as $p\bar{p}\to \bar{Y} Y^\ast$. Compared to photoproduction, the associated 
production of kaons is not required in $p\bar{p}$ reactions and the predicted $\Xi$ 
and $\Omega$ cross sections are large.

It has been well known that rescattering effects play an important role in 
photoproduction processes. This means that in a particular reaction,
the baryon and meson in an intermediate state are not necessarily the
same as those in the final state and the reaction can proceed through
different resonances. The ideal analysis of experimental data needs to 
be performed in the framework of a coupled-channel analysis since
different channels feed into each other. For the same reason, it is
necessary to carry out a combined search for nucleon resonances using 
both hadronic and photoproduction data. A coupled-channel analysis is 
theoretically very challenging and has led to the formation of the Excited 
Baryon Analysis Center (EBAC) at Jefferson Lab. However, challenges
also remain on the experimental side since the data on $\pi N$ and $K
N$ reactions is scarce. As discussed in Section~\ref{Section:PionNucleon}, a fair 
amount of data is available on $\pi N\to \pi N$, but statistics for
reactions such as $\pi N\to\pi\pi N$ and $\pi p\to K Y$ are very poor 
and stem from a few decades ago. New possibilities for studying reactions 
using hadronic probes are being discussed for the Japan Proton 
Accelerator Research Complex (J-PARC) in Tokai, Japan. 

\ack

We acknowledge useful discussions with Jose Goity, Jozef Dudek, Robert
Edwards, Simon Capstick, Reinhard Beck, Ulrike Thoma, Victor Mokeev, Vitaliy Shklyar, Daniel Mohler
and Igor Strakovsky. Special thanks to Jozef Dudek and Robert Edwards 
for providing the numbers from their simulations. This work was partially 
supported by the Department of Energy, Office of Nuclear Physics, under
contracts DE-AC05-06OR23177 
and DE-FG02-92ER40735. 

\section*{References}

\end{document}